\documentclass[final,5p,times,twocolumn]{elsarticle}
\usepackage[table]{xcolor}
\usepackage{graphicx}
\usepackage{booktabs}
\usepackage{todonotes}
\usepackage{hyperref}
\usepackage{subcaption}
\usepackage{multirow}
\usepackage{array}
\usepackage[flushleft]{threeparttable}
\usepackage{footmisc}
\usepackage{algorithm}
\usepackage[noend]{algorithmic}
\usepackage[cmex10,nosumlimits]{amsmath}
\usepackage{textcomp}
\usepackage{todonotes}
\usepackage{url}
\usepackage{algorithm,algorithmic,xpatch}
\usepackage{color, colortbl}
\definecolor{Gray}{gray}{0.85}
\usepackage[toc,page]{appendix}

\usepackage{fancyhdr}
\newcolumntype{L}[1]{>{\raggedright\let\newline\\\arraybackslash\hspace{0pt}}m{#1}}
\newcolumntype{C}[1]{>{\centering\let\newline\\\arraybackslash\hspace{0pt}}m{#1}}
\newcolumntype{R}[1]{>{\raggedleft\let\newline\\\arraybackslash\hspace{0pt}}m{#1}}

\def\arraystretch{1.5}%
\graphicspath{{figures/}}

\begin{document}
	
\begin{frontmatter}
\title{Quantitative mapping of the brain's structural connectivity using diffusion MRI tractography: a review}

\author[1]{Fan Zhang}
\author[2]{Alessandro Daducci}
\author[3,4,5,6]{Yong He}
\author[2]{Simona Schiavi}
\author[7,8]{Caio Seguin}
\author[9,10]{Robert Smith}
\author[11]{Chun-Hung Yeh}
\author[3,4,5]{Tengda Zhao}
\author[1]{Lauren J. O'Donnell}
		
\address[1]{Brigham and Women's Hospital, Harvard Medical School, Boston, USA}
\address[2]{Department of Computer Science, University of Verona, Verona, Italy}
\address[3]{State Key Laboratory of Cognitive Neuroscience and Learning, Beijing Normal University, Beijing, China}
\address[4]{Beijing Key Laboratory of Brain Imaging and Connectomics, Beijing Normal University, Beijing, China}
\address[5]{IDG/McGovern Institute for Brain Research, Beijing Normal University, Beijing, China}
\address[6]{Chinese Institute for Brain Research, Beijing, China}
\address[7]{Melbourne Neuropsychiatry Centre, University of Melbourne and Melbourne Health, Melbourne, Australia}
\address[8]{The University of Sydney, School of Biomedical Engineering, Sydney, Australia}
\address[9]{The Florey Institute of Neuroscience and Mental Health, Melbourne, Australia}
\address[10]{The University of Melbourne, Melbourne, Australia}
\address[11]{Institute for Radiological Research, Chang Gung University and Chang Gung Memorial Hospital, Taoyuan, Taiwan}

\begin{abstract}
	
Diffusion magnetic resonance imaging (dMRI) tractography is an advanced imaging technique that enables \textit{in~vivo} mapping of the brain's white matter connections at macro scale. Over the last two decades, the study of brain connectivity using dMRI tractography has played a prominent role in the neuroimaging research landscape. In this paper, we provide a high-level overview of how tractography is used to enable quantitative analysis of the brain's structural connectivity in health and disease. We first provide a review of methodology involved in three main processing steps that are common across most approaches for quantitative analysis of tractography, including methods for tractography \textit{correction}, \textit{segmentation} and \textit{quantification}. For each step, we aim to describe methodological choices, their popularity, and potential pros and cons. We then review studies that have used quantitative tractography approaches to study the brain's white matter, focusing on applications in neurodevelopment, aging, neurological disorders, mental disorders, and neurosurgery. We conclude that, while there have been considerable advancements in methodological technologies and breadth of applications, there nevertheless remains no consensus about the ``best'' methodology in quantitative analysis of tractography, and researchers should remain cautious when interpreting results in research and clinical applications. 

\end{abstract}

\end{frontmatter}
	
\section{Introduction}
\label{secIntroduction}

Diffusion magnetic resonance imaging (dMRI) tractography is an imaging method that uniquely enables \textit{in~vivo} mapping of the brain's white matter connections at macro scale \cite{basser2000vivo}. Since the first dMRI tractography methods were proposed in 1998-2000 \cite{basser1998fiber,conturo1999tracking,mori1999three,westin1999image,basser2000vivo}, tractography has enabled mapping of the brain's structural connectivity in many neurological applications such as aging, development and disease \cite{nucifora2007diffusion,ciccarelli2008diffusion,assaf2008diffusion,yamada2009mr,essayed2017white,shi2017connectome}. In this paper, we provide a high-level overview of how tractography is used to enable \textit{quantitative} analysis of the brain's structural connectivity. This review is intended to be useful to researchers studying the white matter, developers of quantitative analysis methods, and clinicians interpreting results related to tractography.

\begin{figure*}[!t]
	\centering
	\includegraphics[width=0.92\linewidth]{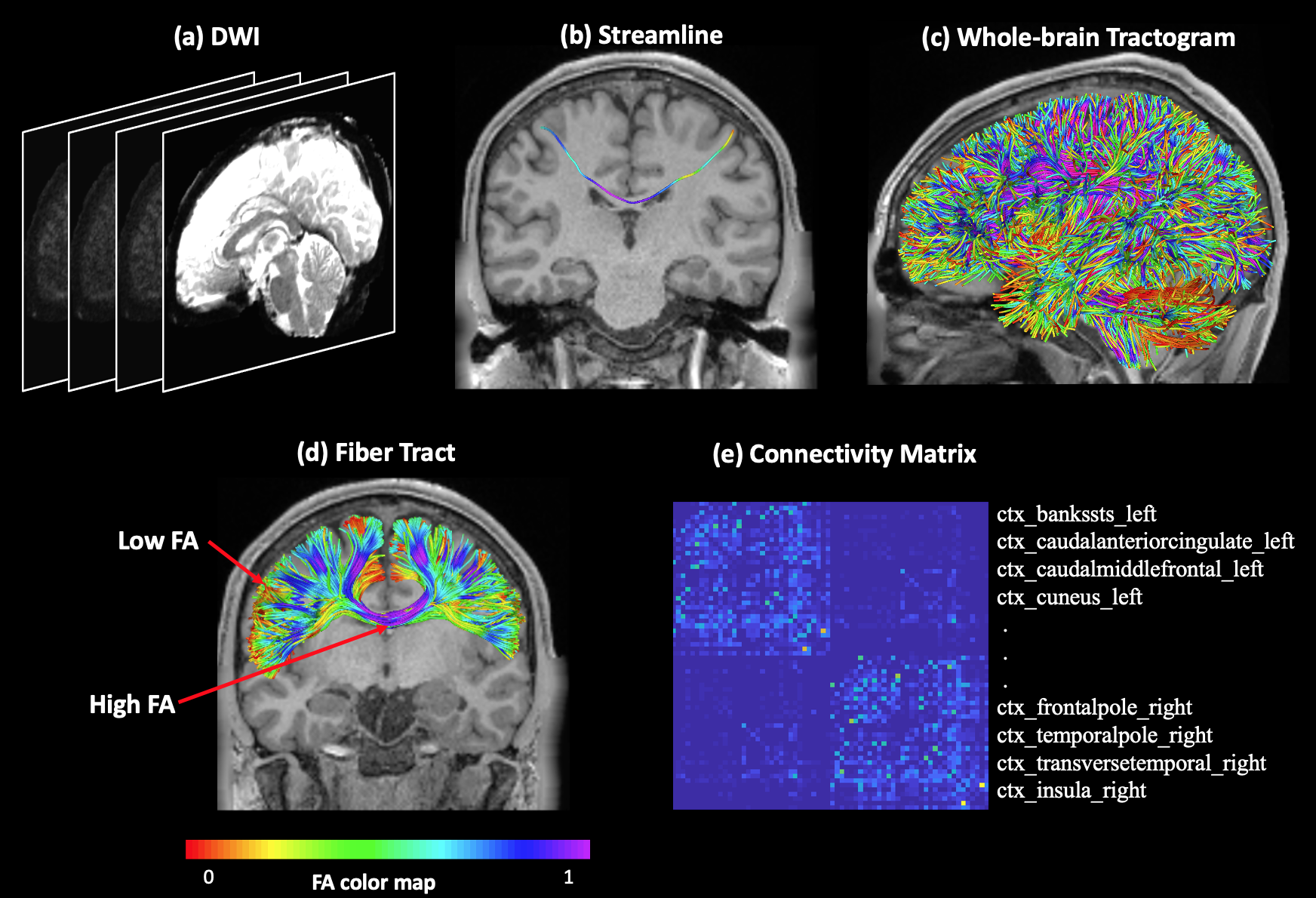}
	\caption{Graphic illustration of tractography. (a) Example dMRI data, also known as diffusion weighted imaging (DWI) data. (b) An individual streamline computed after performing tractography. (c) An example whole-brain tractogram that consists of streamlines covering the entire white matter. (d) A fiber bundle that consists of a set of streamlines, representing an anatomical fiber tract named the corpus callosum. The streamlines are colored by a microstructural measure, i.e., fractional anisotropy~(FA) that measures the anisotropy of water diffusion. A low FA can be seen at the endpoint region of the streamlines (near the cortex) and a high FA can be seen in the middle of the streamlines (in the deep white matter). (e) An example brain structural connectivity matrix, which is constructed based on white matter tractography from the whole brain. Each row (and column) represents a brain gray matter ROI (See Figure~\ref{fig_segmentation}(c) for an example brain gray matter parcellation), and the value in an element of the matrix is the strength of the white matter connection between the two corresponding ROIs (quantified as the number of streamlines in this case).} 
	\label{fig_definition}
\end{figure*}

Due to the large number of proposed quantitative approaches and the evolution of the field over two decades, there is a proliferation of terminology related to the quantitative analysis of dMRI tractography. Therefore, we begin with a listing of common terms used throughout this paper and their definitions, as provided in Table~\ref{table_definition}. We also provide a visualization of many key concepts that will be used in the paper (Figure~\ref{fig_definition}).

\begin{table*}[!t] 
	\caption{Common terms used throughout this paper and their definitions. We note that there are traditional terms that are widely used, but are not technically or biologically precise; in this table, we emphasize such terms and encourage the avoidance of their usage in future studies.}
	\label{table_definition}
	\renewcommand{\arraystretch}{1.4}
	\centering
	\begin{tabular}{ L{3cm} | L{14cm} }
		\hline
		\hline
		\textit{Term} & \textit{Definition} \\
		\hline	
		Tractography / \newline Fiber tracking & Any computational process that estimates the anatomical trajectories of the white matter fiber pathways from dMRI data. \newline
		\textit{Note}: The term ``Fiber tracing''  is also widely used. However, we strongly suggest avoiding this term because ``tracing'' is frequently used in the context of \textit{ex~vivo} tracer injections. \\
		\hline
		Streamline & A set of ordered points in 3D space, encoding a trajectory estimated through performing tractography (see Figure \ref{fig_definition}(b)). \newline \textit{Note}: The term ``fiber '' is also widely used. However, we strongly suggest avoiding this term because ``fiber'' is in reference to biology, while ``streamline'' implies the digital reconstruction of such. \\
		\hline
		Tractogram & A set of streamlines, often generated in such a way as to cover the entire white matter in order to capture any possible white matter connections. This can be referred to as a ``whole-brain tractogram'' (see Figure \ref{fig_definition}(c)). \\
		\hline
		Fiber Bundle / \newline Fiber Tract / \newline Fiber Fasciculus / \newline Fiber Pathway & These terms have biological meanings as a set of white matter fibers (axons) forming a corticocortical or corticosubcortical connection in the brain \cite{schmahmann2009fiber}. 
		In the neuroimaging literature, these terms are commonly used \textit{instead} to refer to white matter connections reconstructed using tractography. 
		For clarity, in this paper, we will use the term ``fiber pathway'' to refer to a set of streamlines resulting from the subdivision of a tractogram, while we will use the terms ``fiber tract'' or ``fiber bundle'' to refer to a fiber pathway that additionally corresponds to known anatomy with a traditional name (e.g., the corpus callosum or the corticospinal tract) (see Figure~\ref{fig_definition}(d)). \\
		\hline
		Brain Connectivity & In neuroimaging, the somewhat elusive and ambiguous concept of brain connectivity refers to measures of the structural and/or functional relationship between different brain regions \cite{sakkalis2011review,rossini2019methods,horwitz2003elusive,uddin2013complex}. \\
		\hline
		Structural Connectivity & A specific type of brain connectivity. Two brain regions are structurally connected if a fiber tract physically interconnects them. This is typically measured \textit{in~vivo} in humans using dMRI. However, there is no consensus on how this should be best quantified. Structural connectivity measures (also called ``weights'' or ``strengths'') can include a variety of quantitative connectivity measures computed from a specific set of streamlines corresponding to a pathway of interest (e.g. those connecting two specific endpoints). The goal is often to approximate the true underlying fiber density or number of axons \cite{jones2010challenges,Smith2020Quantitative}. \\
		\hline
		Connectome / \newline Brain Connectivity Matrix & A two-dimensional matrix wherein the rows and columns correspond to specific brain gray matter regions of interest (ROIs), and the value stored within each element of the matrix is the computed connectivity `` strength'' between those regions corresponding to that row \& column \cite{sporns2005human} (see Figure~\ref{fig_definition}(e)). Such data are described mathematically as a graph. This matrix representation is directly inspired by invasive axonal tract-tracing experiments in animals, where the results of multiple studies are expressed as a quantitative connectivity matrix \cite{bakker2012cocomac}. Without further specification, we use ``connectome'' throughout to implicitly refer to the \textit{structural} connectome constructed using dMRI tractography, as opposed to those derived through other imaging modalities (e.g., functional MRI). \\
		\hline
		Diffusion Model & A theoretical model that connects the dMRI signal to salient features of tissue microstructure at the cellular level \cite{novikov2018modeling,yablonskiy2010theoretical,panagiotaki2012compartment}. This includes tissue microstructure or biophysical models such as Neurite Orientation Dispersion and Density Imaging~(NODDI) \cite{Zhang2012} and Free Water~(FW) \cite{Pasternak2009,Pasternak2012}, as well as diffusion signal representations that include the diffusion tensor \cite{Basser1996_dti}, diffusion kurtosis \cite{Jensen2010a} and many others \cite{alexander2006introduction,afzali2020sensitivity}. \\
		\hline
		Microstructural Measure & Any parameter extracted from a diffusion model fit in each voxel that provides information regarding the underlying tissue microstructure (e.g., the fractional anisotropy~(FA) that describes water diffusion anisotropy \cite{basser2011microstructural}). \\
		\hline
		\hline
	\end{tabular}
\end{table*}

Many quantitative tractography approaches can be applied to study the brain's structural connectivity in health and disease. The fundamental goal of these analyses is to estimate quantitative measures of connectivity (or microstructure) of some pathway (or pathways) of interest. Quantitative analyses of tractography can be categorized into two main categories or styles: \textit{tract-specific} analyses and \textit{connectome-based} analyses (this categorization is helpful but imperfect, as some approaches blend aspects of both analysis styles). Tract-specific analysis refers to research that is typically hypothesis-driven and studies particular anatomical fiber tracts \cite{alexander2007diffusion,yeo2014different,shany2017diffusion,levitt2020miswiring}. Tract-specific analysis has been increasingly adopted, particularly for the study of local white matter regions in health and disease \cite{alexander2007diffusion,yeo2014different,shany2017diffusion}. Connectome-based analysis refers to research that is more data-driven and generally studies the structural connectivity of the entire brain \cite{sporns2005human,zalesky2012connectivity,ingalhalikar2014sex,bastiani_human_2012}. This type of analysis aims to understand patterns of whole-brain anatomical connectivity, and therefore relies on tractography performed across the entire white matter. 

In this paper, we first provide a brief introduction to tractography (Section~\ref{secTractography}), then a review of methodology for quantitative tractography analysis (Sections~\ref{secTractCorrection}--\ref{secTractMeasurement}), followed by a review of studies that use quantitative tractography analysis to study the brain in health and disease (Section~\ref{secApplication}). For the methodology review, we organize the paper according to three main processing steps that are common across most approaches for quantitative analysis of tractography. The first common step corrects potential biases in tractography reconstruction that may otherwise be detrimental to, or outright preclude, quantitative analysis of such data; we refer to this as tractography \textit{correction} (Section~\ref{secTractCorrection}). The second common processing step identifies white matter pathways (e.g., subdivisions of the tractogram) that are meaningful for quantification of brain connectivity; we refer to this as tractography \textit{segmentation} (Section~\ref{secTractSegmentation}). The third common step extracts quantitative indices that describe the microstructure and/or the ``strength'' of the brain connections; we refer to this as tractography \textit{quantification} (Section~\ref{secTractMeasurement}). In each section, we aim to describe methodological choices, their popularity, and their potential pros and cons. Finally, we review studies that have used quantitative tractography approaches to study the brain's white matter, including applications in development, aging, neurological disorders, mental disorders, and neurosurgery (Section~\ref{secApplication}).

\section{Brief introduction to tractography}
\label{secTractography}
``Tractography'' can refer to any computational process that estimates the anatomical trajectories of white matter fiber pathways from dMRI data. There are many methods available to perform tractography, including traditional streamline tractography based on the diffusion tensor model \cite{Basser1996_dti,basser2000vivo}, which was originally proposed (as ``hyperstreamlines'') for the visualization of tensor fields in the computer graphics community in 1993 \cite{delmarcelle1993visualizing}. More recently, many streamline tractography methods using advanced diffusion models have been proposed and are under active development \cite{jeurissen2014multi,tournier2010improved,descoteaux2008deterministic,behrens2007probabilistic,friman2006bayesian,lazar2005bootstrap,tournier2003diffusion,jackowski2005white,fernandez2012high,malcolm2010filtered,reddy2016joint,wasserthal2019combined,feng2020asymmetric}. Furthermore, many alternative tractography algorithms have been proposed, such as front evolution \cite{tournier2003diffusion,pichon_hamilton-jacobi-bellman_2005,sepasian_multivalued_2012}, geodesic (identifying likely connection trajectories based on assumed endpoints) \cite{odonnell2002new,jbabdi_bayesian_2007,jbabdi_accurate_2008,wu_genetic_2009,li_knowledge-based_2014,schreiber_plausibility_2014}, atlas-based \cite{yendiki2011automated,wasserthal2019combined}, and various forms of ``global'' tractography that simultaneously fit the entire tractography reconstruction to all image data (either by progressively forming linked chains from a large set of randomly-initialized short segments \cite{mangin_framework_2002,kreher_gibbs_2008,fillard_novel_2009,reisert_global_2011,mangin_toward_2013,christiaens_global_2015,teillac2017novel}, or iteratively perturbing a set of candidate trajectories \cite{wu_globally_2012,aattocchio2020improving,lemkaddem_global_2014,close_fourier_2015}). Work is additionally underway to improve tractography using machine learning \cite{poulin2019tractography,neher2017fiber,wegmayr2018data,sarwar2020towards,wasserthal2019combined}. 

While a detailed introduction of tractography methods is out of the scope of this review, we refer the readers to these review papers \cite{neher2015strengths,jeurissen2019diffusion,Rheault2020,smith2020diffusion} that are specific to tractography algorithms. 

Many software packages are available to perform tractography research, including  
ANIMA \cite{descoteaux2008deterministic}, 
BrainSUITE \cite{shattuck2002brainsuite}, 
Camino \cite{cook2005camino}, 
COMMIT \cite{daducci2013convex,daducci2014commit,schiavi2019reducing, ocampo2021hierarchical},
Diffusion toolkit \cite{wang2007diffusion}, 
Dipy \cite{garyfallidis2014dipy}, 
DMIPY \cite{fick2019dmipy}, 
DSI studio \cite{yeh2013deterministic},  
DTIstudio \cite{jiang2006dtistudio},  
ExploreDTI \cite{leemans2009exploredti}, 
FiberNavigator \cite{chamberland2014real}, 
FSL \cite{jenkinson2012fsl}, 
MITK \cite{wolf2005medical}, 
MRtrix3 \cite{tournier2012mrtrix,tournier2019mrtrix3},  
PANDA \cite{cui2013panda}, 
SlicerDMRI \cite{norton2017slicerdmri,zhang2020slicerdmri}, 
TractSeg \cite{wasserthal2018tractseg},
and Tracula \cite{yendiki2011automated}. 
It is important to note that tractography is known to be sensitive to the choice of the underlying fiber tracking algorithms, which can further affect any subsequent quantitative analyses using the tractography data. On one hand, tractography results may vary across the large set of available diffusion models \cite{yablonskiy2010theoretical,panagiotaki2012compartment,alexander2006introduction,afzali2020sensitivity} and tractography methods \cite{basser2000vivo,jeurissen2014multi,tournier2010improved,descoteaux2008deterministic,behrens2007probabilistic,friman2006bayesian,lazar2005bootstrap,tournier2003diffusion,jackowski2005white,fernandez2012high,malcolm2010filtered,reddy2016joint,wasserthal2019combined,feng2020asymmetric}. On the other hand, tractography results can also be sensitive to the choice of parameters (such as seeding and stopping thresholds) within a certain algorithm \cite{xie2020anatomical,gong2018free,cote2013tractometer,chamberland2014real,moldrich2010comparative}. While many research studies have been performed to compare different tractography algorithms \cite{bastiani_human_2012,pujol2015dti,sinke2018diffusion,wilkins2015fiber,petrov2017evaluating,zhan2015comparison,fillard2011quantitative}, there is no consensus on ``the best algorithm.''

\section{Improving tractography quality: methods for tractography correction}
\label{secTractCorrection}

The goal of tractography correction methods is to tackle potential biases and errors that are produced in conventional or raw tractography algorithms. This is important to improve the biological accuracy of the reconstructed white matter fiber pathways for the extraction of anatomical fiber tracts (Section \ref{ref_tractidentification}) and the construction of quantitative structural connectivity matrices (Section \ref{secConstructionMatrix}). In this section, we consider the near-ubiquitous ``streamlines'' paradigm for dMRI tractography \cite{basser1998fiber,conturo1999tracking,mori1999three,westin1999image,basser2000vivo,tournier2010improved,descoteaux2008deterministic,behrens2007probabilistic,friman2006bayesian,lazar2005bootstrap,tournier2003diffusion,jackowski2005white,fernandez2012high,malcolm2010filtered,reddy2016joint,wasserthal2019combined,feng2020asymmetric}, and the ways in which specific undesirable behaviors or biases may be ameliorated through specific targeted mechanisms.

\subsection{Curvature overshoot bias}

Possibly the earliest bias established for dMRI tractography was the fact that if the streamline algorithm simply takes a step of finite size in the direction of the local fiber orientation at the current vertex (a so-called ``first-order'' method),  streamlines in curved bundles will tend to under-estimate the curvature, resulting in erroneous trajectories \cite{tournier_limitations_2002}. While use of a step size that is small relative to the image voxel size mitigates the issue, use of a higher-order integration method during tracking that directly accounts for such curvature is a more direct solution \cite{basser2000vivo}. However, doing so in a manner that is compatible with diffusion models that account for crossing fibers can be technically difficult \cite{tournier2010improved,cherifi_combining_2018,aydogan_parallel_2020}.

\subsection{Termination bias}

``Termination'' refers to the location at which the propagation of a streamline is ceased. Just as a tractography algorithm should produce streamlines whose tangents are faithful to the underlying fiber orientations, those streamlines should also terminate at locations corresponding to the endpoints of the underlying fiber tracts. Biases or inadequacies in such can result in partial fiber streamlines that stop prematurely in the white matter --- despite the fact that white matter fibers synapse in the gray matter \cite{daducci2016microstructure} --- or even that enter fluid-filled regions or cross sulcal banks. Such issues can be prevalent as tractography algorithms often operate using only the fitted diffusion model (e.g., diffusion tensor \cite{Basser1996_dti} or constrained spherical deconvolution~(CSD) \cite{tournier2007robust,jeurissen2014multi}) in each image voxel. While these models provide strong evidence regarding fiber \textit{orientations} to inform the direction of streamline propagation, the evidence they provide regarding where such fibers \textit{stop} (and hence where streamlines should ideally be terminated) is weak (indeed, the diffusion-weighted signal itself does not provide direct evidence of fiber terminations \cite{smith_chapter_2020}). Most fiber tracking algorithms exploit heuristic thresholds on features such as diffusion model anisotropy and streamlines curvature to serve as tractography termination criteria. However, the indirect nature of these metrics for this task, combined with the inferior spatial resolution of diffusion-weighted imaging, leads to such errors being highly prevalent \cite{smith_anatomically-constrained_2012}. These common termination criteria are therefore not adequate to ensure biologically plausible streamline generation, resulting in a substantial proportion of streamlines from whole-brain tractography being unreasonable for quantification of white matter fiber connectivity \cite{yeh2016correction}.

The ill-posed nature of streamline terminations can be addressed by utilizing anatomical reference data to impose relevant prior information to ensure that the reconstructed streamlines fulfill the basic characteristics of how neuronal fibers are organized in the brain. Unlike conventional tractography algorithms, in which termination points can be distributed almost anywhere in the brain, the Anatomically-Constrained Tractography~(ACT) framework \cite{smith_anatomically-constrained_2012} and similar methods \cite{girard_anatomical_2012,yeh_mesh-based_2017} ensure that both streamline propagation and termination are constrained based on knowledge of where neuronal fibers locate, e.g., connecting between gray matter areas via the white matter. Although these methods cannot guarantee that every streamline accurately traces the complete trajectory of an underlying fiber connection, they do prevent the generation of streamlines that \textit{cannot possibly} represent biological connectivity within the brain.

\subsection{Connection density biases} 
\label{secDensityBias}

While streamline tractography provides estimates of structural connection trajectories that are consistent with the underlying fiber orientations, it does not provide any guarantees regarding consistency between the \textit{number} of such reconstructed connections with the density of those underlying fibers (i.e., the actual number of axons in a white matter region) \cite{jones2010challenges,Jones2013}. This limitation is often expressed as the mantra ``streamline count is not quantitative.'' Because each streamline is generated independently of all others, and is treated as an infinitesimally thin line with no volume, the number of streamlines (whether traversing a voxel or ascribed to some pathway of interest) is not a quantitative measure, being influenced by myriad non-biological influences including streamline seeding \cite{li_effects_2012} and streamlines algorithm \cite{bastiani_human_2012}. As such, the relative streamline count within different pathways cannot be naively used as a marker of relative connection strength.

What would give greater confidence to the interpretation of tractography-based connection density as a quantitative measure is if, throughout the white matter, there were correspondence between the local density of reconstructed connections and the local density of fibers as evidenced by the diffusion image data \cite{Smith2020Quantitative}. Initially, aspiration for achieving this was limited to ``global'' tractography algorithms as mentioned in Section~\ref{secTractography}. More recently, methods have been devised to address this task by instead utilizing a pre-generated whole-brain streamline tractogram and modulating the contribution of each streamline toward the reconstructed white matter fiber density. Initial methods achieved this by selecting a subset of streamlines that best fit the diffusion image data, i.e., BlueMatter~\cite{sherbondy_think_2009}, MicroTrack~\cite{sherbondy_microtrack:_2010} and SIFT~\cite{smith_sift:_2013}, while later methods estimate a multiplier to be applied to each streamline, i.e., COMMIT~\cite{daducci2013convex,daducci2014commit}, LiFE~\cite{pestilli_evaluation_2014}, SIFT2~\cite{smith_sift2:_2015}, COMMIT2~\cite{schiavi2019reducing} and COMMIT2\textsubscript{tree}~\cite{ocampo2021hierarchical}; these ``weights'' represent an effective cross-sectional area of each streamline, which can be utilized as a direct measure of connection density \cite{Smith2020Quantitative} (or indeed to modulate streamline contributions toward other aggregate measures of the connectivity ``strength''; see Section~\ref{secTractMeasurement}).

\subsection{Gyral bias}

Gyral bias in tractography is the bias towards generating streamlines that terminate in gyri rather than sulci \cite{schilling2018confirmation,wu2020mitigating}, which reduces the agreement of the initiations or terminations of fiber streamlines with known ground truth from tracer studies and histology \cite{aydogan2018tractography,budde2013quantification}. The source of the gyral bias is multifactorial, such as the complexity of axon arrangement at the junction of cortical grey matter and superficial white matter \cite{vanessen_ch16_2014}. The partial volume effect from the limited MRI spatial resolution introduces difficulties in distinguishing complex fiber configurations based on the reconstructed fiber orientation distributions (FODs). Increasing the image resolution is beneficial to mitigate the gyral bias \cite{heidemann_zooppa_2012,sotiropoulos_fusion_2016}, but is constrained by the signal-to-noise ratio of the acquired dMRI data. Even operating on high-resolution dMRI data, current tractography algorithms have been shown to deviate from the ground-truth fiber projections derived from histological staining, with greater streamline densities at gyral crowns than sulcal banks \cite{schilling_gyral_bias_2018}. The complex folding and convolutions of cortical gyri also pose challenges for the reconstruction of long-range connections \cite{reveley_superficial_wm_2015}.

A surface flow approach has been proposed to model the arrangement of superficial axonal fibers at gyral crowns, thereby improving the consistency between streamlines' initiations or terminations and the expectations from histological data \cite{stonge_set_2018}. More recently, the gyral bias has been shown to be alleviated using an asymmetric FOD technique \cite{wu_afods_2020}, which can depict the highly-curved fiber geometry appearing at the superficial white matter \cite{bastiani_afods_2017,wu_afods_2020}.

\subsection{False positive connections}
\label{secFalseFiltering}

Several orders of magnitude separate the resolution of dMRI acquisitions (cubic millimeters) from the typical size of the axons (few micrometers). This discrepancy introduces ambiguities in the white matter that can be compatible with multiple streamline configurations and are \textit{difficult for tractography to resolve} \cite{guevara2012automatic, Maier-Hein.2017, Girard2020}. This ill-posed nature of tractography has received a renewed interest in the past few years, and several studies have raised serious concerns about the anatomical accuracy of the reconstructions \cite{Thomas.2014, Maier-Hein.2017}. In particular, it has been shown that tractography techniques suffer from a large number of false positives, i.e., streamlines that are reconstructed but do not correspond to real anatomical bundles, and that these erroneous connections can severely bias the estimation of connectivity \cite{Drakesmith.2015, Zalesky.2016, Maier-Hein.2017}. The anatomical accuracy of the reconstructions can be improved by \textit{manually filtering the streamlines} with inclusion/exclusion ROIs, as seen in Section~\ref{ref_tractidentification}, but this requires prior  knowledge of ``where white matter pathways start, where they end, and where they do not go'' \cite{Schilling.2020}. However, more automated procedures are desirable especially when analyzing large cohorts of subjects.

A widely-used approach to attempt to alleviate false positive connections when building a structural connectivity matrix is \textit{thresholding} \cite{fornito:2016}, which refers to the removal of edges whose strength is smaller that a certain cut-off value. This strategy implicitly assumes that all weak connections are spurious (and thus implicitly that strong connections are reliable), but it was demonstrated that this indiscriminate filtering (thresholding) does not differentiate between true and spurious connections, and its effectiveness is still debated \cite{Civier.2019, Drakesmith.2015, schiavi2019reducing,BUCHANAN.2020}.
\textit{Alternative solutions} have been proposed, which are \textit{either knowledge- or data-driven}. The former are described in Section~\ref{ref_tractidentification}, and typically use fully- or semi-automated algorithms, e.g., clustering, as a means to detect and discard streamlines considered outliers based on anatomical priors or geometrical properties, e.g., length and shape. On the other hand, the methods described in Section~\ref{secDensityBias} for addressing density biases can also be used as filtering procedures, in which the streamlines that are not supported by the acquired dMRI data, i.e., whose estimated weight is zero, are discarded. While these approaches achieve good results in terms of removing duplicate streamlines or isolated ones that are likely spurious, actually none proved effective in discriminating between true and spurious connections in the connectome networks \cite{Maier-Hein.2017, schiavi2019reducing}.

It was recently demonstrated that the performance of filtering methods can be significantly boosted by \textit{combining knowledge- and data-driven strategies}. These new formulations \cite{schiavi2019reducing,ocampo2021hierarchical} allow taking explicitly into account the fundamental anatomical knowledge that fibers in the brain are naturally organized in bundles, and use this prior to seek for solutions which explain the measured signal using the minimum number of bundles. This is achieved by organizing the streamlines into groups and using the Group Lasso regularization \cite{yuan2006model} to promote sparsity in the space of connections, rather than in the space of individual streamlines as implicitly done in previous filtering methods. This approach helps resolving some of the ambiguities present in the data and, as a consequence, allows improving significantly the anatomical accuracy of the connectome \cite{schiavi2019reducing,ocampo2021hierarchical}.

Research is still underway to determine the best method(s) for filtering, with some disagreement on the beneficial versus detrimental effect of these methods on the structural connectome \cite{Civier.2019, Drakesmith.2015,yeh2016correction,BUCHANAN.2020,schiavi2019reducing,frigo2020diffusion}.

\section{Defining regions for quantitation: methods for tractography segmentation }
\label{secTractSegmentation}

\begin{figure*}[!t]
	\centering
	\includegraphics[width=0.9\linewidth]{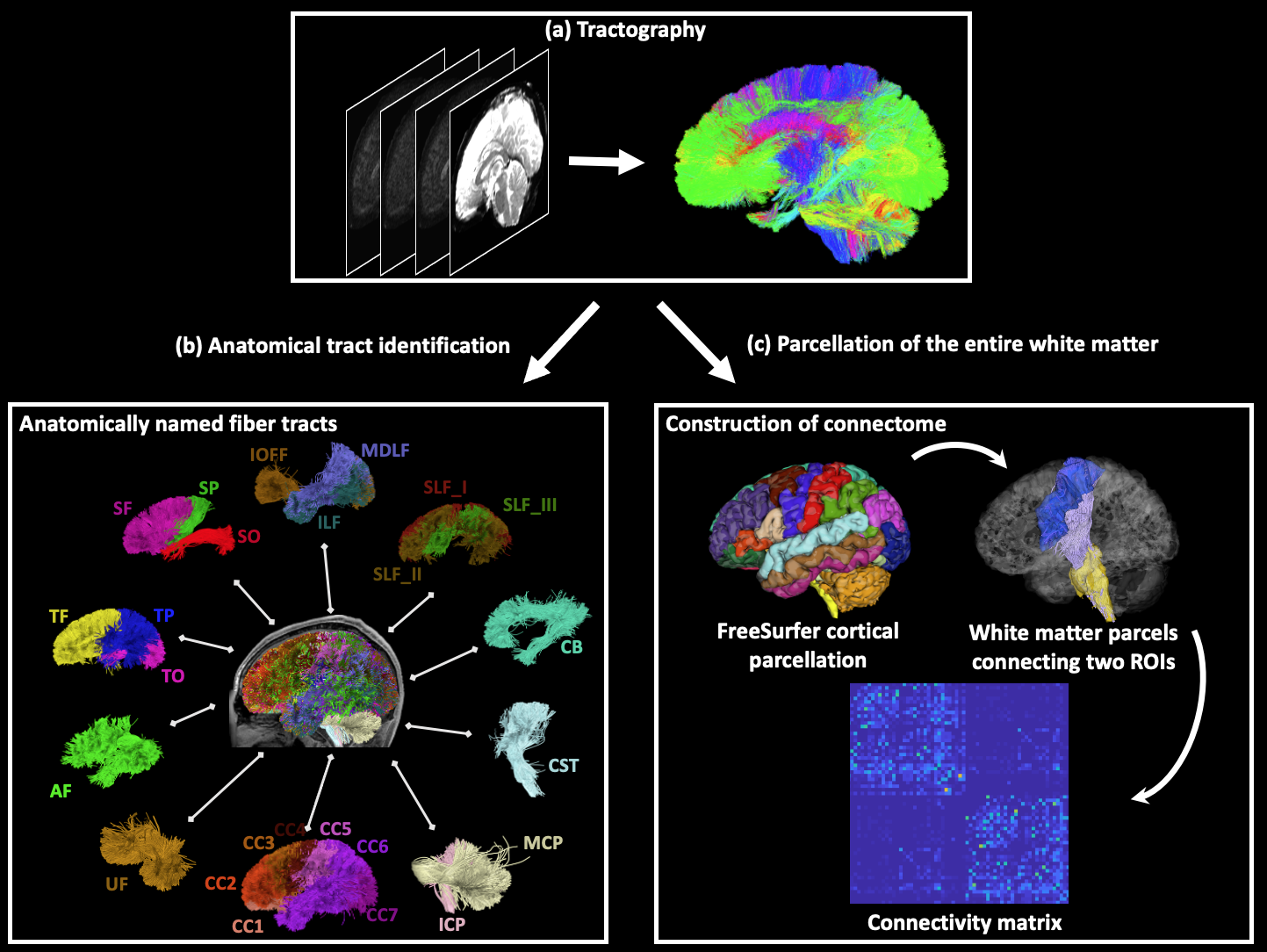}
	\caption{Tractography segmentation: (a) Example whole-brain tractogram computed by performing tractography in DWI data; (b) Example anatomical tracts extracted from the tractogram; and (c) Example structural connectivity matrix constructed by performing whole brain tractography segmentation between all pairs of FreeSurfer cortical regions.} 
	\label{fig_segmentation}
\end{figure*}

The goal of white matter tractography segmentation is to identify white matter pathways (e.g., subdivisions of the tractogram) that are meaningful for quantification of the brain's structural connectivity. (Tractography segmentation is also critical for qualitative applications such as visualization of fiber tracts, e.g., for neurosurgical white matter mapping; see Section~\ref{secTractoInNeurosurgery}.) Tractography segmentation methods can be generally grouped into two categories, related to the tract-specific and connectome-based analysis approaches, as illustrated in Figure~\ref{fig_segmentation}. The first category of methods identify anatomical white matter fiber tracts and label them with traditional names (such as the arcuate fasciculus or the corticospinal tract) (Section~\ref{ref_tractidentification}). The second category of methods parcellate, or subdivide, the entire white matter into many white matter parcels based on information about the streamline trajectories and/or endpoints (Section~\ref{ref_wmparcellation}).

\subsection{Anatomical tract identification}
\label{ref_tractidentification}

Anatomical tract identification aims to identify tractography streamlines that correspond to anatomically named white matter fiber tracts. This task is non-trivial, given the complex structures of the white matter anatomy and the large number of streamlines in the tractogram.

Conventional anatomical tract identification methods rely on \textit{manual streamline selection}, also referred to as \textit{virtual dissection}, where experts in anatomy interactively select tractography streamlines using manually drawn ROIs in the brain \cite{stieltjes2001diffusion,catani2002virtual,wakana2007reproducibility,de2011atlasing}. Usually, inclusion ROIs are placed in the gray matter (cortical and subcortical) to define where the streamlines should terminate and in the white matter to define where the streamlines should pass, and exclusion ROIs are placed in other regions to exclude undesired streamlines \cite{wakana2007reproducibility,Rheault2020_tractostorm}. Manual streamline selection is considered to be the gold standard to delineate anatomical fiber tracts in tractography and has been widely used to validate other anatomical tract identification techniques \cite{pujol2015dti,poulin2019tractography,xie2020anatomical}. Manual selection is also used in studies where specific clinical expertise is needed, e.g., for presurgical white matter mapping in tumor patients where the tumor and lesion can largely displace white matter fiber tracts \cite{radmanesh2015comparison}.

Due to the fact that manual tract selection is time-consuming and has high clinical and expert labor costs, modern studies increasingly focus on automated tract identification methods. These methods can be categorized into three categories: \textit{ROI-based}, \textit{streamline labeling}, and \textit{direct segmentation}. 

The \textit{ROI-based} methods are the most commonly used. They automate ROI computation and perform streamline selection based on the ROIs a streamline terminates in and/or passes through. The majority of the ROI-based methods leverage a brain ROI atlas and use an image registration to automatically align the ROIs in the atlas space to the subject space \cite{wassermann2016white,hua2008tract,zhang2008automated,zhang2010atlas,yeatman2012tract,yendiki2011automated,zollei2019tracts,astolfi2020stem,verhoeven2010construction,lawes2008atlas,schurr2019tractography,allan2020parcellation}. Currently, popularly used brain ROI atlases include those provided in Freesurfer \cite{desikan2006automated,fischl2012freesurfer}, MNI--ICBM152 \cite{mazziotta2001probabilistic,mori2008stereotaxic,oishi2008human,oishi2009atlas}, and JHU-DTI \cite{wakana2004fiber,wakana2007reproducibility} (see \cite{hansen2020pandora} for a review of more atlases). There are also ROI-based methods that directly predict ROIs in subject space using machine learning \cite{astolfi2020stem,li2020two}. 

The second category of automated tract identification methods, \textit{streamline labeling}, assigns an anatomical label to each individual streamline. Often, streamline labeling is done by computing the geometric distance of each streamline to labeled streamlines in a reference tract segmentation, and then assigning a streamline label based on the closest reference tract \cite{maddah2005automated,clayden2007probabilistic,labra2017fast,gupta2017fibernet,gupta2018fibernet,lam2018trafic,xu2019objective,kumar2019white,liu2019deepbundle,berto2020classifyber,yang2020functional,zhang2020deep}. There are also learning-based segmentation approaches that train a model from the reference tract segmentation data and predict an anatomical label for each streamline in a new subject \cite{gupta2017fibernet,lam2018trafic,gupta2018fibernet,kumar2019white,yang2020functional,zhang2020deep,liu2019deepbundle,xu2019objective}. To reduce the amount of labeling for each streamline, many streamline labeling methods first group streamlines into clusters (known as fiber clustering), followed by assigning an anatomical label for each cluster, thus labeling each streamline \cite{o2007automatic,ziyan2009consistency,li2010hybrid,guevara2012automatic,ros2013atlas,chekir2014hybrid,tuncc2014automated,yoo2015example,roman2017clustering,yeh2018population,garyfallidis2018recognition,zhang2018anatomically,siless2018anatomicuts,avila2019inference,vazquez2020automatic,siless2020registration,wu2020tract}. 

The third category of automated tract identification is called \textit{direct segmentation} in the literature \cite{eckstein2009active,bazin2011direct,ratnarajah2014multi,reisert2018hamlet,wasserthal2018tractseg,wasserthal2019combined,dong2019multimodality,rheault2019bundle,lu2020white,li2020neuro4neuro}. Unlike the aforementioned methods that work on tractography streamline data, the \textit{direct segmentation} methods operated on volumetric image data (e.g., FOD maps \cite{wasserthal2018tractseg} and streamline density maps \cite{lu2020white}) to predict the tract location in the dMRI data, followed by fiber tracking in the segmented regions. Although the direct segmentation methods are relatively less used than the ROI-based and streamline labeling methods, recently proposed direct segmentation methods such as TractSeg \cite{wasserthal2018tractseg} and Bundle-specific tractography (BST) \cite{rheault2019bundle} have shown highly promising tract segmentation performance without the need for generation of streamlines in advance.

There are several points that should be considered when selecting a method for anatomical tract identification. First, given the fact that there is no ground truth in tractography, validation of a tract identification method is difficult. One commonly used strategy to assess a tract identification method is to evaluate its reproducibility (in terms of intra- and inter-rater as well as test-retest reproducibility) \cite{zhang2019test,tong2019reproducibility,Rheault2020_tractostorm}. Methods using manual streamline selection have been shown to be highly reproducible for specific tracts and segmentation protocols using the diffusion tensor tractography \cite{wakana2007reproducibility}, but with more modern multi-fiber tractography, reproducibility is lower, potentially due to the increase in spurious streamlines \cite{Rheault2020_tractostorm}. However, manual streamline selection is still considered to be the gold standard for benchmarked comparisons and algorithm training \cite{kreilkamp2019comparison,wasserthal2018tractseg,zhang2010atlas}. For automated tract identification, one study indicates that streamline labeling has higher test-retest reproducibility than ROI-based segmentation \cite{zhang2019test}; however, no consensus has been reached given that many existing tract identification methods have not been compared. The second point is that a good tract identification method should be highly consistent across different populations and acquisitions \cite{zhang2018anatomically,sydnor2018comparison,wasserthal2018tractseg,yendiki2011automated}. This is particularly important for automated tract identification methods, which are essential to perform across-lifespan and multi-site studies, which usually involve a large number of datasets \cite{alexander2017open,casey2018adolescent,thompson2017enigma,harms2018extending,cetin2020white}. One factor that can affect the consistency of methods that segment tractography streamline data is the underlying tractography method that is employed. In general, studies indicate that modern multi-fiber tractography is more sensitive when performing fiber tracking and thus is more consistent across subjects and timepoints than traditional single-fiber DTI tractography \cite{bucci2013quantifying,prvckovska2016reproducibility}. The third point is that anatomical tract identification results can be highly variable across studies due to a lack of consistent fiber tract definitions, where different methods may give different results for the same target anatomical fiber tracts \cite{schilling2020tractography,Rheault2020_tractostorm,dick2012beyond}. Work is underway to provide sets of consistent rules for fiber tract definitions (e.g. the White Matter Query Language \cite{wassermann2016white}) and tractography-based fiber tract atlases \cite{mori2008stereotaxic,zhang2018anatomically,yeh2018population} that are provided in an open fashion for use and possible extension or modification by the community. Lastly, deep learning techniques have been increasingly applied for anatomical tract identification, showing high potential for future work and improvements in computational speed \cite{reisert2018hamlet,wasserthal2018tractseg,wasserthal2019combined,gupta2017brainsegnet,lam2018trafic,gupta2017fibernet,xu2019objective,gupta2018fibernet,zhang2020deep}. 

\subsection{Parcellation of the whole-brain tractogram}
\label{ref_wmparcellation}

Parcellation of the entire white matter (as represented by a whole-brain tractogram) aims to enable quantitative analysis of all possible white matter connections in the whole brain. There are generally two categories of methods: \textit{cortical-parcellation-based} methods and \textit{fiber clustering} methods \cite{o2013fiber}. The cortical-parcellation-based methods are more widely used
as they enable construction of a connectivity matrix and its subsequent analysis using techniques from graph theory (as described in Section~\ref{secGraph}) \cite{bullmore2009complex,sporns2005human,gong2009mapping,zalesky2012connectivity,ingalhalikar2014sex,yeh2016connectometry,bassett2017small}. Though relatively less used, fiber clustering tractography parcellation methods are increasingly applied to study the brain's structural connectivity in applications such as disease classification and between-population statistical analysis \cite{zhang2018suprathreshold,feng2020local,zhang2018whole,ji2019increased,wu2018investigation}. (Note that these methods could perhaps more correctly be called ``streamline clustering,'' but the term ``fiber clustering'' is very established in the literature.)

The \textit{cortical-parcellation-based} methods work from a gray-matter-centric perspective. They parcellate tractography according to a cortical (and sometimes a subcortical) gray matter parcellation, focusing on the structural connectivity among different gray matter ROIs \cite{bullmore2009complex,sporns2005human,gong2009mapping,zalesky2012connectivity,ingalhalikar2014sex,yeh2016connectometry,bassett2017small}. Specifically, tractography segmentation is performed by extracting streamlines that connect pairs of ROIs. Therefore the resulting tractography segmentation is mainly determined by the selection of a cortical parcellation scheme. The majority of methods adopt a cortical parcellation that is computed from T1-weighted or T2-weighted MRI. The most popularly used cortical parcellation is the Freesurfer Desikan-Killiany cortical parcellation \cite{desikan2006automated,fischl2012freesurfer}, while many other cortical parcellation schemes have also been widely used \cite{tzourio2002automated,destrieux2010automatic,shattuck2008construction}. Many cortical-parcellation-based methods have also used a functional cortical parcellation computed using functional MRI data \cite{eickhoff2005new,glasser2016multi,schaefer2018local,yeo2011organization}. Finally, an alternative approach is to use a vertex-wise cortical ``parcellation,'' i.e., by identifying streamlines that connect pairs of vertices in the cortical surface \cite{besson2014intra,fellner2020frequent,tian2021high}. Currently, there is no consensus about which brain parcellation technique could be most useful \cite{de2013parcellation,yeh2020mapping,zalesky2012connectivity}.

The \textit{fiber clustering} methods work instead from a white-matter-centric perspective. They group tractography streamlines according to their geometric trajectories, describing the structural connectivity according to the white matter anatomy \cite{o2007automatic,ziyan2009consistency,li2010hybrid,guevara2012automatic,ros2013atlas,chekir2014hybrid,tuncc2014automated,yoo2015example,roman2017clustering,yeh2018population,garyfallidis2018recognition,zhang2018anatomically,siless2018anatomicuts,avila2019inference,vazquez2020automatic,siless2020registration,wu2020tract}. In general, fiber clustering methods start with a computation of pairwise streamline geometric similarities, followed by a computational clustering method to group similar streamlines into clusters. Compared to the cortical-parcellation-based methods that focus on streamline terminal regions, fiber clustering methods leverage the full lengths of the streamline trajectories. As a result, the cortical-parcellation-based methods can erroneously group streamlines that follow completely different trajectories through the white matter but nevertheless end up at the same gray matter endpoints, whereas fiber clustering can be insensitive to streamlines diverging at the cortex if they pass through the same deep white matter regions. In addition, unlike the cortical-parcellation-based methods that leverage additional cortical parcellation information, the majority of fiber clustering methods work on tractography data \textit{only}, such that there is no need for an inter-MR-modality registration, e.g., an image registration between dMRI and T1w images that can be affected by differences in image resolution \cite{malinsky2013registration} and echo-planar imaging~(EPI) distortion in dMRI data \cite{albi2018image}. Due to these factors, several studies have demonstrated advantages of fiber clustering methods over the cortical-parcellation-based methods, including more consistent parcellation across subjects \cite{zhang2017comparison,sydnor2018comparison} and a higher reproducibility between test-retest scans \cite{zhang2019test}.

When choosing a method to perform parcellation of the entire white matter, the previously described points related to reproducibility and consistency of anatomical tract identification across different populations and acquisitions (as described in Section~\ref{ref_tractidentification}) are also important facts to consider \cite{Smith:2015aa,buchanan2014test,zhang2019test,zhang2018mapping}.

An additional point to be highlighted for parcellation of the entire white matter is the scale or the granularity of the parcellation, i.e., how many white matter parcels are to be obtained after parcellation. For constructing the connectome, this is related to the choice of the cortical parcellation \cite{messe2020parcellation,hagmann:2008}, while for fiber clustering this is related to the number of fiber clusters obtained by the clustering algorithms \cite{zhang2018anatomically,wu2020tract}. Various parcellation scales have been applied in different studies, ranging from tens to millions of white matter parcels \cite{rodrigues2013evaluating,liu2017fine,zhang2018whole,osmanliouglu2020connectomic,zalesky2010whole,besson2014intra}. Choosing parcellation scales depends on the target application, e.g., studies have suggested a fine scale white matter parcellation (over 2000 parcels) can be beneficial for machine learning and statistical analysis \cite{zhang2018whole,liu2017fine} while a coarse scale parcellation (fewer than 200 parcels) can improve the connectome consistency across individuals \cite{osmanliouglu2020connectomic,rodrigues2013evaluating}.

\section{Performing quantitative analysis: methods for tractography quantification}
\label{secTractMeasurement}

The goal of tractography quantification is to extract quantitative measures that are useful to assess the structural connectivity of the brain's white matter pathways. In this section, we will first introduce the quantitative measures that can be computed from tractography (Section~\ref{secQuantitativeFromTractography}). Then we will focus on how to extract these measures within an individual white matter fiber pathway (Section~\ref{secQuantitativePathways}) and how to perform filtering techniques to reduce potential biases in the extracted measures (Section~\ref{secAdjustmentFiltering}). Next, we introduce how the extracted measures can be used for tract-specific analysis of anatomical white matter tracts (Section~\ref{secTractspecificAnalysis}) and for computing the edge weights to construct a connectivity matrix (Section~\ref{secConstructionMatrix}). Finally, we will introduce topological analyses of whole-brain connectivity based on graph theoretical measures and more advanced connectome measurements (Section~\ref{secGraph}).

\subsection{Quantitative measures computed from tractography}
\label{secQuantitativeFromTractography}

There are many quantitative measures that can be computed from tractography, which can be subdivided into two categories based on the source of data used to compute the measures. The first category of quantitative measures are based on \textit{the tractography data alone}, which are often used to define the connectivity strength \cite{jones2010challenges, SotiropoulosZalesky2019, yeh2020mapping, Smith2020Quantitative}. The number of streamlines~(NOS) is one widely used measure to quantify the connectivity strength \cite{van2018minimum,sporns2005human,ingalhalikar2014sex,roberts2016contribution,hagmann2008mapping}. Several studies conducted on animals using invasive tract‐tracing techniques have shown good agreement between the NOS and the biological connectivity as measured using \textit{ex~vivo} tract‐tracing data \cite{vanDenHeuvel2015,Delettre2019_tracers,Girard2020}. However, because of the difference in resolution between the measured dMRI signal and actual axon dimensions (as described in Section~\ref{secDensityBias}: ``Connection density biases''), many studies have emphasized that NOS does not provide a truly quantitative measure of connection strength \cite{Jones2013, jones2010challenges, SotiropoulosZalesky2019, yeh2020mapping}. Recent work aims to provide alternative, more quantitative measures of connectivity that better estimate the underlying connection density \cite{Smith2020Quantitative} or modulate streamline contributions toward aggregate measures of connectivity (see Section~\ref{secAdjustmentFiltering}). Other quantitative measures can be computed directly from tractography, such as the volume and the length of a fiber tract \cite{Colby2012,yeatman2012tract,Chandio2020,Rheault2020_tractostorm,catani2007symmetries,song2015asymmetry,bajada2019fiber,behrman2015fiber,grinberg2018microstructure}, have also been used. 

The second category of quantitative measures computed from tractography leverages microstructural measures that are usually computed from a diffusion model (e.g., diffusion tensor model) or another quantitative imaging modality (e.g., myelin imaging). In this category of measures, tractography is generally used to define the locations in which microstructural measures are sampled. (This sampling may be done after performing tractography, or during fiber tracking as part of sampling or computing a diffusion model). The most widely used microstructural measures are based on computational modeling of the dMRI signal. Popularly used diffusion-modeling-based microstructural measures include the fractional anisotropy, axial, radial and mean diffusivities (FA, AD, RD and MD respectively) derived from traditional diffusion tensor model \cite{Basser1996_dti}. More advanced measures include those reflecting inter- and intra-cellular signal fractions derived from Diffusion Kurtosis Imaging~(DKI) \cite{Jensen2010a}, Free Water modeling \cite{Pasternak2009,Pasternak2012}, Neurite Orientation Dispersion and Density Imaging~(NODDI) \cite{Zhang2012}, Spherical Mean Technique~(SMT) \cite{Kaden2016_SMTmultiComp}, or Apparent Fiber Density~(AFD) \cite{Raffelt2012}. It is important to be aware that changes in dMRI microstructural measures can be non-specific, in part due to the difference in scales between voxel-averaged dMRI signals (mm scale) and the scale of the individual axons and cells that are probed by the diffusing water (micrometer scale). For example, many factors (e.g., cell death, edema, gliosis, inflammation, change in myelination, increase in connectivity of crossing fibers, increase in extracellular or intracellular water, etc.) may cause changes in FA \cite{le2012diffusion,assaf2008diffusion,Jones2013}. More details about sensitivity, specificity and interpretability of these microstructural measures are out of the scope of the present review but the reader can refer to the following studies \cite{novikov2019quantifying,novikov2018modeling,jelescu2017design,alexander2019imaging}. Other approaches for tractography quantification leverage information from imaging modalities other than dMRI, such as T1-weighted~(T1w) imaging \cite{yeatman2014lifespan,boshkovski2020r1}, myelin sensitive maps \citep{Lee2020, Mancini2020, Piredda2020} and g-ratio \citep{Campbell2018}. For example, researchers have used the longitudinal relaxation rate~(R1), a measure sensitive to myelin, along the fiber tracts \cite{boshkovski2020r1,yeatman2012tract}.

These microstructural measures can be computed along fiber pathways using several approaches. Usually, this is done by computing a model in each voxel to derive a 3D microstructural map (e.g., an FA image) and then using tractography to define the locations from which the values of this microstructural measure should be sampled. This sampling may be done by first generating a binary mask that defines the voxels through which streamlines pass and then sampling from an underlying microstructure image within this mask \cite{ciccarelli2003diffusion,kunimatsu2004optimal,voineskos2010diffusion,heiervang2006between}. The sampling may alternatively be done by sampling microstructural measures at each point of the streamlines from a pre-calculated microstructure image \cite{heiervang2006between,ciccarelli2003diffusion,kunimatsu2004optimal,voineskos2010diffusion}. Instead of using a pre-calculated microstructure image, the diffusion or fiber model can be simultaneously estimated during fiber tracking so that the microstructural measures are directly computed at points along each streamline \cite{malcolm2010filtered,reddy2016joint,girard2017ax,olszewski2017social,gong2018free}.

\subsection{Domain of analysis: extraction of quantitative measures within white matter fiber pathways}
\label{secQuantitativePathways}

After choosing a quantitative measure of interest (Section~\ref{secQuantitativeFromTractography}) and a method for tractography segmentation (Section~\ref{secTractSegmentation}), then there are many methods available to extract the measure within an individual white matter fiber pathway. This is important to enable tract-specific analysis (Section~\ref{secTractspecificAnalysis}) and construction of a connectivity matrix (Section~\ref{secConstructionMatrix}). Two main strategies can be employed: using a scalar value as a summary statistic, or using data along the length of the fiber pathway. 

The most commonly adopted strategy for extraction of quantitative measures is to compute a \textit{scalar value} as a summary statistic of the fiber pathway. (This is required for connectome-based analysis and is a popular approach for tract-specific analysis). While some quantitative measures intrinsically provide a single scalar value per pathway, e.g. NOS or tract volume, others (e.g. sampled values of a quantitative metric) necessitate calculation of some statistic in order to produce such a scalar. For microstructural measures, the \textit{mean measure within the fiber pathway} is the most widely used. Other summary statistics such as the \textit{median}, \textit{maximum} and \textit{minimum} have also been employed \cite{Boshkovski2020,zhang2018suprathreshold,zhang2018whole}. A summary statistic can be obtained in several ways according to how microstructural measures are computed along fiber pathways (as described in Section~\ref{secQuantitativeFromTractography}). It can be computed within a binary mask defining a fiber pathway in a microstructure image (e.g. the mean FA within the mask) \cite{heiervang2006between,ciccarelli2003diffusion,kunimatsu2004optimal,voineskos2010diffusion}. It can also be computed by averaging microstructural measures sampled along each point of the streamlines \cite{olszewski2017social,zhang2018whole,gong2018free,zhang2019test,jones2006age}. The choice of the best summary statistic for data measured within a fiber pathway is still open. For example, compared to the mean statistic, studies have shown that the median can be more robust against outliers and does not rely on the normality assumption for the distribution of the microstructure parameter along a streamline \cite{Boshkovski2020,zhang2018suprathreshold}. One study has suggested that the maxima and minima are more discriminative than the mean value in machine-learning-based disease classification \cite{zhang2018whole}.

The second strategy to quantify an individual fiber pathway is to measure the distribution of the microstructural measures \textit{along} the fiber pathway. This can enable the study of the tissue microstructure in local regions along the fiber pathway. This approach requires definition of a coordinate system, sampling framework, or surface-based representation to define how to sample the microstructural data of interest \cite{yeatman2012tract,odonnell2009tract,bells2011tractometry,jones2005pasta,corouge2006fiber,colby2012along,chen2016maturation,irimia2020mapping}. Most often, data are averaged across corresponding points along each streamline in the pathway, such that data can be analyzed (e.g., an along-tract or along-pathway plot) versus streamline arc length or other parameterization \cite{corouge2006fiber,colby2012along,yeatman2012tract}. Other approaches fit a medial surface representation to the pathway, such that sampled data can be represented in two dimensions and can be analyzed not only along the pathway but also across its cross-section \cite{chen2016maturation,qiu2010surface,yushkevich2009structure}.

\subsection{Quantitative measures using filtering techniques}
\label{secAdjustmentFiltering}

The so called filtering methods discussed in Section~\ref{secDensityBias} can also be used to perform ``microstructure-informed'' tractography with the aim to extract quantitative measures of white matter fiber pathways and reduce potential bias in the estimation of connectivity.
Assuming that the microstructure properties corresponding to a single streamline remain constant along its path, these methods assign a specific microstructural property to each reconstructed pathway. The underling assumption comes from the fact that a streamline reconstructed from tractography cannot represent a single axon, but rather a group of axons following the same trajectory, and thus we can suppose that on average the microstructure properties the magnetic resonance signal is sensitive to at achievable resolution remain constant. 
While in its original implementation SIFT~\cite{smith_sift:_2013} provides only the NOS adjusted based on voxel-wise spherical deconvolution, its evolution SIFT2~\cite{smith_sift2:_2015} optimizes per-streamline cross-section multipliers to match a whole-brain tractogram to fixel-wise fiber densities and thus provides a value for each streamline that can be used (for example) to define the edge weights of the connectome. Similarly, LiFE~\cite{pestilli_evaluation_2014}, COMMIT~\cite{daducci2013convex,daducci2014commit,schomburg2019formulation}, COMMIT2~\cite{schiavi2019reducing} and COMMIT2\textsubscript{tree}~\cite{ocampo2021hierarchical} deconvolve the measured dMRI signal on the streamlines and assign a single contribution, or ``weight'', to each one of them using classical multi-compartment models \cite{panagiotaki2012compartment}. These methods have been used to investigate properties of healthy and pathological brains, showing better performance than the standard NOS \citep{Yeatman2014_occipitalFasciculus,Smith:2015aa,Schiavi2020_commit}. In particular, because of their flexible implementations, COMMIT, COMMIT2 and COMMIT2\textsubscript{tree} allow complementing tractography with biophysical models of the tissue microstructure and, thus, enable one to access more quantitative and biologically informative features of individual bundles, such as average axon diameter \citep{barakovic2018}, myelin content \citep{Schiavi2019ISMRM} and bundle-specific T2 \citep{Barakovic2020}.

\subsection{Tract-specific analysis using statistical or machine learning techniques}
\label{secTractspecificAnalysis}

Once a quantitative measure has been extracted for individual fiber pathways, several options exist for tract-specific analysis, which may be hypothesis-driven using statistical analysis or data-driven using machine learning. These methods may rely on a summary statistic per tract, or analyze data along a tract. 

One widely used analysis is based on a hypothesis-driven strategy to assess if there are statistical differences of tract(s)-of-interest between groups (e.g., between health and disease, or between different subtypes of a disease). Usually, given a quantitative measure of interest, a selected summary statistic (e.g. mean of the chosen microstructural measure) is extracted from the tract(s)-of-interest and is used to compute the level of group differences (e.g., the $p$-value) using statistical group-wise comparison methods such as Student's t-test, ANOVA, or other more advanced statistical analysis methods \cite{habeck2010multivariate,marti2020survey,ombao2016handbook,zhang2018suprathreshold}. Another hypothesis-driven analysis uses a regression model (e.g., a generalized linear model \cite{dobson2018introduction} or support vector regression \cite{zhang2020support}) to assess correlation of the tract quantitative measurement and a behavioral or a disease symptom score. Such analyses have been applied to study, e.g., how white matter fiber tracts are affected in individuals with different disease severity \cite{cha2015neural,price2007abnormal,chiang2016different}, and how tracts are developing during human neurodevelopment and aging \cite{gertheiss2013longitudinal,michielse2010selective,hasan2009development}. One important point to highlight is that correction for multiple comparisons is needed if there are multiple tracts and/or multiple quantitative measurements analyzed. Commonly used multiple comparison correction methods include false discovery rate (FDR) \cite{benjamini1995controlling} and Bonferroni \cite{holm1979simple} methods. 

Another tract-specific analysis strategy is data-driven and uses machine learning techniques to perform tasks such as disease classification and prediction \cite{zhang2018whole,deng2019tractography,payabvash2019diffusion,o2012white}. In machine learning analysis, the quantitative measures computed from individual fiber pathways are treated as feature descriptors, which are input to a machine learning algorithm (e.g., a support vector machine) to train a model from a set of training samples with known information (e.g., labels such as disease or healthy control). Then the trained model can be used to predict new samples. Unlike statistical analysis methods that usually aim to find white matter structures with statistical group differences, the machine-learning-based methods aim to offer predictive relevance.

Instead of using a single scalar value to summarize the entire pathway, many methods perform along-tract analysis to investigate the distribution of the microstructural measures along the fiber pathway (variously called ``tractometry,'' ``profilometry,'' and ``tract-based morphometry'' ) \cite{Bells2011,yeatman2012tract,odonnell2009tract}. Along-tract analysis enables the study of local regions along the tract. After mapping microstructural measures along each point of the streamlines along the fiber pathway, a widely used strategy is to analyze data that are averaged across corresponding points along each streamline in the pathway \cite{wang2016tractography,colby2012along,yeatman2012tract,chandio2020bundle}, while other studies have analyzed tracts represented as surfaces \cite{qiu2010surface,chen2016maturation,yushkevich2009structure} or performed data reduction techniques to reduce the dimensionality of the point-wise measurements along tracts \cite{geeraert2020multimodal,chamberland2019dimensionality,ceschin2015regional}. Several studies have shown that along-tract analysis can be more sensitive to detect differences that are not apparent in the mean value within the tract \cite{odonnell2009tract,colby2012along,st2019reducing}. 

\subsection{Construction of a structural connectivity matrix}
\label{secConstructionMatrix}

Quantitative measures extracted for individual fiber pathways can also be used to construct a structural connectivity matrix to map the whole-brain network of structural connectivity between all pairs of gray matter regions \cite{sporns2005human, hagmann:2008}. This approach is motivated by a general shift of focus away from the roles of isolated regions towards understanding the brain as a networked system \cite{friston:2002,barch2016introduction}. Construction of the connectivity matrix has been described in several reviews \cite{yeh2020mapping,sotiropoulos_fusion_2016} and includes two major steps. First, parcellation of the whole-brain tractogram is needed to compute all possible fiber pathways in the brain, i.e., white matter parcels connecting all pairs of gray matter regions (as described in Section~\ref{ref_wmparcellation}). Then, a metric of ``connectivity'' is needed to define quantitative edge weights in the connectivity matrix (see Figure~\ref{fig_definition}(e) for a graphic illustration). This is usually done by computing a scalar value for each white matter parcel (as described in Section \ref{secQuantitativePathways}). The most popular scalar value for connectivity matrix analysis is the NOS \cite{van2018minimum,sporns2005human,ingalhalikar2014sex,roberts2016contribution,hagmann2008mapping}, where higher values are considered to indicate stronger connectivity between pairs of gray matter regions. Other scalar values that are generally interpreted as measures of connectivity ``strength'' include microstructural measures, e.g., FA \cite{bathelt2017global} and R1 \cite{boshkovski2020r1}. For scalar values that are interpreted as inversely related to connectivity ``strength,'' such as RD, MD, or maps reflecting extracellular or isotropic/free water components of the signal, a transformation of the connectivity matrix may be performed using the reciprocal or the log function prior to any subsequent connectivity analysis \cite{fornito:2016}.

While the construction of the connectivity matrix is straightforward, multiple factors that need careful consideration have been pointed out by researchers. Selection of a gray matter parcellation scheme (as described in Section~\ref{ref_wmparcellation}) is the first essential consideration when constructing a structural connectivity matrix, and it has been shown to be highly consequential in connectome analysis \cite{de2013parcellation,yeh2020mapping,zalesky2012connectivity}. Myriad components of the processing pipeline can also influence the resulting connectome matrix data, even down to the level of detail of the mechanism by which streamlines are assigned to those parcels \cite{yeh2019connectomes}. In addition, filtering false positive connections is usually needed to reduce potential bias in the estimation of connectivity (as described in Section~\ref{secAdjustmentFiltering}). Selection of tractography filtering strategies is important and has been shown to affect the analysis of the topology of brain networks \cite{frigo2020diffusion,Civier.2019,yeh2016correction} and affect sensitivity and specificity when analyzing the connectomes \cite{zalesky2016connectome}. Moreover, as concluded in several studies \cite{qi2015influence,bastiani_human_2012,yeh2016correction}, several other factors (such as MRI gradient schemes, fiber orientation models, and tractography algorithms) that are not specifically reviewed in our paper, have also been shown to affect most network measures derived from the connectivity matrix.

\subsection{Graph theoretical measures and advanced structural connectome measurements}
\label{secGraph}

A structural connectivity matrix defines a network (or \textit{graph}) of whole-brain anatomical connectivity \cite{sporns2005human,hagmann:2008}. Graph theory and network science are disciplines dedicated to the quantitative analysis of networked systems and connectivity data. Measures and methods from these domains provide a mathematical framework to investigate high-order organizational properties of the brain's structural wiring. Over the last two decades, application of these measures has become increasingly popular in the neuroscience and neuroimaging communities, giving rise to the new fields of connectomics and network neuroscience \cite{fornito:2015, fornito:2016, bassett:2017a, fornito:2013}.

A daunting number of graph measures have been proposed to study the connectome \cite{bullmore:2009,rubinov:2010}. The choice of which measures to focus on is dependent on the research questions and hypotheses at hand. Graph measures can be classified based on their resolution or scope---from local measures that quantify properties of individual ROIs, to mesoscale measures describing clusters of interconnected ROIs, to global measures that describe whole-brain connectivity properties \cite{betzel:2017a,fornito:2016}. In addition, a second useful categorization divides graph measures into models of integration or segregation \cite{bullmore:2012,heuvel:2019}. Measures of integration seek to understand how information is communicated via structural connections to facilitate functional interactions between different ROIs and neural systems. In contrast, measures of segregation aim to characterize how the brain's anatomical wiring contributes to segregated information processing and clusters of functionally specialized regions.

At the regional level, node centrality measures provide a useful characterization of the integrative importance of individual brain regions \cite{sporns:2007}. Degree or strength centrality are the simplest and most popular node centralities, quantifying, respectively, the number of connections and the sum of connection weights of individual regions. Other advanced centrality measures go beyond a region's direct connectivity, and instead consider the importance of ROIs as mediators of polysynaptic communication \cite{oldham:2019}. Centrality measures find particular utility in identifying hub regions that may play a central role in integrative brain function \cite{heuvel:2013a}. Indeed, hubs identified with these measures have been found to entail high metabolic expenditures and be disproportionately implicated in brain disorders \cite{crossley:2014,fornito:2015}. A subset of hub regions has been found to form the brain's rich club---a densely interconnected core of highly connected regions \cite{heuvel:2012}. The rich club is conjectured to be the connectome's structural backbone that facilitates efficient propagation of signals across distant parts of the brain \cite{heuvel:2011, griffa:2018}.

Structural connectivity can be partitioned into modules (also known as communities) that reflect the presence of mesoscale connectivity clusters embedded in the brain's wiring (\cite{sporns:2016,betzel:2020}). Community detection summarizes the complexity of structural networks into coarse-grained blocks, aiding the identification of sub-networks and large-scale connectivity motifs relevant to specific research questions. This modular structure constrains the flow of information through the network in a manner thought to promote functional segregation and clusters of specialized information processing \cite{betzel:2013}. Popular methods for community detection in structural brain networks include the Louvain algorithm and stochastic block models, though determining best practices around the utilization of these and other approaches remain an active topic of research \cite{fortunato:2016}.

At a global scale, network communication measures quantify how the structural connectome provides support for the integration of information between ROIs and cognitive systems \cite{misic:2015,avena:2018}. Traditionally, connectome communication has been quantified using graph measures based on the notion that neural information transfer occurs via topological shortest paths. Shortest-path-based measures include, e.g., the popular characteristic path length and network efficiency \cite{rubinov:2010}. However, algorithmically, the identification of shortest paths demands individual regions to possess global knowledge of structural connectivity, a requirement unlikely to be met in decentralized systems such as the brain \cite{goni:2014,seguin:2018,avena:2019}.

In light of this shortcoming, recent work on connectome communication has started to focus on communication measures based on decentralized models of network propagation that are more likely to approximate biological neural signalling mechanisms \cite{avena:2018}. Examples of these methods include navigation efficiency \cite{seguin:2018}, communicability \cite{estrada:2008}, search information \cite{goni:2014}, and linear transmission models \cite{misic:2015}. Each of these methods proposes a different conceptualization of how signals traverse the brain's structural connectivity, and determining which approaches most accurately reflect underlying biological processes remains a crucial open question \cite{seguin:2020}. Nonetheless, recent work has shown that advanced communication measures can elucidate how the organization of structural connectivity shapes neural information processing and functional dynamics \cite{goni:2014, misic:2018, xwang:2019, imms:2021}. For instance, despite the inherent limitations of dMRI tractography in resolving axonal fiber directionality, navigation efficiency and search information computed on the undirected human connectome can reveal patterns of asymmetric interregional signaling, as evidenced by significant associations to the directionality of effective connectivity computed using dynamic causal modeling \cite{seguin:2019}. Therefore, the use of advanced communication measures enables the study of directional and asymmetric functional interactions from structural connectivity mapped with dMRI.

\section{Applications of quantitative tractography analysis}
\label{secApplication}

In the last two decades, modern dMRI-based tractography that offers quantitative measurements of fiber tracts has extended neuroanatomy methods and become a key technique for the study of the brain's white matter in a wide range of applications. In this section, we focus on several essential areas including development, aging, neurological disorders, mental disorders, and neurosurgery. These studies can be organized into two types that correspond to the above-mentioned tract-specific and connectome-based analyses. The first category concerns major fiber pathways or localized structural connectivity in order to capture fine-grained details of brain circuits. The second category regards the entire white matter as a complex system and thus adopts clustering analysis for feature reduction or utilizes graph theoretical analysis to quantify topological characteristics with respect to information integration and segregation.

\subsection{Developmental tractography of the brain}

Brain white matter undergoes highly ordered changes during the ontogeny from a neural tube to a complex connectome. In the past decade, researchers have implemented dMRI tractography investigations across broad brain developmental stages ranging from the middle fetal stage to adulthood \cite{cao2017developmental,dubois2014early,gilmore2018imaging,huang2010delineating,lebel2018development,lebel2019review,qiu2015diffusion,zhao2019graph}.

For the prenatal stage, most tractography studies aim to identify the migration pathways of neurons and emergence of key tracts of the brain as important developmental milestones \cite{dubois2014early,gilmore2018imaging,huang2010delineating,vasung2019exploring}. For instance, using \textit{ex~vivo} dMRI scans on fetal brain samples, researchers are able to identify the major tract parts of the fornix and cingulum bundles at 13 weeks, but a part of the corpus callosum at 15 weeks \cite{huang2009anatomical}. Similar studies also show that the radial and tangential pathways organization are prominent at 17 weeks but gradually disappear at later gestational ages, and all major structural pathways are already identifiable by term \cite{takahashi2002magnetic}. These results are consistent with classical histological studies \cite{mitter2015validation,vasung2010development}. Due to recent advances in methodological progress such as motion correction algorithms, \textit{in~vivo} tractography on the fetal brain has become available and exhibit highly agreement with \textit{ex~vivo} and neuroanatomical studies \cite{khan2019fetal}, which indicates strong potential for future prenatal brain investigations.

In the postnatal stage, \textit{in~vivo} tractography offers a superior tool to chart the spatiotemporal maturation patterns of the reconstructed tracts by employing quantitative dMRI metrics from the infancy period \cite{dubois2014early,gilmore2018imaging,lebel2018development,qiu2015diffusion,vasung2019exploring} to the adolescent stage \cite{lebel2019review,tamnes2018diffusion}. Despite focusing on different developmental ranges, these studies consistently observe significant age-related increases of FA and MD in widespread fiber tracts during postnatal growth in both cross-sectional and longitudinal population designs \cite{lebel2018development}. Studies also characterize several typical nonlinear trajectories (piecewise, exponential and quadratic) of diffusion metrics for major tracts across large developmental ranges such as infancy to childhood \cite{pecheva2017tract,reynolds2019global,stephens2020white} or adolescence to adulthood \cite{chang2015white,lebel2011longitudinal}. Researchers can further estimate the growth of tracts at the subsystem level by multivariate analysis; for example, one study investigated the maturational calendars of linguistic bundles and found that the dorsal pathway development falls behind the ventral pathway at birth and catches up later at the first postnatal months \cite{dubois2016exploring}.

Using the newly emerging developmental connectomics framework, researchers are able to seek the wiring principles of brain white matter tracts \cite{cao2017developmental,cao2016toward,collin2013ontogeny,gilmore2018imaging,ouyang2019delineation,tymofiyeva2014structural,vertes2015annual,zhao2019graph}. Many studies have discovered largely adult-like topological structures including significant small-world, modular, and rich-club organization during the middle to final trimester of gestation \cite{brown2014structural,song2017human,tymofiyeva2013dti,van2015neonatal,zhao2019graph}, which reveals that the whole brain tract network is already highly efficient and specialized around term age. During prenatal development, the whole network becomes increasingly efficient and segregated with typical system-level refinements such as increased small-worldness \cite{brown2014structural} and increased modularity \cite{van2015neonatal}. After birth until adulthood, the brain network reshapes to improve integration capacity with the reconfigurations of decreased modularity and decreased small-worldness \cite{chen2013graph,dennis2013development,hagmann2010white,huang2015development,tymofiyeva2013dti}. In terms of regional changes, although neonatal brains already show similar structural hub distribution as adult brains, many refinements occur for the topological roles for each node \cite{yap2011development}. The nodal efficiency of provincial hubs and the strength of within-module connections \cite{zhao2019structural} develop fast during the prenatal stage, and hubs expand into the inferior frontal and insula regions at term age \cite{ball2013development,van2015neonatal}. After birth, the left anterior cingulate gyrus and left superior occipital gyrus become hubs in toddler brains \cite{huang2015development}. The centrality of the precuneus and cuneus still increases before pre-adolescence \cite{huang2015development,yap2011development}. 

The interpretation of the above tract-related changes largely depends on the corresponding biological changes during development. In the fetal brain, it is probably caused by pre-myelination phases such as proliferation and maturation of oligodendrocyte and progenitor cells, the glia that form the myelin sheath \cite{jakovcevski2009oligodendrocyte}. In the early postnatal stage, ongoing myelination is a clear cause \cite{paydar2014diffusional} while axonal packing and decreasing water content \cite{neil2002diffusion} may also make a contribution. In later childhood and adolescence, results seem to be attributed to changes of neurite density, particularly axonal packing \cite{lebel2018development}. 

\subsection{Aging and lifespan tractography of the brain}

Aging is associated with broad alterations of axon density and axonal structure. Researchers have performed many tractography investigations to trace the normal aging patterns and lifespan trajectories of brain connectivity across wide age ranges, spanning childhood until old age \cite{collin2013ontogeny,madden2009cerebral,moseley2002diffusion}.

Early tractography studies usually estimated the aging effect on specific fiber bundles such as the corpus callosum \cite{sullivan2006selective}, uncinate fasciculus \cite{hasan2009development}, and fornix \cite{stadlbauer2008quantitative}, which show consistent changes including decreased FA, increased MD and decreased tract volume. By subdividing across or along certain tracts (as described in Section~\ref{secTractspecificAnalysis}), researchers have been able to delineate the elaborate aging patterns within a tract \cite{davis2009assessing,michielse2010selective,sullivan2006selective}. For instance, FA values of the corpus callosum declined with aging only in the genu part while not in the body and splenium parts in participants aged 22–84 years \cite{michielse2010selective}. Using various tract identification methods, many studies have investigated a wide range of tracts \cite{davis2009assessing,stadlbauer2008age,sullivan2010quantitative,westlye2010life}, revealing a typical spatial degeneration pattern across brain systems in which  prefrontal association tracts are the most vulnerable to aging, while cingulum, temporal and parietal-occipital commissural connections are relatively preserved. From a lifespan view \cite{hasan2010quantification,hasan2009diffusion,lebel2010age,lebel2012diffusion,yeatman2014lifespan}, although most tracts show increased FA during childhood and adolescence and decreased FA at older ages, the age when FA reaches its highest peak varies across different white matter tracts. In general, the anterior and posterior parts of tracts peak earlier than the central part of tracts, around 20 to 40 years old \cite{lebel2010age}, supporting a classical ``last-in-first-out'' theory. Recently, with increasing numbers of adult imaging datasets and development of automated tract identification pipelines (as described in Section~\ref{ref_tractidentification}), researchers have begin to assess the effect of aging on the brain's tracts by employing thousands of participants \cite{cox2016ageing,de2015tract,tseng2021microstructural}, which brings us into an unprecedented ``big data'' era \cite{xia2017functional}.

From a connectomics perspective (as described in Section~\ref{secGraph}), tractography studies show that the basic layouts of the brain's structural network are largely preserved until old age, such as the economical small-world character, the modular organization and rich club organization \cite{gong2009age,li2020age,wen2011discrete,wu2012age}. Meanwhile, both global and local alterations of the brain's structural topology occur with aging, including decreased network efficiency \cite{bi2021relationship,gong2009age,li2020age,madden2020influence,shu2018disrupted,wu2012age} and decreased inter-/intra-modularity strength that may due to the widespread degeneration of whole brain tracts \cite{li2020age,wu2012age} and loss of frontal hubs that may be driven by the vulnerable prefrontal tracts \cite{zhao2015age}. These tractography-based network changes may offer potential imaging markers for individualized cognition. As strong evidence, a recent large sample study implicates the global dimensions of variation in the human structural connectome in aging-related cognitive decline \cite{madole2020aging}. Lifespan changes of the connectome also show good consistency with cognitive maturation \cite{zhao2015age}. For instance, following an inverted U-shaped trajectory, the peak ages of network efficiency and small-world measures (30 years) fall within the range of peak ages on general cognitive performance \cite{schaie2005developmental,schroeder2004age}, and brain regions within the default mode network, which is involved broadly in high-level cognition \cite{raichle2015brain,vatansever2018default,vatansever2017default}, display the latest maturation age of peak.

These various tract-related changes could be attributable to different physiological brain changes at older ages. For instance, decreased axonal density and loss of fibers may cause a decline of FA values, while an increased amount of water content along with healthy atrophy may cause increased MD \cite{moseley2002diffusion}. Results from a study that employ an advanced diffusion model (NODDI) indicate that the age-related declines in FA are primarily induced by declines in neurite density rather than changes in tract complexity \cite{cox2016ageing}. Using multi-modal dMRI and PET imaging analysis, researchers further interpreted the metabolic mechanisms underlying aging-related changes of brain tractography, in which they found that the overall connectome efficiency-metabolism coupling across brain regions significantly increased with aging \cite{bi2021relationship}.

\subsection{Tractography in neurological and mental brain disorders}

dMRI tractography has become an indispensable tool for studies on various brain disorders including, but not limited to, multiple sclerosis~(MS) \cite{fleischer2019graph,sbardella2013dti}, amnestic mild cognitive impairment~(aMCI) \cite{bai2009abnormal,zhao2017age}, Alzheimer's disease \cite{jack2013biomarker,lo2010diffusion,mito2018fibre,toga2013connectomics}, stroke \cite{mukherjee2005diffusion}, schizophrenia \cite{collin2016connectomics,goldsmith2018update,wheeler2014review}, depression \cite{de2017white,korgaonkar2014abnormal}, obsessive-compulsive disorder~(OCD) \cite{cao2021effects,chiu2011white,gan2017abnormal,gruner2012white,koch2014diffusion}, attention-deficit hyperactivity disorder~(ADHD) \cite{cao2014imaging,cao2013probabilistic,damatac2020white,hong2014connectomic} and autism \cite{ikuta2014abnormal,langen2012fronto,thomas2011anatomy,zhang2018whole}. In this section, we are condensing the explanation using two exemplar disorders, \textit{multiple sclerosis} and \textit{schizophrenia}, as the methods utilized in the study of such are representative of the work done across a breadth of disorders.

\subsubsection{Multiple sclerosis}
Multiple sclerosis is a demyelinating disease with multiple focal white matter lesions \cite{fleischer2019graph}. The accurate quantification of tract-specific damage is especially important for patients in MS because researchers have found that the location and surrounding damages of lesions contribute more than the number and volume of lesions \cite{lipp2020tractography}. Since the pathological lesions of MS hinder basic tract reconstruction, clinical studies frequently employ advanced tractography methods to obtain fiber streamlines \cite{sbardella2013dti}. After a tract has been identified out with streamlines, specific dMRI metrics can be estimated within the tract (as described in Section~\ref{secQuantitativePathways}. Tractography studies in MS have found that decreased FA or increased MD value in specific tracts, such as the corticospinal tract, the corpus callosum and cerebellar peduncles, are related to individual motor disabilities \cite{anderson2011comprehensive,kern2011corpus,lin2005importance,wilson2003pyramidal} and longitudinal changes of diffusion metrics of these tracts in patients are associated with the rehabilitation of motor function during training \cite{filippi2019brain,prosperini2014multiple}. Importantly, quantitative tractography analysis in MS further highlights that white matter tracts that exhibit normal appearance are also affected pathologically \cite{droby2015impact,fleischer2019graph} and the FA abnormalities in these tracts are significantly correlated with cognitive abilities in pediatric patients at the early phase of the disease \cite{sbardella2013dti}.

MS is considered to be a typical disconnection syndrome due to the massive lesions observed \cite{rocca2015clinical}. Researchers have observed consistent disrupted network integration including reduction of global efficiency and network strength \cite{liu2018disrupted,shu2016disrupted,shu2011diffusion,Schiavi2020_commit}, breakdown of long-range connections that account for remote communication \cite{muthuraman2016structural}, decreased frontal network communicability that represents the parallel communication capacity \cite{li2013diffusion} and impairments of rich-club connectivity that serve as core highways \cite{shu2018progressive} in MS patients compared with healthy controls. Meanwhile, studies find increased clustering and modular connectivity patterns at 6 and 12 months after the first clinical event, which indicates a disease-related abnormality of network segregation ability \cite{fleischer2017increased,muthuraman2016structural}. Addressing such topological network abnormalities in MS benefits the tracking of high-level behavior disruptions and complex disease state at individual level. The disruption of structural networks is associated with impaired cognitive performance, especially involving attention and executive function \cite{llufriu2017structural}. The decrease in global efficiency is significantly correlated with Expanded Disability Status Scale~(EDSS) scores and disease duration \cite{shu2011diffusion}. Researchers have further used network metrics to differentiate clinical subtypes of MS and achieved promising accuracy \cite{kocevar2016graph}.

\subsubsection{Schizophrenia}
Schizophrenia has been considered to be a dysconnectivity disorder with a complex nature since the late 19th century \cite{collin2016connectomics,goldsmith2018update}. Modern tractography techniques provide \textit{in~vivo} evidence for this hypothesis, indicating that schizophrenia may be seen in terms of affected pathways between cortical regions \cite{goldsmith2018update,griffa2013structural}. For instance, the corpus callosum exhibits impairments of decreased FA in anatomical subdivisions in schizophrenia. These impairments are found reproducible across studies \cite{ohoshi2019microstructural,price2007abnormal,shahab2018sex} and are predictive for multiple cognitive domains \cite{sui2018multimodal} in both health and patients across datasets. A general question about schizophrenia is whether it is a disorder of whole brain dysfunction or one of selectively affected neural systems \cite{fornito2017opportunities}. Studies frequently find consistent decreases of FA in tracts among fronto-temporal systems such as the superior/ inferior longitudinal fasciculi, uncinate fasciculi, arcuate fasciculi, and cingulum bundles \cite{peters2015white,wheeler2014review} in both chronic and first episode schizophrenia patients. These tract-specific impairments exhibit significant symptom associations in deficits of working memory \cite{karlsgodt2008diffusion}, hallucinations \cite{shergill2007diffusion} and attention \cite{abdolalizadeh2020white}. However, increasing evidence suggests that the impairments of tracts in schizophrenia are more widespread than classical fronto-temporal regions \cite{fitzsimmons2013review,kelly2018widespread,klauser2017white,zalesky2011disrupted}, which emphasizes the need for whole brain investigation. Furthermore, by combining along-tract profiles of whole brain fiber pathways with an unsupervised clustering framework, researchers were able to identify biologically defined subtypes of schizophrenia rather than symptom-based divisions, which shows great promise for future studies \cite{sun2015two}.

Using the graph-theoretical framework (as described in \ref{secGraph}), structural connectome studies \cite{collin2016connectomics,fornito2017opportunities} have shown altered global network topology such as reduced network efﬁciency and increased characteristic path length in chronic \cite{van2010aberrant,wheeler2014review,zalesky2011disrupted} or first episode patients \cite{zhang2015disrupted}, unaffected siblings \cite{collin2014structural} and high-risk infants \cite{shi2012altered}, compared with healthy controls. These results support that the impaired network integration may be foundational to schizophrenia's etiology and possibly reflect genetic vulnerability. Meanwhile, the disproportionate deﬁcits of long-distance association edges, disruptions of anatomical brain hubs and rich club connections \cite{collin2014impaired,crossley2017connectomic,klauser2017white,van2013abnormal,yeo2016graph} in both patients and their unaffected siblings also suggest damage of topological infrastructures for inter-domain functional integration in schizophrenia \cite{collin2016connectomics}. Studies also find evidence of abnormally enhanced network segregation such as increased modularity in schizophrenia patients and individuals with high-risk syndrome \cite{schmidt2017structural,van2013abnormal}. These deficiencies of network topology are generally associated with the cognitive impairments and treatment responses during disease. The reduced global efficiency of patients is correlated with Positive and Negative Syndrome Scale~(PANSS) scores \cite{wang2012anatomical} while higher global efﬁciency trend to be found in first-onset patients who subsequently respond to treatment, compared to non-responders \cite{crossley2017connectomic}. The disruption of rich-club organization is also associated with PANSS scores and shows potential for classification between schizophrenia patients and healthy controls \cite{cui2019connectome,zhao2017abnormal}. Tractography-based networks are also a valuable framework to enable cross-species comparisons for pathogenesis; for instance, by comparing brain network connectivity between humans and chimpanzees, researchers find evidence of evolutionary modiﬁcations of human white matter connectivity to signiﬁcantly overlap with the cortical pattern of schizophrenia-related dysconnectivity \cite{van2019evolutionary}.

\subsection{Tractography in neurosurgery}
\label{secTractoInNeurosurgery}

For neurosurgeons, \textit{in~vivo} tractography is invaluable for the planning and implementation of surgery, allowing for the visualization and localization of white matter tracts that are displaced or otherwise affected by the tumor \cite{jellison2004diffusion}. Although there remain technical considerations \cite{essayed2017white}, tractography has shown to be beneficial for the accurate targeting of lesions, which is important for quality of life and overall survival of neurosurgical patients \cite{costabile2019current,essayed2017white,fortin2012tractography,panesar2019tractography}.

Although quantitative image-based metrics of local tracts are of limited use in the field of neurosurgery, \textit{in~vivo} tractography offers essential spatial information about individual anatomical pathways for surgery planning. In neoplastic lesion surgery, balancing the trade-off between function preservation and maximized resection is highly significant. The classical gold standard is the intraoperative identification of eloquent areas by direct electrical stimulation (DES) \cite{duffau2005lessons,duffau2015stimulation,essayed2017white,szelenyi2010intraoperative}. Recent applications of \textit{in~vivo} tractography from intraoperative MRI show relatively high correlation between distances to reconstructed tracts and positive DES intensities \cite{essayed2017white}. Using intraoperative MRI has also shown to provide a more comprehensive surgical white matter mapping by identifying fiber tracts that were not detected in preoperative tractography \cite{javadi2017evaluation}. Studies have shown that when combined with navigated transcranial magnetic stimulation (nTMS), intraoperative tractography aids in preserving speech for patients with tumors adjacent to the superior longitudinal fasciculus \cite{negwer2017language,sollmann2018setup}. Additionally, in language area surgery, when awake surgery is impossible or unavailable, tractography-based estimation may become the only available solution for assessing involved white matter tracts \cite{d2016safe}. The applications of \textit{in~vivo} tractography have been widely used in various lesions located in both supratentorial and infratentorial areas \cite{ellis2012corticospinal,fernandes2016high,kovanlikaya2011assessment}, involving not only the major tracts but also smaller fiber structures \cite{meola2016human}.

The brain connectomic concept benefits traditional surgical strategy by a potential shift from focusing on local tumor topography to a network-guided ``oncological disconnection surgery'' \cite{duffau2021brain,hart2020connections}. In a classical neurosurgical view, the inflexible definition of ``eloquent'' brain regions represented by the sensorimotor, language and visual cortices serves as an irrefutable principle: tumors involving ``eloquent'' cortex would be not selected for resection while tumors locating in ``non-eloquent'' areas could be operated without considerations \cite{spetzler1986proposed}. However, considering the potential for functional compensation and the presence of individual variation, a common location of ``eloquent'' areas across patients seems inaccurate \cite{southwell2017resection}. By combining tractography atlases with intraoperative electrical stimulation for functional mapping, researchers have constructed probabilistic atlases of white matter pathways and structural hubs that define a minimal common brain connectome with a low inter-individual variability and a low potential of post-lesional compensation \cite{herbet2016mapping,sarubbo2020mapping}. Such prior ``structural and functional skeletons'' provide a whole brain knowledge map of essential brain sub-networks, which may be helpful in informing the resection of certain neuronal circuits during tumor surgery \cite{duffau2021brain}. For instance, connectomic studies have shown that the lexical access speed, which is related to the return to professional work after awake surgery for patients with low-grade gliomas, may be completely disrupted only when the left inferior longitudinal fasciculus~(ILF) is damaged \cite{moritz2012lexical}. An elaborate surgical management can now take into account the white matter circuits involved with functional reconfiguration that affected by tumors/lesion\cite{van2012mapping}. Researchers also have tried to develop a ``connectomic risk signature'' for lesion surgery using brain function simulation based on tractography results \cite{aerts2018modeling}. In this study, the authors employed neural dynamic models on empirical tractography data of a cohort of patients with gliomas and meningiomas to create a virtual brain functional network and observed distinct individual signatures depending whether the brain regions were directly affected by tumors. These studies indicate the potential of using individual quantitative tractography to generate neurosurgical prognostic markers at the brain system level in the future.

\section{Discussion and conclusion}
\label{secDiscussion}

In this review paper, we have provided a high-level overview of how tractography can be used to enable quantitative analysis of the brain's structural connectivity in health and disease. We reviewed methods that are involved in the main processing steps for quantitative analysis of tractography. We also reviewed studies that have used quantitative tractography to study the brain's white matter.

For research and clinical applications using tractography, we note that researchers should be extremely cautious when making biological interpretations about quantitative results. The reconstructed streamlines are only simulated entities that do not correspond directly to nerve fibers \cite{jeurissen2019diffusion,yeh2020mapping}, and basic diffusion metrics are only inferences based on local diffusion properties, which are not direct measures of tissue properties \cite{assaf2019role}. Multiple review papers illustrate pitfalls to be avoided when performing quantitative dMRI analysis and studying the brain’s connectivity using tractography \cite{o2015does,daducci2016microstructure,rheault2020common,jones2010challenges}. Research is underway to improve biological specificity to the type of tissue change, by improving the information that is obtained at the acquisition level \cite{westin2016q,hutter2018integrated,barakovic2021resolving,shemesh2016conventions,henriques2020correlation,ning2019joint}, and by proposing advanced mathematical modeling and machine learning techniques \cite{ning2021probing,wu2017image,pizzolato2020acquiring}.

The anatomical accuracy of tractography is an ongoing issue in quantitative tractography analysis, where both false positive and false negative tracking results pose challenges \cite{Maier-Hein.2017,Thomas.2014}. While technical improvements (e.g., using fiber filtering techniques as introduced in Section~\ref{secFalseFiltering}) can improve anatomical accuracy, researchers should be cautious when interpreting results in research and clinical applications by considering prior anatomical knowledge learned from studies in histology \cite{huttenlocher1984synapse,miller2012prolonged,sidman1973neuronal,yakovlev1967myelogenetic} and animals \cite{song2003diffusion,song2002dysmyelination}.

Tractography has enabled the study of the brain's white matter connections across the lifespan in health and disease. Overall, we conclude that, while there have been considerable advancements in methodological technologies and breadth of applications, there nevertheless remains no consensus about the ``best'' methodology in quantitative analysis of tractography, and researchers should remain cautious when interpreting results in research and clinical applications. 

\section*{Acknowledgements}

We acknowledge the following funding.
FZ and LJO acknowledge funding provided by the following National Institutes of Health~(NIH) grants: P41EB015902, P41EB015898, R01MH074794, R01MH125860, P41EB028741 and R01MH119222.
YH and TDZ are supported by the Natural Science Foundation of China (Grant Nos. 81620108016, 81801783). 
RS is supported by fellowship funding from the National Imaging Facility~(NIF), an Australian Government National Collaborative Research Infrastructure Strategy~(NCRIS) capability. 
CHY is grateful to the Ministry of Science and Technology of Taiwan (MOST 109-2222-E-182-001-MY3) for the support.

\bibliography{NeuroImage-review}

\begin{thebibliography}{552}
\expandafter\ifx\csname natexlab\endcsname\relax\def\natexlab#1{#1}\fi
\providecommand{\url}[1]{\texttt{#1}}
\providecommand{\href}[2]{#2}
\providecommand{\path}[1]{#1}
\providecommand{\DOIprefix}{doi:}
\providecommand{\ArXivprefix}{arXiv:}
\providecommand{\URLprefix}{URL: }
\providecommand{\Pubmedprefix}{pmid:}
\providecommand{\doi}[1]{\href{http://dx.doi.org/#1}{\path{#1}}}
\providecommand{\Pubmed}[1]{\href{pmid:#1}{\path{#1}}}
\providecommand{\bibinfo}[2]{#2}
\ifx\xfnm\relax \def\xfnm[#1]{\unskip,\space#1}\fi
\bibitem[{Abdolalizadeh et~al.(2020)Abdolalizadeh, Ostadrahimi, Mohajer,
  Darvishi, Sattarian, Ershadi and Abbasi}]{abdolalizadeh2020white}
\bibinfo{author}{Abdolalizadeh, A.}, \bibinfo{author}{Ostadrahimi, H.},
  \bibinfo{author}{Mohajer, B.}, \bibinfo{author}{Darvishi, A.},
  \bibinfo{author}{Sattarian, M.}, \bibinfo{author}{Ershadi, A.S.B.},
  \bibinfo{author}{Abbasi, N.}, \bibinfo{year}{2020}.
\newblock \bibinfo{title}{White matter microstructural properties associated
  with impaired attention in chronic schizophrenia: a multi-center study}.
\newblock \bibinfo{journal}{Psychiatry Research: Neuroimaging}
  \bibinfo{volume}{302}, \bibinfo{pages}{111105}.
\bibitem[{Aerts et~al.(2018)Aerts, Schirner, Jeurissen, Van~Roost, Achten,
  Ritter and Marinazzo}]{aerts2018modeling}
\bibinfo{author}{Aerts, H.}, \bibinfo{author}{Schirner, M.},
  \bibinfo{author}{Jeurissen, B.}, \bibinfo{author}{Van~Roost, D.},
  \bibinfo{author}{Achten, E.}, \bibinfo{author}{Ritter, P.},
  \bibinfo{author}{Marinazzo, D.}, \bibinfo{year}{2018}.
\newblock \bibinfo{title}{Modeling brain dynamics in brain tumor patients using
  the virtual brain}.
\newblock \bibinfo{journal}{Eneuro} \bibinfo{volume}{5}.
\bibitem[{Afzali et~al.(2020)Afzali, Pieciak, Newman, Garifallidis,
  {\"O}zarslan, Cheng and Jones}]{afzali2020sensitivity}
\bibinfo{author}{Afzali, M.}, \bibinfo{author}{Pieciak, T.},
  \bibinfo{author}{Newman, S.}, \bibinfo{author}{Garifallidis, E.},
  \bibinfo{author}{{\"O}zarslan, E.}, \bibinfo{author}{Cheng, H.},
  \bibinfo{author}{Jones, D.K.}, \bibinfo{year}{2020}.
\newblock \bibinfo{title}{The sensitivity of diffusion {MRI} to microstructural
  properties and experimental factors}.
\newblock \bibinfo{journal}{Journal of Neuroscience Methods} ,
  \bibinfo{pages}{108951}.
\bibitem[{Albi et~al.(2018)Albi, Meola, Zhang, Kahali, Rigolo, Tax, Ciris,
  Essayed, Unadkat, Norton et~al.}]{albi2018image}
\bibinfo{author}{Albi, A.}, \bibinfo{author}{Meola, A.},
  \bibinfo{author}{Zhang, F.}, \bibinfo{author}{Kahali, P.},
  \bibinfo{author}{Rigolo, L.}, \bibinfo{author}{Tax, C.M.},
  \bibinfo{author}{Ciris, P.A.}, \bibinfo{author}{Essayed, W.I.},
  \bibinfo{author}{Unadkat, P.}, \bibinfo{author}{Norton, I.}, et~al.,
  \bibinfo{year}{2018}.
\newblock \bibinfo{title}{Image registration to compensate for epi distortion
  in patients with brain tumors: An evaluation of tract-specific effects}.
\newblock \bibinfo{journal}{Journal of Neuroimaging} \bibinfo{volume}{28},
  \bibinfo{pages}{173--182}.
\bibitem[{Alexander et~al.(2007)Alexander, Lee, Lazar, Boudos, DuBray, Oakes,
  Miller, Lu, Jeong, McMahon et~al.}]{alexander2007diffusion}
\bibinfo{author}{Alexander, A.L.}, \bibinfo{author}{Lee, J.E.},
  \bibinfo{author}{Lazar, M.}, \bibinfo{author}{Boudos, R.},
  \bibinfo{author}{DuBray, M.B.}, \bibinfo{author}{Oakes, T.R.},
  \bibinfo{author}{Miller, J.N.}, \bibinfo{author}{Lu, J.},
  \bibinfo{author}{Jeong, E.K.}, \bibinfo{author}{McMahon, W.M.}, et~al.,
  \bibinfo{year}{2007}.
\newblock \bibinfo{title}{Diffusion tensor imaging of the corpus callosum in
  autism}.
\newblock \bibinfo{journal}{Neuroimage} \bibinfo{volume}{34},
  \bibinfo{pages}{61--73}.
\bibitem[{Alexander(2006)}]{alexander2006introduction}
\bibinfo{author}{Alexander, D.C.}, \bibinfo{year}{2006}.
\newblock \bibinfo{title}{An introduction to computational diffusion {MRI}: the
  diffusion tensor and beyond}, in: \bibinfo{booktitle}{Visualization and
  processing of tensor fields}. \bibinfo{publisher}{Springer}, pp.
  \bibinfo{pages}{83--106}.
\bibitem[{Alexander et~al.(2019)Alexander, Dyrby, Nilsson and
  Zhang}]{alexander2019imaging}
\bibinfo{author}{Alexander, D.C.}, \bibinfo{author}{Dyrby, T.B.},
  \bibinfo{author}{Nilsson, M.}, \bibinfo{author}{Zhang, H.},
  \bibinfo{year}{2019}.
\newblock \bibinfo{title}{Imaging brain microstructure with diffusion mri:
  practicality and applications}.
\newblock \bibinfo{journal}{NMR in Biomedicine} \bibinfo{volume}{32},
  \bibinfo{pages}{e3841}.
\bibitem[{Alexander et~al.(2017)Alexander, Escalera, Ai, Andreotti, Febre,
  Mangone, Vega-Potler, Langer, Alexander, Kovacs et~al.}]{alexander2017open}
\bibinfo{author}{Alexander, L.M.}, \bibinfo{author}{Escalera, J.},
  \bibinfo{author}{Ai, L.}, \bibinfo{author}{Andreotti, C.},
  \bibinfo{author}{Febre, K.}, \bibinfo{author}{Mangone, A.},
  \bibinfo{author}{Vega-Potler, N.}, \bibinfo{author}{Langer, N.},
  \bibinfo{author}{Alexander, A.}, \bibinfo{author}{Kovacs, M.}, et~al.,
  \bibinfo{year}{2017}.
\newblock \bibinfo{title}{An open resource for transdiagnostic research in
  pediatric mental health and learning disorders}.
\newblock \bibinfo{journal}{Scientific data} \bibinfo{volume}{4},
  \bibinfo{pages}{1--26}.
\bibitem[{Allan et~al.(2020)Allan, Briggs, Conner, O'Neal, Bonney, Maxwell,
  Baker, Burks, Sali, Glenn et~al.}]{allan2020parcellation}
\bibinfo{author}{Allan, P.G.}, \bibinfo{author}{Briggs, R.G.},
  \bibinfo{author}{Conner, A.K.}, \bibinfo{author}{O'Neal, C.M.},
  \bibinfo{author}{Bonney, P.A.}, \bibinfo{author}{Maxwell, B.D.},
  \bibinfo{author}{Baker, C.M.}, \bibinfo{author}{Burks, J.D.},
  \bibinfo{author}{Sali, G.}, \bibinfo{author}{Glenn, C.A.}, et~al.,
  \bibinfo{year}{2020}.
\newblock \bibinfo{title}{Parcellation-based tractographic modeling of the
  ventral attention network}.
\newblock \bibinfo{journal}{Journal of the neurological sciences}
  \bibinfo{volume}{408}, \bibinfo{pages}{116548}.
\bibitem[{Anderson et~al.(2011)Anderson, Wheeler-Kingshott, Abdel-Aziz, Miller,
  Toosy, Thompson and Ciccarelli}]{anderson2011comprehensive}
\bibinfo{author}{Anderson, V.}, \bibinfo{author}{Wheeler-Kingshott, C.},
  \bibinfo{author}{Abdel-Aziz, K.}, \bibinfo{author}{Miller, D.},
  \bibinfo{author}{Toosy, A.}, \bibinfo{author}{Thompson, A.},
  \bibinfo{author}{Ciccarelli, O.}, \bibinfo{year}{2011}.
\newblock \bibinfo{title}{A comprehensive assessment of cerebellar damage in
  multiple sclerosis using diffusion tractography and volumetric analysis}.
\newblock \bibinfo{journal}{Multiple Sclerosis Journal} \bibinfo{volume}{17},
  \bibinfo{pages}{1079--1087}.
\bibitem[{Assaf et~al.(2019)Assaf, Johansen-Berg and Thiebaut~de
  Schotten}]{assaf2019role}
\bibinfo{author}{Assaf, Y.}, \bibinfo{author}{Johansen-Berg, H.},
  \bibinfo{author}{Thiebaut~de Schotten, M.}, \bibinfo{year}{2019}.
\newblock \bibinfo{title}{The role of diffusion {MRI} in neuroscience}.
\newblock \bibinfo{journal}{NMR in Biomedicine} \bibinfo{volume}{32},
  \bibinfo{pages}{e3762}.
\bibitem[{Assaf and Pasternak(2008)}]{assaf2008diffusion}
\bibinfo{author}{Assaf, Y.}, \bibinfo{author}{Pasternak, O.},
  \bibinfo{year}{2008}.
\newblock \bibinfo{title}{Diffusion tensor imaging (dti)-based white matter
  mapping in brain research: a review}.
\newblock \bibinfo{journal}{Journal of molecular neuroscience}
  \bibinfo{volume}{34}, \bibinfo{pages}{51--61}.
\bibitem[{Astolfi et~al.(2020)Astolfi, De~Benedictis, Sarubbo, Bert{\'o},
  Olivetti, Sona and Avesani}]{astolfi2020stem}
\bibinfo{author}{Astolfi, P.}, \bibinfo{author}{De~Benedictis, A.},
  \bibinfo{author}{Sarubbo, S.}, \bibinfo{author}{Bert{\'o}, G.},
  \bibinfo{author}{Olivetti, E.}, \bibinfo{author}{Sona, D.},
  \bibinfo{author}{Avesani, P.}, \bibinfo{year}{2020}.
\newblock \bibinfo{title}{A stem-based dissection of inferior fronto-occipital
  fasciculus with a deep learning model}, in: \bibinfo{booktitle}{2020 IEEE
  17th International Symposium on Biomedical Imaging (ISBI)},
  \bibinfo{organization}{IEEE}. pp. \bibinfo{pages}{267--270}.
\bibitem[{Avena-Koenigsberger et~al.(2018)Avena-Koenigsberger, Misic and
  Sporns}]{avena:2018}
\bibinfo{author}{Avena-Koenigsberger, A.}, \bibinfo{author}{Misic, B.},
  \bibinfo{author}{Sporns, O.}, \bibinfo{year}{2018}.
\newblock \bibinfo{title}{Communication dynamics in complex brain networks}.
\newblock \bibinfo{journal}{Nat Rev Neurosci} \bibinfo{volume}{19},
  \bibinfo{pages}{17--33}.
\bibitem[{Avena-Koenigsberger et~al.(2019)Avena-Koenigsberger, Yan, Kolchinsky,
  van~den Heuvel, Hagmann and Sporns}]{avena:2019}
\bibinfo{author}{Avena-Koenigsberger, A.}, \bibinfo{author}{Yan, X.},
  \bibinfo{author}{Kolchinsky, A.}, \bibinfo{author}{van~den Heuvel, M.},
  \bibinfo{author}{Hagmann, P.}, \bibinfo{author}{Sporns, O.},
  \bibinfo{year}{2019}.
\newblock \bibinfo{title}{A spectrum of routing strategies for brain networks}.
\newblock \bibinfo{journal}{PLoS Comput Biol} \bibinfo{volume}{15},
  \bibinfo{pages}{e1006833}.
\bibitem[{Avila et~al.(2019)Avila, Lebenberg, Rivi{\`e}re, Auzias, Fischer,
  Poupon, Guevara, Poupon and Mangin}]{avila2019inference}
\bibinfo{author}{Avila, N.L.}, \bibinfo{author}{Lebenberg, J.},
  \bibinfo{author}{Rivi{\`e}re, D.}, \bibinfo{author}{Auzias, G.},
  \bibinfo{author}{Fischer, C.}, \bibinfo{author}{Poupon, F.},
  \bibinfo{author}{Guevara, P.}, \bibinfo{author}{Poupon, C.},
  \bibinfo{author}{Mangin, J.F.}, \bibinfo{year}{2019}.
\newblock \bibinfo{title}{Inference of an extended short fiber bundle atlas
  using sulcus-based constraints for a diffeomorphic inter-subject alignment},
  in: \bibinfo{booktitle}{International Conference on Medical Image Computing
  and Computer-Assisted Intervention}, \bibinfo{organization}{Springer}. pp.
  \bibinfo{pages}{323--333}.
\bibitem[{Aydogan et~al.(2018)Aydogan, Jacobs, Dulawa, Thompson, Francois,
  Toga, Dong, Knowles and Shi}]{aydogan2018tractography}
\bibinfo{author}{Aydogan, D.B.}, \bibinfo{author}{Jacobs, R.},
  \bibinfo{author}{Dulawa, S.}, \bibinfo{author}{Thompson, S.L.},
  \bibinfo{author}{Francois, M.C.}, \bibinfo{author}{Toga, A.W.},
  \bibinfo{author}{Dong, H.}, \bibinfo{author}{Knowles, J.A.},
  \bibinfo{author}{Shi, Y.}, \bibinfo{year}{2018}.
\newblock \bibinfo{title}{When tractography meets tracer injections: a
  systematic study of trends and variation sources of diffusion-based
  connectivity}.
\newblock \bibinfo{journal}{Brain Structure and Function}
  \bibinfo{volume}{223}, \bibinfo{pages}{2841--2858}.
\bibitem[{Aydogan and Shi(2020)}]{aydogan_parallel_2020}
\bibinfo{author}{Aydogan, D.B.}, \bibinfo{author}{Shi, Y.},
  \bibinfo{year}{2020}.
\newblock \bibinfo{title}{Parallel transport tractography}.
\newblock \bibinfo{journal}{IEEE Transactions on Medical Imaging}
  \bibinfo{volume}{40}, \bibinfo{pages}{635--647}.
\bibitem[{Bai et~al.(2009)Bai, Zhang, Watson, Yu, Shi, Yuan, Qian and
  Jia}]{bai2009abnormal}
\bibinfo{author}{Bai, F.}, \bibinfo{author}{Zhang, Z.},
  \bibinfo{author}{Watson, D.R.}, \bibinfo{author}{Yu, H.},
  \bibinfo{author}{Shi, Y.}, \bibinfo{author}{Yuan, Y.}, \bibinfo{author}{Qian,
  Y.}, \bibinfo{author}{Jia, J.}, \bibinfo{year}{2009}.
\newblock \bibinfo{title}{Abnormal integrity of association fiber tracts in
  amnestic mild cognitive impairment}.
\newblock \bibinfo{journal}{Journal of the neurological sciences}
  \bibinfo{volume}{278}, \bibinfo{pages}{102--106}.
\bibitem[{Bajada et~al.(2019)Bajada, Schreiber and Caspers}]{bajada2019fiber}
\bibinfo{author}{Bajada, C.J.}, \bibinfo{author}{Schreiber, J.},
  \bibinfo{author}{Caspers, S.}, \bibinfo{year}{2019}.
\newblock \bibinfo{title}{Fiber length profiling: A novel approach to
  structural brain organization}.
\newblock \bibinfo{journal}{Neuroimage} \bibinfo{volume}{186},
  \bibinfo{pages}{164--173}.
\bibitem[{Bakker et~al.(2012)Bakker, Wachtler and Diesmann}]{bakker2012cocomac}
\bibinfo{author}{Bakker, R.}, \bibinfo{author}{Wachtler, T.},
  \bibinfo{author}{Diesmann, M.}, \bibinfo{year}{2012}.
\newblock \bibinfo{title}{Cocomac 2.0 and the future of tract-tracing
  databases}.
\newblock \bibinfo{journal}{Frontiers in neuroinformatics} \bibinfo{volume}{6},
  \bibinfo{pages}{30}.
\bibitem[{Ball et~al.(2013)Ball, Srinivasan, Aljabar, Counsell, Durighel,
  Hajnal, Rutherford and Edwards}]{ball2013development}
\bibinfo{author}{Ball, G.}, \bibinfo{author}{Srinivasan, L.},
  \bibinfo{author}{Aljabar, P.}, \bibinfo{author}{Counsell, S.J.},
  \bibinfo{author}{Durighel, G.}, \bibinfo{author}{Hajnal, J.V.},
  \bibinfo{author}{Rutherford, M.A.}, \bibinfo{author}{Edwards, A.D.},
  \bibinfo{year}{2013}.
\newblock \bibinfo{title}{Development of cortical microstructure in the preterm
  human brain}.
\newblock \bibinfo{journal}{Proceedings of the National Academy of Sciences}
  \bibinfo{volume}{110}, \bibinfo{pages}{9541--9546}.
\bibitem[{Barakovic et~al.(2018)Barakovic, Girard, Romascano, Patino~Lopez,
  Descoteaux, Innocenti, Jones, Thiran and Daducci}]{barakovic2018}
\bibinfo{author}{Barakovic, M.}, \bibinfo{author}{Girard, G.},
  \bibinfo{author}{Romascano, D.P.R.}, \bibinfo{author}{Patino~Lopez, J.R.},
  \bibinfo{author}{Descoteaux, M.}, \bibinfo{author}{Innocenti, G.},
  \bibinfo{author}{Jones, D.K.}, \bibinfo{author}{Thiran, J.P.},
  \bibinfo{author}{Daducci, A.}, \bibinfo{year}{2018}.
\newblock \bibinfo{title}{{Assessing feasibility and reproducibility of a
  bundle-specific framework on in vivo axon diameter estimates at 300mT/m}},
  in: \bibinfo{booktitle}{26th annual meeting of the International Society for
  Magnetic Resonance in Medicine (ISMRM)}.
\bibitem[{Barakovic et~al.(2021a)Barakovic, Tax, Rudrapatna, Chamberland,
  Rafael-Patino, Granziera, Thiran, Daducci, Canales-Rodr{\'\i}guez and
  Jones}]{barakovic2021resolving}
\bibinfo{author}{Barakovic, M.}, \bibinfo{author}{Tax, C.M.},
  \bibinfo{author}{Rudrapatna, U.}, \bibinfo{author}{Chamberland, M.},
  \bibinfo{author}{Rafael-Patino, J.}, \bibinfo{author}{Granziera, C.},
  \bibinfo{author}{Thiran, J.P.}, \bibinfo{author}{Daducci, A.},
  \bibinfo{author}{Canales-Rodr{\'\i}guez, E.J.}, \bibinfo{author}{Jones,
  D.K.}, \bibinfo{year}{2021}a.
\newblock \bibinfo{title}{Resolving bundle-specific intra-axonal t2 values
  within a voxel using diffusion-relaxation tract-based estimation}.
\newblock \bibinfo{journal}{NeuroImage} \bibinfo{volume}{227},
  \bibinfo{pages}{117617}.
\bibitem[{Barakovic et~al.(2021b)Barakovic, Tax, Rudrapatna, Chamberland,
  Rafael-Patino, Granziera, Thiran, Daducci, Canales-Rodr{\'{i}}guez and
  Jones}]{Barakovic2020}
\bibinfo{author}{Barakovic, M.}, \bibinfo{author}{Tax, C.M.W.},
  \bibinfo{author}{Rudrapatna, U.}, \bibinfo{author}{Chamberland, M.},
  \bibinfo{author}{Rafael-Patino, J.}, \bibinfo{author}{Granziera, C.},
  \bibinfo{author}{Thiran, J.P.}, \bibinfo{author}{Daducci, A.},
  \bibinfo{author}{Canales-Rodr{\'{i}}guez, E.J.}, \bibinfo{author}{Jones,
  D.K.}, \bibinfo{year}{2021}b.
\newblock \bibinfo{title}{{Resolving bundle-specific intra-axonal T2 values
  within a voxel using diffusion-relaxation tract-based estimation}}.
\newblock \bibinfo{journal}{NeuroImage} \bibinfo{volume}{227},
  \bibinfo{pages}{117617}.
\bibitem[{Barch et~al.(2016)Barch, Verfaellie and Rao}]{barch2016introduction}
\bibinfo{author}{Barch, D.M.}, \bibinfo{author}{Verfaellie, M.},
  \bibinfo{author}{Rao, S.M.}, \bibinfo{year}{2016}.
\newblock \bibinfo{title}{Introduction to jins special issue on human brain
  connectivity in the modern era: Relevance to understanding health and
  disease}.
\newblock \bibinfo{journal}{Journal of the International Neuropsychological
  Society: JINS} \bibinfo{volume}{22}, \bibinfo{pages}{101}.
\bibitem[{Basser(1998)}]{basser1998fiber}
\bibinfo{author}{Basser, P.J.}, \bibinfo{year}{1998}.
\newblock \bibinfo{title}{{Fiber-tractography via diffusion tensor MRI
  (DT-MRI)}}, in: \bibinfo{booktitle}{Proceedings of the 6th Annual Meeting
  ISMRM, Sydney, Australia}, p. \bibinfo{pages}{1226}.
\bibitem[{Basser et~al.(2000)Basser, Pajevic, Pierpaoli, Duda and
  Aldroubi}]{basser2000vivo}
\bibinfo{author}{Basser, P.J.}, \bibinfo{author}{Pajevic, S.},
  \bibinfo{author}{Pierpaoli, C.}, \bibinfo{author}{Duda, J.},
  \bibinfo{author}{Aldroubi, A.}, \bibinfo{year}{2000}.
\newblock \bibinfo{title}{In vivo fiber tractography using dt-mri data}.
\newblock \bibinfo{journal}{Magnetic resonance in medicine}
  \bibinfo{volume}{44}, \bibinfo{pages}{625--632}.
\bibitem[{Basser and Pierpaoli(1996)}]{Basser1996_dti}
\bibinfo{author}{Basser, P.J.}, \bibinfo{author}{Pierpaoli, C.},
  \bibinfo{year}{1996}.
\newblock \bibinfo{title}{{Microstructural and Physiological Features of
  Tissues Elucidated by Quantitative-Diffusion-Tensor MRI}}.
\newblock \bibinfo{journal}{Journal of Magnetic Resonance, Series B}
  \bibinfo{volume}{111}, \bibinfo{pages}{209--219}.
\newblock \DOIprefix\doi{https://doi.org/10.1006/jmrb.1996.0086}.
\bibitem[{Basser and Pierpaoli(2011)}]{basser2011microstructural}
\bibinfo{author}{Basser, P.J.}, \bibinfo{author}{Pierpaoli, C.},
  \bibinfo{year}{2011}.
\newblock \bibinfo{title}{Microstructural and physiological features of tissues
  elucidated by quantitative-diffusion-tensor {MRI}}.
\newblock \bibinfo{journal}{Journal of magnetic resonance}
  \bibinfo{volume}{213}, \bibinfo{pages}{560--570}.
\bibitem[{Bassett and Bullmore(2017)}]{bassett2017small}
\bibinfo{author}{Bassett, D.S.}, \bibinfo{author}{Bullmore, E.T.},
  \bibinfo{year}{2017}.
\newblock \bibinfo{title}{Small-world brain networks revisited}.
\newblock \bibinfo{journal}{The Neuroscientist} \bibinfo{volume}{23},
  \bibinfo{pages}{499--516}.
\bibitem[{Bassett and Sporns(2017)}]{bassett:2017a}
\bibinfo{author}{Bassett, D.S.}, \bibinfo{author}{Sporns, O.},
  \bibinfo{year}{2017}.
\newblock \bibinfo{title}{Network neuroscience}.
\newblock \bibinfo{journal}{Nat Neurosci} \bibinfo{volume}{20},
  \bibinfo{pages}{353--364}.
\bibitem[{Bastiani et~al.(2017)Bastiani, Cottaar, Dikranian, Ghosh, Zhang,
  Alexander, Behrens, Jbabdi and Sotiropoulos}]{bastiani_afods_2017}
\bibinfo{author}{Bastiani, M.}, \bibinfo{author}{Cottaar, M.},
  \bibinfo{author}{Dikranian, K.}, \bibinfo{author}{Ghosh, A.},
  \bibinfo{author}{Zhang, H.}, \bibinfo{author}{Alexander, D.C.},
  \bibinfo{author}{Behrens, T.E.}, \bibinfo{author}{Jbabdi, S.},
  \bibinfo{author}{Sotiropoulos, S.N.}, \bibinfo{year}{2017}.
\newblock \bibinfo{title}{Improved tractography using asymmetric fibre
  orientation distributions}.
\newblock \bibinfo{journal}{NeuroImage} \bibinfo{volume}{158},
  \bibinfo{pages}{205--218}.
\bibitem[{Bastiani et~al.(2012)Bastiani, Shah, Goebel and
  Roebroeck}]{bastiani_human_2012}
\bibinfo{author}{Bastiani, M.}, \bibinfo{author}{Shah, N.J.},
  \bibinfo{author}{Goebel, R.}, \bibinfo{author}{Roebroeck, A.},
  \bibinfo{year}{2012}.
\newblock \bibinfo{title}{Human cortical connectome reconstruction from
  diffusion weighted {MRI}: {The} effect of tractography algorithm}.
\newblock \bibinfo{journal}{NeuroImage} \bibinfo{volume}{62},
  \bibinfo{pages}{1732--1749}.
\bibitem[{Bathelt et~al.(2017)Bathelt, Barnes, Raymond, Baker and
  Astle}]{bathelt2017global}
\bibinfo{author}{Bathelt, J.}, \bibinfo{author}{Barnes, J.},
  \bibinfo{author}{Raymond, F.L.}, \bibinfo{author}{Baker, K.},
  \bibinfo{author}{Astle, D.}, \bibinfo{year}{2017}.
\newblock \bibinfo{title}{Global and local connectivity differences converge
  with gene expression in a neurodevelopmental disorder of known genetic
  origin}.
\newblock \bibinfo{journal}{Cerebral Cortex} \bibinfo{volume}{27},
  \bibinfo{pages}{3806--3817}.
\bibitem[{Battocchio et~al.(2020)Battocchio, Schiavi, Descoteaux and
  Daducci}]{aattocchio2020improving}
\bibinfo{author}{Battocchio, M.}, \bibinfo{author}{Schiavi, S.},
  \bibinfo{author}{Descoteaux, M.}, \bibinfo{author}{Daducci, A.},
  \bibinfo{year}{2020}.
\newblock \bibinfo{title}{Improving tractography accuracy using dynamic
  filtering}, in: \bibinfo{booktitle}{CDMRI}, p. \bibinfo{pages}{accepted to
  appear}.
\bibitem[{Bazin et~al.(2011)Bazin, Ye, Bogovic, Shiee, Reich, Prince and
  Pham}]{bazin2011direct}
\bibinfo{author}{Bazin, P.L.}, \bibinfo{author}{Ye, C.},
  \bibinfo{author}{Bogovic, J.A.}, \bibinfo{author}{Shiee, N.},
  \bibinfo{author}{Reich, D.S.}, \bibinfo{author}{Prince, J.L.},
  \bibinfo{author}{Pham, D.L.}, \bibinfo{year}{2011}.
\newblock \bibinfo{title}{Direct segmentation of the major white matter tracts
  in diffusion tensor images}.
\newblock \bibinfo{journal}{NeuroImage} \bibinfo{volume}{58},
  \bibinfo{pages}{458--468}.
\bibitem[{Behrens et~al.(2007)Behrens, Berg, Jbabdi, Rushworth and
  Woolrich}]{behrens2007probabilistic}
\bibinfo{author}{Behrens, T.E.}, \bibinfo{author}{Berg, H.J.},
  \bibinfo{author}{Jbabdi, S.}, \bibinfo{author}{Rushworth, M.F.},
  \bibinfo{author}{Woolrich, M.W.}, \bibinfo{year}{2007}.
\newblock \bibinfo{title}{Probabilistic diffusion tractography with multiple
  fibre orientations: What can we gain?}
\newblock \bibinfo{journal}{Neuroimage} \bibinfo{volume}{34},
  \bibinfo{pages}{144--155}.
\bibitem[{Behrman-Lay et~al.(2015)Behrman-Lay, Usher, Conturo, Correia,
  Laidlaw, Lane, Bolzenius, Heaps, Salminen, Baker et~al.}]{behrman2015fiber}
\bibinfo{author}{Behrman-Lay, A.M.}, \bibinfo{author}{Usher, C.},
  \bibinfo{author}{Conturo, T.E.}, \bibinfo{author}{Correia, S.},
  \bibinfo{author}{Laidlaw, D.H.}, \bibinfo{author}{Lane, E.M.},
  \bibinfo{author}{Bolzenius, J.}, \bibinfo{author}{Heaps, J.M.},
  \bibinfo{author}{Salminen, L.E.}, \bibinfo{author}{Baker, L.M.}, et~al.,
  \bibinfo{year}{2015}.
\newblock \bibinfo{title}{Fiber bundle length and cognition: a length-based
  tractography {MRI} study}.
\newblock \bibinfo{journal}{Brain imaging and behavior} \bibinfo{volume}{9},
  \bibinfo{pages}{765--775}.
\bibitem[{Bells et~al.(2011a)Bells, Cercignani, Deoni, Assaf, Pasternak, Evans,
  Leemans and Jones}]{bells2011tractometry}
\bibinfo{author}{Bells, S.}, \bibinfo{author}{Cercignani, M.},
  \bibinfo{author}{Deoni, S.}, \bibinfo{author}{Assaf, Y.},
  \bibinfo{author}{Pasternak, O.}, \bibinfo{author}{Evans, C.},
  \bibinfo{author}{Leemans, A.}, \bibinfo{author}{Jones, D.},
  \bibinfo{year}{2011}a.
\newblock \bibinfo{title}{Tractometry--comprehensive multi-modal quantitative
  assessment of white matter along specific tracts}, in:
  \bibinfo{booktitle}{Proc. ISMRM}, p.~\bibinfo{pages}{1}.
\bibitem[{Bells et~al.(2011b)Bells, Cercignani, Deoni, Assaf, Pasternak, Evans,
  Leemans and Jones}]{Bells2011}
\bibinfo{author}{Bells, S.}, \bibinfo{author}{Cercignani, M.},
  \bibinfo{author}{Deoni, S.}, \bibinfo{author}{Assaf, Y.},
  \bibinfo{author}{Pasternak, O.}, \bibinfo{author}{Evans, C.J.},
  \bibinfo{author}{Leemans, A.}, \bibinfo{author}{Jones, D.K.},
  \bibinfo{year}{2011}b.
\newblock \bibinfo{title}{{Tractometry--comprehensive multi-modal quantitative
  assessment of white matter along specific tracts}}, in:
  \bibinfo{booktitle}{Proc. ISMRM}, p.~\bibinfo{pages}{1}.
\bibitem[{Benjamini and Hochberg(1995)}]{benjamini1995controlling}
\bibinfo{author}{Benjamini, Y.}, \bibinfo{author}{Hochberg, Y.},
  \bibinfo{year}{1995}.
\newblock \bibinfo{title}{Controlling the false discovery rate: a practical and
  powerful approach to multiple testing}.
\newblock \bibinfo{journal}{Journal of the Royal statistical society: series B
  (Methodological)} \bibinfo{volume}{57}, \bibinfo{pages}{289--300}.
\bibitem[{Bert{\`o} et~al.(2020)Bert{\`o}, Bullock, Astolfi, Hayashi, Zigiotto,
  Annicchiarico, Corsini, De~Benedictis, Sarubbo, Pestilli
  et~al.}]{berto2020classifyber}
\bibinfo{author}{Bert{\`o}, G.}, \bibinfo{author}{Bullock, D.},
  \bibinfo{author}{Astolfi, P.}, \bibinfo{author}{Hayashi, S.},
  \bibinfo{author}{Zigiotto, L.}, \bibinfo{author}{Annicchiarico, L.},
  \bibinfo{author}{Corsini, F.}, \bibinfo{author}{De~Benedictis, A.},
  \bibinfo{author}{Sarubbo, S.}, \bibinfo{author}{Pestilli, F.}, et~al.,
  \bibinfo{year}{2020}.
\newblock \bibinfo{title}{Classifyber, a robust streamline-based linear
  classifier for white matter bundle segmentation}.
\newblock \bibinfo{journal}{BioRxiv} .
\bibitem[{Besson et~al.(2014)Besson, Lopes, Leclerc, Derambure and
  Tyvaert}]{besson2014intra}
\bibinfo{author}{Besson, P.}, \bibinfo{author}{Lopes, R.},
  \bibinfo{author}{Leclerc, X.}, \bibinfo{author}{Derambure, P.},
  \bibinfo{author}{Tyvaert, L.}, \bibinfo{year}{2014}.
\newblock \bibinfo{title}{Intra-subject reliability of the high-resolution
  whole-brain structural connectome}.
\newblock \bibinfo{journal}{NeuroImage} \bibinfo{volume}{102},
  \bibinfo{pages}{283--293}.
\bibitem[{Betzel(2020)}]{betzel:2020}
\bibinfo{author}{Betzel, R.F.}, \bibinfo{year}{2020}.
\newblock \bibinfo{title}{Community detection in network neuroscience}.
\newblock \bibinfo{journal}{arXiv preprint arXiv:2011.06723} .
\bibitem[{Betzel and Bassett(2017)}]{betzel:2017a}
\bibinfo{author}{Betzel, R.F.}, \bibinfo{author}{Bassett, D.S.},
  \bibinfo{year}{2017}.
\newblock \bibinfo{title}{Multi-scale brain networks}.
\newblock \bibinfo{journal}{Neuroimage} \bibinfo{volume}{160},
  \bibinfo{pages}{73--83}.
\bibitem[{Betzel et~al.(2013)Betzel, Griffa, Avena-Koenigsberger, Go{\~n}i,
  Thiran, Hagmann and Sporns}]{betzel:2013}
\bibinfo{author}{Betzel, R.F.}, \bibinfo{author}{Griffa, A.},
  \bibinfo{author}{Avena-Koenigsberger, A.}, \bibinfo{author}{Go{\~n}i, J.},
  \bibinfo{author}{Thiran, J.P.}, \bibinfo{author}{Hagmann, P.},
  \bibinfo{author}{Sporns, O.}, \bibinfo{year}{2013}.
\newblock \bibinfo{title}{Multi-scale community organization of the human
  structural connectome and its relationship with resting-state functional
  connectivity}.
\newblock \bibinfo{journal}{Network Science} \bibinfo{volume}{1},
  \bibinfo{pages}{353--373}.
\bibitem[{Bi et~al.(2021)Bi, Wang, Niu, Li, Wang, Huang, Chen, Xu, Zhang, Chen
  et~al.}]{bi2021relationship}
\bibinfo{author}{Bi, Q.}, \bibinfo{author}{Wang, W.}, \bibinfo{author}{Niu,
  N.}, \bibinfo{author}{Li, H.}, \bibinfo{author}{Wang, Y.},
  \bibinfo{author}{Huang, W.}, \bibinfo{author}{Chen, K.}, \bibinfo{author}{Xu,
  K.}, \bibinfo{author}{Zhang, J.}, \bibinfo{author}{Chen, Y.}, et~al.,
  \bibinfo{year}{2021}.
\newblock \bibinfo{title}{Relationship between the disrupted topological
  efficiency of the structural brain connectome and glucose hypometabolism in
  normal aging}.
\newblock \bibinfo{journal}{NeuroImage} \bibinfo{volume}{226},
  \bibinfo{pages}{117591}.
\bibitem[{Boshkovski et~al.(2020a)Boshkovski, Kocarev, Cohen-Adad,
  Mi{\v{s}}i{\'c}, Leh{\'e}ricy, Stikov and Mancini}]{boshkovski2020r1}
\bibinfo{author}{Boshkovski, T.}, \bibinfo{author}{Kocarev, L.},
  \bibinfo{author}{Cohen-Adad, J.}, \bibinfo{author}{Mi{\v{s}}i{\'c}, B.},
  \bibinfo{author}{Leh{\'e}ricy, S.}, \bibinfo{author}{Stikov, N.},
  \bibinfo{author}{Mancini, M.}, \bibinfo{year}{2020}a.
\newblock \bibinfo{title}{The r1-weighted connectome: complementing brain
  networks with a myelin-sensitive measure}.
\newblock \bibinfo{journal}{Network Neuroscience} , \bibinfo{pages}{1--34}.
\bibitem[{Boshkovski et~al.(2020b)Boshkovski, Kocarev, Cohen-Adad,
  Mi{\v{s}}i{\'{c}}, Leh{\'{e}}ricy, Stikov and Mancini}]{Boshkovski2020}
\bibinfo{author}{Boshkovski, T.}, \bibinfo{author}{Kocarev, L.},
  \bibinfo{author}{Cohen-Adad, J.}, \bibinfo{author}{Mi{\v{s}}i{\'{c}}, B.},
  \bibinfo{author}{Leh{\'{e}}ricy, S.}, \bibinfo{author}{Stikov, N.},
  \bibinfo{author}{Mancini, M.}, \bibinfo{year}{2020}b.
\newblock \bibinfo{title}{{The R1-weighted connectome: complementing brain
  networks with a myelin-sensitive measure}}.
\newblock \bibinfo{journal}{Network Neuroscience} ,
  \bibinfo{pages}{1--34}\DOIprefix\doi{10.1101/2020.08.06.237941}.
\bibitem[{Brown et~al.(2014)Brown, Miller, Booth, Andrews, Chau, Poskitt and
  Hamarneh}]{brown2014structural}
\bibinfo{author}{Brown, C.J.}, \bibinfo{author}{Miller, S.P.},
  \bibinfo{author}{Booth, B.G.}, \bibinfo{author}{Andrews, S.},
  \bibinfo{author}{Chau, V.}, \bibinfo{author}{Poskitt, K.J.},
  \bibinfo{author}{Hamarneh, G.}, \bibinfo{year}{2014}.
\newblock \bibinfo{title}{Structural network analysis of brain development in
  young preterm neonates}.
\newblock \bibinfo{journal}{Neuroimage} \bibinfo{volume}{101},
  \bibinfo{pages}{667--680}.
\bibitem[{Bucci et~al.(2013)Bucci, Mandelli, Berman, Amirbekian, Nguyen, Berger
  and Henry}]{bucci2013quantifying}
\bibinfo{author}{Bucci, M.}, \bibinfo{author}{Mandelli, M.L.},
  \bibinfo{author}{Berman, J.I.}, \bibinfo{author}{Amirbekian, B.},
  \bibinfo{author}{Nguyen, C.}, \bibinfo{author}{Berger, M.S.},
  \bibinfo{author}{Henry, R.G.}, \bibinfo{year}{2013}.
\newblock \bibinfo{title}{Quantifying diffusion {MRI} tractography of the
  corticospinal tract in brain tumors with deterministic and probabilistic
  methods}.
\newblock \bibinfo{journal}{NeuroImage: Clinical} \bibinfo{volume}{3},
  \bibinfo{pages}{361--368}.
\bibitem[{Buchanan et~al.(2020)Buchanan, Bastin, Ritchie, Liewald, Madole,
  Tucker-Drob, Deary and Cox}]{BUCHANAN.2020}
\bibinfo{author}{Buchanan, C.R.}, \bibinfo{author}{Bastin, M.E.},
  \bibinfo{author}{Ritchie, S.J.}, \bibinfo{author}{Liewald, D.C.},
  \bibinfo{author}{Madole, J.W.}, \bibinfo{author}{Tucker-Drob, E.M.},
  \bibinfo{author}{Deary, I.J.}, \bibinfo{author}{Cox, S.R.},
  \bibinfo{year}{2020}.
\newblock \bibinfo{title}{The effect of network thresholding and weighting on
  structural brain networks in the uk biobank}.
\newblock \bibinfo{journal}{NeuroImage} \bibinfo{volume}{211},
  \bibinfo{pages}{116443}.
\newblock \DOIprefix\doi{https://doi.org/10.1016/j.neuroimage.2019.116443}.
\bibitem[{Buchanan et~al.(2014)Buchanan, Pernet, Gorgolewski, Storkey and
  Bastin}]{buchanan2014test}
\bibinfo{author}{Buchanan, C.R.}, \bibinfo{author}{Pernet, C.R.},
  \bibinfo{author}{Gorgolewski, K.J.}, \bibinfo{author}{Storkey, A.J.},
  \bibinfo{author}{Bastin, M.E.}, \bibinfo{year}{2014}.
\newblock \bibinfo{title}{Test--retest reliability of structural brain networks
  from diffusion {MRI}}.
\newblock \bibinfo{journal}{Neuroimage} \bibinfo{volume}{86},
  \bibinfo{pages}{231--243}.
\bibitem[{Budde and Annese(2013)}]{budde2013quantification}
\bibinfo{author}{Budde, M.D.}, \bibinfo{author}{Annese, J.},
  \bibinfo{year}{2013}.
\newblock \bibinfo{title}{Quantification of anisotropy and fiber orientation in
  human brain histological sections}.
\newblock \bibinfo{journal}{Frontiers in integrative neuroscience}
  \bibinfo{volume}{7}, \bibinfo{pages}{3}.
\bibitem[{Bullmore and Sporns(2009a)}]{bullmore2009complex}
\bibinfo{author}{Bullmore, E.}, \bibinfo{author}{Sporns, O.},
  \bibinfo{year}{2009}a.
\newblock \bibinfo{title}{Complex brain networks: graph theoretical analysis of
  structural and functional systems}.
\newblock \bibinfo{journal}{Nature reviews neuroscience} \bibinfo{volume}{10},
  \bibinfo{pages}{186--198}.
\bibitem[{Bullmore and Sporns(2009b)}]{bullmore:2009}
\bibinfo{author}{Bullmore, E.}, \bibinfo{author}{Sporns, O.},
  \bibinfo{year}{2009}b.
\newblock \bibinfo{title}{Complex brain networks: graph theoretical analysis of
  structural and functional systems}.
\newblock \bibinfo{journal}{Nature Reviews Neuroscience} \bibinfo{volume}{10},
  \bibinfo{pages}{186--198}.
\bibitem[{Bullmore and Sporns(2012)}]{bullmore:2012}
\bibinfo{author}{Bullmore, E.}, \bibinfo{author}{Sporns, O.},
  \bibinfo{year}{2012}.
\newblock \bibinfo{title}{The economy of brain network organization}.
\newblock \bibinfo{journal}{Nat Rev Neurosci} \bibinfo{volume}{13},
  \bibinfo{pages}{336--49}.
\bibitem[{Campbell et~al.(2018)Campbell, Leppert, Narayanan, Boudreau, Duval,
  Cohen-Adad, Pike and Stikov}]{Campbell2018}
\bibinfo{author}{Campbell, J.S.W.}, \bibinfo{author}{Leppert, I.R.},
  \bibinfo{author}{Narayanan, S.}, \bibinfo{author}{Boudreau, M.},
  \bibinfo{author}{Duval, T.}, \bibinfo{author}{Cohen-Adad, J.},
  \bibinfo{author}{Pike, G.B.}, \bibinfo{author}{Stikov, N.},
  \bibinfo{year}{2018}.
\newblock \bibinfo{title}{{Promise and pitfalls of g-ratio estimation with
  MRI.}}
\newblock \bibinfo{journal}{NeuroImage} \bibinfo{volume}{182},
  \bibinfo{pages}{80--96}.
\newblock \DOIprefix\doi{10.1016/j.neuroimage.2017.08.038}.
\bibitem[{Cao et~al.(2017)Cao, Huang and He}]{cao2017developmental}
\bibinfo{author}{Cao, M.}, \bibinfo{author}{Huang, H.}, \bibinfo{author}{He,
  Y.}, \bibinfo{year}{2017}.
\newblock \bibinfo{title}{Developmental connectomics from infancy through early
  childhood}.
\newblock \bibinfo{journal}{Trends in neurosciences} \bibinfo{volume}{40},
  \bibinfo{pages}{494--506}.
\bibitem[{Cao et~al.(2016)Cao, Huang, Peng, Dong and He}]{cao2016toward}
\bibinfo{author}{Cao, M.}, \bibinfo{author}{Huang, H.}, \bibinfo{author}{Peng,
  Y.}, \bibinfo{author}{Dong, Q.}, \bibinfo{author}{He, Y.},
  \bibinfo{year}{2016}.
\newblock \bibinfo{title}{Toward developmental connectomics of the human
  brain}.
\newblock \bibinfo{journal}{Frontiers in neuroanatomy} \bibinfo{volume}{10},
  \bibinfo{pages}{25}.
\bibitem[{Cao et~al.(2014)Cao, Shu, Cao, Wang and He}]{cao2014imaging}
\bibinfo{author}{Cao, M.}, \bibinfo{author}{Shu, N.}, \bibinfo{author}{Cao,
  Q.}, \bibinfo{author}{Wang, Y.}, \bibinfo{author}{He, Y.},
  \bibinfo{year}{2014}.
\newblock \bibinfo{title}{Imaging functional and structural brain connectomics
  in attention-deficit/hyperactivity disorder}.
\newblock \bibinfo{journal}{Molecular neurobiology} \bibinfo{volume}{50},
  \bibinfo{pages}{1111--1123}.
\bibitem[{Cao et~al.(2013)Cao, Shu, An, Wang, Sun, Xia, Wang, Gong, Zang, Wang
  et~al.}]{cao2013probabilistic}
\bibinfo{author}{Cao, Q.}, \bibinfo{author}{Shu, N.}, \bibinfo{author}{An, L.},
  \bibinfo{author}{Wang, P.}, \bibinfo{author}{Sun, L.}, \bibinfo{author}{Xia,
  M.R.}, \bibinfo{author}{Wang, J.H.}, \bibinfo{author}{Gong, G.L.},
  \bibinfo{author}{Zang, Y.F.}, \bibinfo{author}{Wang, Y.F.}, et~al.,
  \bibinfo{year}{2013}.
\newblock \bibinfo{title}{Probabilistic diffusion tractography and graph theory
  analysis reveal abnormal white matter structural connectivity networks in
  drug-naive boys with attention deficit/hyperactivity disorder}.
\newblock \bibinfo{journal}{Journal of Neuroscience} \bibinfo{volume}{33},
  \bibinfo{pages}{10676--10687}.
\bibitem[{Cao et~al.(2021)Cao, Yang, Luo, Wang, Meng, Xia, He, Zhao and
  Li}]{cao2021effects}
\bibinfo{author}{Cao, R.}, \bibinfo{author}{Yang, X.}, \bibinfo{author}{Luo,
  J.}, \bibinfo{author}{Wang, P.}, \bibinfo{author}{Meng, F.},
  \bibinfo{author}{Xia, M.}, \bibinfo{author}{He, Y.}, \bibinfo{author}{Zhao,
  T.}, \bibinfo{author}{Li, Z.}, \bibinfo{year}{2021}.
\newblock \bibinfo{title}{The effects of cognitive behavioral therapy on the
  whole brain structural connectome in unmedicated patients with
  obsessive-compulsive disorder}.
\newblock \bibinfo{journal}{Progress in Neuro-Psychopharmacology and Biological
  Psychiatry} \bibinfo{volume}{104}, \bibinfo{pages}{110037}.
\bibitem[{Casey et~al.(2018)Casey, Cannonier, Conley, Cohen, Barch, Heitzeg,
  Soules, Teslovich, Dellarco, Garavan et~al.}]{casey2018adolescent}
\bibinfo{author}{Casey, B.}, \bibinfo{author}{Cannonier, T.},
  \bibinfo{author}{Conley, M.I.}, \bibinfo{author}{Cohen, A.O.},
  \bibinfo{author}{Barch, D.M.}, \bibinfo{author}{Heitzeg, M.M.},
  \bibinfo{author}{Soules, M.E.}, \bibinfo{author}{Teslovich, T.},
  \bibinfo{author}{Dellarco, D.V.}, \bibinfo{author}{Garavan, H.}, et~al.,
  \bibinfo{year}{2018}.
\newblock \bibinfo{title}{The adolescent brain cognitive development (abcd)
  study: imaging acquisition across 21 sites}.
\newblock \bibinfo{journal}{Developmental cognitive neuroscience}
  \bibinfo{volume}{32}, \bibinfo{pages}{43--54}.
\bibitem[{Catani et~al.(2007)Catani, Allin, Husain, Pugliese, Mesulam, Murray
  and Jones}]{catani2007symmetries}
\bibinfo{author}{Catani, M.}, \bibinfo{author}{Allin, M.P.},
  \bibinfo{author}{Husain, M.}, \bibinfo{author}{Pugliese, L.},
  \bibinfo{author}{Mesulam, M.M.}, \bibinfo{author}{Murray, R.M.},
  \bibinfo{author}{Jones, D.K.}, \bibinfo{year}{2007}.
\newblock \bibinfo{title}{Symmetries in human brain language pathways correlate
  with verbal recall}.
\newblock \bibinfo{journal}{Proceedings of the National Academy of Sciences}
  \bibinfo{volume}{104}, \bibinfo{pages}{17163--17168}.
\bibitem[{Catani et~al.(2002)Catani, Howard, Pajevic and
  Jones}]{catani2002virtual}
\bibinfo{author}{Catani, M.}, \bibinfo{author}{Howard, R.J.},
  \bibinfo{author}{Pajevic, S.}, \bibinfo{author}{Jones, D.K.},
  \bibinfo{year}{2002}.
\newblock \bibinfo{title}{Virtual in vivo interactive dissection of white
  matter fasciculi in the human brain}.
\newblock \bibinfo{journal}{Neuroimage} \bibinfo{volume}{17},
  \bibinfo{pages}{77--94}.
\bibitem[{Ceschin et~al.(2015)Ceschin, Lee, Schmithorst and
  Panigrahy}]{ceschin2015regional}
\bibinfo{author}{Ceschin, R.}, \bibinfo{author}{Lee, V.K.},
  \bibinfo{author}{Schmithorst, V.}, \bibinfo{author}{Panigrahy, A.},
  \bibinfo{year}{2015}.
\newblock \bibinfo{title}{Regional vulnerability of longitudinal cortical
  association connectivity: associated with structural network topology
  alterations in preterm children with cerebral palsy}.
\newblock \bibinfo{journal}{NeuroImage: Clinical} \bibinfo{volume}{9},
  \bibinfo{pages}{322--337}.
\bibitem[{Cetin-Karayumak et~al.(2020)Cetin-Karayumak, Di~Biase, Chunga, Reid,
  Somes, Lyall, Kelly, Solgun, Pasternak, Vangel et~al.}]{cetin2020white}
\bibinfo{author}{Cetin-Karayumak, S.}, \bibinfo{author}{Di~Biase, M.A.},
  \bibinfo{author}{Chunga, N.}, \bibinfo{author}{Reid, B.},
  \bibinfo{author}{Somes, N.}, \bibinfo{author}{Lyall, A.E.},
  \bibinfo{author}{Kelly, S.}, \bibinfo{author}{Solgun, B.},
  \bibinfo{author}{Pasternak, O.}, \bibinfo{author}{Vangel, M.}, et~al.,
  \bibinfo{year}{2020}.
\newblock \bibinfo{title}{White matter abnormalities across the lifespan of
  schizophrenia: a harmonized multi-site diffusion {MRI} study}.
\newblock \bibinfo{journal}{Molecular psychiatry} \bibinfo{volume}{25},
  \bibinfo{pages}{3208--3219}.
\bibitem[{Cha et~al.(2015)Cha, Fekete, Siciliano, Biezonski, Greenhill,
  Pliszka, Blader, Roy, Leibenluft and Posner}]{cha2015neural}
\bibinfo{author}{Cha, J.}, \bibinfo{author}{Fekete, T.},
  \bibinfo{author}{Siciliano, F.}, \bibinfo{author}{Biezonski, D.},
  \bibinfo{author}{Greenhill, L.}, \bibinfo{author}{Pliszka, S.R.},
  \bibinfo{author}{Blader, J.C.}, \bibinfo{author}{Roy, A.K.},
  \bibinfo{author}{Leibenluft, E.}, \bibinfo{author}{Posner, J.},
  \bibinfo{year}{2015}.
\newblock \bibinfo{title}{Neural correlates of aggression in medication-naive
  children with adhd: multivariate analysis of morphometry and tractography}.
\newblock \bibinfo{journal}{Neuropsychopharmacology} \bibinfo{volume}{40},
  \bibinfo{pages}{1717--1725}.
\bibitem[{Chamberland et~al.(2019)Chamberland, Raven, Genc, Duffy, Descoteaux,
  Parker, Tax and Jones}]{chamberland2019dimensionality}
\bibinfo{author}{Chamberland, M.}, \bibinfo{author}{Raven, E.P.},
  \bibinfo{author}{Genc, S.}, \bibinfo{author}{Duffy, K.},
  \bibinfo{author}{Descoteaux, M.}, \bibinfo{author}{Parker, G.D.},
  \bibinfo{author}{Tax, C.M.}, \bibinfo{author}{Jones, D.K.},
  \bibinfo{year}{2019}.
\newblock \bibinfo{title}{Dimensionality reduction of diffusion {MRI} measures
  for improved tractometry of the human brain}.
\newblock \bibinfo{journal}{NeuroImage} \bibinfo{volume}{200},
  \bibinfo{pages}{89--100}.
\bibitem[{Chamberland et~al.(2014)Chamberland, Whittingstall, Fortin, Mathieu
  and Descoteaux}]{chamberland2014real}
\bibinfo{author}{Chamberland, M.}, \bibinfo{author}{Whittingstall, K.},
  \bibinfo{author}{Fortin, D.}, \bibinfo{author}{Mathieu, D.},
  \bibinfo{author}{Descoteaux, M.}, \bibinfo{year}{2014}.
\newblock \bibinfo{title}{Real-time multi-peak tractography for instantaneous
  connectivity display}.
\newblock \bibinfo{journal}{Frontiers in neuroinformatics} \bibinfo{volume}{8},
  \bibinfo{pages}{59}.
\bibitem[{Chandio et~al.(2020a)Chandio, Risacher, Pestilli, Bullock, Yeh,
  Koudoro, Rokem, Harezlak and Garyfallidis}]{Chandio2020}
\bibinfo{author}{Chandio, B.Q.}, \bibinfo{author}{Risacher, S.L.},
  \bibinfo{author}{Pestilli, F.}, \bibinfo{author}{Bullock, D.},
  \bibinfo{author}{Yeh, F.C.}, \bibinfo{author}{Koudoro, S.},
  \bibinfo{author}{Rokem, A.}, \bibinfo{author}{Harezlak, J.},
  \bibinfo{author}{Garyfallidis, E.}, \bibinfo{year}{2020}a.
\newblock \bibinfo{title}{{Bundle analytics, a computational framework for
  investigating the shapes and profiles of brain pathways across populations}}.
\newblock \bibinfo{journal}{Scientific Reports} \bibinfo{volume}{10},
  \bibinfo{pages}{17149}.
\newblock \DOIprefix\doi{10.1038/s41598-020-74054-4}.
\bibitem[{Chandio et~al.(2020b)Chandio, Risacher, Pestilli, Bullock, Yeh,
  Koudoro, Rokem, Harezlak and Garyfallidis}]{chandio2020bundle}
\bibinfo{author}{Chandio, B.Q.}, \bibinfo{author}{Risacher, S.L.},
  \bibinfo{author}{Pestilli, F.}, \bibinfo{author}{Bullock, D.},
  \bibinfo{author}{Yeh, F.C.}, \bibinfo{author}{Koudoro, S.},
  \bibinfo{author}{Rokem, A.}, \bibinfo{author}{Harezlak, J.},
  \bibinfo{author}{Garyfallidis, E.}, \bibinfo{year}{2020}b.
\newblock \bibinfo{title}{Bundle analytics, a computational framework for
  investigating the shapes and profiles of brain pathways across populations}.
\newblock \bibinfo{journal}{Scientific reports} \bibinfo{volume}{10},
  \bibinfo{pages}{1--18}.
\bibitem[{Chang et~al.(2015)Chang, Owen, Pojman, Thieu, Bukshpun, Wakahiro,
  Berman, Roberts, Nagarajan, Sherr et~al.}]{chang2015white}
\bibinfo{author}{Chang, Y.S.}, \bibinfo{author}{Owen, J.P.},
  \bibinfo{author}{Pojman, N.J.}, \bibinfo{author}{Thieu, T.},
  \bibinfo{author}{Bukshpun, P.}, \bibinfo{author}{Wakahiro, M.L.},
  \bibinfo{author}{Berman, J.I.}, \bibinfo{author}{Roberts, T.P.},
  \bibinfo{author}{Nagarajan, S.S.}, \bibinfo{author}{Sherr, E.H.}, et~al.,
  \bibinfo{year}{2015}.
\newblock \bibinfo{title}{White matter changes of neurite density and fiber
  orientation dispersion during human brain maturation}.
\newblock \bibinfo{journal}{PloS one} \bibinfo{volume}{10},
  \bibinfo{pages}{e0123656}.
\bibitem[{Chekir et~al.(2014)Chekir, Descoteaux, Garyfallidis, C{\^o}t{\'e} and
  Boumghar}]{chekir2014hybrid}
\bibinfo{author}{Chekir, A.}, \bibinfo{author}{Descoteaux, M.},
  \bibinfo{author}{Garyfallidis, E.}, \bibinfo{author}{C{\^o}t{\'e}, M.A.},
  \bibinfo{author}{Boumghar, F.O.}, \bibinfo{year}{2014}.
\newblock \bibinfo{title}{A hybrid approach for optimal automatic segmentation
  of white matter tracts in hardi}, in: \bibinfo{booktitle}{2014 IEEE
  Conference on Biomedical Engineering and Sciences (IECBES)},
  \bibinfo{organization}{IEEE}. pp. \bibinfo{pages}{177--180}.
\bibitem[{Chen et~al.(2013)Chen, Liu, Gross and Beaulieu}]{chen2013graph}
\bibinfo{author}{Chen, Z.}, \bibinfo{author}{Liu, M.}, \bibinfo{author}{Gross,
  D.W.}, \bibinfo{author}{Beaulieu, C.}, \bibinfo{year}{2013}.
\newblock \bibinfo{title}{Graph theoretical analysis of developmental patterns
  of the white matter network}.
\newblock \bibinfo{journal}{Frontiers in human neuroscience}
  \bibinfo{volume}{7}, \bibinfo{pages}{716}.
\bibitem[{Chen et~al.(2016)Chen, Zhang, Yushkevich, Liu and
  Beaulieu}]{chen2016maturation}
\bibinfo{author}{Chen, Z.}, \bibinfo{author}{Zhang, H.},
  \bibinfo{author}{Yushkevich, P.A.}, \bibinfo{author}{Liu, M.},
  \bibinfo{author}{Beaulieu, C.}, \bibinfo{year}{2016}.
\newblock \bibinfo{title}{Maturation along white matter tracts in human brain
  using a diffusion tensor surface model tract-specific analysis}.
\newblock \bibinfo{journal}{Frontiers in neuroanatomy} \bibinfo{volume}{10},
  \bibinfo{pages}{9}.
\bibitem[{Cherifi et~al.(2018)Cherifi, Boudjada, Morsli, Girard and
  Deriche}]{cherifi_combining_2018}
\bibinfo{author}{Cherifi, D.}, \bibinfo{author}{Boudjada, M.},
  \bibinfo{author}{Morsli, A.}, \bibinfo{author}{Girard, G.},
  \bibinfo{author}{Deriche, R.}, \bibinfo{year}{2018}.
\newblock \bibinfo{title}{Combining {Improved} {Euler} and {Runge}-{Kutta} 4th
  order for {Tractography} in {Diffusion}-{Weighted} {MRI}}.
\newblock \bibinfo{journal}{Biomedical Signal Processing and Control}
  \bibinfo{volume}{41}, \bibinfo{pages}{90--99}.
\bibitem[{Chiang et~al.(2016)Chiang, Chen, Shang, Tseng and
  Gau}]{chiang2016different}
\bibinfo{author}{Chiang, H.}, \bibinfo{author}{Chen, Y.},
  \bibinfo{author}{Shang, C.}, \bibinfo{author}{Tseng, W.I.},
  \bibinfo{author}{Gau, S.S.}, \bibinfo{year}{2016}.
\newblock \bibinfo{title}{Different neural substrates for executive functions
  in youths with adhd: a diffusion spectrum imaging tractography study}.
\newblock \bibinfo{journal}{Psychological medicine} \bibinfo{volume}{46},
  \bibinfo{pages}{1225}.
\bibitem[{Chiu et~al.(2011)Chiu, Lo, Tang, Liu, Chiang, Yeh, Jaw and
  Tseng}]{chiu2011white}
\bibinfo{author}{Chiu, C.H.}, \bibinfo{author}{Lo, Y.C.},
  \bibinfo{author}{Tang, H.S.}, \bibinfo{author}{Liu, I.C.},
  \bibinfo{author}{Chiang, W.Y.}, \bibinfo{author}{Yeh, F.C.},
  \bibinfo{author}{Jaw, F.S.}, \bibinfo{author}{Tseng, W.Y.I.},
  \bibinfo{year}{2011}.
\newblock \bibinfo{title}{White matter abnormalities of fronto-striato-thalamic
  circuitry in obsessive--compulsive disorder: a study using diffusion spectrum
  imaging tractography}.
\newblock \bibinfo{journal}{Psychiatry Research: Neuroimaging}
  \bibinfo{volume}{192}, \bibinfo{pages}{176--182}.
\bibitem[{Christiaens et~al.(2015)Christiaens, Reisert, Dhollander, Sunaert,
  Suetens and Maes}]{christiaens_global_2015}
\bibinfo{author}{Christiaens, D.}, \bibinfo{author}{Reisert, M.},
  \bibinfo{author}{Dhollander, T.}, \bibinfo{author}{Sunaert, S.},
  \bibinfo{author}{Suetens, P.}, \bibinfo{author}{Maes, F.},
  \bibinfo{year}{2015}.
\newblock \bibinfo{title}{Global tractography of multi-shell diffusion-weighted
  imaging data using a multi-tissue model}.
\newblock \bibinfo{journal}{NeuroImage} \bibinfo{volume}{123},
  \bibinfo{pages}{89--101}.
\bibitem[{Ciccarelli et~al.(2008)Ciccarelli, Catani, Johansen-Berg, Clark and
  Thompson}]{ciccarelli2008diffusion}
\bibinfo{author}{Ciccarelli, O.}, \bibinfo{author}{Catani, M.},
  \bibinfo{author}{Johansen-Berg, H.}, \bibinfo{author}{Clark, C.},
  \bibinfo{author}{Thompson, A.}, \bibinfo{year}{2008}.
\newblock \bibinfo{title}{Diffusion-based tractography in neurological
  disorders: concepts, applications, and future developments}.
\newblock \bibinfo{journal}{The Lancet Neurology} \bibinfo{volume}{7},
  \bibinfo{pages}{715--727}.
\bibitem[{Ciccarelli et~al.(2003)Ciccarelli, Parker, Toosy, Wheeler-Kingshott,
  Barker, Boulby, Miller and Thompson}]{ciccarelli2003diffusion}
\bibinfo{author}{Ciccarelli, O.}, \bibinfo{author}{Parker, G.},
  \bibinfo{author}{Toosy, A.}, \bibinfo{author}{Wheeler-Kingshott, C.},
  \bibinfo{author}{Barker, G.}, \bibinfo{author}{Boulby, P.},
  \bibinfo{author}{Miller, D.}, \bibinfo{author}{Thompson, A.},
  \bibinfo{year}{2003}.
\newblock \bibinfo{title}{From diffusion tractography to quantitative white
  matter tract measures: a reproducibility study}.
\newblock \bibinfo{journal}{Neuroimage} \bibinfo{volume}{18},
  \bibinfo{pages}{348--359}.
\bibitem[{Civier et~al.(2019)Civier, Smith, Yeh, Connelly and
  Calamante}]{Civier.2019}
\bibinfo{author}{Civier, O.}, \bibinfo{author}{Smith, R.E.},
  \bibinfo{author}{Yeh, C.H.}, \bibinfo{author}{Connelly, A.},
  \bibinfo{author}{Calamante, F.}, \bibinfo{year}{2019}.
\newblock \bibinfo{title}{{Is removal of weak connections necessary for
  graph-theoretical analysis of dense weighted structural connectomes from
  diffusion MRI?}}
\newblock \bibinfo{journal}{NeuroImage} \bibinfo{volume}{194},
  \bibinfo{pages}{68--81}.
\newblock \DOIprefix\doi{10.1016/j.neuroimage.2019.02.039}.
\bibitem[{Clayden et~al.(2007)Clayden, Storkey and
  Bastin}]{clayden2007probabilistic}
\bibinfo{author}{Clayden, J.D.}, \bibinfo{author}{Storkey, A.J.},
  \bibinfo{author}{Bastin, M.E.}, \bibinfo{year}{2007}.
\newblock \bibinfo{title}{A probabilistic model-based approach to consistent
  white matter tract segmentation}.
\newblock \bibinfo{journal}{IEEE transactions on medical imaging}
  \bibinfo{volume}{26}, \bibinfo{pages}{1555--1561}.
\bibitem[{Close et~al.(2015)Close, Tournier, Johnston, Calamante, Mareels and
  Connelly}]{close_fourier_2015}
\bibinfo{author}{Close, T.G.}, \bibinfo{author}{Tournier, J.D.},
  \bibinfo{author}{Johnston, L.A.}, \bibinfo{author}{Calamante, F.},
  \bibinfo{author}{Mareels, I.}, \bibinfo{author}{Connelly, A.},
  \bibinfo{year}{2015}.
\newblock \bibinfo{title}{Fourier {Tract} {Sampling} ({FouTS}): {A} framework
  for improved inference of white matter tracts from diffusion {MRI} by
  explicitly modelling tract volume}.
\newblock \bibinfo{journal}{NeuroImage} \bibinfo{volume}{120},
  \bibinfo{pages}{412--427}.
\bibitem[{Colby et~al.(2012a)Colby, Soderberg, Lebel, Dinov, Thompson and
  Sowell}]{Colby2012}
\bibinfo{author}{Colby, J.B.}, \bibinfo{author}{Soderberg, L.},
  \bibinfo{author}{Lebel, C.}, \bibinfo{author}{Dinov, I.D.},
  \bibinfo{author}{Thompson, P.M.}, \bibinfo{author}{Sowell, E.R.},
  \bibinfo{year}{2012}a.
\newblock \bibinfo{title}{{Along-tract statistics allow for enhanced
  tractography analysis.}}
\newblock \bibinfo{journal}{NeuroImage} \bibinfo{volume}{59},
  \bibinfo{pages}{3227--3242}.
\newblock \DOIprefix\doi{10.1016/j.neuroimage.2011.11.004}.
\bibitem[{Colby et~al.(2012b)Colby, Soderberg, Lebel, Dinov, Thompson and
  Sowell}]{colby2012along}
\bibinfo{author}{Colby, J.B.}, \bibinfo{author}{Soderberg, L.},
  \bibinfo{author}{Lebel, C.}, \bibinfo{author}{Dinov, I.D.},
  \bibinfo{author}{Thompson, P.M.}, \bibinfo{author}{Sowell, E.R.},
  \bibinfo{year}{2012}b.
\newblock \bibinfo{title}{Along-tract statistics allow for enhanced
  tractography analysis}.
\newblock \bibinfo{journal}{Neuroimage} \bibinfo{volume}{59},
  \bibinfo{pages}{3227--3242}.
\bibitem[{Collin et~al.(2014a)Collin, Kahn, De~Reus, Cahn and Van
  Den~Heuvel}]{collin2014impaired}
\bibinfo{author}{Collin, G.}, \bibinfo{author}{Kahn, R.S.},
  \bibinfo{author}{De~Reus, M.A.}, \bibinfo{author}{Cahn, W.},
  \bibinfo{author}{Van Den~Heuvel, M.P.}, \bibinfo{year}{2014}a.
\newblock \bibinfo{title}{Impaired rich club connectivity in unaffected
  siblings of schizophrenia patients}.
\newblock \bibinfo{journal}{Schizophrenia bulletin} \bibinfo{volume}{40},
  \bibinfo{pages}{438--448}.
\bibitem[{Collin et~al.(2014b)Collin, Sporns, Mandl and Van
  Den~Heuvel}]{collin2014structural}
\bibinfo{author}{Collin, G.}, \bibinfo{author}{Sporns, O.},
  \bibinfo{author}{Mandl, R.C.}, \bibinfo{author}{Van Den~Heuvel, M.P.},
  \bibinfo{year}{2014}b.
\newblock \bibinfo{title}{Structural and functional aspects relating to cost
  and benefit of rich club organization in the human cerebral cortex}.
\newblock \bibinfo{journal}{Cerebral cortex} \bibinfo{volume}{24},
  \bibinfo{pages}{2258--2267}.
\bibitem[{Collin et~al.(2016)Collin, Turk and Van
  Den~Heuvel}]{collin2016connectomics}
\bibinfo{author}{Collin, G.}, \bibinfo{author}{Turk, E.}, \bibinfo{author}{Van
  Den~Heuvel, M.P.}, \bibinfo{year}{2016}.
\newblock \bibinfo{title}{Connectomics in schizophrenia: from early pioneers to
  recent brain network findings}.
\newblock \bibinfo{journal}{Biological Psychiatry: Cognitive Neuroscience and
  Neuroimaging} \bibinfo{volume}{1}, \bibinfo{pages}{199--208}.
\bibitem[{Collin and Van Den~Heuvel(2013)}]{collin2013ontogeny}
\bibinfo{author}{Collin, G.}, \bibinfo{author}{Van Den~Heuvel, M.P.},
  \bibinfo{year}{2013}.
\newblock \bibinfo{title}{The ontogeny of the human connectome: development and
  dynamic changes of brain connectivity across the life span}.
\newblock \bibinfo{journal}{The Neuroscientist} \bibinfo{volume}{19},
  \bibinfo{pages}{616--628}.
\bibitem[{Conturo et~al.(1999)Conturo, Lori, Cull, Akbudak, Snyder, Shimony,
  McKinstry, Burton and Raichle}]{conturo1999tracking}
\bibinfo{author}{Conturo, T.E.}, \bibinfo{author}{Lori, N.F.},
  \bibinfo{author}{Cull, T.S.}, \bibinfo{author}{Akbudak, E.},
  \bibinfo{author}{Snyder, A.Z.}, \bibinfo{author}{Shimony, J.S.},
  \bibinfo{author}{McKinstry, R.C.}, \bibinfo{author}{Burton, H.},
  \bibinfo{author}{Raichle, M.E.}, \bibinfo{year}{1999}.
\newblock \bibinfo{title}{Tracking neuronal fiber pathways in the living human
  brain}.
\newblock \bibinfo{journal}{Proceedings of the National Academy of Sciences}
  \bibinfo{volume}{96}, \bibinfo{pages}{10422--10427}.
\bibitem[{Cook et~al.(2006)Cook, Bai, Hall, Nedjati-Gilani, Seunarine and
  Alexander}]{cook2005camino}
\bibinfo{author}{Cook, P.A.}, \bibinfo{author}{Bai, Y.}, \bibinfo{author}{Hall,
  M.G.}, \bibinfo{author}{Nedjati-Gilani, S.}, \bibinfo{author}{Seunarine,
  K.K.}, \bibinfo{author}{Alexander, D.C.}, \bibinfo{year}{2006}.
\newblock \bibinfo{title}{Camino: Diffusion {MRI} reconstruction and
  processing}, in: \bibinfo{booktitle}{International Society for Magnetic
  Resonance in Medicine (ISMRM)}, p. \bibinfo{pages}{2759}.
\bibitem[{Corouge et~al.(2006)Corouge, Fletcher, Joshi, Gouttard and
  Gerig}]{corouge2006fiber}
\bibinfo{author}{Corouge, I.}, \bibinfo{author}{Fletcher, P.T.},
  \bibinfo{author}{Joshi, S.}, \bibinfo{author}{Gouttard, S.},
  \bibinfo{author}{Gerig, G.}, \bibinfo{year}{2006}.
\newblock \bibinfo{title}{Fiber tract-oriented statistics for quantitative
  diffusion tensor {MRI} analysis}.
\newblock \bibinfo{journal}{Medical image analysis} \bibinfo{volume}{10},
  \bibinfo{pages}{786--798}.
\bibitem[{Costabile et~al.(2019)Costabile, Alaswad, D'Souza, Thompson and
  Ormond}]{costabile2019current}
\bibinfo{author}{Costabile, J.D.}, \bibinfo{author}{Alaswad, E.},
  \bibinfo{author}{D'Souza, S.}, \bibinfo{author}{Thompson, J.A.},
  \bibinfo{author}{Ormond, D.R.}, \bibinfo{year}{2019}.
\newblock \bibinfo{title}{Current applications of diffusion tensor imaging and
  tractography in intracranial tumor resection}.
\newblock \bibinfo{journal}{Frontiers in oncology} \bibinfo{volume}{9},
  \bibinfo{pages}{426}.
\bibitem[{C{\^o}t{\'e} et~al.(2013)C{\^o}t{\'e}, Girard, Bor{\'e},
  Garyfallidis, Houde and Descoteaux}]{cote2013tractometer}
\bibinfo{author}{C{\^o}t{\'e}, M.A.}, \bibinfo{author}{Girard, G.},
  \bibinfo{author}{Bor{\'e}, A.}, \bibinfo{author}{Garyfallidis, E.},
  \bibinfo{author}{Houde, J.C.}, \bibinfo{author}{Descoteaux, M.},
  \bibinfo{year}{2013}.
\newblock \bibinfo{title}{Tractometer: towards validation of tractography
  pipelines}.
\newblock \bibinfo{journal}{Medical image analysis} \bibinfo{volume}{17},
  \bibinfo{pages}{844--857}.
\bibitem[{Cox et~al.(2016)Cox, Ritchie, Tucker-Drob, Liewald, Hagenaars,
  Davies, Wardlaw, Gale, Bastin and Deary}]{cox2016ageing}
\bibinfo{author}{Cox, S.R.}, \bibinfo{author}{Ritchie, S.J.},
  \bibinfo{author}{Tucker-Drob, E.M.}, \bibinfo{author}{Liewald, D.C.},
  \bibinfo{author}{Hagenaars, S.P.}, \bibinfo{author}{Davies, G.},
  \bibinfo{author}{Wardlaw, J.M.}, \bibinfo{author}{Gale, C.R.},
  \bibinfo{author}{Bastin, M.E.}, \bibinfo{author}{Deary, I.J.},
  \bibinfo{year}{2016}.
\newblock \bibinfo{title}{{Ageing and brain white matter structure in 3,513 UK
  Biobank participants}}.
\newblock \bibinfo{journal}{Nature communications} \bibinfo{volume}{7},
  \bibinfo{pages}{1--13}.
\bibitem[{Crossley et~al.(2017)Crossley, Marques, Taylor, Chaddock,
  Dell’Acqua, Reinders, Mondelli, DiForti, Simmons, David
  et~al.}]{crossley2017connectomic}
\bibinfo{author}{Crossley, N.A.}, \bibinfo{author}{Marques, T.R.},
  \bibinfo{author}{Taylor, H.}, \bibinfo{author}{Chaddock, C.},
  \bibinfo{author}{Dell’Acqua, F.}, \bibinfo{author}{Reinders, A.A.},
  \bibinfo{author}{Mondelli, V.}, \bibinfo{author}{DiForti, M.},
  \bibinfo{author}{Simmons, A.}, \bibinfo{author}{David, A.S.}, et~al.,
  \bibinfo{year}{2017}.
\newblock \bibinfo{title}{Connectomic correlates of response to treatment in
  first-episode psychosis}.
\newblock \bibinfo{journal}{Brain} \bibinfo{volume}{140},
  \bibinfo{pages}{487--496}.
\bibitem[{Crossley et~al.(2014)Crossley, Mechelli, Scott, Carletti, Fox,
  McGuire and Bullmore}]{crossley:2014}
\bibinfo{author}{Crossley, N.A.}, \bibinfo{author}{Mechelli, A.},
  \bibinfo{author}{Scott, J.}, \bibinfo{author}{Carletti, F.},
  \bibinfo{author}{Fox, P.T.}, \bibinfo{author}{McGuire, P.},
  \bibinfo{author}{Bullmore, E.T.}, \bibinfo{year}{2014}.
\newblock \bibinfo{title}{The hubs of the human connectome are generally
  implicated in the anatomy of brain disorders}.
\newblock \bibinfo{journal}{Brain} \bibinfo{volume}{137},
  \bibinfo{pages}{2382--2395}.
\bibitem[{Cui et~al.(2019)Cui, Wei, Xi, Griffa, De~Lange, Kahn, Yin and Van~den
  Heuvel}]{cui2019connectome}
\bibinfo{author}{Cui, L.B.}, \bibinfo{author}{Wei, Y.}, \bibinfo{author}{Xi,
  Y.B.}, \bibinfo{author}{Griffa, A.}, \bibinfo{author}{De~Lange, S.C.},
  \bibinfo{author}{Kahn, R.S.}, \bibinfo{author}{Yin, H.},
  \bibinfo{author}{Van~den Heuvel, M.P.}, \bibinfo{year}{2019}.
\newblock \bibinfo{title}{Connectome-based patterns of first-episode
  medication-naive patients with schizophrenia}.
\newblock \bibinfo{journal}{Schizophrenia bulletin} \bibinfo{volume}{45},
  \bibinfo{pages}{1291--1299}.
\bibitem[{Cui et~al.(2013)Cui, Zhong, Xu, Gong and He}]{cui2013panda}
\bibinfo{author}{Cui, Z.}, \bibinfo{author}{Zhong, S.}, \bibinfo{author}{Xu,
  P.}, \bibinfo{author}{Gong, G.}, \bibinfo{author}{He, Y.},
  \bibinfo{year}{2013}.
\newblock \bibinfo{title}{Panda: a pipeline toolbox for analyzing brain
  diffusion images}.
\newblock \bibinfo{journal}{Frontiers in human neuroscience}
  \bibinfo{volume}{7}, \bibinfo{pages}{42}.
\bibitem[{Daducci et~al.(2016)Daducci, Dal~Pal{\'u}, Descoteaux and
  Thiran}]{daducci2016microstructure}
\bibinfo{author}{Daducci, A.}, \bibinfo{author}{Dal~Pal{\'u}, A.},
  \bibinfo{author}{Descoteaux, M.}, \bibinfo{author}{Thiran, J.P.},
  \bibinfo{year}{2016}.
\newblock \bibinfo{title}{Microstructure informed tractography: pitfalls and
  open challenges}.
\newblock \bibinfo{journal}{Frontiers in neuroscience} \bibinfo{volume}{10},
  \bibinfo{pages}{247}.
\bibitem[{Daducci et~al.(2013)Daducci, Dal~Palu, Lemkaddem and
  Thiran}]{daducci2013convex}
\bibinfo{author}{Daducci, A.}, \bibinfo{author}{Dal~Palu, A.},
  \bibinfo{author}{Lemkaddem, A.}, \bibinfo{author}{Thiran, J.P.},
  \bibinfo{year}{2013}.
\newblock \bibinfo{title}{A convex optimization framework for global
  tractography}, in: \bibinfo{booktitle}{2013 IEEE 10th International Symposium
  on Biomedical Imaging}, \bibinfo{organization}{IEEE}. pp.
  \bibinfo{pages}{524--527}.
\bibitem[{Daducci et~al.(2014)Daducci, Dal~Pal{\`u}, Lemkaddem and
  Thiran}]{daducci2014commit}
\bibinfo{author}{Daducci, A.}, \bibinfo{author}{Dal~Pal{\`u}, A.},
  \bibinfo{author}{Lemkaddem, A.}, \bibinfo{author}{Thiran, J.P.},
  \bibinfo{year}{2014}.
\newblock \bibinfo{title}{{COMMIT}: convex optimization modeling for
  microstructure informed tractography}.
\newblock \bibinfo{journal}{IEEE transactions on medical imaging}
  \bibinfo{volume}{34}, \bibinfo{pages}{246--257}.
\bibitem[{Damatac et~al.(2020)Damatac, Chauvin, Zwiers, van Rooij, Akkermans,
  Naaijen, Hoekstra, Hartman, Oosterlaan, Franke et~al.}]{damatac2020white}
\bibinfo{author}{Damatac, C.G.}, \bibinfo{author}{Chauvin, R.J.},
  \bibinfo{author}{Zwiers, M.P.}, \bibinfo{author}{van Rooij, D.},
  \bibinfo{author}{Akkermans, S.E.}, \bibinfo{author}{Naaijen, J.},
  \bibinfo{author}{Hoekstra, P.J.}, \bibinfo{author}{Hartman, C.A.},
  \bibinfo{author}{Oosterlaan, J.}, \bibinfo{author}{Franke, B.}, et~al.,
  \bibinfo{year}{2020}.
\newblock \bibinfo{title}{White matter microstructure in
  attention-deficit/hyperactivity disorder: a systematic tractography study in
  654 individuals}.
\newblock \bibinfo{journal}{Biological Psychiatry: Cognitive Neuroscience and
  Neuroimaging} .
\bibitem[{D'Andrea et~al.(2016)D'Andrea, Familiari, Di~Lauro, Angelini and
  Sessa}]{d2016safe}
\bibinfo{author}{D'Andrea, G.}, \bibinfo{author}{Familiari, P.},
  \bibinfo{author}{Di~Lauro, A.}, \bibinfo{author}{Angelini, A.},
  \bibinfo{author}{Sessa, G.}, \bibinfo{year}{2016}.
\newblock \bibinfo{title}{Safe resection of gliomas of the dominant angular
  gyrus availing of preoperative {FMRI} and intraoperative dti: preliminary
  series and surgical technique}.
\newblock \bibinfo{journal}{World neurosurgery} \bibinfo{volume}{87},
  \bibinfo{pages}{627--639}.
\bibitem[{Davis et~al.(2009)Davis, Dennis, Buchler, White, Madden and
  Cabeza}]{davis2009assessing}
\bibinfo{author}{Davis, S.W.}, \bibinfo{author}{Dennis, N.A.},
  \bibinfo{author}{Buchler, N.G.}, \bibinfo{author}{White, L.E.},
  \bibinfo{author}{Madden, D.J.}, \bibinfo{author}{Cabeza, R.},
  \bibinfo{year}{2009}.
\newblock \bibinfo{title}{Assessing the effects of age on long white matter
  tracts using diffusion tensor tractography}.
\newblock \bibinfo{journal}{Neuroimage} \bibinfo{volume}{46},
  \bibinfo{pages}{530--541}.
\bibitem[{De~Witte and Mueller(2017)}]{de2017white}
\bibinfo{author}{De~Witte, N.A.}, \bibinfo{author}{Mueller, S.C.},
  \bibinfo{year}{2017}.
\newblock \bibinfo{title}{White matter integrity in brain networks relevant to
  anxiety and depression: evidence from the human connectome project dataset}.
\newblock \bibinfo{journal}{Brain Imaging and Behavior} \bibinfo{volume}{11},
  \bibinfo{pages}{1604--1615}.
\bibitem[{Delettre et~al.(2019)Delettre, Mess{\'{e}}, Dell, Foubet, Heuer,
  Larrat, Meriaux, Mangin, Reillo, {de Juan Romero}, Borrell, Toro and
  Hilgetag}]{Delettre2019_tracers}
\bibinfo{author}{Delettre, C.}, \bibinfo{author}{Mess{\'{e}}, A.},
  \bibinfo{author}{Dell, L.A.}, \bibinfo{author}{Foubet, O.},
  \bibinfo{author}{Heuer, K.}, \bibinfo{author}{Larrat, B.},
  \bibinfo{author}{Meriaux, S.}, \bibinfo{author}{Mangin, J.F.},
  \bibinfo{author}{Reillo, I.}, \bibinfo{author}{{de Juan Romero}, C.},
  \bibinfo{author}{Borrell, V.}, \bibinfo{author}{Toro, R.},
  \bibinfo{author}{Hilgetag, C.C.}, \bibinfo{year}{2019}.
\newblock \bibinfo{title}{{Comparison between diffusion MRI tractography and
  histological tract-tracing of cortico-cortical structural connectivity in the
  ferret brain.}}
\newblock \bibinfo{journal}{Network neuroscience (Cambridge, Mass.)}
  \bibinfo{volume}{3}, \bibinfo{pages}{1038--1050}.
\newblock \DOIprefix\doi{10.1162/netn_a_00098}.
\bibitem[{van Dellen et~al.(2018)van Dellen, Sommer, Bohlken, Tewarie,
  Draaisma, Zalesky, Di~Biase, Brown, Douw, Otte et~al.}]{van2018minimum}
\bibinfo{author}{van Dellen, E.}, \bibinfo{author}{Sommer, I.E.},
  \bibinfo{author}{Bohlken, M.M.}, \bibinfo{author}{Tewarie, P.},
  \bibinfo{author}{Draaisma, L.}, \bibinfo{author}{Zalesky, A.},
  \bibinfo{author}{Di~Biase, M.}, \bibinfo{author}{Brown, J.A.},
  \bibinfo{author}{Douw, L.}, \bibinfo{author}{Otte, W.M.}, et~al.,
  \bibinfo{year}{2018}.
\newblock \bibinfo{title}{Minimum spanning tree analysis of the human
  connectome}.
\newblock \bibinfo{journal}{Human brain mapping} \bibinfo{volume}{39},
  \bibinfo{pages}{2455--2471}.
\bibitem[{Delmarcelle and Hesselink(1993)}]{delmarcelle1993visualizing}
\bibinfo{author}{Delmarcelle, T.}, \bibinfo{author}{Hesselink, L.},
  \bibinfo{year}{1993}.
\newblock \bibinfo{title}{Visualizing second-order tensor fields with
  hyperstreamlines}.
\newblock \bibinfo{journal}{IEEE Computer Graphics and Applications}
  \bibinfo{volume}{13}, \bibinfo{pages}{25--33}.
\bibitem[{Deng et~al.(2019)Deng, Hung, Lui, Chui, Lee, Wang, Li, Mak, Sham,
  Chan et~al.}]{deng2019tractography}
\bibinfo{author}{Deng, Y.}, \bibinfo{author}{Hung, K.S.}, \bibinfo{author}{Lui,
  S.S.}, \bibinfo{author}{Chui, W.W.}, \bibinfo{author}{Lee, J.C.},
  \bibinfo{author}{Wang, Y.}, \bibinfo{author}{Li, Z.}, \bibinfo{author}{Mak,
  H.K.}, \bibinfo{author}{Sham, P.C.}, \bibinfo{author}{Chan, R.C.}, et~al.,
  \bibinfo{year}{2019}.
\newblock \bibinfo{title}{Tractography-based classification in distinguishing
  patients with first-episode schizophrenia from healthy individuals}.
\newblock \bibinfo{journal}{Progress in Neuro-Psychopharmacology and Biological
  Psychiatry} \bibinfo{volume}{88}, \bibinfo{pages}{66--73}.
\bibitem[{Dennis et~al.(2013)Dennis, Jahanshad, McMahon, de~Zubicaray, Martin,
  Hickie, Toga, Wright and Thompson}]{dennis2013development}
\bibinfo{author}{Dennis, E.L.}, \bibinfo{author}{Jahanshad, N.},
  \bibinfo{author}{McMahon, K.L.}, \bibinfo{author}{de~Zubicaray, G.I.},
  \bibinfo{author}{Martin, N.G.}, \bibinfo{author}{Hickie, I.B.},
  \bibinfo{author}{Toga, A.W.}, \bibinfo{author}{Wright, M.J.},
  \bibinfo{author}{Thompson, P.M.}, \bibinfo{year}{2013}.
\newblock \bibinfo{title}{{Development of brain structural connectivity between
  ages 12 and 30: a 4-Tesla diffusion imaging study in 439 adolescents and
  adults}}.
\newblock \bibinfo{journal}{Neuroimage} \bibinfo{volume}{64},
  \bibinfo{pages}{671--684}.
\bibitem[{Descoteaux et~al.(2008)Descoteaux, Deriche, Knosche and
  Anwander}]{descoteaux2008deterministic}
\bibinfo{author}{Descoteaux, M.}, \bibinfo{author}{Deriche, R.},
  \bibinfo{author}{Knosche, T.R.}, \bibinfo{author}{Anwander, A.},
  \bibinfo{year}{2008}.
\newblock \bibinfo{title}{Deterministic and probabilistic tractography based on
  complex fibre orientation distributions}.
\newblock \bibinfo{journal}{IEEE transactions on medical imaging}
  \bibinfo{volume}{28}, \bibinfo{pages}{269--286}.
\bibitem[{Desikan et~al.(2006)Desikan, S{\'e}gonne, Fischl, Quinn, Dickerson,
  Blacker, Buckner, Dale, Maguire, Hyman et~al.}]{desikan2006automated}
\bibinfo{author}{Desikan, R.S.}, \bibinfo{author}{S{\'e}gonne, F.},
  \bibinfo{author}{Fischl, B.}, \bibinfo{author}{Quinn, B.T.},
  \bibinfo{author}{Dickerson, B.C.}, \bibinfo{author}{Blacker, D.},
  \bibinfo{author}{Buckner, R.L.}, \bibinfo{author}{Dale, A.M.},
  \bibinfo{author}{Maguire, R.P.}, \bibinfo{author}{Hyman, B.T.}, et~al.,
  \bibinfo{year}{2006}.
\newblock \bibinfo{title}{An automated labeling system for subdividing the
  human cerebral cortex on {MRI} scans into gyral based regions of interest}.
\newblock \bibinfo{journal}{Neuroimage} \bibinfo{volume}{31},
  \bibinfo{pages}{968--980}.
\bibitem[{Destrieux et~al.(2010)Destrieux, Fischl, Dale and
  Halgren}]{destrieux2010automatic}
\bibinfo{author}{Destrieux, C.}, \bibinfo{author}{Fischl, B.},
  \bibinfo{author}{Dale, A.}, \bibinfo{author}{Halgren, E.},
  \bibinfo{year}{2010}.
\newblock \bibinfo{title}{Automatic parcellation of human cortical gyri and
  sulci using standard anatomical nomenclature}.
\newblock \bibinfo{journal}{Neuroimage} \bibinfo{volume}{53},
  \bibinfo{pages}{1--15}.
\bibitem[{Dick and Tremblay(2012)}]{dick2012beyond}
\bibinfo{author}{Dick, A.S.}, \bibinfo{author}{Tremblay, P.},
  \bibinfo{year}{2012}.
\newblock \bibinfo{title}{Beyond the arcuate fasciculus: consensus and
  controversy in the connectional anatomy of language}.
\newblock \bibinfo{journal}{Brain} \bibinfo{volume}{135},
  \bibinfo{pages}{3529--3550}.
\bibitem[{Dobson and Barnett(2018)}]{dobson2018introduction}
\bibinfo{author}{Dobson, A.J.}, \bibinfo{author}{Barnett, A.G.},
  \bibinfo{year}{2018}.
\newblock \bibinfo{title}{An introduction to generalized linear models}.
\newblock \bibinfo{publisher}{CRC press}.
\bibitem[{Dong et~al.(2019)Dong, Yang, Peng and Wu}]{dong2019multimodality}
\bibinfo{author}{Dong, X.}, \bibinfo{author}{Yang, Z.}, \bibinfo{author}{Peng,
  J.}, \bibinfo{author}{Wu, X.}, \bibinfo{year}{2019}.
\newblock \bibinfo{title}{Multimodality white matter tract segmentation using
  cnn}, in: \bibinfo{booktitle}{Proceedings of the ACM Turing Celebration
  Conference-China}, pp. \bibinfo{pages}{1--8}.
\bibitem[{Drakesmith et~al.(2015)Drakesmith, Caeyenberghs, Dutt, Lewis, David
  and Jones}]{Drakesmith.2015}
\bibinfo{author}{Drakesmith, M.}, \bibinfo{author}{Caeyenberghs, K.},
  \bibinfo{author}{Dutt, A.}, \bibinfo{author}{Lewis, G.},
  \bibinfo{author}{David, A.}, \bibinfo{author}{Jones, D.},
  \bibinfo{year}{2015}.
\newblock \bibinfo{title}{{Overcoming the effects of false positives and
  threshold bias in graph theoretical analyses of neuroimaging data}}.
\newblock \bibinfo{journal}{NeuroImage} \bibinfo{volume}{118},
  \bibinfo{pages}{313--333}.
\newblock \DOIprefix\doi{10.1016/j.neuroimage.2015.05.011}.
\bibitem[{Droby et~al.(2015)Droby, Fleischer, Carnini, Zimmermann, Siffrin,
  Gawehn, Erb, Hildebrandt, Baier and Zipp}]{droby2015impact}
\bibinfo{author}{Droby, A.}, \bibinfo{author}{Fleischer, V.},
  \bibinfo{author}{Carnini, M.}, \bibinfo{author}{Zimmermann, H.},
  \bibinfo{author}{Siffrin, V.}, \bibinfo{author}{Gawehn, J.},
  \bibinfo{author}{Erb, M.}, \bibinfo{author}{Hildebrandt, A.},
  \bibinfo{author}{Baier, B.}, \bibinfo{author}{Zipp, F.},
  \bibinfo{year}{2015}.
\newblock \bibinfo{title}{The impact of isolated lesions on white-matter fiber
  tracts in multiple sclerosis patients}.
\newblock \bibinfo{journal}{NeuroImage: Clinical} \bibinfo{volume}{8},
  \bibinfo{pages}{110--116}.
\bibitem[{Dubois et~al.(2014)Dubois, Dehaene-Lambertz, Kulikova, Poupon,
  H{\"u}ppi and Hertz-Pannier}]{dubois2014early}
\bibinfo{author}{Dubois, J.}, \bibinfo{author}{Dehaene-Lambertz, G.},
  \bibinfo{author}{Kulikova, S.}, \bibinfo{author}{Poupon, C.},
  \bibinfo{author}{H{\"u}ppi, P.S.}, \bibinfo{author}{Hertz-Pannier, L.},
  \bibinfo{year}{2014}.
\newblock \bibinfo{title}{The early development of brain white matter: a review
  of imaging studies in fetuses, newborns and infants}.
\newblock \bibinfo{journal}{Neuroscience} \bibinfo{volume}{276},
  \bibinfo{pages}{48--71}.
\bibitem[{Dubois et~al.(2016)Dubois, Poupon, Thirion, Simonnet, Kulikova,
  Leroy, Hertz-Pannier and Dehaene-Lambertz}]{dubois2016exploring}
\bibinfo{author}{Dubois, J.}, \bibinfo{author}{Poupon, C.},
  \bibinfo{author}{Thirion, B.}, \bibinfo{author}{Simonnet, H.},
  \bibinfo{author}{Kulikova, S.}, \bibinfo{author}{Leroy, F.},
  \bibinfo{author}{Hertz-Pannier, L.}, \bibinfo{author}{Dehaene-Lambertz, G.},
  \bibinfo{year}{2016}.
\newblock \bibinfo{title}{Exploring the early organization and maturation of
  linguistic pathways in the human infant brain}.
\newblock \bibinfo{journal}{Cerebral Cortex} \bibinfo{volume}{26},
  \bibinfo{pages}{2283--2298}.
\bibitem[{Duffau(2005)}]{duffau2005lessons}
\bibinfo{author}{Duffau, H.}, \bibinfo{year}{2005}.
\newblock \bibinfo{title}{Lessons from brain mapping in surgery for low-grade
  glioma: insights into associations between tumour and brain plasticity}.
\newblock \bibinfo{journal}{The Lancet Neurology} \bibinfo{volume}{4},
  \bibinfo{pages}{476--486}.
\bibitem[{Duffau(2015)}]{duffau2015stimulation}
\bibinfo{author}{Duffau, H.}, \bibinfo{year}{2015}.
\newblock \bibinfo{title}{Stimulation mapping of white matter tracts to study
  brain functional connectivity}.
\newblock \bibinfo{journal}{Nature Reviews Neurology} \bibinfo{volume}{11},
  \bibinfo{pages}{255}.
\bibitem[{Duffau(2021)}]{duffau2021brain}
\bibinfo{author}{Duffau, H.}, \bibinfo{year}{2021}.
\newblock \bibinfo{title}{Brain connectomics applied to oncological
  neuroscience: from a traditional surgical strategy focusing on glioma
  topography to a meta-network approach}.
\newblock \bibinfo{journal}{Acta Neurochirurgica} , \bibinfo{pages}{1--13}.
\bibitem[{Eckstein et~al.(2009)Eckstein, Shattuck, Stein, McMahon,
  de~Zubicaray, Wright, Thompson and Toga}]{eckstein2009active}
\bibinfo{author}{Eckstein, I.}, \bibinfo{author}{Shattuck, D.W.},
  \bibinfo{author}{Stein, J.L.}, \bibinfo{author}{McMahon, K.L.},
  \bibinfo{author}{de~Zubicaray, G.}, \bibinfo{author}{Wright, M.J.},
  \bibinfo{author}{Thompson, P.M.}, \bibinfo{author}{Toga, A.W.},
  \bibinfo{year}{2009}.
\newblock \bibinfo{title}{Active fibers: Matching deformable tract templates to
  diffusion tensor images}.
\newblock \bibinfo{journal}{Neuroimage} \bibinfo{volume}{47},
  \bibinfo{pages}{T82--T89}.
\bibitem[{Eickhoff et~al.(2005)Eickhoff, Stephan, Mohlberg, Grefkes, Fink,
  Amunts and Zilles}]{eickhoff2005new}
\bibinfo{author}{Eickhoff, S.B.}, \bibinfo{author}{Stephan, K.E.},
  \bibinfo{author}{Mohlberg, H.}, \bibinfo{author}{Grefkes, C.},
  \bibinfo{author}{Fink, G.R.}, \bibinfo{author}{Amunts, K.},
  \bibinfo{author}{Zilles, K.}, \bibinfo{year}{2005}.
\newblock \bibinfo{title}{A new spm toolbox for combining probabilistic
  cytoarchitectonic maps and functional imaging data}.
\newblock \bibinfo{journal}{Neuroimage} \bibinfo{volume}{25},
  \bibinfo{pages}{1325--1335}.
\bibitem[{Ellis et~al.(2012)Ellis, Rutka, Kulkarni, Dirks and
  Widjaja}]{ellis2012corticospinal}
\bibinfo{author}{Ellis, M.J.}, \bibinfo{author}{Rutka, J.T.},
  \bibinfo{author}{Kulkarni, A.V.}, \bibinfo{author}{Dirks, P.B.},
  \bibinfo{author}{Widjaja, E.}, \bibinfo{year}{2012}.
\newblock \bibinfo{title}{Corticospinal tract mapping in children with ruptured
  arteriovenous malformations using functionally guided diffusion-tensor
  imaging: Report of 3 cases}.
\newblock \bibinfo{journal}{Journal of Neurosurgery: Pediatrics}
  \bibinfo{volume}{9}, \bibinfo{pages}{505--510}.
\bibitem[{Essayed et~al.(2017)Essayed, Zhang, Unadkat, Cosgrove, Golby and
  O'Donnell}]{essayed2017white}
\bibinfo{author}{Essayed, W.I.}, \bibinfo{author}{Zhang, F.},
  \bibinfo{author}{Unadkat, P.}, \bibinfo{author}{Cosgrove, G.R.},
  \bibinfo{author}{Golby, A.J.}, \bibinfo{author}{O'Donnell, L.J.},
  \bibinfo{year}{2017}.
\newblock \bibinfo{title}{White matter tractography for neurosurgical planning:
  A topography-based review of the current state of the art}.
\newblock \bibinfo{journal}{NeuroImage: Clinical} \bibinfo{volume}{15},
  \bibinfo{pages}{659--672}.
\bibitem[{Estrada and Hatano(2008)}]{estrada:2008}
\bibinfo{author}{Estrada, E.}, \bibinfo{author}{Hatano, N.},
  \bibinfo{year}{2008}.
\newblock \bibinfo{title}{Communicability in complex networks}.
\newblock \bibinfo{journal}{Phys Rev E Stat Nonlin Soft Matter Phys}
  \bibinfo{volume}{77}, \bibinfo{pages}{036111}.
\bibitem[{Fellner et~al.(2020)Fellner, Varga and
  Grolmusz}]{fellner2020frequent}
\bibinfo{author}{Fellner, M.}, \bibinfo{author}{Varga, B.},
  \bibinfo{author}{Grolmusz, V.}, \bibinfo{year}{2020}.
\newblock \bibinfo{title}{The frequent complete subgraphs in the human
  connectome}.
\newblock \bibinfo{journal}{PloS one} \bibinfo{volume}{15},
  \bibinfo{pages}{e0236883}.
\bibitem[{Feng and He(2020)}]{feng2020asymmetric}
\bibinfo{author}{Feng, Y.}, \bibinfo{author}{He, J.}, \bibinfo{year}{2020}.
\newblock \bibinfo{title}{Asymmetric fiber trajectory distribution estimation
  using streamline differential equation}.
\newblock \bibinfo{journal}{Medical image analysis} \bibinfo{volume}{63},
  \bibinfo{pages}{101686}.
\bibitem[{Feng et~al.(2020)Feng, Yan, Wang, Song, Zeng and
  Zhao}]{feng2020local}
\bibinfo{author}{Feng, Y.}, \bibinfo{author}{Yan, W.}, \bibinfo{author}{Wang,
  J.}, \bibinfo{author}{Song, J.}, \bibinfo{author}{Zeng, Q.},
  \bibinfo{author}{Zhao, C.}, \bibinfo{year}{2020}.
\newblock \bibinfo{title}{Local white matter fiber clustering differentiates
  parkinson’s disease diagnoses}.
\newblock \bibinfo{journal}{Neuroscience} \bibinfo{volume}{435},
  \bibinfo{pages}{146--160}.
\bibitem[{Fernandes-Cabral et~al.(2016)Fernandes-Cabral, Zenonos, Hamilton,
  Panesar and Fernandez-Miranda}]{fernandes2016high}
\bibinfo{author}{Fernandes-Cabral, D.T.}, \bibinfo{author}{Zenonos, G.A.},
  \bibinfo{author}{Hamilton, R.L.}, \bibinfo{author}{Panesar, S.S.},
  \bibinfo{author}{Fernandez-Miranda, J.C.}, \bibinfo{year}{2016}.
\newblock \bibinfo{title}{High-definition fiber tractography in the evaluation
  and surgical planning of lhermitte-duclos disease: a case report}.
\newblock \bibinfo{journal}{World neurosurgery} \bibinfo{volume}{92},
  \bibinfo{pages}{587.e9--587.e13}.
\bibitem[{Fernandez-Miranda et~al.(2012)Fernandez-Miranda, Pathak, Engh, Jarbo,
  Verstynen, Yeh, Wang, Mintz, Boada, Schneider et~al.}]{fernandez2012high}
\bibinfo{author}{Fernandez-Miranda, J.C.}, \bibinfo{author}{Pathak, S.},
  \bibinfo{author}{Engh, J.}, \bibinfo{author}{Jarbo, K.},
  \bibinfo{author}{Verstynen, T.}, \bibinfo{author}{Yeh, F.C.},
  \bibinfo{author}{Wang, Y.}, \bibinfo{author}{Mintz, A.},
  \bibinfo{author}{Boada, F.}, \bibinfo{author}{Schneider, W.}, et~al.,
  \bibinfo{year}{2012}.
\newblock \bibinfo{title}{High-definition fiber tractography of the human
  brain: neuroanatomical validation and neurosurgical applications}.
\newblock \bibinfo{journal}{Neurosurgery} \bibinfo{volume}{71},
  \bibinfo{pages}{430--453}.
\bibitem[{Fick et~al.(2019)Fick, Wassermann and Deriche}]{fick2019dmipy}
\bibinfo{author}{Fick, R.H.}, \bibinfo{author}{Wassermann, D.},
  \bibinfo{author}{Deriche, R.}, \bibinfo{year}{2019}.
\newblock \bibinfo{title}{The dmipy toolbox: Diffusion {MRI} multi-compartment
  modeling and microstructure recovery made easy}.
\newblock \bibinfo{journal}{Frontiers in neuroinformatics}
  \bibinfo{volume}{13}, \bibinfo{pages}{64}.
\bibitem[{Filippi et~al.(2019)Filippi, Preziosa and Rocca}]{filippi2019brain}
\bibinfo{author}{Filippi, M.}, \bibinfo{author}{Preziosa, P.},
  \bibinfo{author}{Rocca, M.A.}, \bibinfo{year}{2019}.
\newblock \bibinfo{title}{Brain mapping in multiple sclerosis: Lessons learned
  about the human brain}.
\newblock \bibinfo{journal}{Neuroimage} \bibinfo{volume}{190},
  \bibinfo{pages}{32--45}.
\bibitem[{Fillard et~al.(2011)Fillard, Descoteaux, Goh, Gouttard, Jeurissen,
  Malcolm, Ramirez-Manzanares, Reisert, Sakaie, Tensaouti
  et~al.}]{fillard2011quantitative}
\bibinfo{author}{Fillard, P.}, \bibinfo{author}{Descoteaux, M.},
  \bibinfo{author}{Goh, A.}, \bibinfo{author}{Gouttard, S.},
  \bibinfo{author}{Jeurissen, B.}, \bibinfo{author}{Malcolm, J.},
  \bibinfo{author}{Ramirez-Manzanares, A.}, \bibinfo{author}{Reisert, M.},
  \bibinfo{author}{Sakaie, K.}, \bibinfo{author}{Tensaouti, F.}, et~al.,
  \bibinfo{year}{2011}.
\newblock \bibinfo{title}{Quantitative evaluation of 10 tractography algorithms
  on a realistic diffusion mr phantom}.
\newblock \bibinfo{journal}{Neuroimage} \bibinfo{volume}{56},
  \bibinfo{pages}{220--234}.
\bibitem[{Fillard et~al.(2009)Fillard, Poupon and Mangin}]{fillard_novel_2009}
\bibinfo{author}{Fillard, P.}, \bibinfo{author}{Poupon, C.},
  \bibinfo{author}{Mangin, J.F.}, \bibinfo{year}{2009}.
\newblock \bibinfo{title}{A {Novel} {Global} {Tractography} {Algorithm} {Based}
  on an {Adaptive} {Spin} {Glass} {Model}}, in: \bibinfo{booktitle}{Medical
  {Image} {Computing} and {Computer}-{Assisted} {Intervention}},
  \bibinfo{publisher}{Springer}. pp. \bibinfo{pages}{927--934}.
\bibitem[{Fischl(2012)}]{fischl2012freesurfer}
\bibinfo{author}{Fischl, B.}, \bibinfo{year}{2012}.
\newblock \bibinfo{title}{Freesurfer}.
\newblock \bibinfo{journal}{Neuroimage} \bibinfo{volume}{62},
  \bibinfo{pages}{774--781}.
\bibitem[{Fitzsimmons et~al.(2013)Fitzsimmons, Kubicki and
  Shenton}]{fitzsimmons2013review}
\bibinfo{author}{Fitzsimmons, J.}, \bibinfo{author}{Kubicki, M.},
  \bibinfo{author}{Shenton, M.E.}, \bibinfo{year}{2013}.
\newblock \bibinfo{title}{Review of functional and anatomical brain
  connectivity findings in schizophrenia}.
\newblock \bibinfo{journal}{Current opinion in psychiatry}
  \bibinfo{volume}{26}, \bibinfo{pages}{172--187}.
\bibitem[{Fleischer et~al.(2017)Fleischer, Gr{\"o}ger, Koirala, Droby,
  Muthuraman, Kolber, Reuter, Meuth, Zipp and Groppa}]{fleischer2017increased}
\bibinfo{author}{Fleischer, V.}, \bibinfo{author}{Gr{\"o}ger, A.},
  \bibinfo{author}{Koirala, N.}, \bibinfo{author}{Droby, A.},
  \bibinfo{author}{Muthuraman, M.}, \bibinfo{author}{Kolber, P.},
  \bibinfo{author}{Reuter, E.}, \bibinfo{author}{Meuth, S.G.},
  \bibinfo{author}{Zipp, F.}, \bibinfo{author}{Groppa, S.},
  \bibinfo{year}{2017}.
\newblock \bibinfo{title}{Increased structural white and grey matter network
  connectivity compensates for functional decline in early multiple sclerosis}.
\newblock \bibinfo{journal}{Multiple Sclerosis Journal} \bibinfo{volume}{23},
  \bibinfo{pages}{432--441}.
\bibitem[{Fleischer et~al.(2019)Fleischer, Radetz, Ciolac, Muthuraman,
  Gonzalez-Escamilla, Zipp and Groppa}]{fleischer2019graph}
\bibinfo{author}{Fleischer, V.}, \bibinfo{author}{Radetz, A.},
  \bibinfo{author}{Ciolac, D.}, \bibinfo{author}{Muthuraman, M.},
  \bibinfo{author}{Gonzalez-Escamilla, G.}, \bibinfo{author}{Zipp, F.},
  \bibinfo{author}{Groppa, S.}, \bibinfo{year}{2019}.
\newblock \bibinfo{title}{Graph theoretical framework of brain networks in
  multiple sclerosis: a review of concepts}.
\newblock \bibinfo{journal}{Neuroscience} \bibinfo{volume}{403},
  \bibinfo{pages}{35--53}.
\bibitem[{Fornito et~al.(2017)Fornito, Bullmore and
  Zalesky}]{fornito2017opportunities}
\bibinfo{author}{Fornito, A.}, \bibinfo{author}{Bullmore, E.T.},
  \bibinfo{author}{Zalesky, A.}, \bibinfo{year}{2017}.
\newblock \bibinfo{title}{Opportunities and challenges for psychiatry in the
  connectomic era}.
\newblock \bibinfo{journal}{Biological Psychiatry: Cognitive Neuroscience and
  Neuroimaging} \bibinfo{volume}{2}, \bibinfo{pages}{9--19}.
\bibitem[{Fornito et~al.(2013)Fornito, Zalesky and Breakspear}]{fornito:2013}
\bibinfo{author}{Fornito, A.}, \bibinfo{author}{Zalesky, A.},
  \bibinfo{author}{Breakspear, M.}, \bibinfo{year}{2013}.
\newblock \bibinfo{title}{Graph analysis of the human connectome: promise,
  progress, and pitfalls}.
\newblock \bibinfo{journal}{Neuroimage} \bibinfo{volume}{80},
  \bibinfo{pages}{426--44}.
\bibitem[{Fornito et~al.(2015)Fornito, Zalesky and Breakspear}]{fornito:2015}
\bibinfo{author}{Fornito, A.}, \bibinfo{author}{Zalesky, A.},
  \bibinfo{author}{Breakspear, M.}, \bibinfo{year}{2015}.
\newblock \bibinfo{title}{The connectomics of brain disorders}.
\newblock \bibinfo{journal}{Nat Rev Neurosci} \bibinfo{volume}{16},
  \bibinfo{pages}{159--72}.
\bibitem[{Fornito et~al.(2016)Fornito, Zalesky and Bullmore}]{fornito:2016}
\bibinfo{author}{Fornito, A.}, \bibinfo{author}{Zalesky, A.},
  \bibinfo{author}{Bullmore, E.T.}, \bibinfo{year}{2016}.
\newblock \bibinfo{title}{Fundamentals of brain network analysis}.
\bibitem[{Fortin et~al.(2012)Fortin, Aubin-Lemay, Bor{\'e}, Girard, Houde,
  Whittingstall and Descoteaux}]{fortin2012tractography}
\bibinfo{author}{Fortin, D.}, \bibinfo{author}{Aubin-Lemay, C.},
  \bibinfo{author}{Bor{\'e}, A.}, \bibinfo{author}{Girard, G.},
  \bibinfo{author}{Houde, J.C.}, \bibinfo{author}{Whittingstall, K.},
  \bibinfo{author}{Descoteaux, M.}, \bibinfo{year}{2012}.
\newblock \bibinfo{title}{Tractography in the study of the human brain: a
  neurosurgical perspective}.
\newblock \bibinfo{journal}{Canadian journal of neurological sciences}
  \bibinfo{volume}{39}, \bibinfo{pages}{747--756}.
\bibitem[{Fortunato and Hric(2016)}]{fortunato:2016}
\bibinfo{author}{Fortunato, S.}, \bibinfo{author}{Hric, D.},
  \bibinfo{year}{2016}.
\newblock \bibinfo{title}{Community detection in networks: A user guide}.
\newblock \bibinfo{journal}{Physics reports} \bibinfo{volume}{659},
  \bibinfo{pages}{1--44}.
\bibitem[{Frigo et~al.(2020)Frigo, Deslauriers-Gauthier, Parker, Ismail, Kim,
  Verma and Deriche}]{frigo2020diffusion}
\bibinfo{author}{Frigo, M.}, \bibinfo{author}{Deslauriers-Gauthier, S.},
  \bibinfo{author}{Parker, D.}, \bibinfo{author}{Ismail, A.A.O.},
  \bibinfo{author}{Kim, J.J.}, \bibinfo{author}{Verma, R.},
  \bibinfo{author}{Deriche, R.}, \bibinfo{year}{2020}.
\newblock \bibinfo{title}{Diffusion {MRI} tractography filtering techniques
  change the topology of structural connectomes}.
\newblock \bibinfo{journal}{Journal of Neural Engineering}
  \bibinfo{volume}{17}, \bibinfo{pages}{065002}.
\bibitem[{Friman et~al.(2006)Friman, Farneback and Westin}]{friman2006bayesian}
\bibinfo{author}{Friman, O.}, \bibinfo{author}{Farneback, G.},
  \bibinfo{author}{Westin, C.F.}, \bibinfo{year}{2006}.
\newblock \bibinfo{title}{A bayesian approach for stochastic white matter
  tractography}.
\newblock \bibinfo{journal}{IEEE transactions on medical imaging}
  \bibinfo{volume}{25}, \bibinfo{pages}{965--978}.
\bibitem[{Friston(2002)}]{friston:2002}
\bibinfo{author}{Friston, K.}, \bibinfo{year}{2002}.
\newblock \bibinfo{title}{Beyond phrenology: what can neuroimaging tell us
  about distributed circuitry?}
\newblock \bibinfo{journal}{Annu Rev Neurosci} \bibinfo{volume}{25},
  \bibinfo{pages}{221--50}.
\bibitem[{Gan et~al.(2017)Gan, Zhong, Fan, Liu, Niu, Cai, Zou, Wang, Wang, Tan
  et~al.}]{gan2017abnormal}
\bibinfo{author}{Gan, J.}, \bibinfo{author}{Zhong, M.}, \bibinfo{author}{Fan,
  J.}, \bibinfo{author}{Liu, W.}, \bibinfo{author}{Niu, C.},
  \bibinfo{author}{Cai, S.}, \bibinfo{author}{Zou, L.}, \bibinfo{author}{Wang,
  Y.}, \bibinfo{author}{Wang, Y.}, \bibinfo{author}{Tan, C.}, et~al.,
  \bibinfo{year}{2017}.
\newblock \bibinfo{title}{Abnormal white matter structural connectivity in
  adults with obsessive-compulsive disorder}.
\newblock \bibinfo{journal}{Translational psychiatry} \bibinfo{volume}{7},
  \bibinfo{pages}{e1062}.
\bibitem[{Garyfallidis et~al.(2014)Garyfallidis, Brett, Amirbekian, Rokem, Van
  Der~Walt, Descoteaux and Nimmo-Smith}]{garyfallidis2014dipy}
\bibinfo{author}{Garyfallidis, E.}, \bibinfo{author}{Brett, M.},
  \bibinfo{author}{Amirbekian, B.}, \bibinfo{author}{Rokem, A.},
  \bibinfo{author}{Van Der~Walt, S.}, \bibinfo{author}{Descoteaux, M.},
  \bibinfo{author}{Nimmo-Smith, I.}, \bibinfo{year}{2014}.
\newblock \bibinfo{title}{Dipy, a library for the analysis of diffusion {MRI}
  data}.
\newblock \bibinfo{journal}{Frontiers in neuroinformatics} \bibinfo{volume}{8},
  \bibinfo{pages}{8}.
\bibitem[{Garyfallidis et~al.(2018)Garyfallidis, C{\^o}t{\'e}, Rheault, Sidhu,
  Hau, Petit, Fortin, Cunanne and Descoteaux}]{garyfallidis2018recognition}
\bibinfo{author}{Garyfallidis, E.}, \bibinfo{author}{C{\^o}t{\'e}, M.A.},
  \bibinfo{author}{Rheault, F.}, \bibinfo{author}{Sidhu, J.},
  \bibinfo{author}{Hau, J.}, \bibinfo{author}{Petit, L.},
  \bibinfo{author}{Fortin, D.}, \bibinfo{author}{Cunanne, S.},
  \bibinfo{author}{Descoteaux, M.}, \bibinfo{year}{2018}.
\newblock \bibinfo{title}{Recognition of white matter bundles using local and
  global streamline-based registration and clustering}.
\newblock \bibinfo{journal}{NeuroImage} \bibinfo{volume}{170},
  \bibinfo{pages}{283--295}.
\bibitem[{Geeraert et~al.(2020)Geeraert, Chamberland, Lebel and
  Lebel}]{geeraert2020multimodal}
\bibinfo{author}{Geeraert, B.L.}, \bibinfo{author}{Chamberland, M.},
  \bibinfo{author}{Lebel, R.M.}, \bibinfo{author}{Lebel, C.},
  \bibinfo{year}{2020}.
\newblock \bibinfo{title}{Multimodal principal component analysis to identify
  major features of white matter structure and links to reading}.
\newblock \bibinfo{journal}{PloS one} \bibinfo{volume}{15},
  \bibinfo{pages}{e0233244}.
\bibitem[{Gertheiss et~al.(2013)Gertheiss, Goldsmith, Crainiceanu and
  Greven}]{gertheiss2013longitudinal}
\bibinfo{author}{Gertheiss, J.}, \bibinfo{author}{Goldsmith, J.},
  \bibinfo{author}{Crainiceanu, C.}, \bibinfo{author}{Greven, S.},
  \bibinfo{year}{2013}.
\newblock \bibinfo{title}{Longitudinal scalar-on-functions regression with
  application to tractography data}.
\newblock \bibinfo{journal}{Biostatistics} \bibinfo{volume}{14},
  \bibinfo{pages}{447--461}.
\bibitem[{Gilmore et~al.(2018)Gilmore, Knickmeyer and Gao}]{gilmore2018imaging}
\bibinfo{author}{Gilmore, J.H.}, \bibinfo{author}{Knickmeyer, R.C.},
  \bibinfo{author}{Gao, W.}, \bibinfo{year}{2018}.
\newblock \bibinfo{title}{Imaging structural and functional brain development
  in early childhood}.
\newblock \bibinfo{journal}{Nature Reviews Neuroscience} \bibinfo{volume}{19},
  \bibinfo{pages}{123}.
\bibitem[{Girard et~al.(2020)Girard, Caminiti, Battaglia-Mayer, St-Onge,
  Ambrosen, Eskildsen, Krug, Dyrby, Descoteaux, Thiran and
  Innocenti}]{Girard2020}
\bibinfo{author}{Girard, G.}, \bibinfo{author}{Caminiti, R.},
  \bibinfo{author}{Battaglia-Mayer, A.}, \bibinfo{author}{St-Onge, E.},
  \bibinfo{author}{Ambrosen, K.S.}, \bibinfo{author}{Eskildsen, S.F.},
  \bibinfo{author}{Krug, K.}, \bibinfo{author}{Dyrby, T.B.},
  \bibinfo{author}{Descoteaux, M.}, \bibinfo{author}{Thiran, J.P.},
  \bibinfo{author}{Innocenti, G.M.}, \bibinfo{year}{2020}.
\newblock \bibinfo{title}{{On the cortical connectivity in the macaque brain: A
  comparison of diffusion tractography and histological tracing data}}.
\newblock \bibinfo{journal}{NeuroImage} \bibinfo{volume}{221},
  \bibinfo{pages}{117201}.
\bibitem[{Girard et~al.(2017)Girard, Daducci, Petit, Thiran, Whittingstall,
  Deriche, Wassermann and Descoteaux}]{girard2017ax}
\bibinfo{author}{Girard, G.}, \bibinfo{author}{Daducci, A.},
  \bibinfo{author}{Petit, L.}, \bibinfo{author}{Thiran, J.P.},
  \bibinfo{author}{Whittingstall, K.}, \bibinfo{author}{Deriche, R.},
  \bibinfo{author}{Wassermann, D.}, \bibinfo{author}{Descoteaux, M.},
  \bibinfo{year}{2017}.
\newblock \bibinfo{title}{Ax t ract: Toward microstructure informed
  tractography}.
\newblock \bibinfo{journal}{Human brain mapping} \bibinfo{volume}{38},
  \bibinfo{pages}{5485--5500}.
\bibitem[{Girard and Descoteaux(2012)}]{girard_anatomical_2012}
\bibinfo{author}{Girard, G.}, \bibinfo{author}{Descoteaux, M.},
  \bibinfo{year}{2012}.
\newblock \bibinfo{title}{Anatomical {Tissue} {Probability} {Priors} for
  {Tractography}}, in: \bibinfo{booktitle}{{CDMRI}}, pp.
  \bibinfo{pages}{174--185}.
\bibitem[{Glasser et~al.(2016)Glasser, Coalson, Robinson, Hacker, Harwell,
  Yacoub, Ugurbil, Andersson, Beckmann, Jenkinson et~al.}]{glasser2016multi}
\bibinfo{author}{Glasser, M.F.}, \bibinfo{author}{Coalson, T.S.},
  \bibinfo{author}{Robinson, E.C.}, \bibinfo{author}{Hacker, C.D.},
  \bibinfo{author}{Harwell, J.}, \bibinfo{author}{Yacoub, E.},
  \bibinfo{author}{Ugurbil, K.}, \bibinfo{author}{Andersson, J.},
  \bibinfo{author}{Beckmann, C.F.}, \bibinfo{author}{Jenkinson, M.}, et~al.,
  \bibinfo{year}{2016}.
\newblock \bibinfo{title}{A multi-modal parcellation of human cerebral cortex}.
\newblock \bibinfo{journal}{Nature} \bibinfo{volume}{536},
  \bibinfo{pages}{171--178}.
\bibitem[{Goldsmith et~al.(2018)Goldsmith, Crooks, Walker and
  Cotes}]{goldsmith2018update}
\bibinfo{author}{Goldsmith, D.R.}, \bibinfo{author}{Crooks, C.L.},
  \bibinfo{author}{Walker, E.F.}, \bibinfo{author}{Cotes, R.O.},
  \bibinfo{year}{2018}.
\newblock \bibinfo{title}{An update on promising biomarkers in schizophrenia}.
\newblock \bibinfo{journal}{Focus} \bibinfo{volume}{16},
  \bibinfo{pages}{153--163}.
\bibitem[{Gong et~al.(2009a)Gong, He, Concha, Lebel, Gross, Evans and
  Beaulieu}]{gong2009mapping}
\bibinfo{author}{Gong, G.}, \bibinfo{author}{He, Y.}, \bibinfo{author}{Concha,
  L.}, \bibinfo{author}{Lebel, C.}, \bibinfo{author}{Gross, D.W.},
  \bibinfo{author}{Evans, A.C.}, \bibinfo{author}{Beaulieu, C.},
  \bibinfo{year}{2009}a.
\newblock \bibinfo{title}{Mapping anatomical connectivity patterns of human
  cerebral cortex using in vivo diffusion tensor imaging tractography}.
\newblock \bibinfo{journal}{Cerebral cortex} \bibinfo{volume}{19},
  \bibinfo{pages}{524--536}.
\bibitem[{Gong et~al.(2009b)Gong, Rosa-Neto, Carbonell, Chen, He and
  Evans}]{gong2009age}
\bibinfo{author}{Gong, G.}, \bibinfo{author}{Rosa-Neto, P.},
  \bibinfo{author}{Carbonell, F.}, \bibinfo{author}{Chen, Z.J.},
  \bibinfo{author}{He, Y.}, \bibinfo{author}{Evans, A.C.},
  \bibinfo{year}{2009}b.
\newblock \bibinfo{title}{Age- and gender-related differences in the cortical
  anatomical network}.
\newblock \bibinfo{journal}{Journal of Neuroscience} \bibinfo{volume}{29},
  \bibinfo{pages}{15684--15693}.
\bibitem[{Gong et~al.(2018)Gong, Zhang, Norton, Essayed, Unadkat, Rigolo,
  Pasternak, Rathi, Hou, Golby et~al.}]{gong2018free}
\bibinfo{author}{Gong, S.}, \bibinfo{author}{Zhang, F.},
  \bibinfo{author}{Norton, I.}, \bibinfo{author}{Essayed, W.I.},
  \bibinfo{author}{Unadkat, P.}, \bibinfo{author}{Rigolo, L.},
  \bibinfo{author}{Pasternak, O.}, \bibinfo{author}{Rathi, Y.},
  \bibinfo{author}{Hou, L.}, \bibinfo{author}{Golby, A.J.}, et~al.,
  \bibinfo{year}{2018}.
\newblock \bibinfo{title}{Free water modeling of peritumoral edema using
  multi-fiber tractography: Application to tracking the arcuate fasciculus for
  neurosurgical planning}.
\newblock \bibinfo{journal}{PloS one} \bibinfo{volume}{13},
  \bibinfo{pages}{e0197056}.
\bibitem[{Go{\~n}i et~al.(2014)Go{\~n}i, van~den Heuvel, Avena-Koenigsberger,
  Velez~de Mendizabal, Betzel, Griffa, Hagmann, Corominas-Murtra, Thiran and
  Sporns}]{goni:2014}
\bibinfo{author}{Go{\~n}i, J.}, \bibinfo{author}{van~den Heuvel, M.P.},
  \bibinfo{author}{Avena-Koenigsberger, A.}, \bibinfo{author}{Velez~de
  Mendizabal, N.}, \bibinfo{author}{Betzel, R.F.}, \bibinfo{author}{Griffa,
  A.}, \bibinfo{author}{Hagmann, P.}, \bibinfo{author}{Corominas-Murtra, B.},
  \bibinfo{author}{Thiran, J.P.}, \bibinfo{author}{Sporns, O.},
  \bibinfo{year}{2014}.
\newblock \bibinfo{title}{Resting-brain functional connectivity predicted by
  analytic measures of network communication}.
\newblock \bibinfo{journal}{Proc Natl Acad Sci U S A} \bibinfo{volume}{111},
  \bibinfo{pages}{833--8}.
\bibitem[{Griffa et~al.(2013)Griffa, Baumann, Thiran and
  Hagmann}]{griffa2013structural}
\bibinfo{author}{Griffa, A.}, \bibinfo{author}{Baumann, P.S.},
  \bibinfo{author}{Thiran, J.P.}, \bibinfo{author}{Hagmann, P.},
  \bibinfo{year}{2013}.
\newblock \bibinfo{title}{Structural connectomics in brain diseases}.
\newblock \bibinfo{journal}{Neuroimage} \bibinfo{volume}{80},
  \bibinfo{pages}{515--526}.
\bibitem[{Griffa and Van~den Heuvel(2018)}]{griffa:2018}
\bibinfo{author}{Griffa, A.}, \bibinfo{author}{Van~den Heuvel, M.P.},
  \bibinfo{year}{2018}.
\newblock \bibinfo{title}{Rich-club neurocircuitry: function, evolution, and
  vulnerability}.
\newblock \bibinfo{journal}{Dialogues Clin Neurosci} \bibinfo{volume}{20},
  \bibinfo{pages}{121--132}.
\bibitem[{Grinberg et~al.(2018)Grinberg, Maximov, Farrher and
  Shah}]{grinberg2018microstructure}
\bibinfo{author}{Grinberg, F.}, \bibinfo{author}{Maximov, I.I.},
  \bibinfo{author}{Farrher, E.}, \bibinfo{author}{Shah, N.J.},
  \bibinfo{year}{2018}.
\newblock \bibinfo{title}{Microstructure-informed slow diffusion tractography
  in humans enhances visualisation of fibre pathways}.
\newblock \bibinfo{journal}{Magnetic resonance imaging} \bibinfo{volume}{45},
  \bibinfo{pages}{7--17}.
\bibitem[{de~Groot et~al.(2015)de~Groot, Ikram, Akoudad, Krestin, Hofman,
  van~der Lugt, Niessen and Vernooij}]{de2015tract}
\bibinfo{author}{de~Groot, M.}, \bibinfo{author}{Ikram, M.A.},
  \bibinfo{author}{Akoudad, S.}, \bibinfo{author}{Krestin, G.P.},
  \bibinfo{author}{Hofman, A.}, \bibinfo{author}{van~der Lugt, A.},
  \bibinfo{author}{Niessen, W.J.}, \bibinfo{author}{Vernooij, M.W.},
  \bibinfo{year}{2015}.
\newblock \bibinfo{title}{Tract-specific white matter degeneration in aging:
  the {Rotterdam} study}.
\newblock \bibinfo{journal}{Alzheimer's \& Dementia} \bibinfo{volume}{11},
  \bibinfo{pages}{321--330}.
\bibitem[{Gruner et~al.(2012)Gruner, Vo, Ikuta, Mahon, Peters, Malhotra,
  Ulu{\u{g}} and Szeszko}]{gruner2012white}
\bibinfo{author}{Gruner, P.}, \bibinfo{author}{Vo, A.}, \bibinfo{author}{Ikuta,
  T.}, \bibinfo{author}{Mahon, K.}, \bibinfo{author}{Peters, B.D.},
  \bibinfo{author}{Malhotra, A.K.}, \bibinfo{author}{Ulu{\u{g}}, A.M.},
  \bibinfo{author}{Szeszko, P.R.}, \bibinfo{year}{2012}.
\newblock \bibinfo{title}{White matter abnormalities in pediatric
  obsessive-compulsive disorder}.
\newblock \bibinfo{journal}{Neuropsychopharmacology} \bibinfo{volume}{37},
  \bibinfo{pages}{2730--2739}.
\bibitem[{Guevara et~al.(2012)Guevara, Duclap, Poupon, Marrakchi-Kacem,
  Fillard, Le~Bihan, Leboyer, Houenou and Mangin}]{guevara2012automatic}
\bibinfo{author}{Guevara, P.}, \bibinfo{author}{Duclap, D.},
  \bibinfo{author}{Poupon, C.}, \bibinfo{author}{Marrakchi-Kacem, L.},
  \bibinfo{author}{Fillard, P.}, \bibinfo{author}{Le~Bihan, D.},
  \bibinfo{author}{Leboyer, M.}, \bibinfo{author}{Houenou, J.},
  \bibinfo{author}{Mangin, J.F.}, \bibinfo{year}{2012}.
\newblock \bibinfo{title}{Automatic fiber bundle segmentation in massive
  tractography datasets using a multi-subject bundle atlas}.
\newblock \bibinfo{journal}{Neuroimage} \bibinfo{volume}{61},
  \bibinfo{pages}{1083--1099}.
\bibitem[{Gupta et~al.(2017a)Gupta, Patil, Tailor, Thapar and
  Nigam}]{gupta2017brainsegnet}
\bibinfo{author}{Gupta, T.}, \bibinfo{author}{Patil, S.M.},
  \bibinfo{author}{Tailor, M.}, \bibinfo{author}{Thapar, D.},
  \bibinfo{author}{Nigam, A.}, \bibinfo{year}{2017}a.
\newblock \bibinfo{title}{Brainsegnet: a segmentation network for human brain
  fiber tractography data into anatomically meaningful clusters}.
\newblock \bibinfo{journal}{arXiv preprint arXiv:1710.05158} .
\bibitem[{Gupta et~al.(2018)Gupta, Thomopoulos, Corbin, Rashid and
  Thompson}]{gupta2018fibernet}
\bibinfo{author}{Gupta, V.}, \bibinfo{author}{Thomopoulos, S.I.},
  \bibinfo{author}{Corbin, C.K.}, \bibinfo{author}{Rashid, F.},
  \bibinfo{author}{Thompson, P.M.}, \bibinfo{year}{2018}.
\newblock \bibinfo{title}{Fibernet 2.0: An automatic neural network based tool
  for clustering white matter fibers in the brain}, in:
  \bibinfo{booktitle}{2018 IEEE 15th International Symposium on Biomedical
  Imaging (ISBI 2018)}, \bibinfo{organization}{IEEE}. pp.
  \bibinfo{pages}{708--711}.
\bibitem[{Gupta et~al.(2017b)Gupta, Thomopoulos, Rashid and
  Thompson}]{gupta2017fibernet}
\bibinfo{author}{Gupta, V.}, \bibinfo{author}{Thomopoulos, S.I.},
  \bibinfo{author}{Rashid, F.M.}, \bibinfo{author}{Thompson, P.M.},
  \bibinfo{year}{2017}b.
\newblock \bibinfo{title}{Fibernet: An ensemble deep learning framework for
  clustering white matter fibers}, in: \bibinfo{booktitle}{International
  Conference on Medical Image Computing and Computer-Assisted Intervention},
  \bibinfo{organization}{Springer}. pp. \bibinfo{pages}{548--555}.
\bibitem[{Habeck and Stern(2010)}]{habeck2010multivariate}
\bibinfo{author}{Habeck, C.}, \bibinfo{author}{Stern, Y.},
  \bibinfo{year}{2010}.
\newblock \bibinfo{title}{Multivariate data analysis for neuroimaging data:
  overview and application to alzheimer’s disease}.
\newblock \bibinfo{journal}{Cell biochemistry and biophysics}
  \bibinfo{volume}{58}, \bibinfo{pages}{53--67}.
\bibitem[{Hagmann et~al.(2008a)Hagmann, Cammoun, Gigandet, Meuli, Honey, Wedeen
  and Sporns}]{hagmann:2008}
\bibinfo{author}{Hagmann, P.}, \bibinfo{author}{Cammoun, L.},
  \bibinfo{author}{Gigandet, X.}, \bibinfo{author}{Meuli, R.},
  \bibinfo{author}{Honey, C.J.}, \bibinfo{author}{Wedeen, V.J.},
  \bibinfo{author}{Sporns, O.}, \bibinfo{year}{2008}a.
\newblock \bibinfo{title}{Mapping the structural core of human cerebral
  cortex}.
\newblock \bibinfo{journal}{PLoS Biol} \bibinfo{volume}{6},
  \bibinfo{pages}{e159}.
\bibitem[{Hagmann et~al.(2008b)Hagmann, Cammoun, Gigandet, Meuli, Honey, Wedeen
  and Sporns}]{hagmann2008mapping}
\bibinfo{author}{Hagmann, P.}, \bibinfo{author}{Cammoun, L.},
  \bibinfo{author}{Gigandet, X.}, \bibinfo{author}{Meuli, R.},
  \bibinfo{author}{Honey, C.J.}, \bibinfo{author}{Wedeen, V.J.},
  \bibinfo{author}{Sporns, O.}, \bibinfo{year}{2008}b.
\newblock \bibinfo{title}{Mapping the structural core of human cerebral
  cortex}.
\newblock \bibinfo{journal}{PLoS Biol} \bibinfo{volume}{6},
  \bibinfo{pages}{e159}.
\bibitem[{Hagmann et~al.(2010)Hagmann, Sporns, Madan, Cammoun, Pienaar, Wedeen,
  Meuli, Thiran and Grant}]{hagmann2010white}
\bibinfo{author}{Hagmann, P.}, \bibinfo{author}{Sporns, O.},
  \bibinfo{author}{Madan, N.}, \bibinfo{author}{Cammoun, L.},
  \bibinfo{author}{Pienaar, R.}, \bibinfo{author}{Wedeen, V.J.},
  \bibinfo{author}{Meuli, R.}, \bibinfo{author}{Thiran, J.P.},
  \bibinfo{author}{Grant, P.}, \bibinfo{year}{2010}.
\newblock \bibinfo{title}{White matter maturation reshapes structural
  connectivity in the late developing human brain}.
\newblock \bibinfo{journal}{Proceedings of the National Academy of Sciences}
  \bibinfo{volume}{107}, \bibinfo{pages}{19067--19072}.
\bibitem[{Hansen et~al.(2020)Hansen, Yang, Lyu, Rheault, Kerley, Chandio,
  Fadnavis, Williams, Shafer, Resnick et~al.}]{hansen2020pandora}
\bibinfo{author}{Hansen, C.B.}, \bibinfo{author}{Yang, Q.},
  \bibinfo{author}{Lyu, I.}, \bibinfo{author}{Rheault, F.},
  \bibinfo{author}{Kerley, C.}, \bibinfo{author}{Chandio, B.Q.},
  \bibinfo{author}{Fadnavis, S.}, \bibinfo{author}{Williams, O.},
  \bibinfo{author}{Shafer, A.T.}, \bibinfo{author}{Resnick, S.M.}, et~al.,
  \bibinfo{year}{2020}.
\newblock \bibinfo{title}{Pandora: 4-{D} white matter bundle population-based
  atlases derived from diffusion {MRI} fiber tractography}.
\newblock \bibinfo{journal}{Neuroinformatics} ,
  \bibinfo{pages}{1}\DOIprefix\doi{https://doi.org/10.1007/s12021-020-09497-1}.
\bibitem[{Harms et~al.(2018)Harms, Somerville, Ances, Andersson, Barch,
  Bastiani, Bookheimer, Brown, Buckner, Burgess et~al.}]{harms2018extending}
\bibinfo{author}{Harms, M.P.}, \bibinfo{author}{Somerville, L.H.},
  \bibinfo{author}{Ances, B.M.}, \bibinfo{author}{Andersson, J.},
  \bibinfo{author}{Barch, D.M.}, \bibinfo{author}{Bastiani, M.},
  \bibinfo{author}{Bookheimer, S.Y.}, \bibinfo{author}{Brown, T.B.},
  \bibinfo{author}{Buckner, R.L.}, \bibinfo{author}{Burgess, G.C.}, et~al.,
  \bibinfo{year}{2018}.
\newblock \bibinfo{title}{Extending the human connectome project across ages:
  Imaging protocols for the lifespan development and aging projects}.
\newblock \bibinfo{journal}{Neuroimage} \bibinfo{volume}{183},
  \bibinfo{pages}{972--984}.
\bibitem[{Hart et~al.(2020)Hart, Romero-Garcia, Price, Santarius and
  Suckling}]{hart2020connections}
\bibinfo{author}{Hart, M.G.}, \bibinfo{author}{Romero-Garcia, R.},
  \bibinfo{author}{Price, S.J.}, \bibinfo{author}{Santarius, T.},
  \bibinfo{author}{Suckling, J.}, \bibinfo{year}{2020}.
\newblock \bibinfo{title}{Connections, tracts, fractals, and the rest: A
  working guide to network and connectivity studies in neurosurgery}.
\newblock \bibinfo{journal}{World neurosurgery} \bibinfo{volume}{140},
  \bibinfo{pages}{389--400}.
\bibitem[{Hasan et~al.(2009a)Hasan, Iftikhar, Kamali, Kramer, Ashtari, Cirino,
  Papanicolaou, Fletcher and Ewing-Cobbs}]{hasan2009development}
\bibinfo{author}{Hasan, K.M.}, \bibinfo{author}{Iftikhar, A.},
  \bibinfo{author}{Kamali, A.}, \bibinfo{author}{Kramer, L.A.},
  \bibinfo{author}{Ashtari, M.}, \bibinfo{author}{Cirino, P.T.},
  \bibinfo{author}{Papanicolaou, A.C.}, \bibinfo{author}{Fletcher, J.M.},
  \bibinfo{author}{Ewing-Cobbs, L.}, \bibinfo{year}{2009}a.
\newblock \bibinfo{title}{Development and aging of the healthy human brain
  uncinate fasciculus across the lifespan using diffusion tensor tractography}.
\newblock \bibinfo{journal}{Brain research} \bibinfo{volume}{1276},
  \bibinfo{pages}{67--76}.
\bibitem[{Hasan et~al.(2010)Hasan, Kamali, Abid, Kramer, Fletcher and
  Ewing-Cobbs}]{hasan2010quantification}
\bibinfo{author}{Hasan, K.M.}, \bibinfo{author}{Kamali, A.},
  \bibinfo{author}{Abid, H.}, \bibinfo{author}{Kramer, L.A.},
  \bibinfo{author}{Fletcher, J.M.}, \bibinfo{author}{Ewing-Cobbs, L.},
  \bibinfo{year}{2010}.
\newblock \bibinfo{title}{Quantification of the spatiotemporal microstructural
  organization of the human brain association, projection and commissural
  pathways across the lifespan using diffusion tensor tractography}.
\newblock \bibinfo{journal}{Brain Structure and Function}
  \bibinfo{volume}{214}, \bibinfo{pages}{361--373}.
\bibitem[{Hasan et~al.(2009b)Hasan, Kamali, Iftikhar, Kramer, Papanicolaou,
  Fletcher and Ewing-Cobbs}]{hasan2009diffusion}
\bibinfo{author}{Hasan, K.M.}, \bibinfo{author}{Kamali, A.},
  \bibinfo{author}{Iftikhar, A.}, \bibinfo{author}{Kramer, L.A.},
  \bibinfo{author}{Papanicolaou, A.C.}, \bibinfo{author}{Fletcher, J.M.},
  \bibinfo{author}{Ewing-Cobbs, L.}, \bibinfo{year}{2009}b.
\newblock \bibinfo{title}{Diffusion tensor tractography quantification of the
  human corpus callosum fiber pathways across the lifespan}.
\newblock \bibinfo{journal}{Brain research} \bibinfo{volume}{1249},
  \bibinfo{pages}{91--100}.
\bibitem[{Heidemann et~al.(2012)Heidemann, Anwander, Feiweier, Knösche and
  Turner}]{heidemann_zooppa_2012}
\bibinfo{author}{Heidemann, R.M.}, \bibinfo{author}{Anwander, A.},
  \bibinfo{author}{Feiweier, T.}, \bibinfo{author}{Knösche, T.R.},
  \bibinfo{author}{Turner, R.}, \bibinfo{year}{2012}.
\newblock \bibinfo{title}{k-space and q-space: {Combining} ultra-high spatial
  and angular resolution in diffusion imaging using zooppa at 7t}.
\newblock \bibinfo{journal}{NeuroImage} \bibinfo{volume}{60},
  \bibinfo{pages}{967--978}.
\newblock \DOIprefix\doi{10.1016/j.neuroimage.2011.12.081}.
\bibitem[{Heiervang et~al.(2006)Heiervang, Behrens, Mackay, Robson and
  Johansen-Berg}]{heiervang2006between}
\bibinfo{author}{Heiervang, E.}, \bibinfo{author}{Behrens, T.},
  \bibinfo{author}{Mackay, C.E.}, \bibinfo{author}{Robson, M.D.},
  \bibinfo{author}{Johansen-Berg, H.}, \bibinfo{year}{2006}.
\newblock \bibinfo{title}{Between session reproducibility and between subject
  variability of diffusion mr and tractography measures}.
\newblock \bibinfo{journal}{Neuroimage} \bibinfo{volume}{33},
  \bibinfo{pages}{867--877}.
\bibitem[{Henriques et~al.(2020)Henriques, Jespersen and
  Shemesh}]{henriques2020correlation}
\bibinfo{author}{Henriques, R.N.}, \bibinfo{author}{Jespersen, S.N.},
  \bibinfo{author}{Shemesh, N.}, \bibinfo{year}{2020}.
\newblock \bibinfo{title}{Correlation tensor magnetic resonance imaging}.
\newblock \bibinfo{journal}{NeuroImage} \bibinfo{volume}{211},
  \bibinfo{pages}{116605}.
\bibitem[{Herbet et~al.(2016)Herbet, Maheu, Costi, Lafargue and
  Duffau}]{herbet2016mapping}
\bibinfo{author}{Herbet, G.}, \bibinfo{author}{Maheu, M.},
  \bibinfo{author}{Costi, E.}, \bibinfo{author}{Lafargue, G.},
  \bibinfo{author}{Duffau, H.}, \bibinfo{year}{2016}.
\newblock \bibinfo{title}{Mapping neuroplastic potential in brain-damaged
  patients}.
\newblock \bibinfo{journal}{Brain} \bibinfo{volume}{139},
  \bibinfo{pages}{829--844}.
\bibitem[{van~den Heuvel et~al.(2012)van~den Heuvel, Kahn, Go{\~n}i and
  Sporns}]{heuvel:2012}
\bibinfo{author}{van~den Heuvel, M.P.}, \bibinfo{author}{Kahn, R.S.},
  \bibinfo{author}{Go{\~n}i, J.}, \bibinfo{author}{Sporns, O.},
  \bibinfo{year}{2012}.
\newblock \bibinfo{title}{High-cost, high-capacity backbone for global brain
  communication}.
\newblock \bibinfo{journal}{Proc Natl Acad Sci U S A} \bibinfo{volume}{109},
  \bibinfo{pages}{11372--7}.
\bibitem[{van~den Heuvel et~al.(2010)van~den Heuvel, Mandl, Stam, Kahn and
  Pol}]{van2010aberrant}
\bibinfo{author}{van~den Heuvel, M.P.}, \bibinfo{author}{Mandl, R.C.},
  \bibinfo{author}{Stam, C.J.}, \bibinfo{author}{Kahn, R.S.},
  \bibinfo{author}{Pol, H.E.H.}, \bibinfo{year}{2010}.
\newblock \bibinfo{title}{Aberrant frontal and temporal complex network
  structure in schizophrenia: a graph theoretical analysis}.
\newblock \bibinfo{journal}{Journal of Neuroscience} \bibinfo{volume}{30},
  \bibinfo{pages}{15915--15926}.
\bibitem[{van~den Heuvel et~al.(2015)van~den Heuvel, de~Reus, {Feldman
  Barrett}, Scholtens, Coopmans, Schmidt, Preuss, Rilling and
  Li}]{vanDenHeuvel2015}
\bibinfo{author}{van~den Heuvel, M.P.}, \bibinfo{author}{de~Reus, M.A.},
  \bibinfo{author}{{Feldman Barrett}, L.}, \bibinfo{author}{Scholtens, L.H.},
  \bibinfo{author}{Coopmans, F.M.T.}, \bibinfo{author}{Schmidt, R.},
  \bibinfo{author}{Preuss, T.M.}, \bibinfo{author}{Rilling, J.K.},
  \bibinfo{author}{Li, L.}, \bibinfo{year}{2015}.
\newblock \bibinfo{title}{{Comparison of diffusion tractography and
  tract-tracing measures of connectivity strength in rhesus macaque
  connectome}}.
\newblock \bibinfo{journal}{Human Brain Mapping} \bibinfo{volume}{36},
  \bibinfo{pages}{3064--3075}.
\newblock \DOIprefix\doi{10.1002/hbm.22828}.
\bibitem[{van~den Heuvel and Sporns(2011)}]{heuvel:2011}
\bibinfo{author}{van~den Heuvel, M.P.}, \bibinfo{author}{Sporns, O.},
  \bibinfo{year}{2011}.
\newblock \bibinfo{title}{Rich-club organization of the human connectome}.
\newblock \bibinfo{journal}{J Neurosci} \bibinfo{volume}{31},
  \bibinfo{pages}{15775--86}.
\bibitem[{van~den Heuvel and Sporns(2013)}]{heuvel:2013a}
\bibinfo{author}{van~den Heuvel, M.P.}, \bibinfo{author}{Sporns, O.},
  \bibinfo{year}{2013}.
\newblock \bibinfo{title}{Network hubs in the human brain}.
\newblock \bibinfo{journal}{Trends Cogn Sci} \bibinfo{volume}{17},
  \bibinfo{pages}{683--96}.
\bibitem[{van~den Heuvel and Sporns(2019)}]{heuvel:2019}
\bibinfo{author}{van~den Heuvel, M.P.}, \bibinfo{author}{Sporns, O.},
  \bibinfo{year}{2019}.
\newblock \bibinfo{title}{A cross-disorder connectome landscape of brain
  dysconnectivity}.
\newblock \bibinfo{journal}{Nat Rev Neurosci} \bibinfo{volume}{20},
  \bibinfo{pages}{435--446}.
\bibitem[{Holm(1979)}]{holm1979simple}
\bibinfo{author}{Holm, S.}, \bibinfo{year}{1979}.
\newblock \bibinfo{title}{A simple sequentially rejective multiple test
  procedure}.
\newblock \bibinfo{journal}{Scandinavian journal of statistics} ,
  \bibinfo{pages}{65--70}.
\bibitem[{Hong et~al.(2014)Hong, Zalesky, Fornito, Park, Yang, Park, Song,
  Sohn, Shin, Kim et~al.}]{hong2014connectomic}
\bibinfo{author}{Hong, S.B.}, \bibinfo{author}{Zalesky, A.},
  \bibinfo{author}{Fornito, A.}, \bibinfo{author}{Park, S.},
  \bibinfo{author}{Yang, Y.H.}, \bibinfo{author}{Park, M.H.},
  \bibinfo{author}{Song, I.C.}, \bibinfo{author}{Sohn, C.H.},
  \bibinfo{author}{Shin, M.S.}, \bibinfo{author}{Kim, B.N.}, et~al.,
  \bibinfo{year}{2014}.
\newblock \bibinfo{title}{Connectomic disturbances in
  attention-deficit/hyperactivity disorder: a whole-brain tractography
  analysis}.
\newblock \bibinfo{journal}{Biological psychiatry} \bibinfo{volume}{76},
  \bibinfo{pages}{656--663}.
\bibitem[{Horwitz(2003)}]{horwitz2003elusive}
\bibinfo{author}{Horwitz, B.}, \bibinfo{year}{2003}.
\newblock \bibinfo{title}{The elusive concept of brain connectivity}.
\newblock \bibinfo{journal}{Neuroimage} \bibinfo{volume}{19},
  \bibinfo{pages}{466--470}.
\bibitem[{Hua et~al.(2008)Hua, Zhang, Wakana, Jiang, Li, Reich, Calabresi,
  Pekar, van Zijl and Mori}]{hua2008tract}
\bibinfo{author}{Hua, K.}, \bibinfo{author}{Zhang, J.},
  \bibinfo{author}{Wakana, S.}, \bibinfo{author}{Jiang, H.},
  \bibinfo{author}{Li, X.}, \bibinfo{author}{Reich, D.S.},
  \bibinfo{author}{Calabresi, P.A.}, \bibinfo{author}{Pekar, J.J.},
  \bibinfo{author}{van Zijl, P.C.}, \bibinfo{author}{Mori, S.},
  \bibinfo{year}{2008}.
\newblock \bibinfo{title}{Tract probability maps in stereotaxic spaces:
  analyses of white matter anatomy and tract-specific quantification}.
\newblock \bibinfo{journal}{Neuroimage} \bibinfo{volume}{39},
  \bibinfo{pages}{336--347}.
\bibitem[{Huang(2010)}]{huang2010delineating}
\bibinfo{author}{Huang, H.}, \bibinfo{year}{2010}.
\newblock \bibinfo{title}{Delineating neural structures of developmental human
  brains with diffusion tensor imaging}.
\newblock \bibinfo{journal}{TheScientificWorldJOURNAL} \bibinfo{volume}{10},
  \bibinfo{pages}{135--144}.
\bibitem[{Huang et~al.(2015)Huang, Shu, Mishra, Jeon, Chalak, Wang, Rollins,
  Gong, Cheng, Peng et~al.}]{huang2015development}
\bibinfo{author}{Huang, H.}, \bibinfo{author}{Shu, N.},
  \bibinfo{author}{Mishra, V.}, \bibinfo{author}{Jeon, T.},
  \bibinfo{author}{Chalak, L.}, \bibinfo{author}{Wang, Z.J.},
  \bibinfo{author}{Rollins, N.}, \bibinfo{author}{Gong, G.},
  \bibinfo{author}{Cheng, H.}, \bibinfo{author}{Peng, Y.}, et~al.,
  \bibinfo{year}{2015}.
\newblock \bibinfo{title}{Development of human brain structural networks
  through infancy and childhood}.
\newblock \bibinfo{journal}{Cerebral Cortex} \bibinfo{volume}{25},
  \bibinfo{pages}{1389--1404}.
\bibitem[{Huang et~al.(2009)Huang, Xue, Zhang, Ren, Richards, Yarowsky, Miller
  and Mori}]{huang2009anatomical}
\bibinfo{author}{Huang, H.}, \bibinfo{author}{Xue, R.}, \bibinfo{author}{Zhang,
  J.}, \bibinfo{author}{Ren, T.}, \bibinfo{author}{Richards, L.J.},
  \bibinfo{author}{Yarowsky, P.}, \bibinfo{author}{Miller, M.I.},
  \bibinfo{author}{Mori, S.}, \bibinfo{year}{2009}.
\newblock \bibinfo{title}{Anatomical characterization of human fetal brain
  development with diffusion tensor magnetic resonance imaging}.
\newblock \bibinfo{journal}{Journal of Neuroscience} \bibinfo{volume}{29},
  \bibinfo{pages}{4263--4273}.
\bibitem[{Huttenlocher(1984)}]{huttenlocher1984synapse}
\bibinfo{author}{Huttenlocher, P.R.}, \bibinfo{year}{1984}.
\newblock \bibinfo{title}{Synapse elimination and plasticity in developing
  human cerebral cortex.}
\newblock \bibinfo{journal}{American journal of mental deficiency}
  \bibinfo{volume}{88}, \bibinfo{pages}{488–496}.
\bibitem[{Hutter et~al.(2018)Hutter, Slator, Christiaens, Teixeira, Roberts,
  Jackson, Price, Malik and Hajnal}]{hutter2018integrated}
\bibinfo{author}{Hutter, J.}, \bibinfo{author}{Slator, P.J.},
  \bibinfo{author}{Christiaens, D.}, \bibinfo{author}{Teixeira, R.P.A.},
  \bibinfo{author}{Roberts, T.}, \bibinfo{author}{Jackson, L.},
  \bibinfo{author}{Price, A.N.}, \bibinfo{author}{Malik, S.},
  \bibinfo{author}{Hajnal, J.V.}, \bibinfo{year}{2018}.
\newblock \bibinfo{title}{Integrated and efficient diffusion-relaxometry using
  zebra}.
\newblock \bibinfo{journal}{Scientific reports} \bibinfo{volume}{8},
  \bibinfo{pages}{1--13}.
\bibitem[{Ikuta et~al.(2014)Ikuta, Shafritz, Bregman, Peters, Gruner, Malhotra
  and Szeszko}]{ikuta2014abnormal}
\bibinfo{author}{Ikuta, T.}, \bibinfo{author}{Shafritz, K.M.},
  \bibinfo{author}{Bregman, J.}, \bibinfo{author}{Peters, B.D.},
  \bibinfo{author}{Gruner, P.}, \bibinfo{author}{Malhotra, A.K.},
  \bibinfo{author}{Szeszko, P.R.}, \bibinfo{year}{2014}.
\newblock \bibinfo{title}{Abnormal cingulum bundle development in autism: a
  probabilistic tractography study}.
\newblock \bibinfo{journal}{Psychiatry Research: Neuroimaging}
  \bibinfo{volume}{221}, \bibinfo{pages}{63--68}.
\bibitem[{Imms et~al.(2021)Imms, Dom{\'\i}nguez~D, Burmester, Seguin, Clemente,
  Dhollander, Wilson, Poudel and Caeyenberghs}]{imms:2021}
\bibinfo{author}{Imms, P.}, \bibinfo{author}{Dom{\'\i}nguez~D, J.F.},
  \bibinfo{author}{Burmester, A.}, \bibinfo{author}{Seguin, C.},
  \bibinfo{author}{Clemente, A.}, \bibinfo{author}{Dhollander, T.},
  \bibinfo{author}{Wilson, P.H.}, \bibinfo{author}{Poudel, G.},
  \bibinfo{author}{Caeyenberghs, K.}, \bibinfo{year}{2021}.
\newblock \bibinfo{title}{Navigating the link between processing speed and
  network communication in the human brain}.
\newblock \bibinfo{journal}{Brain Struct Funct} \bibinfo{volume}{226},
  \bibinfo{pages}{1281--1302}.
\bibitem[{Ingalhalikar et~al.(2014)Ingalhalikar, Smith, Parker, Satterthwaite,
  Elliott, Ruparel, Hakonarson, Gur, Gur and Verma}]{ingalhalikar2014sex}
\bibinfo{author}{Ingalhalikar, M.}, \bibinfo{author}{Smith, A.},
  \bibinfo{author}{Parker, D.}, \bibinfo{author}{Satterthwaite, T.D.},
  \bibinfo{author}{Elliott, M.A.}, \bibinfo{author}{Ruparel, K.},
  \bibinfo{author}{Hakonarson, H.}, \bibinfo{author}{Gur, R.E.},
  \bibinfo{author}{Gur, R.C.}, \bibinfo{author}{Verma, R.},
  \bibinfo{year}{2014}.
\newblock \bibinfo{title}{Sex differences in the structural connectome of the
  human brain}.
\newblock \bibinfo{journal}{Proceedings of the National Academy of Sciences}
  \bibinfo{volume}{111}, \bibinfo{pages}{823--828}.
\bibitem[{Irimia et~al.(2020)Irimia, Fan, Chaudhari, Ngo, Zhang, Joshi and
  O'Donnell}]{irimia2020mapping}
\bibinfo{author}{Irimia, A.}, \bibinfo{author}{Fan, D.},
  \bibinfo{author}{Chaudhari, N.N.}, \bibinfo{author}{Ngo, V.},
  \bibinfo{author}{Zhang, F.}, \bibinfo{author}{Joshi, S.H.},
  \bibinfo{author}{O'Donnell, L.J.}, \bibinfo{year}{2020}.
\newblock \bibinfo{title}{Mapping cerebral connectivity changes after mild
  traumatic brain injury in older adults using diffusion tensor imaging and
  riemannian matching of elastic curves}, in: \bibinfo{booktitle}{2020 IEEE
  17th International Symposium on Biomedical Imaging (ISBI)},
  \bibinfo{organization}{IEEE}. pp. \bibinfo{pages}{1690--1693}.
\bibitem[{Jack~Jr and Holtzman(2013)}]{jack2013biomarker}
\bibinfo{author}{Jack~Jr, C.R.}, \bibinfo{author}{Holtzman, D.M.},
  \bibinfo{year}{2013}.
\newblock \bibinfo{title}{Biomarker modeling of alzheimer’s disease}.
\newblock \bibinfo{journal}{Neuron} \bibinfo{volume}{80},
  \bibinfo{pages}{1347--1358}.
\bibitem[{Jackowski et~al.(2005)Jackowski, Kao, Qiu, Constable and
  Staib}]{jackowski2005white}
\bibinfo{author}{Jackowski, M.}, \bibinfo{author}{Kao, C.Y.},
  \bibinfo{author}{Qiu, M.}, \bibinfo{author}{Constable, R.T.},
  \bibinfo{author}{Staib, L.H.}, \bibinfo{year}{2005}.
\newblock \bibinfo{title}{White matter tractography by anisotropic wavefront
  evolution and diffusion tensor imaging}.
\newblock \bibinfo{journal}{Medical image analysis} \bibinfo{volume}{9},
  \bibinfo{pages}{427--440}.
\bibitem[{Jakovcevski et~al.(2009)Jakovcevski, Filipovic, Mo, Rakic and
  Zecevic}]{jakovcevski2009oligodendrocyte}
\bibinfo{author}{Jakovcevski, I.}, \bibinfo{author}{Filipovic, R.},
  \bibinfo{author}{Mo, Z.}, \bibinfo{author}{Rakic, S.},
  \bibinfo{author}{Zecevic, N.}, \bibinfo{year}{2009}.
\newblock \bibinfo{title}{Oligodendrocyte development and the onset of
  myelination in the human fetal brain}.
\newblock \bibinfo{journal}{Frontiers in neuroanatomy} \bibinfo{volume}{3},
  \bibinfo{pages}{5}.
\bibitem[{Javadi et~al.(2017)Javadi, Nabavi, Giordano, Faghihzadeh and
  Samii}]{javadi2017evaluation}
\bibinfo{author}{Javadi, S.A.}, \bibinfo{author}{Nabavi, A.},
  \bibinfo{author}{Giordano, M.}, \bibinfo{author}{Faghihzadeh, E.},
  \bibinfo{author}{Samii, A.}, \bibinfo{year}{2017}.
\newblock \bibinfo{title}{Evaluation of diffusion tensor imaging--based
  tractography of the corticospinal tract: a correlative study with
  intraoperative magnetic resonance imaging and direct electrical subcortical
  stimulation}.
\newblock \bibinfo{journal}{Neurosurgery} \bibinfo{volume}{80},
  \bibinfo{pages}{287--299}.
\bibitem[{Jbabdi et~al.(2008)Jbabdi, Bellec, Toro, Daunizeau, Pelegrini-Issac
  and Benali}]{jbabdi_accurate_2008}
\bibinfo{author}{Jbabdi, S.}, \bibinfo{author}{Bellec, P.},
  \bibinfo{author}{Toro, R.}, \bibinfo{author}{Daunizeau, J.},
  \bibinfo{author}{Pelegrini-Issac, M.}, \bibinfo{author}{Benali, H.},
  \bibinfo{year}{2008}.
\newblock \bibinfo{title}{Accurate anisotropic fast marching for
  diffusion-based geodesic tractography}.
\newblock \bibinfo{journal}{Journal of Biomedical Imaging}
  \bibinfo{volume}{2008}, \bibinfo{pages}{1--12}.
\bibitem[{Jbabdi et~al.(2007)Jbabdi, Woolrich, Andersson and
  Behrens}]{jbabdi_bayesian_2007}
\bibinfo{author}{Jbabdi, S.}, \bibinfo{author}{Woolrich, M.},
  \bibinfo{author}{Andersson, J.}, \bibinfo{author}{Behrens, T.},
  \bibinfo{year}{2007}.
\newblock \bibinfo{title}{A {Bayesian} framework for global tractography}.
\newblock \bibinfo{journal}{NeuroImage} \bibinfo{volume}{37},
  \bibinfo{pages}{116--129}.
\bibitem[{Jelescu and Budde(2017)}]{jelescu2017design}
\bibinfo{author}{Jelescu, I.O.}, \bibinfo{author}{Budde, M.D.},
  \bibinfo{year}{2017}.
\newblock \bibinfo{title}{Design and validation of diffusion {MRI} models of
  white matter}.
\newblock \bibinfo{journal}{Frontiers in physics} \bibinfo{volume}{5},
  \bibinfo{pages}{61}.
\bibitem[{Jellison et~al.(2004)Jellison, Field, Medow, Lazar, Salamat and
  Alexander}]{jellison2004diffusion}
\bibinfo{author}{Jellison, B.J.}, \bibinfo{author}{Field, A.S.},
  \bibinfo{author}{Medow, J.}, \bibinfo{author}{Lazar, M.},
  \bibinfo{author}{Salamat, M.S.}, \bibinfo{author}{Alexander, A.L.},
  \bibinfo{year}{2004}.
\newblock \bibinfo{title}{Diffusion tensor imaging of cerebral white matter: a
  pictorial review of physics, fiber tract anatomy, and tumor imaging
  patterns}.
\newblock \bibinfo{journal}{American Journal of Neuroradiology}
  \bibinfo{volume}{25}, \bibinfo{pages}{356--369}.
\bibitem[{Jenkinson et~al.(2012)Jenkinson, Beckmann, Behrens, Woolrich and
  Smith}]{jenkinson2012fsl}
\bibinfo{author}{Jenkinson, M.}, \bibinfo{author}{Beckmann, C.F.},
  \bibinfo{author}{Behrens, T.E.}, \bibinfo{author}{Woolrich, M.W.},
  \bibinfo{author}{Smith, S.M.}, \bibinfo{year}{2012}.
\newblock \bibinfo{title}{Fsl}.
\newblock \bibinfo{journal}{Neuroimage} \bibinfo{volume}{62},
  \bibinfo{pages}{782--790}.
\bibitem[{Jensen and Helpern(2010)}]{Jensen2010a}
\bibinfo{author}{Jensen, J.H.}, \bibinfo{author}{Helpern, J.A.},
  \bibinfo{year}{2010}.
\newblock \bibinfo{title}{{MRI Quantification of Non-Gaussian Water Diffusion
  by Kurtosis Analysis}}.
\newblock \bibinfo{journal}{NMR Biomed.} \bibinfo{volume}{23},
  \bibinfo{pages}{698--710}.
\bibitem[{Jeurissen et~al.(2019)Jeurissen, Descoteaux, Mori and
  Leemans}]{jeurissen2019diffusion}
\bibinfo{author}{Jeurissen, B.}, \bibinfo{author}{Descoteaux, M.},
  \bibinfo{author}{Mori, S.}, \bibinfo{author}{Leemans, A.},
  \bibinfo{year}{2019}.
\newblock \bibinfo{title}{Diffusion {MRI} fiber tractography of the brain}.
\newblock \bibinfo{journal}{NMR in Biomedicine} \bibinfo{volume}{32},
  \bibinfo{pages}{e3785}.
\bibitem[{Jeurissen et~al.(2014)Jeurissen, Tournier, Dhollander, Connelly and
  Sijbers}]{jeurissen2014multi}
\bibinfo{author}{Jeurissen, B.}, \bibinfo{author}{Tournier, J.D.},
  \bibinfo{author}{Dhollander, T.}, \bibinfo{author}{Connelly, A.},
  \bibinfo{author}{Sijbers, J.}, \bibinfo{year}{2014}.
\newblock \bibinfo{title}{Multi-tissue constrained spherical deconvolution for
  improved analysis of multi-shell diffusion {MRI} data}.
\newblock \bibinfo{journal}{NeuroImage} \bibinfo{volume}{103},
  \bibinfo{pages}{411--426}.
\bibitem[{Ji et~al.(2019)Ji, Guevara, Guevara, Grigis, Labra, Sarrazin,
  Hamdani, Bellivier, Delavest, Leboyer et~al.}]{ji2019increased}
\bibinfo{author}{Ji, E.}, \bibinfo{author}{Guevara, P.},
  \bibinfo{author}{Guevara, M.}, \bibinfo{author}{Grigis, A.},
  \bibinfo{author}{Labra, N.}, \bibinfo{author}{Sarrazin, S.},
  \bibinfo{author}{Hamdani, N.}, \bibinfo{author}{Bellivier, F.},
  \bibinfo{author}{Delavest, M.}, \bibinfo{author}{Leboyer, M.}, et~al.,
  \bibinfo{year}{2019}.
\newblock \bibinfo{title}{Increased and decreased superficial white matter
  structural connectivity in schizophrenia and bipolar disorder}.
\newblock \bibinfo{journal}{Schizophrenia bulletin} \bibinfo{volume}{45},
  \bibinfo{pages}{1367--1378}.
\bibitem[{Jiang et~al.(2006)Jiang, Van~Zijl, Kim, Pearlson and
  Mori}]{jiang2006dtistudio}
\bibinfo{author}{Jiang, H.}, \bibinfo{author}{Van~Zijl, P.C.},
  \bibinfo{author}{Kim, J.}, \bibinfo{author}{Pearlson, G.D.},
  \bibinfo{author}{Mori, S.}, \bibinfo{year}{2006}.
\newblock \bibinfo{title}{Dtistudio: resource program for diffusion tensor
  computation and fiber bundle tracking}.
\newblock \bibinfo{journal}{Computer methods and programs in biomedicine}
  \bibinfo{volume}{81}, \bibinfo{pages}{106--116}.
\bibitem[{Jones(2010)}]{jones2010challenges}
\bibinfo{author}{Jones, D.K.}, \bibinfo{year}{2010}.
\newblock \bibinfo{title}{Challenges and limitations of quantifying brain
  connectivity in vivo with diffusion {MRI}}.
\newblock \bibinfo{journal}{Imaging in Medicine} \bibinfo{volume}{2},
  \bibinfo{pages}{341}.
\bibitem[{Jones et~al.(2006)Jones, Catani, Pierpaoli, Reeves, Shergill,
  O'Sullivan, Golesworthy, McGuire, Horsfield, Simmons et~al.}]{jones2006age}
\bibinfo{author}{Jones, D.K.}, \bibinfo{author}{Catani, M.},
  \bibinfo{author}{Pierpaoli, C.}, \bibinfo{author}{Reeves, S.J.},
  \bibinfo{author}{Shergill, S.S.}, \bibinfo{author}{O'Sullivan, M.},
  \bibinfo{author}{Golesworthy, P.}, \bibinfo{author}{McGuire, P.},
  \bibinfo{author}{Horsfield, M.A.}, \bibinfo{author}{Simmons, A.}, et~al.,
  \bibinfo{year}{2006}.
\newblock \bibinfo{title}{Age effects on diffusion tensor magnetic resonance
  imaging tractography measures of frontal cortex connections in
  schizophrenia}.
\newblock \bibinfo{journal}{Human brain mapping} \bibinfo{volume}{27},
  \bibinfo{pages}{230--238}.
\bibitem[{Jones et~al.(2013)Jones, Kn{\"{o}}sche and Turner}]{Jones2013}
\bibinfo{author}{Jones, D.K.}, \bibinfo{author}{Kn{\"{o}}sche, T.R.},
  \bibinfo{author}{Turner, R.}, \bibinfo{year}{2013}.
\newblock \bibinfo{title}{{White matter integrity, fiber count, and other
  fallacies: The do's and don'ts of diffusion MRI}}.
\newblock \DOIprefix\doi{10.1016/j.neuroimage.2012.06.081},
  \href{http://arxiv.org/abs/9905108}{{\tt arXiv:9905108}}.
\bibitem[{Jones et~al.(2005)Jones, Travis, Eden, Pierpaoli and
  Basser}]{jones2005pasta}
\bibinfo{author}{Jones, D.K.}, \bibinfo{author}{Travis, A.R.},
  \bibinfo{author}{Eden, G.}, \bibinfo{author}{Pierpaoli, C.},
  \bibinfo{author}{Basser, P.J.}, \bibinfo{year}{2005}.
\newblock \bibinfo{title}{Pasta: pointwise assessment of streamline
  tractography attributes}.
\newblock \bibinfo{journal}{Magnetic Resonance in Medicine: An Official Journal
  of the International Society for Magnetic Resonance in Medicine}
  \bibinfo{volume}{53}, \bibinfo{pages}{1462--1467}.
\bibitem[{Kaden et~al.(2016)Kaden, Kelm, Carson, Does and
  Alexander}]{Kaden2016_SMTmultiComp}
\bibinfo{author}{Kaden, E.}, \bibinfo{author}{Kelm, N.D.},
  \bibinfo{author}{Carson, R.P.}, \bibinfo{author}{Does, M.D.},
  \bibinfo{author}{Alexander, D.C.}, \bibinfo{year}{2016}.
\newblock \bibinfo{title}{{Multi-compartment microscopic diffusion imaging}}.
\newblock \bibinfo{journal}{NeuroImage} \bibinfo{volume}{139},
  \bibinfo{pages}{346--359}.
\newblock \DOIprefix\doi{https://doi.org/10.1016/j.neuroimage.2016.06.002}.
\bibitem[{Karlsgodt et~al.(2008)Karlsgodt, van Erp, Poldrack, Bearden,
  Nuechterlein and Cannon}]{karlsgodt2008diffusion}
\bibinfo{author}{Karlsgodt, K.H.}, \bibinfo{author}{van Erp, T.G.},
  \bibinfo{author}{Poldrack, R.A.}, \bibinfo{author}{Bearden, C.E.},
  \bibinfo{author}{Nuechterlein, K.H.}, \bibinfo{author}{Cannon, T.D.},
  \bibinfo{year}{2008}.
\newblock \bibinfo{title}{Diffusion tensor imaging of the superior longitudinal
  fasciculus and working memory in recent-onset schizophrenia}.
\newblock \bibinfo{journal}{Biological psychiatry} \bibinfo{volume}{63},
  \bibinfo{pages}{512--518}.
\bibitem[{Kelly et~al.(2018)Kelly, Jahanshad, Zalesky, Kochunov, Agartz,
  Alloza, Andreassen, Arango, Banaj, Bouix et~al.}]{kelly2018widespread}
\bibinfo{author}{Kelly, S.}, \bibinfo{author}{Jahanshad, N.},
  \bibinfo{author}{Zalesky, A.}, \bibinfo{author}{Kochunov, P.},
  \bibinfo{author}{Agartz, I.}, \bibinfo{author}{Alloza, C.},
  \bibinfo{author}{Andreassen, O.}, \bibinfo{author}{Arango, C.},
  \bibinfo{author}{Banaj, N.}, \bibinfo{author}{Bouix, S.}, et~al.,
  \bibinfo{year}{2018}.
\newblock \bibinfo{title}{Widespread white matter microstructural differences
  in schizophrenia across 4322 individuals: results from the enigma
  schizophrenia dti working group}.
\newblock \bibinfo{journal}{Molecular psychiatry} \bibinfo{volume}{23},
  \bibinfo{pages}{1261--1269}.
\bibitem[{Kern et~al.(2011)Kern, Sarcona, Montag, Giesser and
  Sicotte}]{kern2011corpus}
\bibinfo{author}{Kern, K.C.}, \bibinfo{author}{Sarcona, J.},
  \bibinfo{author}{Montag, M.}, \bibinfo{author}{Giesser, B.S.},
  \bibinfo{author}{Sicotte, N.L.}, \bibinfo{year}{2011}.
\newblock \bibinfo{title}{Corpus callosal diffusivity predicts motor impairment
  in relapsing--remitting multiple sclerosis: a tbss and tractography study}.
\newblock \bibinfo{journal}{Neuroimage} \bibinfo{volume}{55},
  \bibinfo{pages}{1169--1177}.
\bibitem[{Khan et~al.(2019)Khan, Vasung, Marami, Rollins, Afacan, Ortinau,
  Yang, Warfield and Gholipour}]{khan2019fetal}
\bibinfo{author}{Khan, S.}, \bibinfo{author}{Vasung, L.},
  \bibinfo{author}{Marami, B.}, \bibinfo{author}{Rollins, C.K.},
  \bibinfo{author}{Afacan, O.}, \bibinfo{author}{Ortinau, C.M.},
  \bibinfo{author}{Yang, E.}, \bibinfo{author}{Warfield, S.K.},
  \bibinfo{author}{Gholipour, A.}, \bibinfo{year}{2019}.
\newblock \bibinfo{title}{Fetal brain growth portrayed by a spatiotemporal
  diffusion tensor {MRI} atlas computed from in utero images}.
\newblock \bibinfo{journal}{NeuroImage} \bibinfo{volume}{185},
  \bibinfo{pages}{593--608}.
\bibitem[{Klauser et~al.(2017)Klauser, Baker, Cropley, Bousman, Fornito,
  Cocchi, Fullerton, Rasser, Schall, Henskens et~al.}]{klauser2017white}
\bibinfo{author}{Klauser, P.}, \bibinfo{author}{Baker, S.T.},
  \bibinfo{author}{Cropley, V.L.}, \bibinfo{author}{Bousman, C.},
  \bibinfo{author}{Fornito, A.}, \bibinfo{author}{Cocchi, L.},
  \bibinfo{author}{Fullerton, J.M.}, \bibinfo{author}{Rasser, P.},
  \bibinfo{author}{Schall, U.}, \bibinfo{author}{Henskens, F.}, et~al.,
  \bibinfo{year}{2017}.
\newblock \bibinfo{title}{White matter disruptions in schizophrenia are
  spatially widespread and topologically converge on brain network hubs}.
\newblock \bibinfo{journal}{Schizophrenia bulletin} \bibinfo{volume}{43},
  \bibinfo{pages}{425--435}.
\bibitem[{Kocevar et~al.(2016)Kocevar, Stamile, Hannoun, Cotton, Vukusic,
  Durand-Dubief and Sappey-Marinier}]{kocevar2016graph}
\bibinfo{author}{Kocevar, G.}, \bibinfo{author}{Stamile, C.},
  \bibinfo{author}{Hannoun, S.}, \bibinfo{author}{Cotton, F.},
  \bibinfo{author}{Vukusic, S.}, \bibinfo{author}{Durand-Dubief, F.},
  \bibinfo{author}{Sappey-Marinier, D.}, \bibinfo{year}{2016}.
\newblock \bibinfo{title}{Graph theory-based brain connectivity for automatic
  classification of multiple sclerosis clinical courses}.
\newblock \bibinfo{journal}{Frontiers in neuroscience} \bibinfo{volume}{10},
  \bibinfo{pages}{478}.
\bibitem[{Koch et~al.(2014)Koch, Ree{\ss}, Rus, Zimmer and
  Zaudig}]{koch2014diffusion}
\bibinfo{author}{Koch, K.}, \bibinfo{author}{Ree{\ss}, T.J.},
  \bibinfo{author}{Rus, O.G.}, \bibinfo{author}{Zimmer, C.},
  \bibinfo{author}{Zaudig, M.}, \bibinfo{year}{2014}.
\newblock \bibinfo{title}{Diffusion tensor imaging (dti) studies in patients
  with obsessive-compulsive disorder (ocd): a review}.
\newblock \bibinfo{journal}{Journal of psychiatric research}
  \bibinfo{volume}{54}, \bibinfo{pages}{26--35}.
\bibitem[{Korgaonkar et~al.(2014)Korgaonkar, Fornito, Williams and
  Grieve}]{korgaonkar2014abnormal}
\bibinfo{author}{Korgaonkar, M.S.}, \bibinfo{author}{Fornito, A.},
  \bibinfo{author}{Williams, L.M.}, \bibinfo{author}{Grieve, S.M.},
  \bibinfo{year}{2014}.
\newblock \bibinfo{title}{Abnormal structural networks characterize major
  depressive disorder: a connectome analysis}.
\newblock \bibinfo{journal}{Biological psychiatry} \bibinfo{volume}{76},
  \bibinfo{pages}{567--574}.
\bibitem[{Kovanlikaya et~al.(2011)Kovanlikaya, Firat, Kovanlikaya, Ulu{\u{g}},
  Cihangiroglu, John, Bingol and Ture}]{kovanlikaya2011assessment}
\bibinfo{author}{Kovanlikaya, I.}, \bibinfo{author}{Firat, Z.},
  \bibinfo{author}{Kovanlikaya, A.}, \bibinfo{author}{Ulu{\u{g}}, A.M.},
  \bibinfo{author}{Cihangiroglu, M.M.}, \bibinfo{author}{John, M.},
  \bibinfo{author}{Bingol, C.A.}, \bibinfo{author}{Ture, U.},
  \bibinfo{year}{2011}.
\newblock \bibinfo{title}{Assessment of the corticospinal tract alterations
  before and after resection of brainstem lesions using diffusion tensor
  imaging (dti) and tractography at 3t}.
\newblock \bibinfo{journal}{European journal of radiology}
  \bibinfo{volume}{77}, \bibinfo{pages}{383--391}.
\bibitem[{Kreher et~al.(2008)Kreher, Mader and Kiselev}]{kreher_gibbs_2008}
\bibinfo{author}{Kreher, B.W.}, \bibinfo{author}{Mader, I.},
  \bibinfo{author}{Kiselev, V.G.}, \bibinfo{year}{2008}.
\newblock \bibinfo{title}{Gibbs tracking: {A} novel approach for the
  reconstruction of neuronal pathways}.
\newblock \bibinfo{journal}{Magnetic Resonance in Medicine}
  \bibinfo{volume}{60}, \bibinfo{pages}{953--963}.
\bibitem[{Kreilkamp et~al.(2019)Kreilkamp, Lisanti, Glenn, Wieshmann, Das,
  Marson and Keller}]{kreilkamp2019comparison}
\bibinfo{author}{Kreilkamp, B.A.}, \bibinfo{author}{Lisanti, L.},
  \bibinfo{author}{Glenn, G.R.}, \bibinfo{author}{Wieshmann, U.C.},
  \bibinfo{author}{Das, K.}, \bibinfo{author}{Marson, A.G.},
  \bibinfo{author}{Keller, S.S.}, \bibinfo{year}{2019}.
\newblock \bibinfo{title}{Comparison of manual and automated fiber
  quantification tractography in patients with temporal lobe epilepsy}.
\newblock \bibinfo{journal}{NeuroImage: Clinical} \bibinfo{volume}{24},
  \bibinfo{pages}{102024}.
\bibitem[{Kumar et~al.(2019)Kumar, Siddiqi and Desrosiers}]{kumar2019white}
\bibinfo{author}{Kumar, K.}, \bibinfo{author}{Siddiqi, K.},
  \bibinfo{author}{Desrosiers, C.}, \bibinfo{year}{2019}.
\newblock \bibinfo{title}{White matter fiber analysis using kernel dictionary
  learning and sparsity priors}.
\newblock \bibinfo{journal}{Pattern Recognition} \bibinfo{volume}{95},
  \bibinfo{pages}{83--95}.
\bibitem[{Kunimatsu et~al.(2004)Kunimatsu, Aoki, Masutani, Abe, Hayashi, Mori,
  Masumoto and Ohtomo}]{kunimatsu2004optimal}
\bibinfo{author}{Kunimatsu, A.}, \bibinfo{author}{Aoki, S.},
  \bibinfo{author}{Masutani, Y.}, \bibinfo{author}{Abe, O.},
  \bibinfo{author}{Hayashi, N.}, \bibinfo{author}{Mori, H.},
  \bibinfo{author}{Masumoto, T.}, \bibinfo{author}{Ohtomo, K.},
  \bibinfo{year}{2004}.
\newblock \bibinfo{title}{The optimal trackability threshold of fractional
  anisotropy for diffusion tensor tractography of the corticospinal tract}.
\newblock \bibinfo{journal}{Magnetic resonance in medical sciences}
  \bibinfo{volume}{3}, \bibinfo{pages}{11--17}.
\bibitem[{Labra et~al.(2017)Labra, Guevara, Duclap, Houenou, Poupon, Mangin and
  Figueroa}]{labra2017fast}
\bibinfo{author}{Labra, N.}, \bibinfo{author}{Guevara, P.},
  \bibinfo{author}{Duclap, D.}, \bibinfo{author}{Houenou, J.},
  \bibinfo{author}{Poupon, C.}, \bibinfo{author}{Mangin, J.F.},
  \bibinfo{author}{Figueroa, M.}, \bibinfo{year}{2017}.
\newblock \bibinfo{title}{Fast automatic segmentation of white matter
  streamlines based on a multi-subject bundle atlas}.
\newblock \bibinfo{journal}{Neuroinformatics} \bibinfo{volume}{15},
  \bibinfo{pages}{71--86}.
\bibitem[{Lam et~al.(2018)Lam, Belhomme, Ferrall, Patterson, Styner and
  Prieto}]{lam2018trafic}
\bibinfo{author}{Lam, P.D.N.}, \bibinfo{author}{Belhomme, G.},
  \bibinfo{author}{Ferrall, J.}, \bibinfo{author}{Patterson, B.},
  \bibinfo{author}{Styner, M.}, \bibinfo{author}{Prieto, J.C.},
  \bibinfo{year}{2018}.
\newblock \bibinfo{title}{Trafic: fiber tract classification using deep
  learning}, in: \bibinfo{booktitle}{Medical Imaging 2018: Image Processing},
  \bibinfo{organization}{International Society for Optics and Photonics}. p.
  \bibinfo{pages}{1057412}.
\bibitem[{Langen et~al.(2012)Langen, Leemans, Johnston, Ecker, Daly, Murphy,
  dell’Acqua, Durston, Murphy, Consortium et~al.}]{langen2012fronto}
\bibinfo{author}{Langen, M.}, \bibinfo{author}{Leemans, A.},
  \bibinfo{author}{Johnston, P.}, \bibinfo{author}{Ecker, C.},
  \bibinfo{author}{Daly, E.}, \bibinfo{author}{Murphy, C.M.},
  \bibinfo{author}{dell’Acqua, F.}, \bibinfo{author}{Durston, S.},
  \bibinfo{author}{Murphy, D.G.}, \bibinfo{author}{Consortium, A.}, et~al.,
  \bibinfo{year}{2012}.
\newblock \bibinfo{title}{Fronto-striatal circuitry and inhibitory control in
  autism: findings from diffusion tensor imaging tractography}.
\newblock \bibinfo{journal}{Cortex} \bibinfo{volume}{48},
  \bibinfo{pages}{183--193}.
\bibitem[{Lawes et~al.(2008)Lawes, Barrick, Murugam, Spierings, Evans, Song and
  Clark}]{lawes2008atlas}
\bibinfo{author}{Lawes, I.N.C.}, \bibinfo{author}{Barrick, T.R.},
  \bibinfo{author}{Murugam, V.}, \bibinfo{author}{Spierings, N.},
  \bibinfo{author}{Evans, D.R.}, \bibinfo{author}{Song, M.},
  \bibinfo{author}{Clark, C.A.}, \bibinfo{year}{2008}.
\newblock \bibinfo{title}{Atlas-based segmentation of white matter tracts of
  the human brain using diffusion tensor tractography and comparison with
  classical dissection}.
\newblock \bibinfo{journal}{Neuroimage} \bibinfo{volume}{39},
  \bibinfo{pages}{62--79}.
\bibitem[{Lazar and Alexander(2005)}]{lazar2005bootstrap}
\bibinfo{author}{Lazar, M.}, \bibinfo{author}{Alexander, A.L.},
  \bibinfo{year}{2005}.
\newblock \bibinfo{title}{Bootstrap white matter tractography (boot-trac)}.
\newblock \bibinfo{journal}{NeuroImage} \bibinfo{volume}{24},
  \bibinfo{pages}{524--532}.
\bibitem[{Le~Bihan and Johansen-Berg(2012)}]{le2012diffusion}
\bibinfo{author}{Le~Bihan, D.}, \bibinfo{author}{Johansen-Berg, H.},
  \bibinfo{year}{2012}.
\newblock \bibinfo{title}{Diffusion {MRI} at 25: exploring brain tissue
  structure and function}.
\newblock \bibinfo{journal}{Neuroimage} \bibinfo{volume}{61},
  \bibinfo{pages}{324--341}.
\bibitem[{Lebel and Beaulieu(2011)}]{lebel2011longitudinal}
\bibinfo{author}{Lebel, C.}, \bibinfo{author}{Beaulieu, C.},
  \bibinfo{year}{2011}.
\newblock \bibinfo{title}{Longitudinal development of human brain wiring
  continues from childhood into adulthood}.
\newblock \bibinfo{journal}{Journal of Neuroscience} \bibinfo{volume}{31},
  \bibinfo{pages}{10937--10947}.
\bibitem[{Lebel et~al.(2010)Lebel, Caverhill-Godkewitsch and
  Beaulieu}]{lebel2010age}
\bibinfo{author}{Lebel, C.}, \bibinfo{author}{Caverhill-Godkewitsch, S.},
  \bibinfo{author}{Beaulieu, C.}, \bibinfo{year}{2010}.
\newblock \bibinfo{title}{Age-related regional variations of the corpus
  callosum identified by diffusion tensor tractography}.
\newblock \bibinfo{journal}{Neuroimage} \bibinfo{volume}{52},
  \bibinfo{pages}{20--31}.
\bibitem[{Lebel and Deoni(2018)}]{lebel2018development}
\bibinfo{author}{Lebel, C.}, \bibinfo{author}{Deoni, S.}, \bibinfo{year}{2018}.
\newblock \bibinfo{title}{The development of brain white matter
  microstructure}.
\newblock \bibinfo{journal}{Neuroimage} \bibinfo{volume}{182},
  \bibinfo{pages}{207--218}.
\bibitem[{Lebel et~al.(2012)Lebel, Gee, Camicioli, Wieler, Martin and
  Beaulieu}]{lebel2012diffusion}
\bibinfo{author}{Lebel, C.}, \bibinfo{author}{Gee, M.},
  \bibinfo{author}{Camicioli, R.}, \bibinfo{author}{Wieler, M.},
  \bibinfo{author}{Martin, W.}, \bibinfo{author}{Beaulieu, C.},
  \bibinfo{year}{2012}.
\newblock \bibinfo{title}{Diffusion tensor imaging of white matter tract
  evolution over the lifespan}.
\newblock \bibinfo{journal}{Neuroimage} \bibinfo{volume}{60},
  \bibinfo{pages}{340--352}.
\bibitem[{Lebel et~al.(2019)Lebel, Treit and Beaulieu}]{lebel2019review}
\bibinfo{author}{Lebel, C.}, \bibinfo{author}{Treit, S.},
  \bibinfo{author}{Beaulieu, C.}, \bibinfo{year}{2019}.
\newblock \bibinfo{title}{A review of diffusion {MRI} of typical white matter
  development from early childhood to young adulthood}.
\newblock \bibinfo{journal}{NMR in Biomedicine} \bibinfo{volume}{32},
  \bibinfo{pages}{e3778}.
\bibitem[{Lee et~al.(2020)Lee, Hyun, Lee, Choi, Shin, Min, Nam, Kim and
  Oh}]{Lee2020}
\bibinfo{author}{Lee, J.}, \bibinfo{author}{Hyun, J.W.}, \bibinfo{author}{Lee,
  J.}, \bibinfo{author}{Choi, E.J.}, \bibinfo{author}{Shin, H.G.},
  \bibinfo{author}{Min, K.}, \bibinfo{author}{Nam, Y.}, \bibinfo{author}{Kim,
  H.J.}, \bibinfo{author}{Oh, S.H.}, \bibinfo{year}{2020}.
\newblock \bibinfo{title}{{So You Want to Image Myelin Using MRI: An Overview
  and Practical Guide for Myelin Water Imaging}}.
\newblock \bibinfo{journal}{Journal of Magnetic Resonance Imaging}
  \bibinfo{volume}{n/a}.
\newblock \DOIprefix\doi{https://doi.org/10.1002/jmri.27059}.
\bibitem[{Leemans et~al.(2009)Leemans, Jeurissen, Sijbers and
  Jones}]{leemans2009exploredti}
\bibinfo{author}{Leemans, A.}, \bibinfo{author}{Jeurissen, B.},
  \bibinfo{author}{Sijbers, J.}, \bibinfo{author}{Jones, D.},
  \bibinfo{year}{2009}.
\newblock \bibinfo{title}{Exploredti: a graphical toolbox for processing,
  analyzing, and visualizing diffusion mr data}, in: \bibinfo{booktitle}{Proc
  Intl Soc Mag Reson Med}, p. \bibinfo{pages}{3537}.
\bibitem[{Lemkaddem et~al.(2014)Lemkaddem, Skiöldebrand, Dal~Palú, Thiran and
  Daducci}]{lemkaddem_global_2014}
\bibinfo{author}{Lemkaddem, A.}, \bibinfo{author}{Skiöldebrand, D.},
  \bibinfo{author}{Dal~Palú, A.}, \bibinfo{author}{Thiran, J.P.},
  \bibinfo{author}{Daducci, A.}, \bibinfo{year}{2014}.
\newblock \bibinfo{title}{Global tractography with embedded anatomical priors
  for quantitative connectivity analysis}.
\newblock \bibinfo{journal}{Frontiers in Neurology} \bibinfo{volume}{5},
  \bibinfo{pages}{232}.
\bibitem[{Levitt et~al.(2020)Levitt, Nestor, Kubicki, Lyall, Zhang,
  Riklin-Raviv, O'Donnell, McCarley, Shenton and Rathi}]{levitt2020miswiring}
\bibinfo{author}{Levitt, J.J.}, \bibinfo{author}{Nestor, P.G.},
  \bibinfo{author}{Kubicki, M.}, \bibinfo{author}{Lyall, A.E.},
  \bibinfo{author}{Zhang, F.}, \bibinfo{author}{Riklin-Raviv, T.},
  \bibinfo{author}{O'Donnell, L.J.}, \bibinfo{author}{McCarley, R.W.},
  \bibinfo{author}{Shenton, M.E.}, \bibinfo{author}{Rathi, Y.},
  \bibinfo{year}{2020}.
\newblock \bibinfo{title}{Miswiring of frontostriatal projections in
  schizophrenia}.
\newblock \bibinfo{journal}{Schizophrenia bulletin} \bibinfo{volume}{46},
  \bibinfo{pages}{990--998}.
\bibitem[{Li et~al.(2020a)Li, de~Groot, Steketee, Meijboom, Smits, Vernooij,
  Ikram, Liu, Niessen and Bron}]{li2020neuro4neuro}
\bibinfo{author}{Li, B.}, \bibinfo{author}{de~Groot, M.},
  \bibinfo{author}{Steketee, R.M.}, \bibinfo{author}{Meijboom, R.},
  \bibinfo{author}{Smits, M.}, \bibinfo{author}{Vernooij, M.W.},
  \bibinfo{author}{Ikram, M.A.}, \bibinfo{author}{Liu, J.},
  \bibinfo{author}{Niessen, W.J.}, \bibinfo{author}{Bron, E.E.},
  \bibinfo{year}{2020}a.
\newblock \bibinfo{title}{Neuro4neuro: A neural network approach for neural
  tract segmentation using large-scale population-based diffusion imaging}.
\newblock \bibinfo{journal}{NeuroImage} , \bibinfo{pages}{116993}.
\bibitem[{Li et~al.(2010)Li, Xue, Guo, Liu, Hunter and Wong}]{li2010hybrid}
\bibinfo{author}{Li, H.}, \bibinfo{author}{Xue, Z.}, \bibinfo{author}{Guo, L.},
  \bibinfo{author}{Liu, T.}, \bibinfo{author}{Hunter, J.},
  \bibinfo{author}{Wong, S.T.}, \bibinfo{year}{2010}.
\newblock \bibinfo{title}{A hybrid approach to automatic clustering of white
  matter fibers}.
\newblock \bibinfo{journal}{NeuroImage} \bibinfo{volume}{49},
  \bibinfo{pages}{1249--1258}.
\bibitem[{Li et~al.(2012)Li, Rilling, Preuss, Glasser and Hu}]{li_effects_2012}
\bibinfo{author}{Li, L.}, \bibinfo{author}{Rilling, J.K.},
  \bibinfo{author}{Preuss, T.M.}, \bibinfo{author}{Glasser, M.F.},
  \bibinfo{author}{Hu, X.}, \bibinfo{year}{2012}.
\newblock \bibinfo{title}{The effects of connection reconstruction method on
  the interregional connectivity of brain networks via diffusion tractography}.
\newblock \bibinfo{journal}{Human Brain Mapping} \bibinfo{volume}{33},
  \bibinfo{pages}{1894--1913}.
\bibitem[{Li et~al.(2014)Li, Ratnanather, Miller and
  Mori}]{li_knowledge-based_2014}
\bibinfo{author}{Li, M.}, \bibinfo{author}{Ratnanather, J.T.},
  \bibinfo{author}{Miller, M.I.}, \bibinfo{author}{Mori, S.},
  \bibinfo{year}{2014}.
\newblock \bibinfo{title}{Knowledge-based automated reconstruction of human
  brain white matter tracts using a path-finding approach with dynamic
  programming}.
\newblock \bibinfo{journal}{NeuroImage} \bibinfo{volume}{88},
  \bibinfo{pages}{271--281}.
\bibitem[{Li et~al.(2020b)Li, Chen, Guo, Zeng and Feng}]{li2020two}
\bibinfo{author}{Li, S.}, \bibinfo{author}{Chen, Z.}, \bibinfo{author}{Guo,
  W.}, \bibinfo{author}{Zeng, Q.}, \bibinfo{author}{Feng, Y.},
  \bibinfo{year}{2020}b.
\newblock \bibinfo{title}{Two parallel stages deep learning network for
  anterior visual pathway segmentation}, in: \bibinfo{booktitle}{CDMRI}, pp.
  \bibinfo{pages}{1--1}.
\bibitem[{Li et~al.(2020c)Li, Wang, Wang, Huang, Chen, Xu, Zhang, Chen, Li, Wei
  et~al.}]{li2020age}
\bibinfo{author}{Li, X.}, \bibinfo{author}{Wang, Y.}, \bibinfo{author}{Wang,
  W.}, \bibinfo{author}{Huang, W.}, \bibinfo{author}{Chen, K.},
  \bibinfo{author}{Xu, K.}, \bibinfo{author}{Zhang, J.}, \bibinfo{author}{Chen,
  Y.}, \bibinfo{author}{Li, H.}, \bibinfo{author}{Wei, D.}, et~al.,
  \bibinfo{year}{2020}c.
\newblock \bibinfo{title}{Age-related decline in the topological efficiency of
  the brain structural connectome and cognitive aging}.
\newblock \bibinfo{journal}{Cerebral Cortex} \bibinfo{volume}{30},
  \bibinfo{pages}{4651--4661}.
\bibitem[{Li et~al.(2013)Li, Jewells, Kim, Chen, Moon, Armao, Troiani,
  Markovic-Plese, Lin and Shen}]{li2013diffusion}
\bibinfo{author}{Li, Y.}, \bibinfo{author}{Jewells, V.}, \bibinfo{author}{Kim,
  M.}, \bibinfo{author}{Chen, Y.}, \bibinfo{author}{Moon, A.},
  \bibinfo{author}{Armao, D.}, \bibinfo{author}{Troiani, L.},
  \bibinfo{author}{Markovic-Plese, S.}, \bibinfo{author}{Lin, W.},
  \bibinfo{author}{Shen, D.}, \bibinfo{year}{2013}.
\newblock \bibinfo{title}{Diffusion tensor imaging based network analysis
  detects alterations of neuroconnectivity in patients with clinically early
  relapsing-remitting multiple sclerosis}.
\newblock \bibinfo{journal}{Human brain mapping} \bibinfo{volume}{34},
  \bibinfo{pages}{3376--3391}.
\bibitem[{Lin et~al.(2005)Lin, Tench, Morgan, Niepel and
  Constantinescu}]{lin2005importance}
\bibinfo{author}{Lin, X.}, \bibinfo{author}{Tench, C.R.},
  \bibinfo{author}{Morgan, P.S.}, \bibinfo{author}{Niepel, G.},
  \bibinfo{author}{Constantinescu, C.S.}, \bibinfo{year}{2005}.
\newblock \bibinfo{title}{‘importance sampling’in ms: use of diffusion
  tensor tractography to quantify pathology related to specific impairment}.
\newblock \bibinfo{journal}{Journal of the neurological sciences}
  \bibinfo{volume}{237}, \bibinfo{pages}{13--19}.
\bibitem[{Lipp et~al.(2020)Lipp, Parker, Tallantyre, Goodall, Grama, Patitucci,
  Heveron, Tomassini and Jones}]{lipp2020tractography}
\bibinfo{author}{Lipp, I.}, \bibinfo{author}{Parker, G.D.},
  \bibinfo{author}{Tallantyre, E.C.}, \bibinfo{author}{Goodall, A.},
  \bibinfo{author}{Grama, S.}, \bibinfo{author}{Patitucci, E.},
  \bibinfo{author}{Heveron, P.}, \bibinfo{author}{Tomassini, V.},
  \bibinfo{author}{Jones, D.K.}, \bibinfo{year}{2020}.
\newblock \bibinfo{title}{Tractography in the presence of multiple sclerosis
  lesions}.
\newblock \bibinfo{journal}{NeuroImage} \bibinfo{volume}{209},
  \bibinfo{pages}{116471}.
\bibitem[{Liu et~al.(2019)Liu, Feng, Chen, Wu, Hong, Yap and
  Shen}]{liu2019deepbundle}
\bibinfo{author}{Liu, F.}, \bibinfo{author}{Feng, J.}, \bibinfo{author}{Chen,
  G.}, \bibinfo{author}{Wu, Y.}, \bibinfo{author}{Hong, Y.},
  \bibinfo{author}{Yap, P.T.}, \bibinfo{author}{Shen, D.},
  \bibinfo{year}{2019}.
\newblock \bibinfo{title}{Deepbundle: Fiber bundle parcellation with graph
  convolution neural networks}, in: \bibinfo{booktitle}{International Workshop
  on Graph Learning in Medical Imaging}, \bibinfo{organization}{Springer}. pp.
  \bibinfo{pages}{88--95}.
\bibitem[{Liu et~al.(2017)Liu, Lauer, Ward, Roberts, Liu, Gollapudy, Rohloff,
  Gross, Xu, Chen et~al.}]{liu2017fine}
\bibinfo{author}{Liu, X.}, \bibinfo{author}{Lauer, K.K.},
  \bibinfo{author}{Ward, B.D.}, \bibinfo{author}{Roberts, C.J.},
  \bibinfo{author}{Liu, S.}, \bibinfo{author}{Gollapudy, S.},
  \bibinfo{author}{Rohloff, R.}, \bibinfo{author}{Gross, W.},
  \bibinfo{author}{Xu, Z.}, \bibinfo{author}{Chen, G.}, et~al.,
  \bibinfo{year}{2017}.
\newblock \bibinfo{title}{Fine-grained parcellation of brain connectivity
  improves differentiation of states of consciousness during graded propofol
  sedation}.
\newblock \bibinfo{journal}{Brain connectivity} \bibinfo{volume}{7},
  \bibinfo{pages}{373--381}.
\bibitem[{Liu et~al.(2018)Liu, Duan, Dong, Barkhof, Li and
  Shu}]{liu2018disrupted}
\bibinfo{author}{Liu, Y.}, \bibinfo{author}{Duan, Y.}, \bibinfo{author}{Dong,
  H.}, \bibinfo{author}{Barkhof, F.}, \bibinfo{author}{Li, K.},
  \bibinfo{author}{Shu, N.}, \bibinfo{year}{2018}.
\newblock \bibinfo{title}{Disrupted module efficiency of structural and
  functional brain connectomes in clinically isolated syndrome and multiple
  sclerosis}.
\newblock \bibinfo{journal}{Frontiers in human neuroscience}
  \bibinfo{volume}{12}, \bibinfo{pages}{138}.
\bibitem[{Llufriu et~al.(2017)Llufriu, Martinez-Heras, Solana, Sola-Valls,
  Sepulveda, Blanco, Martinez-Lapiscina, Andorra, Villoslada, Prats-Galino
  et~al.}]{llufriu2017structural}
\bibinfo{author}{Llufriu, S.}, \bibinfo{author}{Martinez-Heras, E.},
  \bibinfo{author}{Solana, E.}, \bibinfo{author}{Sola-Valls, N.},
  \bibinfo{author}{Sepulveda, M.}, \bibinfo{author}{Blanco, Y.},
  \bibinfo{author}{Martinez-Lapiscina, E.H.}, \bibinfo{author}{Andorra, M.},
  \bibinfo{author}{Villoslada, P.}, \bibinfo{author}{Prats-Galino, A.}, et~al.,
  \bibinfo{year}{2017}.
\newblock \bibinfo{title}{Structural networks involved in attention and
  executive functions in multiple sclerosis}.
\newblock \bibinfo{journal}{NeuroImage: Clinical} \bibinfo{volume}{13},
  \bibinfo{pages}{288--296}.
\bibitem[{Lo et~al.(2010)Lo, Wang, Chou, Wang, He and Lin}]{lo2010diffusion}
\bibinfo{author}{Lo, C.Y.}, \bibinfo{author}{Wang, P.N.},
  \bibinfo{author}{Chou, K.H.}, \bibinfo{author}{Wang, J.},
  \bibinfo{author}{He, Y.}, \bibinfo{author}{Lin, C.P.}, \bibinfo{year}{2010}.
\newblock \bibinfo{title}{Diffusion tensor tractography reveals abnormal
  topological organization in structural cortical networks in alzheimer's
  disease}.
\newblock \bibinfo{journal}{Journal of Neuroscience} \bibinfo{volume}{30},
  \bibinfo{pages}{16876--16885}.
\bibitem[{Lu et~al.(2020)Lu, Li and Ye}]{lu2020white}
\bibinfo{author}{Lu, Q.}, \bibinfo{author}{Li, Y.}, \bibinfo{author}{Ye, C.},
  \bibinfo{year}{2020}.
\newblock \bibinfo{title}{White matter tract segmentation with self-supervised
  learning}, in: \bibinfo{booktitle}{International Conference on Medical Image
  Computing and Computer-Assisted Intervention},
  \bibinfo{organization}{Springer}. pp. \bibinfo{pages}{270--279}.
\bibitem[{Maddah et~al.(2005)Maddah, Mewes, Haker, Grimson and
  Warfield}]{maddah2005automated}
\bibinfo{author}{Maddah, M.}, \bibinfo{author}{Mewes, A.U.},
  \bibinfo{author}{Haker, S.}, \bibinfo{author}{Grimson, W.E.L.},
  \bibinfo{author}{Warfield, S.K.}, \bibinfo{year}{2005}.
\newblock \bibinfo{title}{Automated atlas-based clustering of white matter
  fiber tracts from dtmri}, in: \bibinfo{booktitle}{International Conference on
  Medical Image Computing and Computer-Assisted Intervention},
  \bibinfo{organization}{Springer}. pp. \bibinfo{pages}{188--195}.
\bibitem[{Madden et~al.(2009)Madden, Bennett and Song}]{madden2009cerebral}
\bibinfo{author}{Madden, D.J.}, \bibinfo{author}{Bennett, I.J.},
  \bibinfo{author}{Song, A.W.}, \bibinfo{year}{2009}.
\newblock \bibinfo{title}{Cerebral white matter integrity and cognitive aging:
  contributions from diffusion tensor imaging}.
\newblock \bibinfo{journal}{Neuropsychology review} \bibinfo{volume}{19},
  \bibinfo{pages}{415--435}.
\bibitem[{Madden et~al.(2020)Madden, Jain, Monge, Cook, Lee, Huang, Howard and
  Cohen}]{madden2020influence}
\bibinfo{author}{Madden, D.J.}, \bibinfo{author}{Jain, S.},
  \bibinfo{author}{Monge, Z.A.}, \bibinfo{author}{Cook, A.D.},
  \bibinfo{author}{Lee, A.}, \bibinfo{author}{Huang, H.},
  \bibinfo{author}{Howard, C.M.}, \bibinfo{author}{Cohen, J.R.},
  \bibinfo{year}{2020}.
\newblock \bibinfo{title}{Influence of structural and functional brain
  connectivity on age-related differences in fluid cognition}.
\newblock \bibinfo{journal}{Neurobiology of Aging} \bibinfo{volume}{96},
  \bibinfo{pages}{205--222}.
\bibitem[{Madole et~al.(2020)Madole, Ritchie, Cox, Buchanan, Hern{\'a}ndez,
  Maniega, Wardlaw, Harris, Bastin, Deary et~al.}]{madole2020aging}
\bibinfo{author}{Madole, J.W.}, \bibinfo{author}{Ritchie, S.J.},
  \bibinfo{author}{Cox, S.R.}, \bibinfo{author}{Buchanan, C.R.},
  \bibinfo{author}{Hern{\'a}ndez, M.V.}, \bibinfo{author}{Maniega, S.M.},
  \bibinfo{author}{Wardlaw, J.M.}, \bibinfo{author}{Harris, M.A.},
  \bibinfo{author}{Bastin, M.E.}, \bibinfo{author}{Deary, I.J.}, et~al.,
  \bibinfo{year}{2020}.
\newblock \bibinfo{title}{Aging-sensitive networks within the human structural
  connectome are implicated in late-life cognitive declines}.
\newblock \bibinfo{journal}{Biological Psychiatry} .
\bibitem[{Maier-Hein et~al.(2017)Maier-Hein, Neher, Houde, Côté,
  Garyfallidis, Zhong, Chamberland, Yeh, Lin, Ji, Reddick, Glass, Chen, Feng,
  Gao, Wu, Ma, Renjie, Li, Westin, Deslauriers-Gauthier, González, Paquette,
  St-Jean, Girard, Rheault, Sidhu, Tax, Guo, Mesri, Dávid, Froeling,
  Heemskerk, Leemans, Boré, Pinsard, Bedetti, Desrosiers, Brambati, Doyon,
  Sarica, Vasta, Cerasa, Quattrone, Yeatman, Khan, Hodges, Alexander,
  Romascano, Barakovic, Auría, Esteban, Lemkaddem, Thiran, Cetingul, Odry,
  Mailhe, Nadar, Pizzagalli, Prasad, Villalon-Reina, Galvis, Thompson, Requejo,
  Laguna, Lacerda, Barrett, Dell'Acqua, Catani, Petit, Caruyer, Daducci, Dyrby,
  Holland-Letz, Hilgetag, Stieltjes and Descoteaux}]{Maier-Hein.2017}
\bibinfo{author}{Maier-Hein, K.H.}, \bibinfo{author}{Neher, P.F.},
  \bibinfo{author}{Houde, J.C.}, \bibinfo{author}{Côté, M.A.},
  \bibinfo{author}{Garyfallidis, E.}, \bibinfo{author}{Zhong, J.},
  \bibinfo{author}{Chamberland, M.}, \bibinfo{author}{Yeh, F.C.},
  \bibinfo{author}{Lin, Y.C.}, \bibinfo{author}{Ji, Q.},
  \bibinfo{author}{Reddick, W.E.}, \bibinfo{author}{Glass, J.O.},
  \bibinfo{author}{Chen, D.Q.}, \bibinfo{author}{Feng, Y.},
  \bibinfo{author}{Gao, C.}, \bibinfo{author}{Wu, Y.}, \bibinfo{author}{Ma,
  J.}, \bibinfo{author}{Renjie, H.}, \bibinfo{author}{Li, Q.},
  \bibinfo{author}{Westin, C.F.}, \bibinfo{author}{Deslauriers-Gauthier, S.},
  \bibinfo{author}{González, J.O.O.}, \bibinfo{author}{Paquette, M.},
  \bibinfo{author}{St-Jean, S.}, \bibinfo{author}{Girard, G.},
  \bibinfo{author}{Rheault, F.}, \bibinfo{author}{Sidhu, J.},
  \bibinfo{author}{Tax, C.M.W.}, \bibinfo{author}{Guo, F.},
  \bibinfo{author}{Mesri, H.Y.}, \bibinfo{author}{Dávid, S.},
  \bibinfo{author}{Froeling, M.}, \bibinfo{author}{Heemskerk, A.M.},
  \bibinfo{author}{Leemans, A.}, \bibinfo{author}{Boré, A.},
  \bibinfo{author}{Pinsard, B.}, \bibinfo{author}{Bedetti, C.},
  \bibinfo{author}{Desrosiers, M.}, \bibinfo{author}{Brambati, S.},
  \bibinfo{author}{Doyon, J.}, \bibinfo{author}{Sarica, A.},
  \bibinfo{author}{Vasta, R.}, \bibinfo{author}{Cerasa, A.},
  \bibinfo{author}{Quattrone, A.}, \bibinfo{author}{Yeatman, J.},
  \bibinfo{author}{Khan, A.R.}, \bibinfo{author}{Hodges, W.},
  \bibinfo{author}{Alexander, S.}, \bibinfo{author}{Romascano, D.},
  \bibinfo{author}{Barakovic, M.}, \bibinfo{author}{Auría, A.},
  \bibinfo{author}{Esteban, O.}, \bibinfo{author}{Lemkaddem, A.},
  \bibinfo{author}{Thiran, J.P.}, \bibinfo{author}{Cetingul, H.E.},
  \bibinfo{author}{Odry, B.L.}, \bibinfo{author}{Mailhe, B.},
  \bibinfo{author}{Nadar, M.S.}, \bibinfo{author}{Pizzagalli, F.},
  \bibinfo{author}{Prasad, G.}, \bibinfo{author}{Villalon-Reina, J.E.},
  \bibinfo{author}{Galvis, J.}, \bibinfo{author}{Thompson, P.M.},
  \bibinfo{author}{Requejo, F.D.S.}, \bibinfo{author}{Laguna, P.L.},
  \bibinfo{author}{Lacerda, L.M.}, \bibinfo{author}{Barrett, R.},
  \bibinfo{author}{Dell'Acqua, F.}, \bibinfo{author}{Catani, M.},
  \bibinfo{author}{Petit, L.}, \bibinfo{author}{Caruyer, E.},
  \bibinfo{author}{Daducci, A.}, \bibinfo{author}{Dyrby, T.B.},
  \bibinfo{author}{Holland-Letz, T.}, \bibinfo{author}{Hilgetag, C.C.},
  \bibinfo{author}{Stieltjes, B.}, \bibinfo{author}{Descoteaux, M.},
  \bibinfo{year}{2017}.
\newblock \bibinfo{title}{{The challenge of mapping the human connectome based
  on diffusion tractography}}.
\newblock \bibinfo{journal}{Nature Communications} \bibinfo{volume}{8},
  \bibinfo{pages}{1349}.
\newblock \DOIprefix\doi{10.1038/s41467-017-01285-x}.
\bibitem[{Malcolm et~al.(2010)Malcolm, Shenton and Rathi}]{malcolm2010filtered}
\bibinfo{author}{Malcolm, J.G.}, \bibinfo{author}{Shenton, M.E.},
  \bibinfo{author}{Rathi, Y.}, \bibinfo{year}{2010}.
\newblock \bibinfo{title}{Filtered multitensor tractography}.
\newblock \bibinfo{journal}{IEEE transactions on medical imaging}
  \bibinfo{volume}{29}, \bibinfo{pages}{1664--1675}.
\bibitem[{Malinsky et~al.(2013)Malinsky, Peter, Hodneland, Lundervold,
  Lundervold and Jan}]{malinsky2013registration}
\bibinfo{author}{Malinsky, M.}, \bibinfo{author}{Peter, R.},
  \bibinfo{author}{Hodneland, E.}, \bibinfo{author}{Lundervold, A.J.},
  \bibinfo{author}{Lundervold, A.}, \bibinfo{author}{Jan, J.},
  \bibinfo{year}{2013}.
\newblock \bibinfo{title}{Registration of fa and t1-weighted {MRI} data of
  healthy human brain based on template matching and normalized
  cross-correlation}.
\newblock \bibinfo{journal}{Journal of digital imaging} \bibinfo{volume}{26},
  \bibinfo{pages}{774--785}.
\bibitem[{Mancini et~al.(2020)Mancini, Karakuzu, Cohen-Adad, Cercignani,
  Nichols and Stikov}]{Mancini2020}
\bibinfo{author}{Mancini, M.}, \bibinfo{author}{Karakuzu, A.},
  \bibinfo{author}{Cohen-Adad, J.}, \bibinfo{author}{Cercignani, M.},
  \bibinfo{author}{Nichols, T.E.}, \bibinfo{author}{Stikov, N.},
  \bibinfo{year}{2020}.
\newblock \bibinfo{title}{{An interactive meta-analysis of MRI biomarkers of
  myelin}}.
\newblock \bibinfo{journal}{eLife} \bibinfo{volume}{9},
  \bibinfo{pages}{e61523}.
\newblock \DOIprefix\doi{10.7554/eLife.61523}.
\bibitem[{Mangin et~al.(2013)Mangin, Fillard, Cointepas, Le~Bihan, Frouin and
  Poupon}]{mangin_toward_2013}
\bibinfo{author}{Mangin, J.F.}, \bibinfo{author}{Fillard, P.},
  \bibinfo{author}{Cointepas, Y.}, \bibinfo{author}{Le~Bihan, D.},
  \bibinfo{author}{Frouin, V.}, \bibinfo{author}{Poupon, C.},
  \bibinfo{year}{2013}.
\newblock \bibinfo{title}{Toward global tractography}.
\newblock \bibinfo{journal}{Mapping the Connectome} \bibinfo{volume}{80},
  \bibinfo{pages}{290--296}.
\bibitem[{Mangin et~al.(2002)Mangin, Poupon, Cointepas, Rivière,
  Papadopoulos-Orfanos, Clark, Régis and Le~Bihan}]{mangin_framework_2002}
\bibinfo{author}{Mangin, J.F.}, \bibinfo{author}{Poupon, C.},
  \bibinfo{author}{Cointepas, Y.}, \bibinfo{author}{Rivière, D.},
  \bibinfo{author}{Papadopoulos-Orfanos, D.}, \bibinfo{author}{Clark, C.A.},
  \bibinfo{author}{Régis, J.}, \bibinfo{author}{Le~Bihan, D.},
  \bibinfo{year}{2002}.
\newblock \bibinfo{title}{A framework based on spin glass models for the
  inference of anatomical connectivity from diffusion-weighted {MR} data - a
  technical review}.
\newblock \bibinfo{journal}{NMR in Biomedicine} \bibinfo{volume}{15},
  \bibinfo{pages}{481--492}.
\bibitem[{Mart{\'\i}-Juan et~al.(2020)Mart{\'\i}-Juan, Sanroma-Guell and
  Piella}]{marti2020survey}
\bibinfo{author}{Mart{\'\i}-Juan, G.}, \bibinfo{author}{Sanroma-Guell, G.},
  \bibinfo{author}{Piella, G.}, \bibinfo{year}{2020}.
\newblock \bibinfo{title}{A survey on machine and statistical learning for
  longitudinal analysis of neuroimaging data in alzheimer’s disease}.
\newblock \bibinfo{journal}{Computer methods and programs in biomedicine}
  \bibinfo{volume}{189}, \bibinfo{pages}{105348}.
\bibitem[{Mazziotta et~al.(2001)Mazziotta, Toga, Evans, Fox, Lancaster, Zilles,
  Woods, Paus, Simpson, Pike et~al.}]{mazziotta2001probabilistic}
\bibinfo{author}{Mazziotta, J.}, \bibinfo{author}{Toga, A.},
  \bibinfo{author}{Evans, A.}, \bibinfo{author}{Fox, P.},
  \bibinfo{author}{Lancaster, J.}, \bibinfo{author}{Zilles, K.},
  \bibinfo{author}{Woods, R.}, \bibinfo{author}{Paus, T.},
  \bibinfo{author}{Simpson, G.}, \bibinfo{author}{Pike, B.}, et~al.,
  \bibinfo{year}{2001}.
\newblock \bibinfo{title}{A probabilistic atlas and reference system for the
  human brain: International consortium for brain mapping (icbm)}.
\newblock \bibinfo{journal}{Philosophical Transactions of the Royal Society of
  London. Series B: Biological Sciences} \bibinfo{volume}{356},
  \bibinfo{pages}{1293--1322}.
\bibitem[{Meola et~al.(2016)Meola, Yeh, Fellows-Mayle, Weed and
  Fernandez-Miranda}]{meola2016human}
\bibinfo{author}{Meola, A.}, \bibinfo{author}{Yeh, F.C.},
  \bibinfo{author}{Fellows-Mayle, W.}, \bibinfo{author}{Weed, J.},
  \bibinfo{author}{Fernandez-Miranda, J.C.}, \bibinfo{year}{2016}.
\newblock \bibinfo{title}{Human connectome-based tractographic atlas of the
  brainstem connections and surgical approaches}.
\newblock \bibinfo{journal}{Neurosurgery} \bibinfo{volume}{79},
  \bibinfo{pages}{437--455}.
\bibitem[{Mess{\'e}(2020)}]{messe2020parcellation}
\bibinfo{author}{Mess{\'e}, A.}, \bibinfo{year}{2020}.
\newblock \bibinfo{title}{Parcellation influence on the connectivity-based
  structure--function relationship in the human brain}.
\newblock \bibinfo{journal}{Human brain mapping} \bibinfo{volume}{41},
  \bibinfo{pages}{1167--1180}.
\bibitem[{Michielse et~al.(2010)Michielse, Coupland, Camicioli, Carter, Seres,
  Sabino and Malykhin}]{michielse2010selective}
\bibinfo{author}{Michielse, S.}, \bibinfo{author}{Coupland, N.},
  \bibinfo{author}{Camicioli, R.}, \bibinfo{author}{Carter, R.},
  \bibinfo{author}{Seres, P.}, \bibinfo{author}{Sabino, J.},
  \bibinfo{author}{Malykhin, N.}, \bibinfo{year}{2010}.
\newblock \bibinfo{title}{Selective effects of aging on brain white matter
  microstructure: a diffusion tensor imaging tractography study}.
\newblock \bibinfo{journal}{Neuroimage} \bibinfo{volume}{52},
  \bibinfo{pages}{1190--1201}.
\bibitem[{Miller et~al.(2012)Miller, Duka, Stimpson, Schapiro, Baze, McArthur,
  Fobbs, Sousa, {\v{S}}estan, Wildman et~al.}]{miller2012prolonged}
\bibinfo{author}{Miller, D.J.}, \bibinfo{author}{Duka, T.},
  \bibinfo{author}{Stimpson, C.D.}, \bibinfo{author}{Schapiro, S.J.},
  \bibinfo{author}{Baze, W.B.}, \bibinfo{author}{McArthur, M.J.},
  \bibinfo{author}{Fobbs, A.J.}, \bibinfo{author}{Sousa, A.M.},
  \bibinfo{author}{{\v{S}}estan, N.}, \bibinfo{author}{Wildman, D.E.}, et~al.,
  \bibinfo{year}{2012}.
\newblock \bibinfo{title}{Prolonged myelination in human neocortical
  evolution}.
\newblock \bibinfo{journal}{Proceedings of the National Academy of Sciences}
  \bibinfo{volume}{109}, \bibinfo{pages}{16480--16485}.
\bibitem[{Mi{\v s}ic et~al.(2018)Mi{\v s}ic, Betzel, Griffa, de~Reus, He, Zuo,
  van~den Heuvel, Hagmann, Sporns and Zatorre}]{misic:2018}
\bibinfo{author}{Mi{\v s}ic, B.}, \bibinfo{author}{Betzel, R.F.},
  \bibinfo{author}{Griffa, A.}, \bibinfo{author}{de~Reus, M.A.},
  \bibinfo{author}{He, Y.}, \bibinfo{author}{Zuo, X.N.},
  \bibinfo{author}{van~den Heuvel, M.P.}, \bibinfo{author}{Hagmann, P.},
  \bibinfo{author}{Sporns, O.}, \bibinfo{author}{Zatorre, R.J.},
  \bibinfo{year}{2018}.
\newblock \bibinfo{title}{Network-based asymmetry of the human auditory
  system}.
\newblock \bibinfo{journal}{Cereb Cortex} \bibinfo{volume}{28},
  \bibinfo{pages}{2655--2664}.
\bibitem[{Mi{\v s}i{\'c} et~al.(2015)Mi{\v s}i{\'c}, Betzel, Nematzadeh,
  Go{\~n}i, Griffa, Hagmann, Flammini, Ahn and Sporns}]{misic:2015}
\bibinfo{author}{Mi{\v s}i{\'c}, B.}, \bibinfo{author}{Betzel, R.F.},
  \bibinfo{author}{Nematzadeh, A.}, \bibinfo{author}{Go{\~n}i, J.},
  \bibinfo{author}{Griffa, A.}, \bibinfo{author}{Hagmann, P.},
  \bibinfo{author}{Flammini, A.}, \bibinfo{author}{Ahn, Y.Y.},
  \bibinfo{author}{Sporns, O.}, \bibinfo{year}{2015}.
\newblock \bibinfo{title}{Cooperative and competitive spreading dynamics on the
  human connectome}.
\newblock \bibinfo{journal}{Neuron} \bibinfo{volume}{86},
  \bibinfo{pages}{1518--29}.
\bibitem[{Mito et~al.(2018)Mito, Raffelt, Dhollander, Vaughan, Tournier,
  Salvado, Brodtmann, Rowe, Villemagne and Connelly}]{mito2018fibre}
\bibinfo{author}{Mito, R.}, \bibinfo{author}{Raffelt, D.},
  \bibinfo{author}{Dhollander, T.}, \bibinfo{author}{Vaughan, D.N.},
  \bibinfo{author}{Tournier, J.D.}, \bibinfo{author}{Salvado, O.},
  \bibinfo{author}{Brodtmann, A.}, \bibinfo{author}{Rowe, C.C.},
  \bibinfo{author}{Villemagne, V.L.}, \bibinfo{author}{Connelly, A.},
  \bibinfo{year}{2018}.
\newblock \bibinfo{title}{Fibre-specific white matter reductions in
  alzheimer’s disease and mild cognitive impairment}.
\newblock \bibinfo{journal}{Brain} \bibinfo{volume}{141},
  \bibinfo{pages}{888--902}.
\bibitem[{Mitter et~al.(2015)Mitter, Jakab, Brugger, Ricken, Gruber,
  Bettelheim, Scharrer, Langs, Hainfellner, Prayer
  et~al.}]{mitter2015validation}
\bibinfo{author}{Mitter, C.}, \bibinfo{author}{Jakab, A.},
  \bibinfo{author}{Brugger, P.C.}, \bibinfo{author}{Ricken, G.},
  \bibinfo{author}{Gruber, G.M.}, \bibinfo{author}{Bettelheim, D.},
  \bibinfo{author}{Scharrer, A.}, \bibinfo{author}{Langs, G.},
  \bibinfo{author}{Hainfellner, J.A.}, \bibinfo{author}{Prayer, D.}, et~al.,
  \bibinfo{year}{2015}.
\newblock \bibinfo{title}{Validation of in utero tractography of human fetal
  commissural and internal capsule fibers with histological structure tensor
  analysis}.
\newblock \bibinfo{journal}{Frontiers in neuroanatomy} \bibinfo{volume}{9},
  \bibinfo{pages}{164}.
\bibitem[{Moldrich et~al.()Moldrich, Pannek, Hoch, Rubenstein, Kurniawan and
  Richards}]{moldrich2010comparative}
\bibinfo{author}{Moldrich, R.X.}, \bibinfo{author}{Pannek, K.},
  \bibinfo{author}{Hoch, R.}, \bibinfo{author}{Rubenstein, J.L.},
  \bibinfo{author}{Kurniawan, N.D.}, \bibinfo{author}{Richards, L.J.}, .
\newblock \bibinfo{title}{Comparative mouse brain tractography of diffusion
  magnetic resonance imaging}.
\newblock \bibinfo{journal}{{NeuroImage}} \bibinfo{volume}{51},
  \bibinfo{pages}{1027--1036}.
\bibitem[{Mori et~al.(1999)Mori, Crain, Chacko and Van~Zijl}]{mori1999three}
\bibinfo{author}{Mori, S.}, \bibinfo{author}{Crain, B.J.},
  \bibinfo{author}{Chacko, V.P.}, \bibinfo{author}{Van~Zijl, P.C.},
  \bibinfo{year}{1999}.
\newblock \bibinfo{title}{Three-dimensional tracking of axonal projections in
  the brain by magnetic resonance imaging}.
\newblock \bibinfo{journal}{Annals of Neurology: Official Journal of the
  American Neurological Association and the Child Neurology Society}
  \bibinfo{volume}{45}, \bibinfo{pages}{265--269}.
\bibitem[{Mori et~al.(2008)Mori, Oishi, Jiang, Jiang, Li, Akhter, Hua, Faria,
  Mahmood, Woods et~al.}]{mori2008stereotaxic}
\bibinfo{author}{Mori, S.}, \bibinfo{author}{Oishi, K.},
  \bibinfo{author}{Jiang, H.}, \bibinfo{author}{Jiang, L.},
  \bibinfo{author}{Li, X.}, \bibinfo{author}{Akhter, K.}, \bibinfo{author}{Hua,
  K.}, \bibinfo{author}{Faria, A.V.}, \bibinfo{author}{Mahmood, A.},
  \bibinfo{author}{Woods, R.}, et~al., \bibinfo{year}{2008}.
\newblock \bibinfo{title}{Stereotaxic white matter atlas based on diffusion
  tensor imaging in an icbm template}.
\newblock \bibinfo{journal}{Neuroimage} \bibinfo{volume}{40},
  \bibinfo{pages}{570--582}.
\bibitem[{Moritz-Gasser et~al.(2012)Moritz-Gasser, Herbet, Maldonado and
  Duffau}]{moritz2012lexical}
\bibinfo{author}{Moritz-Gasser, S.}, \bibinfo{author}{Herbet, G.},
  \bibinfo{author}{Maldonado, I.L.}, \bibinfo{author}{Duffau, H.},
  \bibinfo{year}{2012}.
\newblock \bibinfo{title}{Lexical access speed is significantly correlated with
  the return to professional activities after awake surgery for low-grade
  gliomas}.
\newblock \bibinfo{journal}{Journal of neuro-oncology} \bibinfo{volume}{107},
  \bibinfo{pages}{633--641}.
\bibitem[{Moseley(2002)}]{moseley2002diffusion}
\bibinfo{author}{Moseley, M.}, \bibinfo{year}{2002}.
\newblock \bibinfo{title}{Diffusion tensor imaging and aging--a review}.
\newblock \bibinfo{journal}{NMR in Biomedicine: An International Journal
  Devoted to the Development and Application of Magnetic Resonance In Vivo}
  \bibinfo{volume}{15}, \bibinfo{pages}{553--560}.
\bibitem[{Mukherjee(2005)}]{mukherjee2005diffusion}
\bibinfo{author}{Mukherjee, P.}, \bibinfo{year}{2005}.
\newblock \bibinfo{title}{Diffusion tensor imaging and fiber tractography in
  acute stroke}.
\newblock \bibinfo{journal}{Neuroimaging Clinics} \bibinfo{volume}{15},
  \bibinfo{pages}{655--665}.
\bibitem[{Muthuraman et~al.(2016)Muthuraman, Fleischer, Kolber, Luessi, Zipp
  and Groppa}]{muthuraman2016structural}
\bibinfo{author}{Muthuraman, M.}, \bibinfo{author}{Fleischer, V.},
  \bibinfo{author}{Kolber, P.}, \bibinfo{author}{Luessi, F.},
  \bibinfo{author}{Zipp, F.}, \bibinfo{author}{Groppa, S.},
  \bibinfo{year}{2016}.
\newblock \bibinfo{title}{Structural brain network characteristics can
  differentiate cis from early rrms}.
\newblock \bibinfo{journal}{Frontiers in neuroscience} \bibinfo{volume}{10},
  \bibinfo{pages}{14}.
\bibitem[{Negwer et~al.(2017)Negwer, Sollmann, Ille, Hauck, Maurer, Kirschke,
  Ringel, Meyer and Krieg}]{negwer2017language}
\bibinfo{author}{Negwer, C.}, \bibinfo{author}{Sollmann, N.},
  \bibinfo{author}{Ille, S.}, \bibinfo{author}{Hauck, T.},
  \bibinfo{author}{Maurer, S.}, \bibinfo{author}{Kirschke, J.S.},
  \bibinfo{author}{Ringel, F.}, \bibinfo{author}{Meyer, B.},
  \bibinfo{author}{Krieg, S.M.}, \bibinfo{year}{2017}.
\newblock \bibinfo{title}{Language pathway tracking: comparing ntms-based dti
  fiber tracking with a cubic rois-based protocol}.
\newblock \bibinfo{journal}{Journal of neurosurgery} \bibinfo{volume}{126},
  \bibinfo{pages}{1006--1014}.
\bibitem[{Neher et~al.(2017)Neher, C{\^o}t{\'e}, Houde, Descoteaux and
  Maier-Hein}]{neher2017fiber}
\bibinfo{author}{Neher, P.F.}, \bibinfo{author}{C{\^o}t{\'e}, M.A.},
  \bibinfo{author}{Houde, J.C.}, \bibinfo{author}{Descoteaux, M.},
  \bibinfo{author}{Maier-Hein, K.H.}, \bibinfo{year}{2017}.
\newblock \bibinfo{title}{Fiber tractography using machine learning}.
\newblock \bibinfo{journal}{Neuroimage} \bibinfo{volume}{158},
  \bibinfo{pages}{417--429}.
\bibitem[{Neher et~al.(2015)Neher, Descoteaux, Houde, Stieltjes and
  Maier-Hein}]{neher2015strengths}
\bibinfo{author}{Neher, P.F.}, \bibinfo{author}{Descoteaux, M.},
  \bibinfo{author}{Houde, J.C.}, \bibinfo{author}{Stieltjes, B.},
  \bibinfo{author}{Maier-Hein, K.H.}, \bibinfo{year}{2015}.
\newblock \bibinfo{title}{Strengths and weaknesses of state of the art fiber
  tractography pipelines--a comprehensive in-vivo and phantom evaluation study
  using tractometer}.
\newblock \bibinfo{journal}{Medical image analysis} \bibinfo{volume}{26},
  \bibinfo{pages}{287--305}.
\bibitem[{Neil et~al.(2002)Neil, Miller, Mukherjee and
  H{\"u}ppi}]{neil2002diffusion}
\bibinfo{author}{Neil, J.}, \bibinfo{author}{Miller, J.},
  \bibinfo{author}{Mukherjee, P.}, \bibinfo{author}{H{\"u}ppi, P.S.},
  \bibinfo{year}{2002}.
\newblock \bibinfo{title}{Diffusion tensor imaging of normal and injured
  developing human brain-a technical review}.
\newblock \bibinfo{journal}{NMR in Biomedicine: An International Journal
  Devoted to the Development and Application of Magnetic Resonance In Vivo}
  \bibinfo{volume}{15}, \bibinfo{pages}{543--552}.
\bibitem[{Ning et~al.(2019)Ning, Gagoski, Szczepankiewicz, Westin and
  Rathi}]{ning2019joint}
\bibinfo{author}{Ning, L.}, \bibinfo{author}{Gagoski, B.},
  \bibinfo{author}{Szczepankiewicz, F.}, \bibinfo{author}{Westin, C.F.},
  \bibinfo{author}{Rathi, Y.}, \bibinfo{year}{2019}.
\newblock \bibinfo{title}{Joint relaxation-diffusion imaging moments to probe
  neurite microstructure}.
\newblock \bibinfo{journal}{IEEE transactions on medical imaging}
  \bibinfo{volume}{39}, \bibinfo{pages}{668--677}.
\bibitem[{Ning et~al.(2021)Ning, Szczepankiewicz, Nilsson, Rathi and
  Westin}]{ning2021probing}
\bibinfo{author}{Ning, L.}, \bibinfo{author}{Szczepankiewicz, F.},
  \bibinfo{author}{Nilsson, M.}, \bibinfo{author}{Rathi, Y.},
  \bibinfo{author}{Westin, C.F.}, \bibinfo{year}{2021}.
\newblock \bibinfo{title}{Probing tissue microstructure by diffusion skewness
  tensor imaging}.
\newblock \bibinfo{journal}{Scientific Reports} \bibinfo{volume}{11},
  \bibinfo{pages}{1--10}.
\bibitem[{Norton et~al.(2017)Norton, Essayed, Zhang, Pujol, Yarmarkovich,
  Golby, Kindlmann, Wassermann, Estepar, Rathi et~al.}]{norton2017slicerdmri}
\bibinfo{author}{Norton, I.}, \bibinfo{author}{Essayed, W.I.},
  \bibinfo{author}{Zhang, F.}, \bibinfo{author}{Pujol, S.},
  \bibinfo{author}{Yarmarkovich, A.}, \bibinfo{author}{Golby, A.J.},
  \bibinfo{author}{Kindlmann, G.}, \bibinfo{author}{Wassermann, D.},
  \bibinfo{author}{Estepar, R.S.J.}, \bibinfo{author}{Rathi, Y.}, et~al.,
  \bibinfo{year}{2017}.
\newblock \bibinfo{title}{{SlicerDMRI: open source diffusion MRI software for
  brain cancer research}}.
\newblock \bibinfo{journal}{Cancer research} \bibinfo{volume}{77},
  \bibinfo{pages}{e101--e103}.
\bibitem[{Novikov et~al.(2019)Novikov, Fieremans, Jespersen and
  Kiselev}]{novikov2019quantifying}
\bibinfo{author}{Novikov, D.S.}, \bibinfo{author}{Fieremans, E.},
  \bibinfo{author}{Jespersen, S.N.}, \bibinfo{author}{Kiselev, V.G.},
  \bibinfo{year}{2019}.
\newblock \bibinfo{title}{Quantifying brain microstructure with diffusion mri:
  Theory and parameter estimation}.
\newblock \bibinfo{journal}{NMR in Biomedicine} \bibinfo{volume}{32},
  \bibinfo{pages}{e3998}.
\bibitem[{Novikov et~al.(2018)Novikov, Kiselev and
  Jespersen}]{novikov2018modeling}
\bibinfo{author}{Novikov, D.S.}, \bibinfo{author}{Kiselev, V.G.},
  \bibinfo{author}{Jespersen, S.N.}, \bibinfo{year}{2018}.
\newblock \bibinfo{title}{On modeling}.
\newblock \bibinfo{journal}{Magnetic resonance in medicine}
  \bibinfo{volume}{79}, \bibinfo{pages}{3172--3193}.
\bibitem[{Nucifora et~al.(2007)Nucifora, Verma, Lee and
  Melhem}]{nucifora2007diffusion}
\bibinfo{author}{Nucifora, P.G.}, \bibinfo{author}{Verma, R.},
  \bibinfo{author}{Lee, S.K.}, \bibinfo{author}{Melhem, E.R.},
  \bibinfo{year}{2007}.
\newblock \bibinfo{title}{Diffusion-tensor mr imaging and tractography:
  exploring brain microstructure and connectivity}.
\newblock \bibinfo{journal}{Radiology} \bibinfo{volume}{245},
  \bibinfo{pages}{367--384}.
\bibitem[{Ocampo-Pineda et~al.(2021)Ocampo-Pineda, Schiavi, Rheault, Girard,
  Petit, Descoteaux and Daducci}]{ocampo2021hierarchical}
\bibinfo{author}{Ocampo-Pineda, M.}, \bibinfo{author}{Schiavi, S.},
  \bibinfo{author}{Rheault, F.}, \bibinfo{author}{Girard, G.},
  \bibinfo{author}{Petit, L.}, \bibinfo{author}{Descoteaux, M.},
  \bibinfo{author}{Daducci, A.}, \bibinfo{year}{2021}.
\newblock \bibinfo{title}{Hierarchical microstructure informed tractography}.
\newblock \bibinfo{journal}{Brain Connectivity} .
\bibitem[{O'Donnell et~al.(2013)O'Donnell, Golby and Westin}]{o2013fiber}
\bibinfo{author}{O'Donnell, L.J.}, \bibinfo{author}{Golby, A.J.},
  \bibinfo{author}{Westin, C.F.}, \bibinfo{year}{2013}.
\newblock \bibinfo{title}{Fiber clustering versus the parcellation-based
  connectome}.
\newblock \bibinfo{journal}{NeuroImage} \bibinfo{volume}{80},
  \bibinfo{pages}{283--289}.
\bibitem[{O'Donnell and Pasternak(2015)}]{o2015does}
\bibinfo{author}{O'Donnell, L.J.}, \bibinfo{author}{Pasternak, O.},
  \bibinfo{year}{2015}.
\newblock \bibinfo{title}{Does diffusion {MRI} tell us anything about the white
  matter? an overview of methods and pitfalls}.
\newblock \bibinfo{journal}{Schizophrenia research} \bibinfo{volume}{161},
  \bibinfo{pages}{133--141}.
\bibitem[{O'Donnell and Westin(2007)}]{o2007automatic}
\bibinfo{author}{O'Donnell, L.J.}, \bibinfo{author}{Westin, C.F.},
  \bibinfo{year}{2007}.
\newblock \bibinfo{title}{Automatic tractography segmentation using a
  high-dimensional white matter atlas}.
\newblock \bibinfo{journal}{IEEE transactions on medical imaging}
  \bibinfo{volume}{26}, \bibinfo{pages}{1562--1575}.
\bibitem[{O'Donnell et~al.(2009)O'Donnell, Westin and
  Golby}]{odonnell2009tract}
\bibinfo{author}{O'Donnell, L.J.}, \bibinfo{author}{Westin, C.F.},
  \bibinfo{author}{Golby, A.J.}, \bibinfo{year}{2009}.
\newblock \bibinfo{title}{Tract-based morphometry for white matter group
  analysis}.
\newblock \bibinfo{journal}{Neuroimage} \bibinfo{volume}{45},
  \bibinfo{pages}{832--844}.
\bibitem[{O'Dwyer et~al.(2012)O'Dwyer, Lamberton, Matura, Scheibe, Miller,
  Rujescu, Prvulovic and Hampel}]{o2012white}
\bibinfo{author}{O'Dwyer, L.}, \bibinfo{author}{Lamberton, F.},
  \bibinfo{author}{Matura, S.}, \bibinfo{author}{Scheibe, M.},
  \bibinfo{author}{Miller, J.}, \bibinfo{author}{Rujescu, D.},
  \bibinfo{author}{Prvulovic, D.}, \bibinfo{author}{Hampel, H.},
  \bibinfo{year}{2012}.
\newblock \bibinfo{title}{White matter differences between healthy young apoe4
  carriers and non-carriers identified with tractography and support vector
  machines}.
\newblock \bibinfo{journal}{PLoS One} \bibinfo{volume}{7},
  \bibinfo{pages}{e36024}.
\bibitem[{Ohoshi et~al.(2019)Ohoshi, Takahashi, Yamada, Ishida, Tsuda, Tsuji,
  Terada, Shinosaki and Ukai}]{ohoshi2019microstructural}
\bibinfo{author}{Ohoshi, Y.}, \bibinfo{author}{Takahashi, S.},
  \bibinfo{author}{Yamada, S.}, \bibinfo{author}{Ishida, T.},
  \bibinfo{author}{Tsuda, K.}, \bibinfo{author}{Tsuji, T.},
  \bibinfo{author}{Terada, M.}, \bibinfo{author}{Shinosaki, K.},
  \bibinfo{author}{Ukai, S.}, \bibinfo{year}{2019}.
\newblock \bibinfo{title}{Microstructural abnormalities in callosal fibers and
  their relationship with cognitive function in schizophrenia: A tract-specific
  analysis study}.
\newblock \bibinfo{journal}{Brain and behavior} \bibinfo{volume}{9},
  \bibinfo{pages}{e01357}.
\bibitem[{Oishi et~al.(2009)Oishi, Faria, Jiang, Li, Akhter, Zhang, Hsu,
  Miller, van Zijl, Albert et~al.}]{oishi2009atlas}
\bibinfo{author}{Oishi, K.}, \bibinfo{author}{Faria, A.},
  \bibinfo{author}{Jiang, H.}, \bibinfo{author}{Li, X.},
  \bibinfo{author}{Akhter, K.}, \bibinfo{author}{Zhang, J.},
  \bibinfo{author}{Hsu, J.T.}, \bibinfo{author}{Miller, M.I.},
  \bibinfo{author}{van Zijl, P.C.}, \bibinfo{author}{Albert, M.}, et~al.,
  \bibinfo{year}{2009}.
\newblock \bibinfo{title}{Atlas-based whole brain white matter analysis using
  large deformation diffeomorphic metric mapping: application to normal elderly
  and alzheimer's disease participants}.
\newblock \bibinfo{journal}{Neuroimage} \bibinfo{volume}{46},
  \bibinfo{pages}{486--499}.
\bibitem[{Oishi et~al.(2008)Oishi, Zilles, Amunts, Faria, Jiang, Li, Akhter,
  Hua, Woods, Toga et~al.}]{oishi2008human}
\bibinfo{author}{Oishi, K.}, \bibinfo{author}{Zilles, K.},
  \bibinfo{author}{Amunts, K.}, \bibinfo{author}{Faria, A.},
  \bibinfo{author}{Jiang, H.}, \bibinfo{author}{Li, X.},
  \bibinfo{author}{Akhter, K.}, \bibinfo{author}{Hua, K.},
  \bibinfo{author}{Woods, R.}, \bibinfo{author}{Toga, A.W.}, et~al.,
  \bibinfo{year}{2008}.
\newblock \bibinfo{title}{Human brain white matter atlas: identification and
  assignment of common anatomical structures in superficial white matter}.
\newblock \bibinfo{journal}{Neuroimage} \bibinfo{volume}{43},
  \bibinfo{pages}{447--457}.
\bibitem[{Oldham et~al.(2019)Oldham, Fulcher, Parkes, Arnatkevic~I{\=u}t{\.e},
  Suo and Fornito}]{oldham:2019}
\bibinfo{author}{Oldham, S.}, \bibinfo{author}{Fulcher, B.},
  \bibinfo{author}{Parkes, L.}, \bibinfo{author}{Arnatkevic~I{\=u}t{\.e}, A.},
  \bibinfo{author}{Suo, C.}, \bibinfo{author}{Fornito, A.},
  \bibinfo{year}{2019}.
\newblock \bibinfo{title}{Consistency and differences between centrality
  measures across distinct classes of networks}.
\newblock \bibinfo{journal}{PLoS One} \bibinfo{volume}{14},
  \bibinfo{pages}{e0220061}.
\bibitem[{Olszewski et~al.(2017)Olszewski, Kikinis, Gonzalez, Coman, Makris,
  Gong, Rathi, Zhu, Antshel, Fremont et~al.}]{olszewski2017social}
\bibinfo{author}{Olszewski, A.K.}, \bibinfo{author}{Kikinis, Z.},
  \bibinfo{author}{Gonzalez, C.S.}, \bibinfo{author}{Coman, I.L.},
  \bibinfo{author}{Makris, N.}, \bibinfo{author}{Gong, X.},
  \bibinfo{author}{Rathi, Y.}, \bibinfo{author}{Zhu, A.},
  \bibinfo{author}{Antshel, K.M.}, \bibinfo{author}{Fremont, W.}, et~al.,
  \bibinfo{year}{2017}.
\newblock \bibinfo{title}{The social brain network in 22q11. 2 deletion
  syndrome: a diffusion tensor imaging study}.
\newblock \bibinfo{journal}{Behavioral and brain functions}
  \bibinfo{volume}{13}, \bibinfo{pages}{1--17}.
\bibitem[{Ombao et~al.(2016)Ombao, Lindquist, Thompson and
  Aston}]{ombao2016handbook}
\bibinfo{author}{Ombao, H.}, \bibinfo{author}{Lindquist, M.},
  \bibinfo{author}{Thompson, W.}, \bibinfo{author}{Aston, J.},
  \bibinfo{year}{2016}.
\newblock \bibinfo{title}{Handbook of neuroimaging data analysis}.
\newblock \bibinfo{publisher}{CRC Press}.
\bibitem[{Osmanl{\i}o{\u{g}}lu et~al.(2020)Osmanl{\i}o{\u{g}}lu, Alappatt,
  Parker and Verma}]{osmanliouglu2020connectomic}
\bibinfo{author}{Osmanl{\i}o{\u{g}}lu, Y.}, \bibinfo{author}{Alappatt, J.A.},
  \bibinfo{author}{Parker, D.}, \bibinfo{author}{Verma, R.},
  \bibinfo{year}{2020}.
\newblock \bibinfo{title}{Connectomic consistency: a systematic stability
  analysis of structural and functional connectivity}.
\newblock \bibinfo{journal}{Journal of neural engineering}
  \bibinfo{volume}{17}, \bibinfo{pages}{045004}.
\bibitem[{Ouyang et~al.(2019)Ouyang, Dubois, Yu, Mukherjee and
  Huang}]{ouyang2019delineation}
\bibinfo{author}{Ouyang, M.}, \bibinfo{author}{Dubois, J.},
  \bibinfo{author}{Yu, Q.}, \bibinfo{author}{Mukherjee, P.},
  \bibinfo{author}{Huang, H.}, \bibinfo{year}{2019}.
\newblock \bibinfo{title}{Delineation of early brain development from fetuses
  to infants with diffusion {MRI} and beyond}.
\newblock \bibinfo{journal}{Neuroimage} \bibinfo{volume}{185},
  \bibinfo{pages}{836--850}.
\bibitem[{O’Donnell et~al.(2002)O’Donnell, Haker and
  Westin}]{odonnell2002new}
\bibinfo{author}{O’Donnell, L.}, \bibinfo{author}{Haker, S.},
  \bibinfo{author}{Westin, C.F.}, \bibinfo{year}{2002}.
\newblock \bibinfo{title}{New approaches to estimation of white matter
  connectivity in diffusion tensor {MRI}: Elliptic pdes and geodesics in a
  tensor-warped space}, in: \bibinfo{booktitle}{International Conference on
  Medical Image Computing and Computer-Assisted Intervention},
  \bibinfo{organization}{Springer}. pp. \bibinfo{pages}{459--466}.
\bibitem[{Panagiotaki et~al.(2012)Panagiotaki, Schneider, Siow, Hall, Lythgoe
  and Alexander}]{panagiotaki2012compartment}
\bibinfo{author}{Panagiotaki, E.}, \bibinfo{author}{Schneider, T.},
  \bibinfo{author}{Siow, B.}, \bibinfo{author}{Hall, M.G.},
  \bibinfo{author}{Lythgoe, M.F.}, \bibinfo{author}{Alexander, D.C.},
  \bibinfo{year}{2012}.
\newblock \bibinfo{title}{Compartment models of the diffusion mr signal in
  brain white matter: a taxonomy and comparison}.
\newblock \bibinfo{journal}{Neuroimage} \bibinfo{volume}{59},
  \bibinfo{pages}{2241--2254}.
\bibitem[{Panesar et~al.(2019)Panesar, Abhinav, Yeh, Jacquesson, Collins and
  Fernandez-Miranda}]{panesar2019tractography}
\bibinfo{author}{Panesar, S.S.}, \bibinfo{author}{Abhinav, K.},
  \bibinfo{author}{Yeh, F.C.}, \bibinfo{author}{Jacquesson, T.},
  \bibinfo{author}{Collins, M.}, \bibinfo{author}{Fernandez-Miranda, J.},
  \bibinfo{year}{2019}.
\newblock \bibinfo{title}{Tractography for surgical neuro-oncology planning:
  towards a gold standard}.
\newblock \bibinfo{journal}{neurotherapeutics} \bibinfo{volume}{16},
  \bibinfo{pages}{36--51}.
\bibitem[{Pasternak et~al.(2012)Pasternak, Shenton and Westin}]{Pasternak2012}
\bibinfo{author}{Pasternak, O.}, \bibinfo{author}{Shenton, M.E.},
  \bibinfo{author}{Westin, C.F.}, \bibinfo{year}{2012}.
\newblock \bibinfo{title}{{Estimation of extracellular volume from regularized
  multi-shell diffusion MRI.}}
\newblock \bibinfo{journal}{Medical image computing and computer-assisted
  intervention : MICCAI ... International Conference on Medical Image Computing
  and Computer-Assisted Intervention} \bibinfo{volume}{15},
  \bibinfo{pages}{305--312}.
\newblock \DOIprefix\doi{10.1007/978-3-642-33418-4_38}.
\bibitem[{Pasternak et~al.(2009)Pasternak, Sochen, Gur, Intrator and
  Assaf}]{Pasternak2009}
\bibinfo{author}{Pasternak, O.}, \bibinfo{author}{Sochen, N.},
  \bibinfo{author}{Gur, Y.}, \bibinfo{author}{Intrator, N.},
  \bibinfo{author}{Assaf, Y.}, \bibinfo{year}{2009}.
\newblock \bibinfo{title}{{Free water elimination and mapping from diffusion
  MRI}}.
\newblock \bibinfo{journal}{Magnetic Resonance in Medicine}
  \bibinfo{volume}{62}, \bibinfo{pages}{717--730}.
\bibitem[{Payabvash et~al.(2019)Payabvash, Palacios, Owen, Wang, Tavassoli,
  Gerdes, Brandes-Aitken, Marco and Mukherjee}]{payabvash2019diffusion}
\bibinfo{author}{Payabvash, S.}, \bibinfo{author}{Palacios, E.M.},
  \bibinfo{author}{Owen, J.P.}, \bibinfo{author}{Wang, M.B.},
  \bibinfo{author}{Tavassoli, T.}, \bibinfo{author}{Gerdes, M.},
  \bibinfo{author}{Brandes-Aitken, A.}, \bibinfo{author}{Marco, E.J.},
  \bibinfo{author}{Mukherjee, P.}, \bibinfo{year}{2019}.
\newblock \bibinfo{title}{Diffusion tensor tractography in children with
  sensory processing disorder: Potentials for devising machine learning
  classifiers}.
\newblock \bibinfo{journal}{NeuroImage: Clinical} \bibinfo{volume}{23},
  \bibinfo{pages}{101831}.
\bibitem[{Paydar et~al.(2014)Paydar, Fieremans, Nwankwo, Lazar, Sheth,
  Adisetiyo, Helpern, Jensen and Milla}]{paydar2014diffusional}
\bibinfo{author}{Paydar, A.}, \bibinfo{author}{Fieremans, E.},
  \bibinfo{author}{Nwankwo, J.}, \bibinfo{author}{Lazar, M.},
  \bibinfo{author}{Sheth, H.}, \bibinfo{author}{Adisetiyo, V.},
  \bibinfo{author}{Helpern, J.}, \bibinfo{author}{Jensen, J.},
  \bibinfo{author}{Milla, S.}, \bibinfo{year}{2014}.
\newblock \bibinfo{title}{Diffusional kurtosis imaging of the developing
  brain}.
\newblock \bibinfo{journal}{American Journal of Neuroradiology}
  \bibinfo{volume}{35}, \bibinfo{pages}{808--814}.
\bibitem[{Pecheva et~al.(2017)Pecheva, Yushkevich, Batalle, Hughes, Aljabar,
  Wurie, Hajnal, Edwards, Alexander, Counsell et~al.}]{pecheva2017tract}
\bibinfo{author}{Pecheva, D.}, \bibinfo{author}{Yushkevich, P.},
  \bibinfo{author}{Batalle, D.}, \bibinfo{author}{Hughes, E.},
  \bibinfo{author}{Aljabar, P.}, \bibinfo{author}{Wurie, J.},
  \bibinfo{author}{Hajnal, J.V.}, \bibinfo{author}{Edwards, A.D.},
  \bibinfo{author}{Alexander, D.C.}, \bibinfo{author}{Counsell, S.J.}, et~al.,
  \bibinfo{year}{2017}.
\newblock \bibinfo{title}{A tract-specific approach to assessing white matter
  in preterm infants}.
\newblock \bibinfo{journal}{Neuroimage} \bibinfo{volume}{157},
  \bibinfo{pages}{675--694}.
\bibitem[{Pestilli et~al.(2014)Pestilli, Yeatman, Rokem, Kay and
  Wandell}]{pestilli_evaluation_2014}
\bibinfo{author}{Pestilli, F.}, \bibinfo{author}{Yeatman, J.D.},
  \bibinfo{author}{Rokem, A.}, \bibinfo{author}{Kay, K.N.},
  \bibinfo{author}{Wandell, B.A.}, \bibinfo{year}{2014}.
\newblock \bibinfo{title}{Evaluation and statistical inference for human
  connectomes}.
\newblock \bibinfo{journal}{Nat Meth} \bibinfo{volume}{11},
  \bibinfo{pages}{1058--1063}.
\bibitem[{Peters and Karlsgodt(2015)}]{peters2015white}
\bibinfo{author}{Peters, B.D.}, \bibinfo{author}{Karlsgodt, K.H.},
  \bibinfo{year}{2015}.
\newblock \bibinfo{title}{White matter development in the early stages of
  psychosis}.
\newblock \bibinfo{journal}{Schizophrenia research} \bibinfo{volume}{161},
  \bibinfo{pages}{61--69}.
\bibitem[{Petrov et~al.(2017)Petrov, Ivanov, Faskowitz, Gutman, Moyer,
  Villalon, Jahanshad and Thompson}]{petrov2017evaluating}
\bibinfo{author}{Petrov, D.}, \bibinfo{author}{Ivanov, A.},
  \bibinfo{author}{Faskowitz, J.}, \bibinfo{author}{Gutman, B.},
  \bibinfo{author}{Moyer, D.}, \bibinfo{author}{Villalon, J.},
  \bibinfo{author}{Jahanshad, N.}, \bibinfo{author}{Thompson, P.},
  \bibinfo{year}{2017}.
\newblock \bibinfo{title}{Evaluating 35 methods to generate structural
  connectomes using pairwise classification}, in:
  \bibinfo{booktitle}{International Conference on medical Image Computing and
  Computer-Assisted Intervention}, \bibinfo{organization}{Springer}. pp.
  \bibinfo{pages}{515--522}.
\bibitem[{Pichon et~al.(2005)Pichon, Westin and
  Tannenbaum}]{pichon_hamilton-jacobi-bellman_2005}
\bibinfo{author}{Pichon, E.}, \bibinfo{author}{Westin, C.F.},
  \bibinfo{author}{Tannenbaum, A.R.}, \bibinfo{year}{2005}.
\newblock \bibinfo{title}{A {Hamilton}-{Jacobi}-{Bellman} approach to high
  angular resolution diffusion tractography}.
\newblock \bibinfo{journal}{Medical image computing and computer-assisted
  intervention} \bibinfo{volume}{8}, \bibinfo{pages}{180--187}.
\bibitem[{Piredda et~al.(2021)Piredda, Hilbert, Thiran and Kober}]{Piredda2020}
\bibinfo{author}{Piredda, G.F.}, \bibinfo{author}{Hilbert, T.},
  \bibinfo{author}{Thiran, J.P.}, \bibinfo{author}{Kober, T.},
  \bibinfo{year}{2021}.
\newblock \bibinfo{title}{{Probing myelin content of the human brain with MRI:
  A review}}.
\newblock \bibinfo{journal}{Magnetic Resonance in Medicine}
  \bibinfo{volume}{85}, \bibinfo{pages}{627--652}.
\newblock \DOIprefix\doi{https://doi.org/10.1002/mrm.28509}.
\bibitem[{Pizzolato et~al.(2020)Pizzolato, Palombo, Bonet-Carne, Tax, Grussu,
  Ianus, Bogusz, Pieciak, Ning, Larochelle et~al.}]{pizzolato2020acquiring}
\bibinfo{author}{Pizzolato, M.}, \bibinfo{author}{Palombo, M.},
  \bibinfo{author}{Bonet-Carne, E.}, \bibinfo{author}{Tax, C.M.},
  \bibinfo{author}{Grussu, F.}, \bibinfo{author}{Ianus, A.},
  \bibinfo{author}{Bogusz, F.}, \bibinfo{author}{Pieciak, T.},
  \bibinfo{author}{Ning, L.}, \bibinfo{author}{Larochelle, H.}, et~al.,
  \bibinfo{year}{2020}.
\newblock \bibinfo{title}{Acquiring and predicting multidimensional diffusion
  (mudi) data: An open challenge}, in: \bibinfo{booktitle}{Computational
  Diffusion MRI}. \bibinfo{publisher}{Springer}, pp. \bibinfo{pages}{195--208}.
\bibitem[{Poulin et~al.(2019)Poulin, J{\"o}rgens, Jodoin and
  Descoteaux}]{poulin2019tractography}
\bibinfo{author}{Poulin, P.}, \bibinfo{author}{J{\"o}rgens, D.},
  \bibinfo{author}{Jodoin, P.M.}, \bibinfo{author}{Descoteaux, M.},
  \bibinfo{year}{2019}.
\newblock \bibinfo{title}{Tractography and machine learning: Current state and
  open challenges}.
\newblock \bibinfo{journal}{Magnetic resonance imaging} \bibinfo{volume}{64},
  \bibinfo{pages}{37--48}.
\bibitem[{Pr{\v{c}}kovska et~al.(2016)Pr{\v{c}}kovska, Rodrigues,
  Puigdellivol~Sanchez, Ramos, Andorra, Martinez-Heras, Falcon, Prats-Galino
  and Villoslada}]{prvckovska2016reproducibility}
\bibinfo{author}{Pr{\v{c}}kovska, V.}, \bibinfo{author}{Rodrigues, P.},
  \bibinfo{author}{Puigdellivol~Sanchez, A.}, \bibinfo{author}{Ramos, M.},
  \bibinfo{author}{Andorra, M.}, \bibinfo{author}{Martinez-Heras, E.},
  \bibinfo{author}{Falcon, C.}, \bibinfo{author}{Prats-Galino, A.},
  \bibinfo{author}{Villoslada, P.}, \bibinfo{year}{2016}.
\newblock \bibinfo{title}{Reproducibility of the structural connectome
  reconstruction across diffusion methods}.
\newblock \bibinfo{journal}{Journal of Neuroimaging} \bibinfo{volume}{26},
  \bibinfo{pages}{46--57}.
\bibitem[{Price et~al.(2007)Price, Cercignani, Parker, Altmann, Barnes, Barker,
  Joyce and Ron}]{price2007abnormal}
\bibinfo{author}{Price, G.}, \bibinfo{author}{Cercignani, M.},
  \bibinfo{author}{Parker, G.J.}, \bibinfo{author}{Altmann, D.R.},
  \bibinfo{author}{Barnes, T.R.}, \bibinfo{author}{Barker, G.J.},
  \bibinfo{author}{Joyce, E.M.}, \bibinfo{author}{Ron, M.A.},
  \bibinfo{year}{2007}.
\newblock \bibinfo{title}{Abnormal brain connectivity in first-episode
  psychosis: a diffusion {MRI} tractography study of the corpus callosum}.
\newblock \bibinfo{journal}{Neuroimage} \bibinfo{volume}{35},
  \bibinfo{pages}{458--466}.
\bibitem[{Prosperini et~al.(2014)Prosperini, Fanelli, Petsas, Sbardella, Tona,
  Raz, Fortuna, De~Angelis, Pozzilli and Pantano}]{prosperini2014multiple}
\bibinfo{author}{Prosperini, L.}, \bibinfo{author}{Fanelli, F.},
  \bibinfo{author}{Petsas, N.}, \bibinfo{author}{Sbardella, E.},
  \bibinfo{author}{Tona, F.}, \bibinfo{author}{Raz, E.},
  \bibinfo{author}{Fortuna, D.}, \bibinfo{author}{De~Angelis, F.},
  \bibinfo{author}{Pozzilli, C.}, \bibinfo{author}{Pantano, P.},
  \bibinfo{year}{2014}.
\newblock \bibinfo{title}{Multiple sclerosis: changes in microarchitecture of
  white matter tracts after training with a video game balance board}.
\newblock \bibinfo{journal}{Radiology} \bibinfo{volume}{273},
  \bibinfo{pages}{529--538}.
\bibitem[{Pujol et~al.(2015)Pujol, Wells, Pierpaoli, Brun, Gee, Cheng, Vemuri,
  Commowick, Prima, Stamm et~al.}]{pujol2015dti}
\bibinfo{author}{Pujol, S.}, \bibinfo{author}{Wells, W.},
  \bibinfo{author}{Pierpaoli, C.}, \bibinfo{author}{Brun, C.},
  \bibinfo{author}{Gee, J.}, \bibinfo{author}{Cheng, G.},
  \bibinfo{author}{Vemuri, B.}, \bibinfo{author}{Commowick, O.},
  \bibinfo{author}{Prima, S.}, \bibinfo{author}{Stamm, A.}, et~al.,
  \bibinfo{year}{2015}.
\newblock \bibinfo{title}{The dti challenge: toward standardized evaluation of
  diffusion tensor imaging tractography for neurosurgery}.
\newblock \bibinfo{journal}{Journal of Neuroimaging} \bibinfo{volume}{25},
  \bibinfo{pages}{875--882}.
\bibitem[{Qi et~al.(2015)Qi, Meesters, Nicolay, ter Haar~Romeny and
  Ossenblok}]{qi2015influence}
\bibinfo{author}{Qi, S.}, \bibinfo{author}{Meesters, S.},
  \bibinfo{author}{Nicolay, K.}, \bibinfo{author}{ter Haar~Romeny, B.M.},
  \bibinfo{author}{Ossenblok, P.}, \bibinfo{year}{2015}.
\newblock \bibinfo{title}{The influence of construction methodology on
  structural brain network measures: A review}.
\newblock \bibinfo{journal}{Journal of neuroscience methods}
  \bibinfo{volume}{253}, \bibinfo{pages}{170--182}.
\bibitem[{Qiu et~al.(2015)Qiu, Mori and Miller}]{qiu2015diffusion}
\bibinfo{author}{Qiu, A.}, \bibinfo{author}{Mori, S.}, \bibinfo{author}{Miller,
  M.I.}, \bibinfo{year}{2015}.
\newblock \bibinfo{title}{Diffusion tensor imaging for understanding brain
  development in early life}.
\newblock \bibinfo{journal}{Annual review of psychology} \bibinfo{volume}{66},
  \bibinfo{pages}{853--876}.
\bibitem[{Qiu et~al.(2010)Qiu, Oishi, Miller, Lyketsos, Mori and
  Albert}]{qiu2010surface}
\bibinfo{author}{Qiu, A.}, \bibinfo{author}{Oishi, K.},
  \bibinfo{author}{Miller, M.I.}, \bibinfo{author}{Lyketsos, C.G.},
  \bibinfo{author}{Mori, S.}, \bibinfo{author}{Albert, M.},
  \bibinfo{year}{2010}.
\newblock \bibinfo{title}{Surface-based analysis on shape and fractional
  anisotropy of white matter tracts in alzheimer's disease}.
\newblock \bibinfo{journal}{PloS one} \bibinfo{volume}{5},
  \bibinfo{pages}{e9811}.
\bibitem[{Radmanesh et~al.(2015)Radmanesh, Zamani, Whalen, Tie, Suarez and
  Golby}]{radmanesh2015comparison}
\bibinfo{author}{Radmanesh, A.}, \bibinfo{author}{Zamani, A.A.},
  \bibinfo{author}{Whalen, S.}, \bibinfo{author}{Tie, Y.},
  \bibinfo{author}{Suarez, R.O.}, \bibinfo{author}{Golby, A.J.},
  \bibinfo{year}{2015}.
\newblock \bibinfo{title}{Comparison of seeding methods for visualization of
  the corticospinal tracts using single tensor tractography}.
\newblock \bibinfo{journal}{Clinical neurology and neurosurgery}
  \bibinfo{volume}{129}, \bibinfo{pages}{44--49}.
\bibitem[{Raffelt et~al.(2012)Raffelt, Tournier, Rose, Ridgway, Henderson,
  Crozier, Salvado and Connelly}]{Raffelt2012}
\bibinfo{author}{Raffelt, D.}, \bibinfo{author}{Tournier, J.D.},
  \bibinfo{author}{Rose, S.}, \bibinfo{author}{Ridgway, G.R.},
  \bibinfo{author}{Henderson, R.}, \bibinfo{author}{Crozier, S.},
  \bibinfo{author}{Salvado, O.}, \bibinfo{author}{Connelly, A.},
  \bibinfo{year}{2012}.
\newblock \bibinfo{title}{{Apparent Fibre Density: A novel measure for the
  analysis of diffusion-weighted magnetic resonance images}}.
\newblock \bibinfo{journal}{NeuroImage} \bibinfo{volume}{59},
  \bibinfo{pages}{3976--3994}.
\newblock \DOIprefix\doi{https://doi.org/10.1016/j.neuroimage.2011.10.045}.
\bibitem[{Raichle(2015)}]{raichle2015brain}
\bibinfo{author}{Raichle, M.E.}, \bibinfo{year}{2015}.
\newblock \bibinfo{title}{The brain's default mode network}.
\newblock \bibinfo{journal}{Annual review of neuroscience}
  \bibinfo{volume}{38}, \bibinfo{pages}{433--447}.
\bibitem[{Ratnarajah and Qiu(2014)}]{ratnarajah2014multi}
\bibinfo{author}{Ratnarajah, N.}, \bibinfo{author}{Qiu, A.},
  \bibinfo{year}{2014}.
\newblock \bibinfo{title}{Multi-label segmentation of white matter structures:
  Application to neonatal brains}.
\newblock \bibinfo{journal}{NeuroImage} \bibinfo{volume}{102},
  \bibinfo{pages}{913--922}.
\bibitem[{Reddy and Rathi(2016)}]{reddy2016joint}
\bibinfo{author}{Reddy, C.P.}, \bibinfo{author}{Rathi, Y.},
  \bibinfo{year}{2016}.
\newblock \bibinfo{title}{Joint multi-fiber noddi parameter estimation and
  tractography using the unscented information filter}.
\newblock \bibinfo{journal}{Frontiers in neuroscience} \bibinfo{volume}{10},
  \bibinfo{pages}{166}.
\bibitem[{Reisert et~al.(2018)Reisert, Coenen, Kaller, Egger and
  Skibbe}]{reisert2018hamlet}
\bibinfo{author}{Reisert, M.}, \bibinfo{author}{Coenen, V.A.},
  \bibinfo{author}{Kaller, C.}, \bibinfo{author}{Egger, K.},
  \bibinfo{author}{Skibbe, H.}, \bibinfo{year}{2018}.
\newblock \bibinfo{title}{Hamlet: hierarchical harmonic filters for learning
  tracts from diffusion mri}.
\newblock \bibinfo{journal}{arXiv preprint arXiv:1807.01068} .
\bibitem[{Reisert et~al.(2011)Reisert, Mader, Anastasopoulos, Weigel, Schnell
  and Kiselev}]{reisert_global_2011}
\bibinfo{author}{Reisert, M.}, \bibinfo{author}{Mader, I.},
  \bibinfo{author}{Anastasopoulos, C.}, \bibinfo{author}{Weigel, M.},
  \bibinfo{author}{Schnell, S.}, \bibinfo{author}{Kiselev, V.},
  \bibinfo{year}{2011}.
\newblock \bibinfo{title}{Global fiber reconstruction becomes practical}.
\newblock \bibinfo{journal}{NeuroImage} \bibinfo{volume}{54},
  \bibinfo{pages}{955--962}.
\bibitem[{de~Reus and Van~den Heuvel(2013)}]{de2013parcellation}
\bibinfo{author}{de~Reus, M.A.}, \bibinfo{author}{Van~den Heuvel, M.P.},
  \bibinfo{year}{2013}.
\newblock \bibinfo{title}{The parcellation-based connectome: limitations and
  extensions}.
\newblock \bibinfo{journal}{Neuroimage} \bibinfo{volume}{80},
  \bibinfo{pages}{397--404}.
\bibitem[{Reveley et~al.(2015)Reveley, Seth, Pierpaoli, Silva, Yu, Saunders,
  Leopold and Ye}]{reveley_superficial_wm_2015}
\bibinfo{author}{Reveley, C.}, \bibinfo{author}{Seth, A.K.},
  \bibinfo{author}{Pierpaoli, C.}, \bibinfo{author}{Silva, A.C.},
  \bibinfo{author}{Yu, D.}, \bibinfo{author}{Saunders, R.C.},
  \bibinfo{author}{Leopold, D.A.}, \bibinfo{author}{Ye, F.Q.},
  \bibinfo{year}{2015}.
\newblock \bibinfo{title}{Superficial white matter fiber systems impede
  detection of long-range cortical connections in diffusion mr tractography}.
\newblock \bibinfo{journal}{Proceedings of the National Academy of Sciences}
  \bibinfo{volume}{112}, \bibinfo{pages}{E2820--E2828}.
\bibitem[{Reynolds et~al.(2019)Reynolds, Grohs, Dewey and
  Lebel}]{reynolds2019global}
\bibinfo{author}{Reynolds, J.E.}, \bibinfo{author}{Grohs, M.N.},
  \bibinfo{author}{Dewey, D.}, \bibinfo{author}{Lebel, C.},
  \bibinfo{year}{2019}.
\newblock \bibinfo{title}{Global and regional white matter development in early
  childhood}.
\newblock \bibinfo{journal}{Neuroimage} \bibinfo{volume}{196},
  \bibinfo{pages}{49--58}.
\bibitem[{Rheault et~al.(2020a)Rheault, {De Benedictis}, Daducci, Maffei, Tax,
  Romascano, Caverzasi, Morency, Corrivetti, Pestilli, Girard, Theaud,
  Zemmoura, Hau, Glavin, Jordan, Pomiecko, Chamberland, Barakovic, Goyette,
  Poulin, Chenot, Panesar, Sarubbo, Petit and
  Descoteaux}]{Rheault2020_tractostorm}
\bibinfo{author}{Rheault, F.}, \bibinfo{author}{{De Benedictis}, A.},
  \bibinfo{author}{Daducci, A.}, \bibinfo{author}{Maffei, C.},
  \bibinfo{author}{Tax, C.M.W.}, \bibinfo{author}{Romascano, D.},
  \bibinfo{author}{Caverzasi, E.}, \bibinfo{author}{Morency, F.C.},
  \bibinfo{author}{Corrivetti, F.}, \bibinfo{author}{Pestilli, F.},
  \bibinfo{author}{Girard, G.}, \bibinfo{author}{Theaud, G.},
  \bibinfo{author}{Zemmoura, I.}, \bibinfo{author}{Hau, J.},
  \bibinfo{author}{Glavin, K.}, \bibinfo{author}{Jordan, K.M.},
  \bibinfo{author}{Pomiecko, K.}, \bibinfo{author}{Chamberland, M.},
  \bibinfo{author}{Barakovic, M.}, \bibinfo{author}{Goyette, N.},
  \bibinfo{author}{Poulin, P.}, \bibinfo{author}{Chenot, Q.},
  \bibinfo{author}{Panesar, S.S.}, \bibinfo{author}{Sarubbo, S.},
  \bibinfo{author}{Petit, L.}, \bibinfo{author}{Descoteaux, M.},
  \bibinfo{year}{2020}a.
\newblock \bibinfo{title}{{Tractostorm: The what, why, and how of tractography
  dissection reproducibility}}.
\newblock \bibinfo{journal}{Human Brain Mapping} \bibinfo{volume}{41},
  \bibinfo{pages}{1859--1874}.
\newblock \DOIprefix\doi{https://doi.org/10.1002/hbm.24917}.
\bibitem[{Rheault et~al.(2020b)Rheault, Poulin, Caron, St-Onge and
  Descoteaux}]{Rheault2020}
\bibinfo{author}{Rheault, F.}, \bibinfo{author}{Poulin, P.},
  \bibinfo{author}{Caron, A.V.}, \bibinfo{author}{St-Onge, E.},
  \bibinfo{author}{Descoteaux, M.}, \bibinfo{year}{2020}b.
\newblock \bibinfo{title}{{Common misconceptions, hidden biases and modern
  challenges of dMRI tractography}}.
\newblock \bibinfo{journal}{Journal of Neural Engineering}
  \bibinfo{volume}{17}, \bibinfo{pages}{11001}.
\newblock \DOIprefix\doi{10.1088/1741-2552/ab6aad}.
\bibitem[{Rheault et~al.(2020c)Rheault, Poulin, Caron, St-Onge and
  Descoteaux}]{rheault2020common}
\bibinfo{author}{Rheault, F.}, \bibinfo{author}{Poulin, P.},
  \bibinfo{author}{Caron, A.V.}, \bibinfo{author}{St-Onge, E.},
  \bibinfo{author}{Descoteaux, M.}, \bibinfo{year}{2020}c.
\newblock \bibinfo{title}{Common misconceptions, hidden biases and modern
  challenges of dmri tractography}.
\newblock \bibinfo{journal}{Journal of neural engineering}
  \bibinfo{volume}{17}, \bibinfo{pages}{011001}.
\bibitem[{Rheault et~al.(2019)Rheault, St-Onge, Sidhu, Maier-Hein,
  Tzourio-Mazoyer, Petit and Descoteaux}]{rheault2019bundle}
\bibinfo{author}{Rheault, F.}, \bibinfo{author}{St-Onge, E.},
  \bibinfo{author}{Sidhu, J.}, \bibinfo{author}{Maier-Hein, K.},
  \bibinfo{author}{Tzourio-Mazoyer, N.}, \bibinfo{author}{Petit, L.},
  \bibinfo{author}{Descoteaux, M.}, \bibinfo{year}{2019}.
\newblock \bibinfo{title}{Bundle-specific tractography with incorporated
  anatomical and orientational priors}.
\newblock \bibinfo{journal}{NeuroImage} \bibinfo{volume}{186},
  \bibinfo{pages}{382--398}.
\bibitem[{Roberts et~al.(2016)Roberts, Perry, Lord, Roberts, Mitchell, Smith,
  Calamante and Breakspear}]{roberts2016contribution}
\bibinfo{author}{Roberts, J.A.}, \bibinfo{author}{Perry, A.},
  \bibinfo{author}{Lord, A.R.}, \bibinfo{author}{Roberts, G.},
  \bibinfo{author}{Mitchell, P.B.}, \bibinfo{author}{Smith, R.E.},
  \bibinfo{author}{Calamante, F.}, \bibinfo{author}{Breakspear, M.},
  \bibinfo{year}{2016}.
\newblock \bibinfo{title}{The contribution of geometry to the human
  connectome}.
\newblock \bibinfo{journal}{Neuroimage} \bibinfo{volume}{124},
  \bibinfo{pages}{379--393}.
\bibitem[{Rocca et~al.(2015)Rocca, Amato, De~Stefano, Enzinger, Geurts, Penner,
  Rovira, Sumowski, Valsasina, Filippi et~al.}]{rocca2015clinical}
\bibinfo{author}{Rocca, M.A.}, \bibinfo{author}{Amato, M.P.},
  \bibinfo{author}{De~Stefano, N.}, \bibinfo{author}{Enzinger, C.},
  \bibinfo{author}{Geurts, J.J.}, \bibinfo{author}{Penner, I.K.},
  \bibinfo{author}{Rovira, A.}, \bibinfo{author}{Sumowski, J.F.},
  \bibinfo{author}{Valsasina, P.}, \bibinfo{author}{Filippi, M.}, et~al.,
  \bibinfo{year}{2015}.
\newblock \bibinfo{title}{Clinical and imaging assessment of cognitive
  dysfunction in multiple sclerosis}.
\newblock \bibinfo{journal}{The Lancet Neurology} \bibinfo{volume}{14},
  \bibinfo{pages}{302--317}.
\bibitem[{Rodrigues et~al.(2013)Rodrigues, Prats, Gallardo-Pujol, Villoslada,
  Falcon and Pr{\v{c}}kovska}]{rodrigues2013evaluating}
\bibinfo{author}{Rodrigues, P.}, \bibinfo{author}{Prats, A.},
  \bibinfo{author}{Gallardo-Pujol, D.}, \bibinfo{author}{Villoslada, P.},
  \bibinfo{author}{Falcon, C.}, \bibinfo{author}{Pr{\v{c}}kovska, V.},
  \bibinfo{year}{2013}.
\newblock \bibinfo{title}{Evaluating structural connectomics: the effect of the
  cortical parcellation scheme}, in: \bibinfo{booktitle}{Front. Neuroinform.
  Conference Abstract: Imaging the brain at different scales: How to integrate
  multi-scale structural information}.
\bibitem[{Rom{\'a}n et~al.(2017)Rom{\'a}n, Guevara, Valenzuela, Figueroa,
  Houenou, Duclap, Poupon, Mangin and Guevara}]{roman2017clustering}
\bibinfo{author}{Rom{\'a}n, C.}, \bibinfo{author}{Guevara, M.},
  \bibinfo{author}{Valenzuela, R.}, \bibinfo{author}{Figueroa, M.},
  \bibinfo{author}{Houenou, J.}, \bibinfo{author}{Duclap, D.},
  \bibinfo{author}{Poupon, C.}, \bibinfo{author}{Mangin, J.F.},
  \bibinfo{author}{Guevara, P.}, \bibinfo{year}{2017}.
\newblock \bibinfo{title}{Clustering of whole-brain white matter short
  association bundles using hardi data}.
\newblock \bibinfo{journal}{Frontiers in neuroinformatics}
  \bibinfo{volume}{11}, \bibinfo{pages}{73}.
\bibitem[{Ros et~al.(2013)Ros, G{\"u}llmar, Stenzel, Mentzel and
  Reichenbach}]{ros2013atlas}
\bibinfo{author}{Ros, C.}, \bibinfo{author}{G{\"u}llmar, D.},
  \bibinfo{author}{Stenzel, M.}, \bibinfo{author}{Mentzel, H.J.},
  \bibinfo{author}{Reichenbach, J.R.}, \bibinfo{year}{2013}.
\newblock \bibinfo{title}{Atlas-guided cluster analysis of large tractography
  datasets}.
\newblock \bibinfo{journal}{PloS one} \bibinfo{volume}{8},
  \bibinfo{pages}{e83847}.
\bibitem[{Rossini et~al.(2019)Rossini, Di~Iorio, Bentivoglio, Bertini, Ferreri,
  Gerloff, Ilmoniemi, Miraglia, Nitsche, Pestilli et~al.}]{rossini2019methods}
\bibinfo{author}{Rossini, P.}, \bibinfo{author}{Di~Iorio, R.},
  \bibinfo{author}{Bentivoglio, M.}, \bibinfo{author}{Bertini, G.},
  \bibinfo{author}{Ferreri, F.}, \bibinfo{author}{Gerloff, C.},
  \bibinfo{author}{Ilmoniemi, R.}, \bibinfo{author}{Miraglia, F.},
  \bibinfo{author}{Nitsche, M.}, \bibinfo{author}{Pestilli, F.}, et~al.,
  \bibinfo{year}{2019}.
\newblock \bibinfo{title}{Methods for analysis of brain connectivity: An
  ifcn-sponsored review}.
\newblock \bibinfo{journal}{Clinical Neurophysiology} \bibinfo{volume}{130},
  \bibinfo{pages}{1833--1858}.
\bibitem[{Rubinov and Sporns(2010)}]{rubinov:2010}
\bibinfo{author}{Rubinov, M.}, \bibinfo{author}{Sporns, O.},
  \bibinfo{year}{2010}.
\newblock \bibinfo{title}{Complex network measures of brain connectivity: uses
  and interpretations}.
\newblock \bibinfo{journal}{Neuroimage} \bibinfo{volume}{52},
  \bibinfo{pages}{1059--69}.
\bibitem[{Sakkalis(2011)}]{sakkalis2011review}
\bibinfo{author}{Sakkalis, V.}, \bibinfo{year}{2011}.
\newblock \bibinfo{title}{Review of advanced techniques for the estimation of
  brain connectivity measured with eeg/meg}.
\newblock \bibinfo{journal}{Computers in biology and medicine}
  \bibinfo{volume}{41}, \bibinfo{pages}{1110--1117}.
\bibitem[{Sarubbo et~al.(2020)Sarubbo, Tate, De~Benedictis, Merler,
  Moritz-Gasser, Herbet and Duffau}]{sarubbo2020mapping}
\bibinfo{author}{Sarubbo, S.}, \bibinfo{author}{Tate, M.},
  \bibinfo{author}{De~Benedictis, A.}, \bibinfo{author}{Merler, S.},
  \bibinfo{author}{Moritz-Gasser, S.}, \bibinfo{author}{Herbet, G.},
  \bibinfo{author}{Duffau, H.}, \bibinfo{year}{2020}.
\newblock \bibinfo{title}{Mapping critical cortical hubs and white matter
  pathways by direct electrical stimulation: an original functional atlas of
  the human brain}.
\newblock \bibinfo{journal}{Neuroimage} \bibinfo{volume}{205},
  \bibinfo{pages}{116237}.
\bibitem[{Sarwar et~al.(2020)Sarwar, Seguin, Ramamohanarao and
  Zalesky}]{sarwar2020towards}
\bibinfo{author}{Sarwar, T.}, \bibinfo{author}{Seguin, C.},
  \bibinfo{author}{Ramamohanarao, K.}, \bibinfo{author}{Zalesky, A.},
  \bibinfo{year}{2020}.
\newblock \bibinfo{title}{Towards deep learning for connectome mapping: A block
  decomposition framework}.
\newblock \bibinfo{journal}{NeuroImage} \bibinfo{volume}{212},
  \bibinfo{pages}{116654}.
\bibitem[{Sbardella et~al.(2013)Sbardella, Tona, Petsas and
  Pantano}]{sbardella2013dti}
\bibinfo{author}{Sbardella, E.}, \bibinfo{author}{Tona, F.},
  \bibinfo{author}{Petsas, N.}, \bibinfo{author}{Pantano, P.},
  \bibinfo{year}{2013}.
\newblock \bibinfo{title}{Dti measurements in multiple sclerosis: evaluation of
  brain damage and clinical implications}.
\newblock \bibinfo{journal}{Multiple sclerosis international}
  \bibinfo{volume}{2013}.
\bibitem[{Schaefer et~al.(2018)Schaefer, Kong, Gordon, Laumann, Zuo, Holmes,
  Eickhoff and Yeo}]{schaefer2018local}
\bibinfo{author}{Schaefer, A.}, \bibinfo{author}{Kong, R.},
  \bibinfo{author}{Gordon, E.M.}, \bibinfo{author}{Laumann, T.O.},
  \bibinfo{author}{Zuo, X.N.}, \bibinfo{author}{Holmes, A.J.},
  \bibinfo{author}{Eickhoff, S.B.}, \bibinfo{author}{Yeo, B.T.},
  \bibinfo{year}{2018}.
\newblock \bibinfo{title}{Local-global parcellation of the human cerebral
  cortex from intrinsic functional connectivity {MRI}}.
\newblock \bibinfo{journal}{Cerebral cortex} \bibinfo{volume}{28},
  \bibinfo{pages}{3095--3114}.
\bibitem[{Schaie(2005)}]{schaie2005developmental}
\bibinfo{author}{Schaie, K.W.}, \bibinfo{year}{2005}.
\newblock \bibinfo{title}{Developmental influences on adult intelligence: The
  Seattle longitudinal study}.
\newblock \bibinfo{publisher}{Oxford University Press}.
\bibitem[{Schiavi et~al.(2020a)Schiavi, Ocampo-Pineda, Barakovic, Petit,
  Descoteaux, Thiran and Daducci}]{schiavi2019reducing}
\bibinfo{author}{Schiavi, S.}, \bibinfo{author}{Ocampo-Pineda, M.},
  \bibinfo{author}{Barakovic, M.}, \bibinfo{author}{Petit, L.},
  \bibinfo{author}{Descoteaux, M.}, \bibinfo{author}{Thiran, J.P.},
  \bibinfo{author}{Daducci, A.}, \bibinfo{year}{2020}a.
\newblock \bibinfo{title}{A new method for accurate in vivo mapping of human
  brain connections using microstructural and anatomical information}.
\newblock \bibinfo{journal}{Science Advances} \bibinfo{volume}{6}.
\bibitem[{Schiavi et~al.(2020b)Schiavi, Petracca, Battocchio, {El Mendili},
  Paduri, Fleysher, Inglese and Daducci}]{Schiavi2020_commit}
\bibinfo{author}{Schiavi, S.}, \bibinfo{author}{Petracca, M.},
  \bibinfo{author}{Battocchio, M.}, \bibinfo{author}{{El Mendili}, M.M.},
  \bibinfo{author}{Paduri, S.}, \bibinfo{author}{Fleysher, L.},
  \bibinfo{author}{Inglese, M.}, \bibinfo{author}{Daducci, A.},
  \bibinfo{year}{2020}b.
\newblock \bibinfo{title}{{Sensory-motor network topology in multiple
  sclerosis: Structural connectivity analysis accounting for intrinsic density
  discrepancy}}.
\newblock \bibinfo{journal}{Human Brain Mapping} \bibinfo{volume}{n/a}.
\newblock \DOIprefix\doi{10.1002/hbm.24989}.
\bibitem[{Schiavi et~al.(2019)Schiavi, Pizzolato, Ocampo-Pineda,
  Canales-Rodriguez, Thiran and Daducci}]{Schiavi2019ISMRM}
\bibinfo{author}{Schiavi, S.}, \bibinfo{author}{Pizzolato, M.},
  \bibinfo{author}{Ocampo-Pineda, M.}, \bibinfo{author}{Canales-Rodriguez,
  E.J.}, \bibinfo{author}{Thiran, J.P.}, \bibinfo{author}{Daducci, A.},
  \bibinfo{year}{2019}.
\newblock \bibinfo{title}{Is it feasible to directly access the bundle’s
  specific myelin content, instead of averaging? a study with microstructure
  informed tractography}, in: \bibinfo{booktitle}{ISMRM 27th Annual Meeting \&
  Exhibition}.
\bibitem[{Schilling et~al.(2018a)Schilling, Gao, Janve, Stepniewska, Landman
  and Anderson}]{schilling2018confirmation}
\bibinfo{author}{Schilling, K.}, \bibinfo{author}{Gao, Y.},
  \bibinfo{author}{Janve, V.}, \bibinfo{author}{Stepniewska, I.},
  \bibinfo{author}{Landman, B.A.}, \bibinfo{author}{Anderson, A.W.},
  \bibinfo{year}{2018}a.
\newblock \bibinfo{title}{Confirmation of a gyral bias in diffusion {MRI} fiber
  tractography}.
\newblock \bibinfo{journal}{Human brain mapping} \bibinfo{volume}{39},
  \bibinfo{pages}{1449--1466}.
\bibitem[{Schilling et~al.(2018b)Schilling, Gao, Janve, Stepniewska, Landman
  and Anderson}]{schilling_gyral_bias_2018}
\bibinfo{author}{Schilling, K.}, \bibinfo{author}{Gao, Y.},
  \bibinfo{author}{Janve, V.}, \bibinfo{author}{Stepniewska, I.},
  \bibinfo{author}{Landman, B.A.}, \bibinfo{author}{Anderson, A.W.},
  \bibinfo{year}{2018}b.
\newblock \bibinfo{title}{Confirmation of a gyral bias in diffusion {MRI} fiber
  tractography}.
\newblock \bibinfo{journal}{Human Brain Mapping} \bibinfo{volume}{39},
  \bibinfo{pages}{1449--1466}.
\bibitem[{Schilling et~al.(2020a)Schilling, Petit, Rheault, Remedios,
  Pierpaoli, Anderson, Landman and Descoteaux}]{Schilling.2020}
\bibinfo{author}{Schilling, K.G.}, \bibinfo{author}{Petit, L.},
  \bibinfo{author}{Rheault, F.}, \bibinfo{author}{Remedios, S.},
  \bibinfo{author}{Pierpaoli, C.}, \bibinfo{author}{Anderson, A.W.},
  \bibinfo{author}{Landman, B.A.}, \bibinfo{author}{Descoteaux, M.},
  \bibinfo{year}{2020}a.
\newblock \bibinfo{title}{{Brain connections derived from diffusion MRI
  tractography can be highly anatomically accurate—if we know where white
  matter pathways start, where they end, and where they do not go}}.
\newblock \bibinfo{journal}{Brain Structure and Function}
  \bibinfo{volume}{225}, \bibinfo{pages}{2387--2402}.
\newblock \DOIprefix\doi{10.1007/s00429-020-02129-z}.
\bibitem[{Schilling et~al.(2020b)Schilling, Rheault, Petit, Hansen, Nath, Yeh,
  Girard, Barakovic, Rafael-Patino, Yu et~al.}]{schilling2020tractography}
\bibinfo{author}{Schilling, K.G.}, \bibinfo{author}{Rheault, F.},
  \bibinfo{author}{Petit, L.}, \bibinfo{author}{Hansen, C.B.},
  \bibinfo{author}{Nath, V.}, \bibinfo{author}{Yeh, F.C.},
  \bibinfo{author}{Girard, G.}, \bibinfo{author}{Barakovic, M.},
  \bibinfo{author}{Rafael-Patino, J.}, \bibinfo{author}{Yu, T.}, et~al.,
  \bibinfo{year}{2020}b.
\newblock \bibinfo{title}{Tractography dissection variability: what happens
  when 42 groups dissect 14 white matter bundles on the same dataset?}
\newblock \bibinfo{journal}{bioRxiv} .
\bibitem[{Schmahmann et~al.(2009)Schmahmann, Schmahmann and
  Pandya}]{schmahmann2009fiber}
\bibinfo{author}{Schmahmann, J.D.}, \bibinfo{author}{Schmahmann, J.},
  \bibinfo{author}{Pandya, D.}, \bibinfo{year}{2009}.
\newblock \bibinfo{title}{Fiber pathways of the brain}.
\newblock \bibinfo{publisher}{OUP USA}.
\bibitem[{Schmidt et~al.(2017)Schmidt, Crossley, Harrisberger, Smieskova, Lenz,
  Riecher-R{\"o}ssler, Lang, McGuire, Fusar-Poli and
  Borgwardt}]{schmidt2017structural}
\bibinfo{author}{Schmidt, A.}, \bibinfo{author}{Crossley, N.A.},
  \bibinfo{author}{Harrisberger, F.}, \bibinfo{author}{Smieskova, R.},
  \bibinfo{author}{Lenz, C.}, \bibinfo{author}{Riecher-R{\"o}ssler, A.},
  \bibinfo{author}{Lang, U.E.}, \bibinfo{author}{McGuire, P.},
  \bibinfo{author}{Fusar-Poli, P.}, \bibinfo{author}{Borgwardt, S.},
  \bibinfo{year}{2017}.
\newblock \bibinfo{title}{Structural network disorganization in subjects at
  clinical high risk for psychosis}.
\newblock \bibinfo{journal}{Schizophrenia bulletin} \bibinfo{volume}{43},
  \bibinfo{pages}{583--591}.
\bibitem[{Schomburg and Hohage(2019)}]{schomburg2019formulation}
\bibinfo{author}{Schomburg, H.}, \bibinfo{author}{Hohage, T.},
  \bibinfo{year}{2019}.
\newblock \bibinfo{title}{Formulation and efficient computation of l1 - and
  smoothness penalized estimates for microstructure-informed tractography}.
\newblock \bibinfo{journal}{IEEE transactions on medical imaging}
  \bibinfo{volume}{38}, \bibinfo{pages}{1899--1909}.
\bibitem[{de~Schotten et~al.(2011)de~Schotten, Bizzi, Dell'Acqua, Allin,
  Walshe, Murray, Williams, Murphy, Catani et~al.}]{de2011atlasing}
\bibinfo{author}{de~Schotten, M.T.}, \bibinfo{author}{Bizzi, A.},
  \bibinfo{author}{Dell'Acqua, F.}, \bibinfo{author}{Allin, M.},
  \bibinfo{author}{Walshe, M.}, \bibinfo{author}{Murray, R.},
  \bibinfo{author}{Williams, S.C.}, \bibinfo{author}{Murphy, D.G.},
  \bibinfo{author}{Catani, M.}, et~al., \bibinfo{year}{2011}.
\newblock \bibinfo{title}{Atlasing location, asymmetry and inter-subject
  variability of white matter tracts in the human brain with mr diffusion
  tractography}.
\newblock \bibinfo{journal}{Neuroimage} \bibinfo{volume}{54},
  \bibinfo{pages}{49--59}.
\bibitem[{Schreiber et~al.(2014)Schreiber, Riffert, Anwander and
  Knösche}]{schreiber_plausibility_2014}
\bibinfo{author}{Schreiber, J.}, \bibinfo{author}{Riffert, T.},
  \bibinfo{author}{Anwander, A.}, \bibinfo{author}{Knösche, T.R.},
  \bibinfo{year}{2014}.
\newblock \bibinfo{title}{Plausibility {Tracking}: {A} method to evaluate
  anatomical connectivity and microstructural properties along fiber pathways}.
\newblock \bibinfo{journal}{NeuroImage} \bibinfo{volume}{90},
  \bibinfo{pages}{163--178}.
\bibitem[{Schroeder and Salthouse(2004)}]{schroeder2004age}
\bibinfo{author}{Schroeder, D.H.}, \bibinfo{author}{Salthouse, T.A.},
  \bibinfo{year}{2004}.
\newblock \bibinfo{title}{Age-related effects on cognition between 20 and 50
  years of age}.
\newblock \bibinfo{journal}{Personality and individual differences}
  \bibinfo{volume}{36}, \bibinfo{pages}{393--404}.
\bibitem[{Schurr et~al.(2019)Schurr, Filo and Mezer}]{schurr2019tractography}
\bibinfo{author}{Schurr, R.}, \bibinfo{author}{Filo, S.},
  \bibinfo{author}{Mezer, A.A.}, \bibinfo{year}{2019}.
\newblock \bibinfo{title}{Tractography delineation of the vertical occipital
  fasciculus using quantitative t1 mapping}.
\newblock \bibinfo{journal}{NeuroImage} \bibinfo{volume}{202},
  \bibinfo{pages}{116121}.
\bibitem[{Seguin et~al.(2018)Seguin, van~den Heuvel and Zalesky}]{seguin:2018}
\bibinfo{author}{Seguin, C.}, \bibinfo{author}{van~den Heuvel, M.P.},
  \bibinfo{author}{Zalesky, A.}, \bibinfo{year}{2018}.
\newblock \bibinfo{title}{Navigation of brain networks}.
\newblock \bibinfo{journal}{Proc Natl Acad Sci U S A} \bibinfo{volume}{115},
  \bibinfo{pages}{6297--6302}.
\bibitem[{Seguin et~al.(2019)Seguin, Razi and Zalesky}]{seguin:2019}
\bibinfo{author}{Seguin, C.}, \bibinfo{author}{Razi, A.},
  \bibinfo{author}{Zalesky, A.}, \bibinfo{year}{2019}.
\newblock \bibinfo{title}{Inferring neural signalling directionality from
  undirected structural connectomes}.
\newblock \bibinfo{journal}{Nat Commun} \bibinfo{volume}{10},
  \bibinfo{pages}{4289}.
\bibitem[{Seguin et~al.(2020)Seguin, Tian and Zalesky}]{seguin:2020}
\bibinfo{author}{Seguin, C.}, \bibinfo{author}{Tian, Y.},
  \bibinfo{author}{Zalesky, A.}, \bibinfo{year}{2020}.
\newblock \bibinfo{title}{Network communication models improve the behavioral
  and functional predictive utility of the human structural connectome}.
\newblock \bibinfo{journal}{Network Neuroscience} .
\bibitem[{Sepasian et~al.(2012)Sepasian, ten Thije~Boonkkamp, Ter Haar~Romeny
  and Vilanova~Bartroli}]{sepasian_multivalued_2012}
\bibinfo{author}{Sepasian, N.}, \bibinfo{author}{ten Thije~Boonkkamp, J.},
  \bibinfo{author}{Ter Haar~Romeny, B.}, \bibinfo{author}{Vilanova~Bartroli,
  A.}, \bibinfo{year}{2012}.
\newblock \bibinfo{title}{Multivalued geodesic ray-tracing for computing brain
  connections using diffusion tensor imaging}.
\newblock \bibinfo{journal}{SIAM Journal on Imaging Sciences}
  \bibinfo{volume}{5}, \bibinfo{pages}{483--504}.
\bibitem[{Shahab et~al.(2018)Shahab, Stefanik, Foussias, Lai, Anderson and
  Voineskos}]{shahab2018sex}
\bibinfo{author}{Shahab, S.}, \bibinfo{author}{Stefanik, L.},
  \bibinfo{author}{Foussias, G.}, \bibinfo{author}{Lai, M.C.},
  \bibinfo{author}{Anderson, K.K.}, \bibinfo{author}{Voineskos, A.N.},
  \bibinfo{year}{2018}.
\newblock \bibinfo{title}{Sex and diffusion tensor imaging of white matter in
  schizophrenia: A systematic review plus meta-analysis of the corpus
  callosum}.
\newblock \bibinfo{journal}{Schizophrenia bulletin} \bibinfo{volume}{44},
  \bibinfo{pages}{203--221}.
\bibitem[{Shany et~al.(2017)Shany, Inder, Goshen, Lee, Neil, Smyser, Doyle,
  Anderson and Shimony}]{shany2017diffusion}
\bibinfo{author}{Shany, E.}, \bibinfo{author}{Inder, T.E.},
  \bibinfo{author}{Goshen, S.}, \bibinfo{author}{Lee, I.},
  \bibinfo{author}{Neil, J.J.}, \bibinfo{author}{Smyser, C.D.},
  \bibinfo{author}{Doyle, L.W.}, \bibinfo{author}{Anderson, P.J.},
  \bibinfo{author}{Shimony, J.S.}, \bibinfo{year}{2017}.
\newblock \bibinfo{title}{Diffusion tensor tractography of the cerebellar
  peduncles in prematurely born 7-year-old children}.
\newblock \bibinfo{journal}{The Cerebellum} \bibinfo{volume}{16},
  \bibinfo{pages}{314--325}.
\bibitem[{Shattuck and Leahy(2002)}]{shattuck2002brainsuite}
\bibinfo{author}{Shattuck, D.W.}, \bibinfo{author}{Leahy, R.M.},
  \bibinfo{year}{2002}.
\newblock \bibinfo{title}{Brainsuite: an automated cortical surface
  identification tool}.
\newblock \bibinfo{journal}{Medical image analysis} \bibinfo{volume}{6},
  \bibinfo{pages}{129--142}.
\bibitem[{Shattuck et~al.(2008)Shattuck, Mirza, Adisetiyo, Hojatkashani,
  Salamon, Narr, Poldrack, Bilder and Toga}]{shattuck2008construction}
\bibinfo{author}{Shattuck, D.W.}, \bibinfo{author}{Mirza, M.},
  \bibinfo{author}{Adisetiyo, V.}, \bibinfo{author}{Hojatkashani, C.},
  \bibinfo{author}{Salamon, G.}, \bibinfo{author}{Narr, K.L.},
  \bibinfo{author}{Poldrack, R.A.}, \bibinfo{author}{Bilder, R.M.},
  \bibinfo{author}{Toga, A.W.}, \bibinfo{year}{2008}.
\newblock \bibinfo{title}{Construction of a 3d probabilistic atlas of human
  cortical structures}.
\newblock \bibinfo{journal}{Neuroimage} \bibinfo{volume}{39},
  \bibinfo{pages}{1064--1080}.
\bibitem[{Shemesh et~al.(2016)Shemesh, Jespersen, Alexander, Cohen, Drobnjak,
  Dyrby, Finsterbusch, Koch, Kuder, Laun et~al.}]{shemesh2016conventions}
\bibinfo{author}{Shemesh, N.}, \bibinfo{author}{Jespersen, S.N.},
  \bibinfo{author}{Alexander, D.C.}, \bibinfo{author}{Cohen, Y.},
  \bibinfo{author}{Drobnjak, I.}, \bibinfo{author}{Dyrby, T.B.},
  \bibinfo{author}{Finsterbusch, J.}, \bibinfo{author}{Koch, M.A.},
  \bibinfo{author}{Kuder, T.}, \bibinfo{author}{Laun, F.}, et~al.,
  \bibinfo{year}{2016}.
\newblock \bibinfo{title}{Conventions and nomenclature for double diffusion
  encoding nmr and mri}.
\newblock \bibinfo{journal}{Magnetic resonance in medicine}
  \bibinfo{volume}{75}, \bibinfo{pages}{82--87}.
\bibitem[{Sherbondy et~al.(2010)Sherbondy, Rowe and
  Alexander}]{sherbondy_microtrack:_2010}
\bibinfo{author}{Sherbondy, A.}, \bibinfo{author}{Rowe, M.},
  \bibinfo{author}{Alexander, D.}, \bibinfo{year}{2010}.
\newblock \bibinfo{title}{{MicroTrack}: {An} {Algorithm} for {Concurrent}
  {Projectome} and {Microstructure} {Estimation}}, in: \bibinfo{editor}{Jiang,
  T.}, \bibinfo{editor}{Navab, N.}, \bibinfo{editor}{Pluim, J.},
  \bibinfo{editor}{Viergever, M.} (Eds.), \bibinfo{booktitle}{Medical {Image}
  {Computing} and {Computer}-{Assisted} {Intervention}},
  \bibinfo{publisher}{Springer Berlin / Heidelberg}. pp.
  \bibinfo{pages}{183--190}.
\bibitem[{Sherbondy et~al.(2009)Sherbondy, Dougherty, Ananthanarayanan, Modha
  and Wandell}]{sherbondy_think_2009}
\bibinfo{author}{Sherbondy, A.J.}, \bibinfo{author}{Dougherty, R.F.},
  \bibinfo{author}{Ananthanarayanan, R.}, \bibinfo{author}{Modha, D.S.},
  \bibinfo{author}{Wandell, B.A.}, \bibinfo{year}{2009}.
\newblock \bibinfo{title}{Think {Global}, {Act} {Local}; {Projectome}
  {Estimation} with {BlueMatter}}, in: \bibinfo{booktitle}{Medical {Image}
  {Computing} and {Computer}-{Assisted} {Intervention}},
  \bibinfo{publisher}{Springer-Verlag}, \bibinfo{address}{London, UK}. pp.
  \bibinfo{pages}{861--868}.
\bibitem[{Shergill et~al.(2007)Shergill, Kanaan, Chitnis, O’Daly, Jones,
  Frangou, Williams, Howard, Barker, Murray et~al.}]{shergill2007diffusion}
\bibinfo{author}{Shergill, S.S.}, \bibinfo{author}{Kanaan, R.A.},
  \bibinfo{author}{Chitnis, X.A.}, \bibinfo{author}{O’Daly, O.},
  \bibinfo{author}{Jones, D.K.}, \bibinfo{author}{Frangou, S.},
  \bibinfo{author}{Williams, S.C.}, \bibinfo{author}{Howard, R.J.},
  \bibinfo{author}{Barker, G.J.}, \bibinfo{author}{Murray, R.M.}, et~al.,
  \bibinfo{year}{2007}.
\newblock \bibinfo{title}{A diffusion tensor imaging study of fasciculi in
  schizophrenia}.
\newblock \bibinfo{journal}{American Journal of Psychiatry}
  \bibinfo{volume}{164}, \bibinfo{pages}{467--473}.
\bibitem[{Shi et~al.(2012)Shi, Yap, Gao, Lin, Gilmore and
  Shen}]{shi2012altered}
\bibinfo{author}{Shi, F.}, \bibinfo{author}{Yap, P.T.}, \bibinfo{author}{Gao,
  W.}, \bibinfo{author}{Lin, W.}, \bibinfo{author}{Gilmore, J.H.},
  \bibinfo{author}{Shen, D.}, \bibinfo{year}{2012}.
\newblock \bibinfo{title}{Altered structural connectivity in neonates at
  genetic risk for schizophrenia: a combined study using morphological and
  white matter networks}.
\newblock \bibinfo{journal}{Neuroimage} \bibinfo{volume}{62},
  \bibinfo{pages}{1622--1633}.
\bibitem[{Shi and Toga(2017)}]{shi2017connectome}
\bibinfo{author}{Shi, Y.}, \bibinfo{author}{Toga, A.W.}, \bibinfo{year}{2017}.
\newblock \bibinfo{title}{Connectome imaging for mapping human brain pathways}.
\newblock \bibinfo{journal}{Molecular psychiatry} \bibinfo{volume}{22},
  \bibinfo{pages}{1230--1240}.
\bibitem[{Shu et~al.(2018a)Shu, Duan, Huang, Ren, Liu, Dong, Barkhof, Li and
  Liu}]{shu2018progressive}
\bibinfo{author}{Shu, N.}, \bibinfo{author}{Duan, Y.}, \bibinfo{author}{Huang,
  J.}, \bibinfo{author}{Ren, Z.}, \bibinfo{author}{Liu, Z.},
  \bibinfo{author}{Dong, H.}, \bibinfo{author}{Barkhof, F.},
  \bibinfo{author}{Li, K.}, \bibinfo{author}{Liu, Y.}, \bibinfo{year}{2018}a.
\newblock \bibinfo{title}{Progressive brain rich-club network disruption from
  clinically isolated syndrome towards multiple sclerosis}.
\newblock \bibinfo{journal}{NeuroImage: Clinical} \bibinfo{volume}{19},
  \bibinfo{pages}{232--239}.
\bibitem[{Shu et~al.(2016)Shu, Duan, Xia, Schoonheim, Huang, Ren, Sun, Ye,
  Dong, Shi et~al.}]{shu2016disrupted}
\bibinfo{author}{Shu, N.}, \bibinfo{author}{Duan, Y.}, \bibinfo{author}{Xia,
  M.}, \bibinfo{author}{Schoonheim, M.M.}, \bibinfo{author}{Huang, J.},
  \bibinfo{author}{Ren, Z.}, \bibinfo{author}{Sun, Z.}, \bibinfo{author}{Ye,
  J.}, \bibinfo{author}{Dong, H.}, \bibinfo{author}{Shi, F.D.}, et~al.,
  \bibinfo{year}{2016}.
\newblock \bibinfo{title}{Disrupted topological organization of structural and
  functional brain connectomes in clinically isolated syndrome and multiple
  sclerosis}.
\newblock \bibinfo{journal}{Scientific reports} \bibinfo{volume}{6},
  \bibinfo{pages}{1--11}.
\bibitem[{Shu et~al.(2011)Shu, Liu, Li, Duan, Wang, Yu, Dong, Ye and
  He}]{shu2011diffusion}
\bibinfo{author}{Shu, N.}, \bibinfo{author}{Liu, Y.}, \bibinfo{author}{Li, K.},
  \bibinfo{author}{Duan, Y.}, \bibinfo{author}{Wang, J.}, \bibinfo{author}{Yu,
  C.}, \bibinfo{author}{Dong, H.}, \bibinfo{author}{Ye, J.},
  \bibinfo{author}{He, Y.}, \bibinfo{year}{2011}.
\newblock \bibinfo{title}{Diffusion tensor tractography reveals disrupted
  topological efficiency in white matter structural networks in multiple
  sclerosis}.
\newblock \bibinfo{journal}{Cerebral cortex} \bibinfo{volume}{21},
  \bibinfo{pages}{2565--2577}.
\bibitem[{Shu et~al.(2018b)Shu, Wang, Bi, Zhao and Han}]{shu2018disrupted}
\bibinfo{author}{Shu, N.}, \bibinfo{author}{Wang, X.}, \bibinfo{author}{Bi,
  Q.}, \bibinfo{author}{Zhao, T.}, \bibinfo{author}{Han, Y.},
  \bibinfo{year}{2018}b.
\newblock \bibinfo{title}{Disrupted topologic efficiency of white matter
  structural connectome in individuals with subjective cognitive decline}.
\newblock \bibinfo{journal}{Radiology} \bibinfo{volume}{286},
  \bibinfo{pages}{229--238}.
\bibitem[{Sidman and Rakic(1973)}]{sidman1973neuronal}
\bibinfo{author}{Sidman, R.L.}, \bibinfo{author}{Rakic, P.},
  \bibinfo{year}{1973}.
\newblock \bibinfo{title}{Neuronal migration, with special reference to
  developing human brain: a review}.
\newblock \bibinfo{journal}{Brain research} \bibinfo{volume}{62},
  \bibinfo{pages}{1--35}.
\bibitem[{Siless et~al.(2018)Siless, Chang, Fischl and
  Yendiki}]{siless2018anatomicuts}
\bibinfo{author}{Siless, V.}, \bibinfo{author}{Chang, K.},
  \bibinfo{author}{Fischl, B.}, \bibinfo{author}{Yendiki, A.},
  \bibinfo{year}{2018}.
\newblock \bibinfo{title}{Anatomicuts: Hierarchical clustering of tractography
  streamlines based on anatomical similarity}.
\newblock \bibinfo{journal}{NeuroImage} \bibinfo{volume}{166},
  \bibinfo{pages}{32--45}.
\bibitem[{Siless et~al.(2020)Siless, Davidow, Nielsen, Fan, Hedden,
  Hollinshead, Beam, Bustamante, Garrad, Santillana
  et~al.}]{siless2020registration}
\bibinfo{author}{Siless, V.}, \bibinfo{author}{Davidow, J.Y.},
  \bibinfo{author}{Nielsen, J.}, \bibinfo{author}{Fan, Q.},
  \bibinfo{author}{Hedden, T.}, \bibinfo{author}{Hollinshead, M.},
  \bibinfo{author}{Beam, E.}, \bibinfo{author}{Bustamante, C.M.V.},
  \bibinfo{author}{Garrad, M.C.}, \bibinfo{author}{Santillana, R.}, et~al.,
  \bibinfo{year}{2020}.
\newblock \bibinfo{title}{Registration-free analysis of diffusion {MRI}
  tractography data across subjects through the human lifespan}.
\newblock \bibinfo{journal}{NeuroImage} , \bibinfo{pages}{116703}.
\bibitem[{Sinke et~al.(2018)Sinke, Otte, Christiaens, Schmitt, Leemans, van~der
  Toorn, Sarabdjitsingh, Jo{\"e}ls and Dijkhuizen}]{sinke2018diffusion}
\bibinfo{author}{Sinke, M.R.}, \bibinfo{author}{Otte, W.M.},
  \bibinfo{author}{Christiaens, D.}, \bibinfo{author}{Schmitt, O.},
  \bibinfo{author}{Leemans, A.}, \bibinfo{author}{van~der Toorn, A.},
  \bibinfo{author}{Sarabdjitsingh, R.A.}, \bibinfo{author}{Jo{\"e}ls, M.},
  \bibinfo{author}{Dijkhuizen, R.M.}, \bibinfo{year}{2018}.
\newblock \bibinfo{title}{Diffusion {MRI}-based cortical connectome
  reconstruction: dependency on tractography procedures and neuroanatomical
  characteristics}.
\newblock \bibinfo{journal}{Brain Structure and Function}
  \bibinfo{volume}{223}, \bibinfo{pages}{2269--2285}.
\bibitem[{Smith et~al.(2020a)Smith, Raffelt, Tournier and
  Connelly}]{Smith2020Quantitative}
\bibinfo{author}{Smith, R.}, \bibinfo{author}{Raffelt, D.},
  \bibinfo{author}{Tournier, J.D.}, \bibinfo{author}{Connelly, A.},
  \bibinfo{year}{2020}a.
\newblock \bibinfo{title}{Quantitative streamlines tractography: methods and
  inter-subject normalisation}.
\bibitem[{Smith et~al.(2020b)Smith, Connelly and
  Calamante}]{smith_chapter_2020}
\bibinfo{author}{Smith, R.E.}, \bibinfo{author}{Connelly, A.},
  \bibinfo{author}{Calamante, F.}, \bibinfo{year}{2020}b.
\newblock \bibinfo{title}{Chapter 21 - {Diffusion} {MRI} {Fiber}
  {Tractography}}, in: \bibinfo{editor}{Seiberlich, N.},
  \bibinfo{editor}{Gulani, V.}, \bibinfo{editor}{Calamante, F.},
  \bibinfo{editor}{Campbell-Washburn, A.}, \bibinfo{editor}{Doneva, M.},
  \bibinfo{editor}{Hu, H.H.}, \bibinfo{editor}{Sourbron, S.} (Eds.),
  \bibinfo{booktitle}{Advances in {Magnetic} {Resonance} {Technology} and
  {Applications}}. \bibinfo{publisher}{Academic Press}.
  volume~\bibinfo{volume}{1} of \textit{\bibinfo{series}{Quantitative
  {Magnetic} {Resonance} {Imaging}}}, pp. \bibinfo{pages}{533--569}.
\bibitem[{Smith et~al.(2020c)Smith, Connelly and
  Calamante}]{smith2020diffusion}
\bibinfo{author}{Smith, R.E.}, \bibinfo{author}{Connelly, A.},
  \bibinfo{author}{Calamante, F.}, \bibinfo{year}{2020}c.
\newblock \bibinfo{title}{Diffusion {MRI} fiber tractography}, in:
  \bibinfo{booktitle}{Advances in Magnetic Resonance Technology and
  Applications}. \bibinfo{publisher}{Elsevier}. volume~\bibinfo{volume}{1}, pp.
  \bibinfo{pages}{533--569}.
\bibitem[{Smith et~al.(2012)Smith, Tournier, Calamante and
  Connelly}]{smith_anatomically-constrained_2012}
\bibinfo{author}{Smith, R.E.}, \bibinfo{author}{Tournier, J.D.},
  \bibinfo{author}{Calamante, F.}, \bibinfo{author}{Connelly, A.},
  \bibinfo{year}{2012}.
\newblock \bibinfo{title}{Anatomically-constrained tractography: {Improved}
  diffusion {MRI} streamlines tractography through effective use of anatomical
  information}.
\newblock \bibinfo{journal}{NeuroImage} \bibinfo{volume}{62},
  \bibinfo{pages}{1924--1938}.
\bibitem[{Smith et~al.(2013)Smith, Tournier, Calamante and
  Connelly}]{smith_sift:_2013}
\bibinfo{author}{Smith, R.E.}, \bibinfo{author}{Tournier, J.D.},
  \bibinfo{author}{Calamante, F.}, \bibinfo{author}{Connelly, A.},
  \bibinfo{year}{2013}.
\newblock \bibinfo{title}{{SIFT}: {Spherical}-deconvolution informed filtering
  of tractograms}.
\newblock \bibinfo{journal}{NeuroImage} \bibinfo{volume}{67},
  \bibinfo{pages}{298--312}.
\bibitem[{Smith et~al.(2015a)Smith, Tournier, Calamante and
  Connelly}]{smith_sift2:_2015}
\bibinfo{author}{Smith, R.E.}, \bibinfo{author}{Tournier, J.D.},
  \bibinfo{author}{Calamante, F.}, \bibinfo{author}{Connelly, A.},
  \bibinfo{year}{2015}a.
\newblock \bibinfo{title}{{SIFT2}: {Enabling} dense quantitative assessment of
  brain white matter connectivity using streamlines tractography}.
\newblock \bibinfo{journal}{NeuroImage} \bibinfo{volume}{119},
  \bibinfo{pages}{338--351}.
\bibitem[{Smith et~al.(2015b)Smith, Tournier, Calamante and
  Connelly}]{Smith:2015aa}
\bibinfo{author}{Smith, R.E.}, \bibinfo{author}{Tournier, J.D.},
  \bibinfo{author}{Calamante, F.}, \bibinfo{author}{Connelly, A.},
  \bibinfo{year}{2015}b.
\newblock \bibinfo{title}{{The effects of SIFT on the reproducibility and
  biological accuracy of the structural connectome}}.
\newblock \bibinfo{journal}{NeuroImage} \bibinfo{volume}{104},
  \bibinfo{pages}{253--265}.
\bibitem[{Sollmann et~al.(2018)Sollmann, Kelm, Ille, Schr{\"o}der, Zimmer,
  Ringel, Meyer and Krieg}]{sollmann2018setup}
\bibinfo{author}{Sollmann, N.}, \bibinfo{author}{Kelm, A.},
  \bibinfo{author}{Ille, S.}, \bibinfo{author}{Schr{\"o}der, A.},
  \bibinfo{author}{Zimmer, C.}, \bibinfo{author}{Ringel, F.},
  \bibinfo{author}{Meyer, B.}, \bibinfo{author}{Krieg, S.M.},
  \bibinfo{year}{2018}.
\newblock \bibinfo{title}{Setup presentation and clinical outcome analysis of
  treating highly language-eloquent gliomas via preoperative navigated
  transcranial magnetic stimulation and tractography}.
\newblock \bibinfo{journal}{Neurosurgical focus} \bibinfo{volume}{44},
  \bibinfo{pages}{E2}.
\bibitem[{Song et~al.(2015)Song, Mitchell, Kolasinski, Ellen~Grant, Galaburda
  and Takahashi}]{song2015asymmetry}
\bibinfo{author}{Song, J.W.}, \bibinfo{author}{Mitchell, P.D.},
  \bibinfo{author}{Kolasinski, J.}, \bibinfo{author}{Ellen~Grant, P.},
  \bibinfo{author}{Galaburda, A.M.}, \bibinfo{author}{Takahashi, E.},
  \bibinfo{year}{2015}.
\newblock \bibinfo{title}{Asymmetry of white matter pathways in developing
  human brains}.
\newblock \bibinfo{journal}{Cerebral cortex} \bibinfo{volume}{25},
  \bibinfo{pages}{2883--2893}.
\bibitem[{Song et~al.(2017)Song, Mishra, Ouyang, Peng, Slinger, Liu and
  Huang}]{song2017human}
\bibinfo{author}{Song, L.}, \bibinfo{author}{Mishra, V.},
  \bibinfo{author}{Ouyang, M.}, \bibinfo{author}{Peng, Q.},
  \bibinfo{author}{Slinger, M.}, \bibinfo{author}{Liu, S.},
  \bibinfo{author}{Huang, H.}, \bibinfo{year}{2017}.
\newblock \bibinfo{title}{Human fetal brain connectome: structural network
  development from middle fetal stage to birth}.
\newblock \bibinfo{journal}{Frontiers in neuroscience} \bibinfo{volume}{11},
  \bibinfo{pages}{561}.
\bibitem[{Song et~al.(2003)Song, Sun, Ju, Lin, Cross and
  Neufeld}]{song2003diffusion}
\bibinfo{author}{Song, S.K.}, \bibinfo{author}{Sun, S.W.}, \bibinfo{author}{Ju,
  W.K.}, \bibinfo{author}{Lin, S.J.}, \bibinfo{author}{Cross, A.H.},
  \bibinfo{author}{Neufeld, A.H.}, \bibinfo{year}{2003}.
\newblock \bibinfo{title}{Diffusion tensor imaging detects and differentiates
  axon and myelin degeneration in mouse optic nerve after retinal ischemia}.
\newblock \bibinfo{journal}{Neuroimage} \bibinfo{volume}{20},
  \bibinfo{pages}{1714--1722}.
\bibitem[{Song et~al.(2002)Song, Sun, Ramsbottom, Chang, Russell and
  Cross}]{song2002dysmyelination}
\bibinfo{author}{Song, S.K.}, \bibinfo{author}{Sun, S.W.},
  \bibinfo{author}{Ramsbottom, M.J.}, \bibinfo{author}{Chang, C.},
  \bibinfo{author}{Russell, J.}, \bibinfo{author}{Cross, A.H.},
  \bibinfo{year}{2002}.
\newblock \bibinfo{title}{Dysmyelination revealed through {MRI} as increased
  radial (but unchanged axial) diffusion of water}.
\newblock \bibinfo{journal}{Neuroimage} \bibinfo{volume}{17},
  \bibinfo{pages}{1429--1436}.
\bibitem[{Sotiropoulos et~al.(2016)Sotiropoulos, Hernández-Fernández, Vu,
  Andersson, Moeller, Yacoub, Lenglet, Ugurbil, Behrens and
  Jbabdi}]{sotiropoulos_fusion_2016}
\bibinfo{author}{Sotiropoulos, S.N.}, \bibinfo{author}{Hernández-Fernández,
  M.}, \bibinfo{author}{Vu, A.T.}, \bibinfo{author}{Andersson, J.L.},
  \bibinfo{author}{Moeller, S.}, \bibinfo{author}{Yacoub, E.},
  \bibinfo{author}{Lenglet, C.}, \bibinfo{author}{Ugurbil, K.},
  \bibinfo{author}{Behrens, T.E.}, \bibinfo{author}{Jbabdi, S.},
  \bibinfo{year}{2016}.
\newblock \bibinfo{title}{{Fusion in diffusion MRI for improved fibre
  orientation estimation: An application to the 3T and 7T data of the Human
  Connectome Project}}.
\newblock \bibinfo{journal}{NeuroImage} \bibinfo{volume}{134},
  \bibinfo{pages}{396--409}.
\newblock \DOIprefix\doi{10.1016/j.neuroimage.2016.04.014}.
\bibitem[{Sotiropoulos and Zalesky(2019)}]{SotiropoulosZalesky2019}
\bibinfo{author}{Sotiropoulos, S.N.}, \bibinfo{author}{Zalesky, A.},
  \bibinfo{year}{2019}.
\newblock \bibinfo{title}{{Building connectomes using diffusion MRI: why, how
  and but}}.
\newblock \bibinfo{journal}{NMR in Biomedicine} \bibinfo{volume}{32},
  \bibinfo{pages}{e3752}.
\newblock \DOIprefix\doi{10.1002/nbm.3752}.
\bibitem[{Southwell et~al.(2017)Southwell, Birk, Han, Li, Sall and
  Berger}]{southwell2017resection}
\bibinfo{author}{Southwell, D.G.}, \bibinfo{author}{Birk, H.S.},
  \bibinfo{author}{Han, S.J.}, \bibinfo{author}{Li, J.}, \bibinfo{author}{Sall,
  J.W.}, \bibinfo{author}{Berger, M.S.}, \bibinfo{year}{2017}.
\newblock \bibinfo{title}{Resection of gliomas deemed inoperable by
  neurosurgeons based on preoperative imaging studies}.
\newblock \bibinfo{journal}{Journal of neurosurgery} \bibinfo{volume}{129},
  \bibinfo{pages}{567--575}.
\bibitem[{Spetzler and Martin(1986)}]{spetzler1986proposed}
\bibinfo{author}{Spetzler, R.F.}, \bibinfo{author}{Martin, N.A.},
  \bibinfo{year}{1986}.
\newblock \bibinfo{title}{A proposed grading system for arteriovenous
  malformations}.
\newblock \bibinfo{journal}{Journal of neurosurgery} \bibinfo{volume}{65},
  \bibinfo{pages}{476--483}.
\bibitem[{Sporns and Betzel(2016)}]{sporns:2016}
\bibinfo{author}{Sporns, O.}, \bibinfo{author}{Betzel, R.F.},
  \bibinfo{year}{2016}.
\newblock \bibinfo{title}{Modular brain networks}.
\newblock \bibinfo{journal}{Annu Rev Psychol} \bibinfo{volume}{67},
  \bibinfo{pages}{613--40}.
\bibitem[{Sporns et~al.(2007)Sporns, Honey and K{\"o}tter}]{sporns:2007}
\bibinfo{author}{Sporns, O.}, \bibinfo{author}{Honey, C.J.},
  \bibinfo{author}{K{\"o}tter, R.}, \bibinfo{year}{2007}.
\newblock \bibinfo{title}{Identification and classification of hubs in brain
  networks}.
\newblock \bibinfo{journal}{PLoS One} \bibinfo{volume}{2},
  \bibinfo{pages}{e1049}.
\bibitem[{Sporns et~al.(2005)Sporns, Tononi and K{\"o}tter}]{sporns2005human}
\bibinfo{author}{Sporns, O.}, \bibinfo{author}{Tononi, G.},
  \bibinfo{author}{K{\"o}tter, R.}, \bibinfo{year}{2005}.
\newblock \bibinfo{title}{The human connectome: a structural description of the
  human brain}.
\newblock \bibinfo{journal}{PLoS Comput Biol} \bibinfo{volume}{1},
  \bibinfo{pages}{e42}.
\bibitem[{St-Jean et~al.(2019)St-Jean, Chamberland, Viergever and
  Leemans}]{st2019reducing}
\bibinfo{author}{St-Jean, S.}, \bibinfo{author}{Chamberland, M.},
  \bibinfo{author}{Viergever, M.A.}, \bibinfo{author}{Leemans, A.},
  \bibinfo{year}{2019}.
\newblock \bibinfo{title}{Reducing variability in along-tract analysis with
  diffusion profile realignment}.
\newblock \bibinfo{journal}{NeuroImage} \bibinfo{volume}{199},
  \bibinfo{pages}{663--679}.
\bibitem[{St-Onge et~al.(2018)St-Onge, Daducci, Girard and
  Descoteaux}]{stonge_set_2018}
\bibinfo{author}{St-Onge, E.}, \bibinfo{author}{Daducci, A.},
  \bibinfo{author}{Girard, G.}, \bibinfo{author}{Descoteaux, M.},
  \bibinfo{year}{2018}.
\newblock \bibinfo{title}{Surface-enhanced tractography (set)}.
\newblock \bibinfo{journal}{NeuroImage} \bibinfo{volume}{169},
  \bibinfo{pages}{524--539}.
\newblock \DOIprefix\doi{10.1016/j.neuroimage.2017.12.036}.
\bibitem[{Stadlbauer et~al.(2008a)Stadlbauer, Salomonowitz, Strunk, Hammen and
  Ganslandt}]{stadlbauer2008age}
\bibinfo{author}{Stadlbauer, A.}, \bibinfo{author}{Salomonowitz, E.},
  \bibinfo{author}{Strunk, G.}, \bibinfo{author}{Hammen, T.},
  \bibinfo{author}{Ganslandt, O.}, \bibinfo{year}{2008}a.
\newblock \bibinfo{title}{Age-related degradation in the central nervous
  system: assessment with diffusion-tensor imaging and quantitative fiber
  tracking}.
\newblock \bibinfo{journal}{Radiology} \bibinfo{volume}{247},
  \bibinfo{pages}{179--188}.
\bibitem[{Stadlbauer et~al.(2008b)Stadlbauer, Salomonowitz, Strunk, Hammen and
  Ganslandt}]{stadlbauer2008quantitative}
\bibinfo{author}{Stadlbauer, A.}, \bibinfo{author}{Salomonowitz, E.},
  \bibinfo{author}{Strunk, G.}, \bibinfo{author}{Hammen, T.},
  \bibinfo{author}{Ganslandt, O.}, \bibinfo{year}{2008}b.
\newblock \bibinfo{title}{Quantitative diffusion tensor fiber tracking of
  age-related changes in the limbic system}.
\newblock \bibinfo{journal}{European radiology} \bibinfo{volume}{18},
  \bibinfo{pages}{130--137}.
\bibitem[{Stephens et~al.(2020)Stephens, Langworthy, Short, Girault, Styner and
  Gilmore}]{stephens2020white}
\bibinfo{author}{Stephens, R.L.}, \bibinfo{author}{Langworthy, B.W.},
  \bibinfo{author}{Short, S.J.}, \bibinfo{author}{Girault, J.B.},
  \bibinfo{author}{Styner, M.A.}, \bibinfo{author}{Gilmore, J.H.},
  \bibinfo{year}{2020}.
\newblock \bibinfo{title}{White matter development from birth to 6 years of
  age: A longitudinal study}.
\newblock \bibinfo{journal}{Cerebral Cortex} \bibinfo{volume}{30},
  \bibinfo{pages}{6152--6168}.
\bibitem[{Stieltjes et~al.(2001)Stieltjes, Kaufmann, van Zijl, Fredericksen,
  Pearlson, Solaiyappan and Mori}]{stieltjes2001diffusion}
\bibinfo{author}{Stieltjes, B.}, \bibinfo{author}{Kaufmann, W.E.},
  \bibinfo{author}{van Zijl, P.C.}, \bibinfo{author}{Fredericksen, K.},
  \bibinfo{author}{Pearlson, G.D.}, \bibinfo{author}{Solaiyappan, M.},
  \bibinfo{author}{Mori, S.}, \bibinfo{year}{2001}.
\newblock \bibinfo{title}{Diffusion tensor imaging and axonal tracking in the
  human brainstem}.
\newblock \bibinfo{journal}{Neuroimage} \bibinfo{volume}{14},
  \bibinfo{pages}{723--735}.
\bibitem[{Sui et~al.(2018)Sui, Qi, van Erp, Bustillo, Jiang, Lin, Turner,
  Damaraju, Mayer, Cui et~al.}]{sui2018multimodal}
\bibinfo{author}{Sui, J.}, \bibinfo{author}{Qi, S.}, \bibinfo{author}{van Erp,
  T.G.}, \bibinfo{author}{Bustillo, J.}, \bibinfo{author}{Jiang, R.},
  \bibinfo{author}{Lin, D.}, \bibinfo{author}{Turner, J.A.},
  \bibinfo{author}{Damaraju, E.}, \bibinfo{author}{Mayer, A.R.},
  \bibinfo{author}{Cui, Y.}, et~al., \bibinfo{year}{2018}.
\newblock \bibinfo{title}{Multimodal neuromarkers in schizophrenia via
  cognition-guided {MRI} fusion}.
\newblock \bibinfo{journal}{Nature communications} \bibinfo{volume}{9},
  \bibinfo{pages}{1--14}.
\bibitem[{Sullivan et~al.(2006)Sullivan, Adalsteinsson and
  Pfefferbaum}]{sullivan2006selective}
\bibinfo{author}{Sullivan, E.V.}, \bibinfo{author}{Adalsteinsson, E.},
  \bibinfo{author}{Pfefferbaum, A.}, \bibinfo{year}{2006}.
\newblock \bibinfo{title}{Selective age-related degradation of anterior
  callosal fiber bundles quantified in vivo with fiber tracking}.
\newblock \bibinfo{journal}{Cerebral cortex} \bibinfo{volume}{16},
  \bibinfo{pages}{1030--1039}.
\bibitem[{Sullivan et~al.(2010)Sullivan, Rohlfing and
  Pfefferbaum}]{sullivan2010quantitative}
\bibinfo{author}{Sullivan, E.V.}, \bibinfo{author}{Rohlfing, T.},
  \bibinfo{author}{Pfefferbaum, A.}, \bibinfo{year}{2010}.
\newblock \bibinfo{title}{Quantitative fiber tracking of lateral and
  interhemispheric white matter systems in normal aging: relations to timed
  performance}.
\newblock \bibinfo{journal}{Neurobiology of aging} \bibinfo{volume}{31},
  \bibinfo{pages}{464--481}.
\bibitem[{Sun et~al.(2015)Sun, Lui, Yao, Deng, Xiao, Zhang, Huang, Hu, Bi, Li
  et~al.}]{sun2015two}
\bibinfo{author}{Sun, H.}, \bibinfo{author}{Lui, S.}, \bibinfo{author}{Yao,
  L.}, \bibinfo{author}{Deng, W.}, \bibinfo{author}{Xiao, Y.},
  \bibinfo{author}{Zhang, W.}, \bibinfo{author}{Huang, X.},
  \bibinfo{author}{Hu, J.}, \bibinfo{author}{Bi, F.}, \bibinfo{author}{Li, T.},
  et~al., \bibinfo{year}{2015}.
\newblock \bibinfo{title}{Two patterns of white matter abnormalities in
  medication-naive patients with first-episode schizophrenia revealed by
  diffusion tensor imaging and cluster analysis}.
\newblock \bibinfo{journal}{JAMA psychiatry} \bibinfo{volume}{72},
  \bibinfo{pages}{678--686}.
\bibitem[{Sydnor et~al.(2018)Sydnor, Rivas-Grajales, Lyall, Zhang, Bouix,
  Karmacharya, Shenton, Westin, Makris, Wassermann
  et~al.}]{sydnor2018comparison}
\bibinfo{author}{Sydnor, V.J.}, \bibinfo{author}{Rivas-Grajales, A.M.},
  \bibinfo{author}{Lyall, A.E.}, \bibinfo{author}{Zhang, F.},
  \bibinfo{author}{Bouix, S.}, \bibinfo{author}{Karmacharya, S.},
  \bibinfo{author}{Shenton, M.E.}, \bibinfo{author}{Westin, C.F.},
  \bibinfo{author}{Makris, N.}, \bibinfo{author}{Wassermann, D.}, et~al.,
  \bibinfo{year}{2018}.
\newblock \bibinfo{title}{A comparison of three fiber tract delineation methods
  and their impact on white matter analysis}.
\newblock \bibinfo{journal}{NeuroImage} \bibinfo{volume}{178},
  \bibinfo{pages}{318--331}.
\bibitem[{Szel{\'e}nyi et~al.(2010)Szel{\'e}nyi, Bello, Duffau, Fava, Feigl,
  Galanda, Neuloh, Signorelli and Sala}]{szelenyi2010intraoperative}
\bibinfo{author}{Szel{\'e}nyi, A.}, \bibinfo{author}{Bello, L.},
  \bibinfo{author}{Duffau, H.}, \bibinfo{author}{Fava, E.},
  \bibinfo{author}{Feigl, G.C.}, \bibinfo{author}{Galanda, M.},
  \bibinfo{author}{Neuloh, G.}, \bibinfo{author}{Signorelli, F.},
  \bibinfo{author}{Sala, F.}, \bibinfo{year}{2010}.
\newblock \bibinfo{title}{Intraoperative electrical stimulation in awake
  craniotomy: methodological aspects of current practice}.
\newblock \bibinfo{journal}{Neurosurgical focus} \bibinfo{volume}{28},
  \bibinfo{pages}{E7}.
\bibitem[{Takahashi et~al.(2002)Takahashi, Hackney, Zhang, Wehrli, Wright,
  O'Brien, Uematsu, Wehrli and Selzer}]{takahashi2002magnetic}
\bibinfo{author}{Takahashi, M.}, \bibinfo{author}{Hackney, D.B.},
  \bibinfo{author}{Zhang, G.}, \bibinfo{author}{Wehrli, S.L.},
  \bibinfo{author}{Wright, A.C.}, \bibinfo{author}{O'Brien, W.T.},
  \bibinfo{author}{Uematsu, H.}, \bibinfo{author}{Wehrli, F.W.},
  \bibinfo{author}{Selzer, M.E.}, \bibinfo{year}{2002}.
\newblock \bibinfo{title}{Magnetic resonance microimaging of intraaxonal water
  diffusion in live excised lamprey spinal cord}.
\newblock \bibinfo{journal}{Proceedings of the National Academy of Sciences}
  \bibinfo{volume}{99}, \bibinfo{pages}{16192--16196}.
\bibitem[{Tamnes et~al.(2018)Tamnes, Roalf, Goddings and
  Lebel}]{tamnes2018diffusion}
\bibinfo{author}{Tamnes, C.K.}, \bibinfo{author}{Roalf, D.R.},
  \bibinfo{author}{Goddings, A.L.}, \bibinfo{author}{Lebel, C.},
  \bibinfo{year}{2018}.
\newblock \bibinfo{title}{Diffusion {MRI} of white matter microstructure
  development in childhood and adolescence: Methods, challenges and progress}.
\newblock \bibinfo{journal}{Developmental cognitive neuroscience}
  \bibinfo{volume}{33}, \bibinfo{pages}{161--175}.
\bibitem[{Teillac et~al.(2017)Teillac, Beaujoin, Poupon, Mangin and
  Poupon}]{teillac2017novel}
\bibinfo{author}{Teillac, A.}, \bibinfo{author}{Beaujoin, J.},
  \bibinfo{author}{Poupon, F.}, \bibinfo{author}{Mangin, J.F.},
  \bibinfo{author}{Poupon, C.}, \bibinfo{year}{2017}.
\newblock \bibinfo{title}{A novel anatomically-constrained global tractography
  approach to monitor sharp turns in gyri}, in:
  \bibinfo{booktitle}{International Conference on Medical Image Computing and
  Computer-assisted Intervention}, \bibinfo{organization}{Springer}. pp.
  \bibinfo{pages}{532--539}.
\bibitem[{Thomas et~al.(2011)Thomas, Humphreys, Jung, Minshew and
  Behrmann}]{thomas2011anatomy}
\bibinfo{author}{Thomas, C.}, \bibinfo{author}{Humphreys, K.},
  \bibinfo{author}{Jung, K.J.}, \bibinfo{author}{Minshew, N.},
  \bibinfo{author}{Behrmann, M.}, \bibinfo{year}{2011}.
\newblock \bibinfo{title}{The anatomy of the callosal and visual-association
  pathways in high-functioning autism: a dti tractography study}.
\newblock \bibinfo{journal}{Cortex} \bibinfo{volume}{47},
  \bibinfo{pages}{863--873}.
\bibitem[{Thomas et~al.(2014)Thomas, Ye, Irfanoğlu, Modi, Saleem, Leopold and
  Pierpaoli}]{Thomas.2014}
\bibinfo{author}{Thomas, C.}, \bibinfo{author}{Ye, F.Q.},
  \bibinfo{author}{Irfanoğlu, M.O.}, \bibinfo{author}{Modi, P.},
  \bibinfo{author}{Saleem, K.S.}, \bibinfo{author}{Leopold, D.A.},
  \bibinfo{author}{Pierpaoli, C.}, \bibinfo{year}{2014}.
\newblock \bibinfo{title}{{Anatomical accuracy of brain connections derived
  from diffusion MRI tractography is inherently limited.}}
\newblock \bibinfo{journal}{Proceedings of the National Academy of Sciences of
  the United States of America} \bibinfo{volume}{111}, \bibinfo{pages}{16574 --
  16579}.
\newblock \DOIprefix\doi{10.1073/pnas.1405672111}.
\bibitem[{Thompson et~al.(2017)Thompson, Andreassen, Arias-Vasquez, Bearden,
  Boedhoe, Brouwer, Buckner, Buitelaar, Bulayeva, Cannon
  et~al.}]{thompson2017enigma}
\bibinfo{author}{Thompson, P.M.}, \bibinfo{author}{Andreassen, O.A.},
  \bibinfo{author}{Arias-Vasquez, A.}, \bibinfo{author}{Bearden, C.E.},
  \bibinfo{author}{Boedhoe, P.S.}, \bibinfo{author}{Brouwer, R.M.},
  \bibinfo{author}{Buckner, R.L.}, \bibinfo{author}{Buitelaar, J.K.},
  \bibinfo{author}{Bulayeva, K.B.}, \bibinfo{author}{Cannon, D.M.}, et~al.,
  \bibinfo{year}{2017}.
\newblock \bibinfo{title}{Enigma and the individual: Predicting factors that
  affect the brain in 35 countries worldwide}.
\newblock \bibinfo{journal}{Neuroimage} \bibinfo{volume}{145},
  \bibinfo{pages}{389--408}.
\bibitem[{Tian et~al.(2021)Tian, Yeo, Cropley, Zalesky et~al.}]{tian2021high}
\bibinfo{author}{Tian, Y.}, \bibinfo{author}{Yeo, B.T.},
  \bibinfo{author}{Cropley, V.}, \bibinfo{author}{Zalesky, A.}, et~al.,
  \bibinfo{year}{2021}.
\newblock \bibinfo{title}{High-resolution connectomic fingerprints: Mapping
  neural identity and behavior}.
\newblock \bibinfo{journal}{NeuroImage} \bibinfo{volume}{229},
  \bibinfo{pages}{117695}.
\bibitem[{Toga and Thompson(2013)}]{toga2013connectomics}
\bibinfo{author}{Toga, A.W.}, \bibinfo{author}{Thompson, P.M.},
  \bibinfo{year}{2013}.
\newblock \bibinfo{title}{Connectomics sheds new light on alzheimer’s
  disease}.
\newblock \bibinfo{journal}{Biological psychiatry} \bibinfo{volume}{73},
  \bibinfo{pages}{390--392}.
\bibitem[{Tong et~al.(2019)Tong, He, Gong, Li, Liang, Qian, Sun, Ding, Li and
  Zhong}]{tong2019reproducibility}
\bibinfo{author}{Tong, Q.}, \bibinfo{author}{He, H.}, \bibinfo{author}{Gong,
  T.}, \bibinfo{author}{Li, C.}, \bibinfo{author}{Liang, P.},
  \bibinfo{author}{Qian, T.}, \bibinfo{author}{Sun, Y.}, \bibinfo{author}{Ding,
  Q.}, \bibinfo{author}{Li, K.}, \bibinfo{author}{Zhong, J.},
  \bibinfo{year}{2019}.
\newblock \bibinfo{title}{Reproducibility of multi-shell diffusion tractography
  on traveling subjects: a multicenter study prospective}.
\newblock \bibinfo{journal}{Magnetic resonance imaging} \bibinfo{volume}{59},
  \bibinfo{pages}{1--9}.
\bibitem[{Tournier et~al.(2007)Tournier, Calamante and
  Connelly}]{tournier2007robust}
\bibinfo{author}{Tournier, J.D.}, \bibinfo{author}{Calamante, F.},
  \bibinfo{author}{Connelly, A.}, \bibinfo{year}{2007}.
\newblock \bibinfo{title}{Robust determination of the fibre orientation
  distribution in diffusion {MRI}: non-negativity constrained super-resolved
  spherical deconvolution}.
\newblock \bibinfo{journal}{Neuroimage} \bibinfo{volume}{35},
  \bibinfo{pages}{1459--1472}.
\bibitem[{Tournier et~al.(2010)Tournier, Calamante and
  Connelly}]{tournier2010improved}
\bibinfo{author}{Tournier, J.D.}, \bibinfo{author}{Calamante, F.},
  \bibinfo{author}{Connelly, A.}, \bibinfo{year}{2010}.
\newblock \bibinfo{title}{Improved probabilistic streamlines tractography by
  2nd order integration over fibre orientation distributions}, in:
  \bibinfo{booktitle}{Proceedings of the {International} {Society} for
  {Magnetic} {Resonance} in {Medicine}}, p. \bibinfo{pages}{1670}.
\bibitem[{Tournier et~al.(2012)Tournier, Calamante and
  Connelly}]{tournier2012mrtrix}
\bibinfo{author}{Tournier, J.D.}, \bibinfo{author}{Calamante, F.},
  \bibinfo{author}{Connelly, A.}, \bibinfo{year}{2012}.
\newblock \bibinfo{title}{Mrtrix: diffusion tractography in crossing fiber
  regions}.
\newblock \bibinfo{journal}{International journal of imaging systems and
  technology} \bibinfo{volume}{22}, \bibinfo{pages}{53--66}.
\bibitem[{Tournier et~al.(2003)Tournier, Calamante, Gadian and
  Connelly}]{tournier2003diffusion}
\bibinfo{author}{Tournier, J.D.}, \bibinfo{author}{Calamante, F.},
  \bibinfo{author}{Gadian, D.G.}, \bibinfo{author}{Connelly, A.},
  \bibinfo{year}{2003}.
\newblock \bibinfo{title}{Diffusion-weighted magnetic resonance imaging fibre
  tracking using a front evolution algorithm}.
\newblock \bibinfo{journal}{NeuroImage} \bibinfo{volume}{20},
  \bibinfo{pages}{276--288}.
\bibitem[{Tournier et~al.(2002)Tournier, Calamante, King, Gadian and
  Connelly}]{tournier_limitations_2002}
\bibinfo{author}{Tournier, J.D.}, \bibinfo{author}{Calamante, F.},
  \bibinfo{author}{King, M.D.}, \bibinfo{author}{Gadian, D.G.},
  \bibinfo{author}{Connelly, A.}, \bibinfo{year}{2002}.
\newblock \bibinfo{title}{Limitations and requirements of diffusion tensor
  fiber tracking: {An} assessment using simulations}.
\newblock \bibinfo{journal}{Magnetic Resonance in Medicine}
  \bibinfo{volume}{47}, \bibinfo{pages}{701--708}.
\bibitem[{Tournier et~al.(2019)Tournier, Smith, Raffelt, Tabbara, Dhollander,
  Pietsch, Christiaens, Jeurissen, Yeh and Connelly}]{tournier2019mrtrix3}
\bibinfo{author}{Tournier, J.D.}, \bibinfo{author}{Smith, R.},
  \bibinfo{author}{Raffelt, D.}, \bibinfo{author}{Tabbara, R.},
  \bibinfo{author}{Dhollander, T.}, \bibinfo{author}{Pietsch, M.},
  \bibinfo{author}{Christiaens, D.}, \bibinfo{author}{Jeurissen, B.},
  \bibinfo{author}{Yeh, C.H.}, \bibinfo{author}{Connelly, A.},
  \bibinfo{year}{2019}.
\newblock \bibinfo{title}{Mrtrix3: A fast, flexible and open software framework
  for medical image processing and visualisation}.
\newblock \bibinfo{journal}{NeuroImage} \bibinfo{volume}{202},
  \bibinfo{pages}{116137}.
\bibitem[{Tseng et~al.(2021)Tseng, Hsu, Chen, Kang, Kao, Chen and
  Waiter}]{tseng2021microstructural}
\bibinfo{author}{Tseng, W.Y.I.}, \bibinfo{author}{Hsu, Y.C.},
  \bibinfo{author}{Chen, C.L.}, \bibinfo{author}{Kang, Y.J.},
  \bibinfo{author}{Kao, T.W.}, \bibinfo{author}{Chen, P.Y.},
  \bibinfo{author}{Waiter, G.D.}, \bibinfo{year}{2021}.
\newblock \bibinfo{title}{Microstructural differences in white matter tracts
  across middle to late adulthood: a diffusion {MRI} study on 7167 uk biobank
  participants}.
\newblock \bibinfo{journal}{Neurobiology of Aging} \bibinfo{volume}{98},
  \bibinfo{pages}{160--172}.
\bibitem[{Tun{\c{c}} et~al.(2014)Tun{\c{c}}, Parker, Ingalhalikar and
  Verma}]{tuncc2014automated}
\bibinfo{author}{Tun{\c{c}}, B.}, \bibinfo{author}{Parker, W.A.},
  \bibinfo{author}{Ingalhalikar, M.}, \bibinfo{author}{Verma, R.},
  \bibinfo{year}{2014}.
\newblock \bibinfo{title}{Automated tract extraction via atlas based adaptive
  clustering}.
\newblock \bibinfo{journal}{Neuroimage} \bibinfo{volume}{102},
  \bibinfo{pages}{596--607}.
\bibitem[{Tymofiyeva et~al.(2014)Tymofiyeva, Hess, Xu and
  Barkovich}]{tymofiyeva2014structural}
\bibinfo{author}{Tymofiyeva, O.}, \bibinfo{author}{Hess, C.},
  \bibinfo{author}{Xu, D.}, \bibinfo{author}{Barkovich, A.},
  \bibinfo{year}{2014}.
\newblock \bibinfo{title}{Structural {MRI} connectome in development:
  challenges of the changing brain}.
\newblock \bibinfo{journal}{The British journal of radiology}
  \bibinfo{volume}{87}, \bibinfo{pages}{20140086}.
\bibitem[{Tymofiyeva et~al.(2013)Tymofiyeva, Hess, Ziv, Lee, Glass, Ferriero,
  Barkovich and Xu}]{tymofiyeva2013dti}
\bibinfo{author}{Tymofiyeva, O.}, \bibinfo{author}{Hess, C.P.},
  \bibinfo{author}{Ziv, E.}, \bibinfo{author}{Lee, P.N.},
  \bibinfo{author}{Glass, H.C.}, \bibinfo{author}{Ferriero, D.M.},
  \bibinfo{author}{Barkovich, A.J.}, \bibinfo{author}{Xu, D.},
  \bibinfo{year}{2013}.
\newblock \bibinfo{title}{A dti-based template-free cortical connectome study
  of brain maturation}.
\newblock \bibinfo{journal}{PloS one} \bibinfo{volume}{8},
  \bibinfo{pages}{e63310}.
\bibitem[{Tzourio-Mazoyer et~al.(2002)Tzourio-Mazoyer, Landeau, Papathanassiou,
  Crivello, Etard, Delcroix, Mazoyer and Joliot}]{tzourio2002automated}
\bibinfo{author}{Tzourio-Mazoyer, N.}, \bibinfo{author}{Landeau, B.},
  \bibinfo{author}{Papathanassiou, D.}, \bibinfo{author}{Crivello, F.},
  \bibinfo{author}{Etard, O.}, \bibinfo{author}{Delcroix, N.},
  \bibinfo{author}{Mazoyer, B.}, \bibinfo{author}{Joliot, M.},
  \bibinfo{year}{2002}.
\newblock \bibinfo{title}{Automated anatomical labeling of activations in spm
  using a macroscopic anatomical parcellation of the {MNI} {MRI} single-subject
  brain}.
\newblock \bibinfo{journal}{Neuroimage} \bibinfo{volume}{15},
  \bibinfo{pages}{273--289}.
\bibitem[{Uddin(2013)}]{uddin2013complex}
\bibinfo{author}{Uddin, L.Q.}, \bibinfo{year}{2013}.
\newblock \bibinfo{title}{Complex relationships between structural and
  functional brain connectivity}.
\newblock \bibinfo{journal}{Trends in cognitive sciences} \bibinfo{volume}{17},
  \bibinfo{pages}{600--602}.
\bibitem[{Van Den~Heuvel et~al.(2015)Van Den~Heuvel, Kersbergen, De~Reus,
  Keunen, Kahn, Groenendaal, De~Vries and Benders}]{van2015neonatal}
\bibinfo{author}{Van Den~Heuvel, M.P.}, \bibinfo{author}{Kersbergen, K.J.},
  \bibinfo{author}{De~Reus, M.A.}, \bibinfo{author}{Keunen, K.},
  \bibinfo{author}{Kahn, R.S.}, \bibinfo{author}{Groenendaal, F.},
  \bibinfo{author}{De~Vries, L.S.}, \bibinfo{author}{Benders, M.J.},
  \bibinfo{year}{2015}.
\newblock \bibinfo{title}{The neonatal connectome during preterm brain
  development}.
\newblock \bibinfo{journal}{Cerebral cortex} \bibinfo{volume}{25},
  \bibinfo{pages}{3000--3013}.
\bibitem[{Van Den~Heuvel et~al.(2019)Van Den~Heuvel, Scholtens, De~Lange,
  Pijnenburg, Cahn, Van~Haren, Sommer, Bozzali, Koch, Boks
  et~al.}]{van2019evolutionary}
\bibinfo{author}{Van Den~Heuvel, M.P.}, \bibinfo{author}{Scholtens, L.H.},
  \bibinfo{author}{De~Lange, S.C.}, \bibinfo{author}{Pijnenburg, R.},
  \bibinfo{author}{Cahn, W.}, \bibinfo{author}{Van~Haren, N.E.},
  \bibinfo{author}{Sommer, I.E.}, \bibinfo{author}{Bozzali, M.},
  \bibinfo{author}{Koch, K.}, \bibinfo{author}{Boks, M.P.}, et~al.,
  \bibinfo{year}{2019}.
\newblock \bibinfo{title}{Evolutionary modifications in human brain
  connectivity associated with schizophrenia}.
\newblock \bibinfo{journal}{Brain} \bibinfo{volume}{142},
  \bibinfo{pages}{3991--4002}.
\bibitem[{Van Den~Heuvel et~al.(2013)Van Den~Heuvel, Sporns, Collin, Scheewe,
  Mandl, Cahn, Go{\~n}i, Pol and Kahn}]{van2013abnormal}
\bibinfo{author}{Van Den~Heuvel, M.P.}, \bibinfo{author}{Sporns, O.},
  \bibinfo{author}{Collin, G.}, \bibinfo{author}{Scheewe, T.},
  \bibinfo{author}{Mandl, R.C.}, \bibinfo{author}{Cahn, W.},
  \bibinfo{author}{Go{\~n}i, J.}, \bibinfo{author}{Pol, H.E.H.},
  \bibinfo{author}{Kahn, R.S.}, \bibinfo{year}{2013}.
\newblock \bibinfo{title}{Abnormal rich club organization and functional brain
  dynamics in schizophrenia}.
\newblock \bibinfo{journal}{JAMA psychiatry} \bibinfo{volume}{70},
  \bibinfo{pages}{783--792}.
\bibitem[{{Van Essen} et~al.(2014){Van Essen}, Jbabdi, Sotiropoulos, Chen,
  Dikranian, Coalson, Harwell, Behrens and Glasser}]{vanessen_ch16_2014}
\bibinfo{author}{{Van Essen}, D.C.}, \bibinfo{author}{Jbabdi, S.},
  \bibinfo{author}{Sotiropoulos, S.N.}, \bibinfo{author}{Chen, C.},
  \bibinfo{author}{Dikranian, K.}, \bibinfo{author}{Coalson, T.},
  \bibinfo{author}{Harwell, J.}, \bibinfo{author}{Behrens, T.E.},
  \bibinfo{author}{Glasser, M.F.}, \bibinfo{year}{2014}.
\newblock \bibinfo{title}{Chapter 16 - {Mapping} connections in humans and
  non-human primates: {Aspirations} and challenges for diffusion imaging}, in:
  \bibinfo{editor}{Johansen-Berg, H.}, \bibinfo{editor}{Behrens, T.E.} (Eds.),
  \bibinfo{booktitle}{Diffusion MRI (Second Edition)}.
  \bibinfo{publisher}{Academic Press}, \bibinfo{address}{San Diego}, pp.
  \bibinfo{pages}{337--358}.
\newblock \DOIprefix\doi{10.1016/B978-0-12-396460-1.00016-0}.
\bibitem[{Van~Horn et~al.(2012)Van~Horn, Irimia, Torgerson, Chambers, Kikinis
  and Toga}]{van2012mapping}
\bibinfo{author}{Van~Horn, J.D.}, \bibinfo{author}{Irimia, A.},
  \bibinfo{author}{Torgerson, C.M.}, \bibinfo{author}{Chambers, M.C.},
  \bibinfo{author}{Kikinis, R.}, \bibinfo{author}{Toga, A.W.},
  \bibinfo{year}{2012}.
\newblock \bibinfo{title}{Mapping connectivity damage in the case of phineas
  gage}.
\newblock \bibinfo{journal}{PloS one} \bibinfo{volume}{7},
  \bibinfo{pages}{e37454}.
\bibitem[{Vasung et~al.(2010)Vasung, Huang, Jovanov-Milo{\v{s}}evi{\'c},
  Pletikos, Mori and Kostovi{\'c}}]{vasung2010development}
\bibinfo{author}{Vasung, L.}, \bibinfo{author}{Huang, H.},
  \bibinfo{author}{Jovanov-Milo{\v{s}}evi{\'c}, N.}, \bibinfo{author}{Pletikos,
  M.}, \bibinfo{author}{Mori, S.}, \bibinfo{author}{Kostovi{\'c}, I.},
  \bibinfo{year}{2010}.
\newblock \bibinfo{title}{Development of axonal pathways in the human fetal
  fronto-limbic brain: histochemical characterization and diffusion tensor
  imaging}.
\newblock \bibinfo{journal}{Journal of anatomy} \bibinfo{volume}{217},
  \bibinfo{pages}{400--417}.
\bibitem[{Vasung et~al.(2019)Vasung, Turk, Ferradal, Sutin, Stout, Ahtam, Lin
  and Grant}]{vasung2019exploring}
\bibinfo{author}{Vasung, L.}, \bibinfo{author}{Turk, E.A.},
  \bibinfo{author}{Ferradal, S.L.}, \bibinfo{author}{Sutin, J.},
  \bibinfo{author}{Stout, J.N.}, \bibinfo{author}{Ahtam, B.},
  \bibinfo{author}{Lin, P.Y.}, \bibinfo{author}{Grant, P.E.},
  \bibinfo{year}{2019}.
\newblock \bibinfo{title}{Exploring early human brain development with
  structural and physiological neuroimaging}.
\newblock \bibinfo{journal}{Neuroimage} \bibinfo{volume}{187},
  \bibinfo{pages}{226--254}.
\bibitem[{Vatansever et~al.(2018)Vatansever, Manktelow, Sahakian, Menon and
  Stamatakis}]{vatansever2018default}
\bibinfo{author}{Vatansever, D.}, \bibinfo{author}{Manktelow, A.},
  \bibinfo{author}{Sahakian, B.J.}, \bibinfo{author}{Menon, D.K.},
  \bibinfo{author}{Stamatakis, E.A.}, \bibinfo{year}{2018}.
\newblock \bibinfo{title}{Default mode network engagement beyond
  self-referential internal mentation}.
\newblock \bibinfo{journal}{Brain Connectivity} \bibinfo{volume}{8},
  \bibinfo{pages}{245--253}.
\bibitem[{Vatansever et~al.(2017)Vatansever, Menon and
  Stamatakis}]{vatansever2017default}
\bibinfo{author}{Vatansever, D.}, \bibinfo{author}{Menon, D.K.},
  \bibinfo{author}{Stamatakis, E.A.}, \bibinfo{year}{2017}.
\newblock \bibinfo{title}{Default mode contributions to automated information
  processing}.
\newblock \bibinfo{journal}{Proceedings of the National Academy of Sciences}
  \bibinfo{volume}{114}, \bibinfo{pages}{12821--12826}.
\bibitem[{V{\'a}zquez et~al.(2020)V{\'a}zquez, L{\'o}pez-L{\'o}pez, Houenou,
  Poupon, Mangin, Ladra and Guevara}]{vazquez2020automatic}
\bibinfo{author}{V{\'a}zquez, A.}, \bibinfo{author}{L{\'o}pez-L{\'o}pez, N.},
  \bibinfo{author}{Houenou, J.}, \bibinfo{author}{Poupon, C.},
  \bibinfo{author}{Mangin, J.F.}, \bibinfo{author}{Ladra, S.},
  \bibinfo{author}{Guevara, P.}, \bibinfo{year}{2020}.
\newblock \bibinfo{title}{Automatic group-wise whole-brain short association
  fiber bundle labeling based on clustering and cortical surface information}.
\newblock \bibinfo{journal}{BioMedical Engineering OnLine}
  \bibinfo{volume}{19}, \bibinfo{pages}{1--24}.
\bibitem[{Verhoeven et~al.(2010)Verhoeven, Sage, Leemans, Van~Hecke, Callaert,
  Peeters, De~Cock, Lagae and Sunaert}]{verhoeven2010construction}
\bibinfo{author}{Verhoeven, J.S.}, \bibinfo{author}{Sage, C.A.},
  \bibinfo{author}{Leemans, A.}, \bibinfo{author}{Van~Hecke, W.},
  \bibinfo{author}{Callaert, D.}, \bibinfo{author}{Peeters, R.},
  \bibinfo{author}{De~Cock, P.}, \bibinfo{author}{Lagae, L.},
  \bibinfo{author}{Sunaert, S.}, \bibinfo{year}{2010}.
\newblock \bibinfo{title}{Construction of a stereotaxic dti atlas with full
  diffusion tensor information for studying white matter maturation from
  childhood to adolescence using tractography-based segmentations}.
\newblock \bibinfo{journal}{Human brain mapping} \bibinfo{volume}{31},
  \bibinfo{pages}{470--486}.
\bibitem[{V{\'e}rtes and Bullmore(2015)}]{vertes2015annual}
\bibinfo{author}{V{\'e}rtes, P.E.}, \bibinfo{author}{Bullmore, E.T.},
  \bibinfo{year}{2015}.
\newblock \bibinfo{title}{Annual research review: growth connectomics--the
  organization and reorganization of brain networks during normal and abnormal
  development}.
\newblock \bibinfo{journal}{Journal of Child Psychology and Psychiatry}
  \bibinfo{volume}{56}, \bibinfo{pages}{299--320}.
\bibitem[{Voineskos et~al.(2010)Voineskos, Lobaugh, Bouix, Rajji, Miranda,
  Kennedy, Mulsant, Pollock and Shenton}]{voineskos2010diffusion}
\bibinfo{author}{Voineskos, A.N.}, \bibinfo{author}{Lobaugh, N.J.},
  \bibinfo{author}{Bouix, S.}, \bibinfo{author}{Rajji, T.K.},
  \bibinfo{author}{Miranda, D.}, \bibinfo{author}{Kennedy, J.L.},
  \bibinfo{author}{Mulsant, B.H.}, \bibinfo{author}{Pollock, B.G.},
  \bibinfo{author}{Shenton, M.E.}, \bibinfo{year}{2010}.
\newblock \bibinfo{title}{Diffusion tensor tractography findings in
  schizophrenia across the adult lifespan}.
\newblock \bibinfo{journal}{Brain} \bibinfo{volume}{133},
  \bibinfo{pages}{1494--1504}.
\bibitem[{Wakana et~al.(2007)Wakana, Caprihan, Panzenboeck, Fallon, Perry,
  Gollub, Hua, Zhang, Jiang, Dubey et~al.}]{wakana2007reproducibility}
\bibinfo{author}{Wakana, S.}, \bibinfo{author}{Caprihan, A.},
  \bibinfo{author}{Panzenboeck, M.M.}, \bibinfo{author}{Fallon, J.H.},
  \bibinfo{author}{Perry, M.}, \bibinfo{author}{Gollub, R.L.},
  \bibinfo{author}{Hua, K.}, \bibinfo{author}{Zhang, J.},
  \bibinfo{author}{Jiang, H.}, \bibinfo{author}{Dubey, P.}, et~al.,
  \bibinfo{year}{2007}.
\newblock \bibinfo{title}{Reproducibility of quantitative tractography methods
  applied to cerebral white matter}.
\newblock \bibinfo{journal}{Neuroimage} \bibinfo{volume}{36},
  \bibinfo{pages}{630--644}.
\bibitem[{Wakana et~al.(2004)Wakana, Jiang, Nagae-Poetscher, Van~Zijl and
  Mori}]{wakana2004fiber}
\bibinfo{author}{Wakana, S.}, \bibinfo{author}{Jiang, H.},
  \bibinfo{author}{Nagae-Poetscher, L.M.}, \bibinfo{author}{Van~Zijl, P.C.},
  \bibinfo{author}{Mori, S.}, \bibinfo{year}{2004}.
\newblock \bibinfo{title}{Fiber tract--based atlas of human white matter
  anatomy}.
\newblock \bibinfo{journal}{Radiology} \bibinfo{volume}{230},
  \bibinfo{pages}{77--87}.
\bibitem[{Wang et~al.(2016)Wang, Luo, Mok, Chu and Shi}]{wang2016tractography}
\bibinfo{author}{Wang, D.}, \bibinfo{author}{Luo, Y.}, \bibinfo{author}{Mok,
  V.C.}, \bibinfo{author}{Chu, W.C.}, \bibinfo{author}{Shi, L.},
  \bibinfo{year}{2016}.
\newblock \bibinfo{title}{Tractography atlas-based spatial statistics:
  Statistical analysis of diffusion tensor image along fiber pathways}.
\newblock \bibinfo{journal}{NeuroImage} \bibinfo{volume}{125},
  \bibinfo{pages}{301--310}.
\bibitem[{Wang et~al.(2012)Wang, Su, Zhou, Chou, Chen, Jiang and
  Lin}]{wang2012anatomical}
\bibinfo{author}{Wang, Q.}, \bibinfo{author}{Su, T.P.}, \bibinfo{author}{Zhou,
  Y.}, \bibinfo{author}{Chou, K.H.}, \bibinfo{author}{Chen, I.Y.},
  \bibinfo{author}{Jiang, T.}, \bibinfo{author}{Lin, C.P.},
  \bibinfo{year}{2012}.
\newblock \bibinfo{title}{Anatomical insights into disrupted small-world
  networks in schizophrenia}.
\newblock \bibinfo{journal}{Neuroimage} \bibinfo{volume}{59},
  \bibinfo{pages}{1085--1093}.
\bibitem[{Wang et~al.(2007)Wang, Benner, Sorensen and
  Wedeen}]{wang2007diffusion}
\bibinfo{author}{Wang, R.}, \bibinfo{author}{Benner, T.},
  \bibinfo{author}{Sorensen, A.G.}, \bibinfo{author}{Wedeen, V.J.},
  \bibinfo{year}{2007}.
\newblock \bibinfo{title}{Diffusion toolkit: a software package for diffusion
  imaging data processing and tractography}, in: \bibinfo{booktitle}{Proc Intl
  Soc Mag Reson Med}, \bibinfo{organization}{Berlin}. p. \bibinfo{pages}{3720}.
\bibitem[{Wang et~al.(2019)Wang, Seguin, Zalesky, Wong, Chu and
  Tong}]{xwang:2019}
\bibinfo{author}{Wang, X.}, \bibinfo{author}{Seguin, C.},
  \bibinfo{author}{Zalesky, A.}, \bibinfo{author}{Wong, W.W.},
  \bibinfo{author}{Chu, W.C.W.}, \bibinfo{author}{Tong, R.K.Y.},
  \bibinfo{year}{2019}.
\newblock \bibinfo{title}{Synchronization lag in post stroke: relation to motor
  function and structural connectivity}.
\newblock \bibinfo{journal}{Netw Neurosci} \bibinfo{volume}{3},
  \bibinfo{pages}{1121--1140}.
\bibitem[{Wassermann et~al.(2016)Wassermann, Makris, Rathi, Shenton, Kikinis,
  Kubicki and Westin}]{wassermann2016white}
\bibinfo{author}{Wassermann, D.}, \bibinfo{author}{Makris, N.},
  \bibinfo{author}{Rathi, Y.}, \bibinfo{author}{Shenton, M.},
  \bibinfo{author}{Kikinis, R.}, \bibinfo{author}{Kubicki, M.},
  \bibinfo{author}{Westin, C.F.}, \bibinfo{year}{2016}.
\newblock \bibinfo{title}{The white matter query language: a novel approach for
  describing human white matter anatomy}.
\newblock \bibinfo{journal}{Brain Structure and Function}
  \bibinfo{volume}{221}, \bibinfo{pages}{4705--4721}.
\bibitem[{Wasserthal et~al.(2018)Wasserthal, Neher and
  Maier-Hein}]{wasserthal2018tractseg}
\bibinfo{author}{Wasserthal, J.}, \bibinfo{author}{Neher, P.},
  \bibinfo{author}{Maier-Hein, K.H.}, \bibinfo{year}{2018}.
\newblock \bibinfo{title}{Tractseg-fast and accurate white matter tract
  segmentation}.
\newblock \bibinfo{journal}{NeuroImage} \bibinfo{volume}{183},
  \bibinfo{pages}{239--253}.
\bibitem[{Wasserthal et~al.(2019)Wasserthal, Neher, Hirjak and
  Maier-Hein}]{wasserthal2019combined}
\bibinfo{author}{Wasserthal, J.}, \bibinfo{author}{Neher, P.F.},
  \bibinfo{author}{Hirjak, D.}, \bibinfo{author}{Maier-Hein, K.H.},
  \bibinfo{year}{2019}.
\newblock \bibinfo{title}{Combined tract segmentation and orientation mapping
  for bundle-specific tractography}.
\newblock \bibinfo{journal}{Medical image analysis} \bibinfo{volume}{58},
  \bibinfo{pages}{101559}.
\bibitem[{Wegmayr et~al.(2018)Wegmayr, Giuliari, Holdener and
  Buhmann}]{wegmayr2018data}
\bibinfo{author}{Wegmayr, V.}, \bibinfo{author}{Giuliari, G.},
  \bibinfo{author}{Holdener, S.}, \bibinfo{author}{Buhmann, J.},
  \bibinfo{year}{2018}.
\newblock \bibinfo{title}{Data-driven fiber tractography with neural networks},
  in: \bibinfo{booktitle}{2018 IEEE 15th international symposium on biomedical
  imaging (ISBI 2018)}, \bibinfo{organization}{IEEE}. pp.
  \bibinfo{pages}{1030--1033}.
\bibitem[{Wen et~al.(2011)Wen, Zhu, He, Kochan, Reppermund, Slavin, Brodaty,
  Crawford, Xia and Sachdev}]{wen2011discrete}
\bibinfo{author}{Wen, W.}, \bibinfo{author}{Zhu, W.}, \bibinfo{author}{He, Y.},
  \bibinfo{author}{Kochan, N.A.}, \bibinfo{author}{Reppermund, S.},
  \bibinfo{author}{Slavin, M.J.}, \bibinfo{author}{Brodaty, H.},
  \bibinfo{author}{Crawford, J.}, \bibinfo{author}{Xia, A.},
  \bibinfo{author}{Sachdev, P.}, \bibinfo{year}{2011}.
\newblock \bibinfo{title}{Discrete neuroanatomical networks are associated with
  specific cognitive abilities in old age}.
\newblock \bibinfo{journal}{Journal of Neuroscience} \bibinfo{volume}{31},
  \bibinfo{pages}{1204--1212}.
\bibitem[{Westin et~al.(2016)Westin, Knutsson, Pasternak, Szczepankiewicz,
  {\"O}zarslan, van Westen, Mattisson, Bogren, O'Donnell, Kubicki
  et~al.}]{westin2016q}
\bibinfo{author}{Westin, C.F.}, \bibinfo{author}{Knutsson, H.},
  \bibinfo{author}{Pasternak, O.}, \bibinfo{author}{Szczepankiewicz, F.},
  \bibinfo{author}{{\"O}zarslan, E.}, \bibinfo{author}{van Westen, D.},
  \bibinfo{author}{Mattisson, C.}, \bibinfo{author}{Bogren, M.},
  \bibinfo{author}{O'Donnell, L.J.}, \bibinfo{author}{Kubicki, M.}, et~al.,
  \bibinfo{year}{2016}.
\newblock \bibinfo{title}{Q-space trajectory imaging for multidimensional
  diffusion {MRI} of the human brain}.
\newblock \bibinfo{journal}{Neuroimage} \bibinfo{volume}{135},
  \bibinfo{pages}{345--362}.
\bibitem[{Westin et~al.(1999)Westin, Maier, Khidhir, Everett, Jolesz and
  Kikinis}]{westin1999image}
\bibinfo{author}{Westin, C.F.}, \bibinfo{author}{Maier, S.E.},
  \bibinfo{author}{Khidhir, B.}, \bibinfo{author}{Everett, P.},
  \bibinfo{author}{Jolesz, F.A.}, \bibinfo{author}{Kikinis, R.},
  \bibinfo{year}{1999}.
\newblock \bibinfo{title}{Image processing for diffusion tensor magnetic
  resonance imaging}, in: \bibinfo{booktitle}{International Conference on
  Medical Image Computing and Computer-Assisted Intervention},
  \bibinfo{organization}{Springer}. pp. \bibinfo{pages}{441--452}.
\bibitem[{Westlye et~al.(2010)Westlye, Walhovd, Dale, Bj{\o}rnerud,
  Due-T{\o}nnessen, Engvig, Grydeland, Tamnes, {\O}stby and
  Fjell}]{westlye2010life}
\bibinfo{author}{Westlye, L.T.}, \bibinfo{author}{Walhovd, K.B.},
  \bibinfo{author}{Dale, A.M.}, \bibinfo{author}{Bj{\o}rnerud, A.},
  \bibinfo{author}{Due-T{\o}nnessen, P.}, \bibinfo{author}{Engvig, A.},
  \bibinfo{author}{Grydeland, H.}, \bibinfo{author}{Tamnes, C.K.},
  \bibinfo{author}{{\O}stby, Y.}, \bibinfo{author}{Fjell, A.M.},
  \bibinfo{year}{2010}.
\newblock \bibinfo{title}{Life-span changes of the human brain white matter:
  diffusion tensor imaging (dti) and volumetry}.
\newblock \bibinfo{journal}{Cerebral cortex} \bibinfo{volume}{20},
  \bibinfo{pages}{2055--2068}.
\bibitem[{Wheeler and Voineskos(2014)}]{wheeler2014review}
\bibinfo{author}{Wheeler, A.L.}, \bibinfo{author}{Voineskos, A.N.},
  \bibinfo{year}{2014}.
\newblock \bibinfo{title}{A review of structural neuroimaging in schizophrenia:
  from connectivity to connectomics}.
\newblock \bibinfo{journal}{Frontiers in human neuroscience}
  \bibinfo{volume}{8}, \bibinfo{pages}{653}.
\bibitem[{Wilkins et~al.(2015)Wilkins, Lee, Gajawelli, Law and
  Lepor{\'e}}]{wilkins2015fiber}
\bibinfo{author}{Wilkins, B.}, \bibinfo{author}{Lee, N.},
  \bibinfo{author}{Gajawelli, N.}, \bibinfo{author}{Law, M.},
  \bibinfo{author}{Lepor{\'e}, N.}, \bibinfo{year}{2015}.
\newblock \bibinfo{title}{Fiber estimation and tractography in diffusion {MRI}:
  development of simulated brain images and comparison of multi-fiber analysis
  methods at clinical b-values}.
\newblock \bibinfo{journal}{Neuroimage} \bibinfo{volume}{109},
  \bibinfo{pages}{341--356}.
\bibitem[{Wilson et~al.(2003)Wilson, Tench, Morgan and
  Blumhardt}]{wilson2003pyramidal}
\bibinfo{author}{Wilson, M.}, \bibinfo{author}{Tench, C.},
  \bibinfo{author}{Morgan, P.}, \bibinfo{author}{Blumhardt, L.},
  \bibinfo{year}{2003}.
\newblock \bibinfo{title}{Pyramidal tract mapping by diffusion tensor magnetic
  resonance imaging in multiple sclerosis: improving correlations with
  disability}.
\newblock \bibinfo{journal}{Journal of Neurology, Neurosurgery \& Psychiatry}
  \bibinfo{volume}{74}, \bibinfo{pages}{203--207}.
\bibitem[{Wolf et~al.(2005)Wolf, Vetter, Wegner, B{\"o}ttger, Nolden,
  Sch{\"o}binger, Hastenteufel, Kunert and Meinzer}]{wolf2005medical}
\bibinfo{author}{Wolf, I.}, \bibinfo{author}{Vetter, M.},
  \bibinfo{author}{Wegner, I.}, \bibinfo{author}{B{\"o}ttger, T.},
  \bibinfo{author}{Nolden, M.}, \bibinfo{author}{Sch{\"o}binger, M.},
  \bibinfo{author}{Hastenteufel, M.}, \bibinfo{author}{Kunert, T.},
  \bibinfo{author}{Meinzer, H.P.}, \bibinfo{year}{2005}.
\newblock \bibinfo{title}{The medical imaging interaction toolkit}.
\newblock \bibinfo{journal}{Medical image analysis} \bibinfo{volume}{9},
  \bibinfo{pages}{594--604}.
\bibitem[{Wu et~al.(2012a)Wu, Taki, Sato, Kinomura, Goto, Okada, Kawashima, He,
  Evans and Fukuda}]{wu2012age}
\bibinfo{author}{Wu, K.}, \bibinfo{author}{Taki, Y.}, \bibinfo{author}{Sato,
  K.}, \bibinfo{author}{Kinomura, S.}, \bibinfo{author}{Goto, R.},
  \bibinfo{author}{Okada, K.}, \bibinfo{author}{Kawashima, R.},
  \bibinfo{author}{He, Y.}, \bibinfo{author}{Evans, A.C.},
  \bibinfo{author}{Fukuda, H.}, \bibinfo{year}{2012}a.
\newblock \bibinfo{title}{Age-related changes in topological organization of
  structural brain networks in healthy individuals}.
\newblock \bibinfo{journal}{Human brain mapping} \bibinfo{volume}{33},
  \bibinfo{pages}{552--568}.
\bibitem[{Wu and Miller(2017)}]{wu2017image}
\bibinfo{author}{Wu, W.}, \bibinfo{author}{Miller, K.L.}, \bibinfo{year}{2017}.
\newblock \bibinfo{title}{Image formation in diffusion mri: a review of recent
  technical developments}.
\newblock \bibinfo{journal}{Journal of Magnetic Resonance Imaging}
  \bibinfo{volume}{46}, \bibinfo{pages}{646--662}.
\bibitem[{Wu et~al.(2012b)Wu, Xie, Zhou, Anderson, Gore and
  Ding}]{wu_globally_2012}
\bibinfo{author}{Wu, X.}, \bibinfo{author}{Xie, M.}, \bibinfo{author}{Zhou,
  J.}, \bibinfo{author}{Anderson, A.W.}, \bibinfo{author}{Gore, J.C.},
  \bibinfo{author}{Ding, Z.}, \bibinfo{year}{2012}b.
\newblock \bibinfo{title}{Globally optimized fiber tracking and hierarchical
  clustering — a unified framework}.
\newblock \bibinfo{journal}{Magnetic Resonance Imaging} \bibinfo{volume}{30},
  \bibinfo{pages}{485--495}.
\bibitem[{Wu et~al.(2009)Wu, Xu, Xu, Zhou, Anderson and Ding}]{wu_genetic_2009}
\bibinfo{author}{Wu, X.}, \bibinfo{author}{Xu, Q.}, \bibinfo{author}{Xu, L.},
  \bibinfo{author}{Zhou, J.}, \bibinfo{author}{Anderson, A.W.},
  \bibinfo{author}{Ding, Z.}, \bibinfo{year}{2009}.
\newblock \bibinfo{title}{Genetic white matter fiber tractography with global
  optimization}.
\newblock \bibinfo{journal}{Journal of Neuroscience Methods}
  \bibinfo{volume}{184}, \bibinfo{pages}{375--379}.
\bibitem[{Wu et~al.(2020a)Wu, Hong, Ahmad, Lin, Shen, Yap, Consortium
  et~al.}]{wu2020tract}
\bibinfo{author}{Wu, Y.}, \bibinfo{author}{Hong, Y.}, \bibinfo{author}{Ahmad,
  S.}, \bibinfo{author}{Lin, W.}, \bibinfo{author}{Shen, D.},
  \bibinfo{author}{Yap, P.T.}, \bibinfo{author}{Consortium, U.B.C.P.}, et~al.,
  \bibinfo{year}{2020}a.
\newblock \bibinfo{title}{Tract dictionary learning for fast and robust
  recognition of fiber bundles}, in: \bibinfo{booktitle}{International
  Conference on Medical Image Computing and Computer-Assisted Intervention},
  \bibinfo{organization}{Springer}. pp. \bibinfo{pages}{251--259}.
\bibitem[{Wu et~al.(2020b)Wu, Hong, Feng, Shen and Yap}]{wu2020mitigating}
\bibinfo{author}{Wu, Y.}, \bibinfo{author}{Hong, Y.}, \bibinfo{author}{Feng,
  Y.}, \bibinfo{author}{Shen, D.}, \bibinfo{author}{Yap, P.T.},
  \bibinfo{year}{2020}b.
\newblock \bibinfo{title}{Mitigating gyral bias in cortical tractography via
  asymmetric fiber orientation distributions}.
\newblock \bibinfo{journal}{Medical image analysis} \bibinfo{volume}{59},
  \bibinfo{pages}{101543}.
\bibitem[{Wu et~al.(2020c)Wu, Hong, Feng, Shen and Yap}]{wu_afods_2020}
\bibinfo{author}{Wu, Y.}, \bibinfo{author}{Hong, Y.}, \bibinfo{author}{Feng,
  Y.}, \bibinfo{author}{Shen, D.}, \bibinfo{author}{Yap, P.T.},
  \bibinfo{year}{2020}c.
\newblock \bibinfo{title}{Mitigating gyral bias in cortical tractography via
  asymmetric fiber orientation distributions}.
\newblock \bibinfo{journal}{Medical Image Analysis} \bibinfo{volume}{59},
  \bibinfo{pages}{101543}.
\bibitem[{Wu et~al.(2018)Wu, Zhang, Makris, Ning, Norton, She, Peng, Rathi,
  Feng, Wu et~al.}]{wu2018investigation}
\bibinfo{author}{Wu, Y.}, \bibinfo{author}{Zhang, F.}, \bibinfo{author}{Makris,
  N.}, \bibinfo{author}{Ning, Y.}, \bibinfo{author}{Norton, I.},
  \bibinfo{author}{She, S.}, \bibinfo{author}{Peng, H.},
  \bibinfo{author}{Rathi, Y.}, \bibinfo{author}{Feng, Y.}, \bibinfo{author}{Wu,
  H.}, et~al., \bibinfo{year}{2018}.
\newblock \bibinfo{title}{Investigation into local white matter abnormality in
  emotional processing and sensorimotor areas using an automatically annotated
  fiber clustering in major depressive disorder}.
\newblock \bibinfo{journal}{NeuroImage} \bibinfo{volume}{181},
  \bibinfo{pages}{16--29}.
\bibitem[{Xia and He(2017)}]{xia2017functional}
\bibinfo{author}{Xia, M.}, \bibinfo{author}{He, Y.}, \bibinfo{year}{2017}.
\newblock \bibinfo{title}{Functional connectomics from a “big data”
  perspective}.
\newblock \bibinfo{journal}{Neuroimage} \bibinfo{volume}{160},
  \bibinfo{pages}{152--167}.
\bibitem[{Xie et~al.(2020)Xie, Zhang, Leung, Mooney, Epprecht, Norton, Rathi,
  Kikinis, Al-Mefty, Makris et~al.}]{xie2020anatomical}
\bibinfo{author}{Xie, G.}, \bibinfo{author}{Zhang, F.}, \bibinfo{author}{Leung,
  L.}, \bibinfo{author}{Mooney, M.A.}, \bibinfo{author}{Epprecht, L.},
  \bibinfo{author}{Norton, I.}, \bibinfo{author}{Rathi, Y.},
  \bibinfo{author}{Kikinis, R.}, \bibinfo{author}{Al-Mefty, O.},
  \bibinfo{author}{Makris, N.}, et~al., \bibinfo{year}{2020}.
\newblock \bibinfo{title}{Anatomical assessment of trigeminal nerve
  tractography using diffusion {MRI}: A comparison of acquisition b-values and
  single-and multi-fiber tracking strategies}.
\newblock \bibinfo{journal}{NeuroImage: Clinical} \bibinfo{volume}{25},
  \bibinfo{pages}{102160}.
\bibitem[{Xu et~al.(2019)Xu, Dong, Lee, O’Hara, Asano and
  Jeong}]{xu2019objective}
\bibinfo{author}{Xu, H.}, \bibinfo{author}{Dong, M.}, \bibinfo{author}{Lee,
  M.H.}, \bibinfo{author}{O’Hara, N.}, \bibinfo{author}{Asano, E.},
  \bibinfo{author}{Jeong, J.W.}, \bibinfo{year}{2019}.
\newblock \bibinfo{title}{Objective detection of eloquent axonal pathways to
  minimize postoperative deficits in pediatric epilepsy surgery using diffusion
  tractography and convolutional neural networks}.
\newblock \bibinfo{journal}{IEEE Transactions on Medical Imaging}
  \bibinfo{volume}{38}, \bibinfo{pages}{1910--1922}.
\bibitem[{Yablonskiy and Sukstanskii(2010)}]{yablonskiy2010theoretical}
\bibinfo{author}{Yablonskiy, D.A.}, \bibinfo{author}{Sukstanskii, A.L.},
  \bibinfo{year}{2010}.
\newblock \bibinfo{title}{Theoretical models of the diffusion weighted mr
  signal}.
\newblock \bibinfo{journal}{NMR in Biomedicine} \bibinfo{volume}{23},
  \bibinfo{pages}{661--681}.
\bibitem[{Yakovlev and Lecours(1967)}]{yakovlev1967myelogenetic}
\bibinfo{author}{Yakovlev, P.}, \bibinfo{author}{Lecours, A.},
  \bibinfo{year}{1967}.
\newblock \bibinfo{title}{The myelogenetic cycles of regional maturation of the
  brain}.
\newblock \bibinfo{journal}{Regional development of the brain in early life} ,
  \bibinfo{pages}{3--70}.
\bibitem[{Yamada et~al.(2009)Yamada, Sakai, Akazawa, Yuen and
  Nishimura}]{yamada2009mr}
\bibinfo{author}{Yamada, K.}, \bibinfo{author}{Sakai, K.},
  \bibinfo{author}{Akazawa, K.}, \bibinfo{author}{Yuen, S.},
  \bibinfo{author}{Nishimura, T.}, \bibinfo{year}{2009}.
\newblock \bibinfo{title}{Mr tractography: a review of its clinical
  applications}.
\newblock \bibinfo{journal}{Magnetic resonance in medical sciences}
  \bibinfo{volume}{8}, \bibinfo{pages}{165--174}.
\bibitem[{Yang et~al.(2020)Yang, Li, Zhou, Wu and Ding}]{yang2020functional}
\bibinfo{author}{Yang, Z.}, \bibinfo{author}{Li, X.}, \bibinfo{author}{Zhou,
  J.}, \bibinfo{author}{Wu, X.}, \bibinfo{author}{Ding, Z.},
  \bibinfo{year}{2020}.
\newblock \bibinfo{title}{Functional clustering of whole brain white matter
  fibers}.
\newblock \bibinfo{journal}{Journal of Neuroscience Methods}
  \bibinfo{volume}{335}, \bibinfo{pages}{108626}.
\bibitem[{Yap et~al.(2011)Yap, Fan, Chen, Gilmore, Lin and
  Shen}]{yap2011development}
\bibinfo{author}{Yap, P.T.}, \bibinfo{author}{Fan, Y.}, \bibinfo{author}{Chen,
  Y.}, \bibinfo{author}{Gilmore, J.H.}, \bibinfo{author}{Lin, W.},
  \bibinfo{author}{Shen, D.}, \bibinfo{year}{2011}.
\newblock \bibinfo{title}{Development trends of white matter connectivity in
  the first years of life}.
\newblock \bibinfo{journal}{PloS one} \bibinfo{volume}{6},
  \bibinfo{pages}{e24678}.
\bibitem[{Yeatman et~al.(2012)Yeatman, Dougherty, Myall, Wandell and
  Feldman}]{yeatman2012tract}
\bibinfo{author}{Yeatman, J.D.}, \bibinfo{author}{Dougherty, R.F.},
  \bibinfo{author}{Myall, N.J.}, \bibinfo{author}{Wandell, B.A.},
  \bibinfo{author}{Feldman, H.M.}, \bibinfo{year}{2012}.
\newblock \bibinfo{title}{Tract profiles of white matter properties: automating
  fiber-tract quantification}.
\newblock \bibinfo{journal}{PloS one} \bibinfo{volume}{7},
  \bibinfo{pages}{e49790}.
\bibitem[{Yeatman et~al.(2014a)Yeatman, Wandell and
  Mezer}]{yeatman2014lifespan}
\bibinfo{author}{Yeatman, J.D.}, \bibinfo{author}{Wandell, B.A.},
  \bibinfo{author}{Mezer, A.A.}, \bibinfo{year}{2014}a.
\newblock \bibinfo{title}{Lifespan maturation and degeneration of human brain
  white matter}.
\newblock \bibinfo{journal}{Nature communications} \bibinfo{volume}{5},
  \bibinfo{pages}{1--12}.
\bibitem[{Yeatman et~al.(2014b)Yeatman, Weiner, Pestilli, Rokem, Mezer and
  Wandell}]{Yeatman2014_occipitalFasciculus}
\bibinfo{author}{Yeatman, J.D.}, \bibinfo{author}{Weiner, K.S.},
  \bibinfo{author}{Pestilli, F.}, \bibinfo{author}{Rokem, A.},
  \bibinfo{author}{Mezer, A.}, \bibinfo{author}{Wandell, B.A.},
  \bibinfo{year}{2014}b.
\newblock \bibinfo{title}{The vertical occipital fasciculus: A century of
  controversy resolved by in vivo measurements}.
\newblock \bibinfo{journal}{Proceedings of the National Academy of Sciences}
  \bibinfo{volume}{111}, \bibinfo{pages}{E5214--E5223}.
\bibitem[{Yeh et~al.(2020)Yeh, Jones, Liang, Descoteaux and
  Connelly}]{yeh2020mapping}
\bibinfo{author}{Yeh, C.H.}, \bibinfo{author}{Jones, D.K.},
  \bibinfo{author}{Liang, X.}, \bibinfo{author}{Descoteaux, M.},
  \bibinfo{author}{Connelly, A.}, \bibinfo{year}{2020}.
\newblock \bibinfo{title}{{Mapping structural connectivity using diffusion MRI:
  challenges and opportunities}}.
\newblock \bibinfo{journal}{Journal of Magnetic Resonance Imaging} ,
  \bibinfo{pages}{https://doi.org/10.1002/jmri.27188}.
\bibitem[{Yeh et~al.(2019)Yeh, Smith, Dhollander, Calamante and
  Connelly}]{yeh2019connectomes}
\bibinfo{author}{Yeh, C.H.}, \bibinfo{author}{Smith, R.E.},
  \bibinfo{author}{Dhollander, T.}, \bibinfo{author}{Calamante, F.},
  \bibinfo{author}{Connelly, A.}, \bibinfo{year}{2019}.
\newblock \bibinfo{title}{Connectomes from streamlines tractography: Assigning
  streamlines to brain parcellations is not trivial but highly consequential}.
\newblock \bibinfo{journal}{NeuroImage} \bibinfo{volume}{199},
  \bibinfo{pages}{160--171}.
\bibitem[{Yeh et~al.(2017)Yeh, Smith, Dhollander and
  Connelly}]{yeh_mesh-based_2017}
\bibinfo{author}{Yeh, C.H.}, \bibinfo{author}{Smith, R.E.},
  \bibinfo{author}{Dhollander, T.}, \bibinfo{author}{Connelly, A.},
  \bibinfo{year}{2017}.
\newblock \bibinfo{title}{Mesh-based anatomically-constrained tractography for
  effective tracking termination and structural connectome construction}, in:
  \bibinfo{booktitle}{Proceedings of the {ISMRM}}, p.~\bibinfo{pages}{58}.
\bibitem[{Yeh et~al.(2016a)Yeh, Smith, Liang, Calamante and
  Connelly}]{yeh2016correction}
\bibinfo{author}{Yeh, C.H.}, \bibinfo{author}{Smith, R.E.},
  \bibinfo{author}{Liang, X.}, \bibinfo{author}{Calamante, F.},
  \bibinfo{author}{Connelly, A.}, \bibinfo{year}{2016}a.
\newblock \bibinfo{title}{{Correction for diffusion MRI fibre tracking biases:
  The consequences for structural connectomic metrics}}.
\newblock \bibinfo{journal}{Neuroimage} \bibinfo{volume}{142},
  \bibinfo{pages}{150--162}.
\bibitem[{Yeh et~al.(2016b)Yeh, Badre and Verstynen}]{yeh2016connectometry}
\bibinfo{author}{Yeh, F.C.}, \bibinfo{author}{Badre, D.},
  \bibinfo{author}{Verstynen, T.}, \bibinfo{year}{2016}b.
\newblock \bibinfo{title}{Connectometry: a statistical approach harnessing the
  analytical potential of the local connectome}.
\newblock \bibinfo{journal}{Neuroimage} \bibinfo{volume}{125},
  \bibinfo{pages}{162--171}.
\bibitem[{Yeh et~al.(2018)Yeh, Panesar, Fernandes, Meola, Yoshino,
  Fernandez-Miranda, Vettel and Verstynen}]{yeh2018population}
\bibinfo{author}{Yeh, F.C.}, \bibinfo{author}{Panesar, S.},
  \bibinfo{author}{Fernandes, D.}, \bibinfo{author}{Meola, A.},
  \bibinfo{author}{Yoshino, M.}, \bibinfo{author}{Fernandez-Miranda, J.C.},
  \bibinfo{author}{Vettel, J.M.}, \bibinfo{author}{Verstynen, T.},
  \bibinfo{year}{2018}.
\newblock \bibinfo{title}{Population-averaged atlas of the macroscale human
  structural connectome and its network topology}.
\newblock \bibinfo{journal}{NeuroImage} \bibinfo{volume}{178},
  \bibinfo{pages}{57--68}.
\bibitem[{Yeh et~al.(2013)Yeh, Verstynen, Wang, Fern{\'a}ndez-Miranda and
  Tseng}]{yeh2013deterministic}
\bibinfo{author}{Yeh, F.C.}, \bibinfo{author}{Verstynen, T.D.},
  \bibinfo{author}{Wang, Y.}, \bibinfo{author}{Fern{\'a}ndez-Miranda, J.C.},
  \bibinfo{author}{Tseng, W.Y.I.}, \bibinfo{year}{2013}.
\newblock \bibinfo{title}{Deterministic diffusion fiber tracking improved by
  quantitative anisotropy}.
\newblock \bibinfo{journal}{PloS one} \bibinfo{volume}{8},
  \bibinfo{pages}{e80713}.
\bibitem[{Yendiki et~al.(2011)Yendiki, Panneck, Srinivasan, Stevens,
  Z{\"o}llei, Augustinack, Wang, Salat, Ehrlich, Behrens
  et~al.}]{yendiki2011automated}
\bibinfo{author}{Yendiki, A.}, \bibinfo{author}{Panneck, P.},
  \bibinfo{author}{Srinivasan, P.}, \bibinfo{author}{Stevens, A.},
  \bibinfo{author}{Z{\"o}llei, L.}, \bibinfo{author}{Augustinack, J.},
  \bibinfo{author}{Wang, R.}, \bibinfo{author}{Salat, D.},
  \bibinfo{author}{Ehrlich, S.}, \bibinfo{author}{Behrens, T.}, et~al.,
  \bibinfo{year}{2011}.
\newblock \bibinfo{title}{Automated probabilistic reconstruction of
  white-matter pathways in health and disease using an atlas of the underlying
  anatomy}.
\newblock \bibinfo{journal}{Frontiers in neuroinformatics} \bibinfo{volume}{5},
  \bibinfo{pages}{23}.
\bibitem[{Yeo et~al.(2011)Yeo, Krienen, Sepulcre, Sabuncu, Lashkari,
  Hollinshead, Roffman, Smoller, Z{\"o}llei, Polimeni
  et~al.}]{yeo2011organization}
\bibinfo{author}{Yeo, B.T.}, \bibinfo{author}{Krienen, F.M.},
  \bibinfo{author}{Sepulcre, J.}, \bibinfo{author}{Sabuncu, M.R.},
  \bibinfo{author}{Lashkari, D.}, \bibinfo{author}{Hollinshead, M.},
  \bibinfo{author}{Roffman, J.L.}, \bibinfo{author}{Smoller, J.W.},
  \bibinfo{author}{Z{\"o}llei, L.}, \bibinfo{author}{Polimeni, J.R.}, et~al.,
  \bibinfo{year}{2011}.
\newblock \bibinfo{title}{The organization of the human cerebral cortex
  estimated by intrinsic functional connectivity}.
\newblock \bibinfo{journal}{Journal of neurophysiology} .
\bibitem[{Yeo et~al.(2016)Yeo, Ryman, Van Den~Heuvel, De~Reus, Jung, Pommy,
  Mayer, Ehrlich, Schulz, Morrow et~al.}]{yeo2016graph}
\bibinfo{author}{Yeo, R.A.}, \bibinfo{author}{Ryman, S.G.},
  \bibinfo{author}{Van Den~Heuvel, M.P.}, \bibinfo{author}{De~Reus, M.A.},
  \bibinfo{author}{Jung, R.E.}, \bibinfo{author}{Pommy, J.},
  \bibinfo{author}{Mayer, A.R.}, \bibinfo{author}{Ehrlich, S.},
  \bibinfo{author}{Schulz, S.C.}, \bibinfo{author}{Morrow, E.M.}, et~al.,
  \bibinfo{year}{2016}.
\newblock \bibinfo{title}{Graph metrics of structural brain networks in
  individuals with schizophrenia and healthy controls: group differences,
  relationships with intelligence, and genetics}.
\newblock \bibinfo{journal}{Journal of the International Neuropsychological
  Society: JINS} \bibinfo{volume}{22}, \bibinfo{pages}{240}.
\bibitem[{Yeo et~al.(2014)Yeo, Jang and Son}]{yeo2014different}
\bibinfo{author}{Yeo, S.S.}, \bibinfo{author}{Jang, S.H.},
  \bibinfo{author}{Son, S.M.}, \bibinfo{year}{2014}.
\newblock \bibinfo{title}{The different maturation of the corticospinal tract
  and corticoreticular pathway in normal brain development: diffusion tensor
  imaging study}.
\newblock \bibinfo{journal}{Frontiers in human neuroscience}
  \bibinfo{volume}{8}, \bibinfo{pages}{573}.
\bibitem[{Yoo et~al.(2015)Yoo, Guevara, Jeong, Yoo, Shin, Mangin and
  Seong}]{yoo2015example}
\bibinfo{author}{Yoo, S.W.}, \bibinfo{author}{Guevara, P.},
  \bibinfo{author}{Jeong, Y.}, \bibinfo{author}{Yoo, K.},
  \bibinfo{author}{Shin, J.S.}, \bibinfo{author}{Mangin, J.F.},
  \bibinfo{author}{Seong, J.K.}, \bibinfo{year}{2015}.
\newblock \bibinfo{title}{An example-based multi-atlas approach to automatic
  labeling of white matter tracts}.
\newblock \bibinfo{journal}{PloS one} \bibinfo{volume}{10},
  \bibinfo{pages}{e0133337}.
\bibitem[{Yuan and Lin(2006)}]{yuan2006model}
\bibinfo{author}{Yuan, M.}, \bibinfo{author}{Lin, Y.}, \bibinfo{year}{2006}.
\newblock \bibinfo{title}{Model selection and estimation in regression with
  grouped variables}.
\newblock \bibinfo{journal}{Journal of the Royal Statistical Society: Series B
  (Statistical Methodology)} \bibinfo{volume}{68}, \bibinfo{pages}{49--67}.
\bibitem[{Yushkevich et~al.(2009)Yushkevich, Zhang, Simon and
  Gee}]{yushkevich2009structure}
\bibinfo{author}{Yushkevich, P.A.}, \bibinfo{author}{Zhang, H.},
  \bibinfo{author}{Simon, T.J.}, \bibinfo{author}{Gee, J.C.},
  \bibinfo{year}{2009}.
\newblock \bibinfo{title}{Structure-specific statistical mapping of white
  matter tracts}, in: \bibinfo{booktitle}{Visualization and Processing of
  Tensor Fields}. \bibinfo{publisher}{Springer}, pp. \bibinfo{pages}{83--112}.
\bibitem[{Zalesky et~al.(2012)Zalesky, Cocchi, Fornito, Murray and
  Bullmore}]{zalesky2012connectivity}
\bibinfo{author}{Zalesky, A.}, \bibinfo{author}{Cocchi, L.},
  \bibinfo{author}{Fornito, A.}, \bibinfo{author}{Murray, M.M.},
  \bibinfo{author}{Bullmore, E.}, \bibinfo{year}{2012}.
\newblock \bibinfo{title}{Connectivity differences in brain networks}.
\newblock \bibinfo{journal}{Neuroimage} \bibinfo{volume}{60},
  \bibinfo{pages}{1055--1062}.
\bibitem[{Zalesky et~al.(2016a)Zalesky, Fornito, Cocchi, Gollo, van~den Heuvel
  and Breakspear}]{zalesky2016connectome}
\bibinfo{author}{Zalesky, A.}, \bibinfo{author}{Fornito, A.},
  \bibinfo{author}{Cocchi, L.}, \bibinfo{author}{Gollo, L.L.},
  \bibinfo{author}{van~den Heuvel, M.P.}, \bibinfo{author}{Breakspear, M.},
  \bibinfo{year}{2016}a.
\newblock \bibinfo{title}{Connectome sensitivity or specificity: which is more
  important?}
\newblock \bibinfo{journal}{Neuroimage} \bibinfo{volume}{142},
  \bibinfo{pages}{407--420}.
\bibitem[{Zalesky et~al.(2016b)Zalesky, Fornito, Cocchi, Gollo, Heuvel and
  Breakspear}]{Zalesky.2016}
\bibinfo{author}{Zalesky, A.}, \bibinfo{author}{Fornito, A.},
  \bibinfo{author}{Cocchi, L.}, \bibinfo{author}{Gollo, L.L.},
  \bibinfo{author}{Heuvel, M.P.v.d.}, \bibinfo{author}{Breakspear, M.},
  \bibinfo{year}{2016}b.
\newblock \bibinfo{title}{{Connectome sensitivity or specificity: which is more
  important?}}
\newblock \bibinfo{journal}{NeuroImage} \bibinfo{volume}{142},
  \bibinfo{pages}{407 -- 420}.
\bibitem[{Zalesky et~al.(2010)Zalesky, Fornito, Harding, Cocchi, Y{\"u}cel,
  Pantelis and Bullmore}]{zalesky2010whole}
\bibinfo{author}{Zalesky, A.}, \bibinfo{author}{Fornito, A.},
  \bibinfo{author}{Harding, I.H.}, \bibinfo{author}{Cocchi, L.},
  \bibinfo{author}{Y{\"u}cel, M.}, \bibinfo{author}{Pantelis, C.},
  \bibinfo{author}{Bullmore, E.T.}, \bibinfo{year}{2010}.
\newblock \bibinfo{title}{Whole-brain anatomical networks: does the choice of
  nodes matter?}
\newblock \bibinfo{journal}{Neuroimage} \bibinfo{volume}{50},
  \bibinfo{pages}{970--983}.
\bibitem[{Zalesky et~al.(2011)Zalesky, Fornito, Seal, Cocchi, Westin, Bullmore,
  Egan and Pantelis}]{zalesky2011disrupted}
\bibinfo{author}{Zalesky, A.}, \bibinfo{author}{Fornito, A.},
  \bibinfo{author}{Seal, M.L.}, \bibinfo{author}{Cocchi, L.},
  \bibinfo{author}{Westin, C.F.}, \bibinfo{author}{Bullmore, E.T.},
  \bibinfo{author}{Egan, G.F.}, \bibinfo{author}{Pantelis, C.},
  \bibinfo{year}{2011}.
\newblock \bibinfo{title}{Disrupted axonal fiber connectivity in
  schizophrenia}.
\newblock \bibinfo{journal}{Biological psychiatry} \bibinfo{volume}{69},
  \bibinfo{pages}{80--89}.
\bibitem[{Zhan et~al.(2015)Zhan, Zhou, Wang, Jin, Jahanshad, Prasad, Nir,
  Leonardo, Ye, Thompson et~al.}]{zhan2015comparison}
\bibinfo{author}{Zhan, L.}, \bibinfo{author}{Zhou, J.}, \bibinfo{author}{Wang,
  Y.}, \bibinfo{author}{Jin, Y.}, \bibinfo{author}{Jahanshad, N.},
  \bibinfo{author}{Prasad, G.}, \bibinfo{author}{Nir, T.M.},
  \bibinfo{author}{Leonardo, C.D.}, \bibinfo{author}{Ye, J.},
  \bibinfo{author}{Thompson, P.M.}, et~al., \bibinfo{year}{2015}.
\newblock \bibinfo{title}{Comparison of nine tractography algorithms for
  detecting abnormal structural brain networks in alzheimer’s disease}.
\newblock \bibinfo{journal}{Frontiers in aging neuroscience}
  \bibinfo{volume}{7}, \bibinfo{pages}{48}.
\bibitem[{Zhang et~al.(2020a)Zhang, Karayumak, Hoffmann, Rathi, Golby and
  O’Donnell}]{zhang2020deep}
\bibinfo{author}{Zhang, F.}, \bibinfo{author}{Karayumak, S.C.},
  \bibinfo{author}{Hoffmann, N.}, \bibinfo{author}{Rathi, Y.},
  \bibinfo{author}{Golby, A.J.}, \bibinfo{author}{O’Donnell, L.J.},
  \bibinfo{year}{2020}a.
\newblock \bibinfo{title}{Deep white matter analysis (deepwma): Fast and
  consistent tractography segmentation}.
\newblock \bibinfo{journal}{Medical Image Analysis} \bibinfo{volume}{65},
  \bibinfo{pages}{101761}.
\bibitem[{Zhang et~al.(2020b)Zhang, Noh, Juvekar, Frisken, Rigolo, Norton,
  Kapur, Pujol, Wells~III, Yarmarkovich et~al.}]{zhang2020slicerdmri}
\bibinfo{author}{Zhang, F.}, \bibinfo{author}{Noh, T.},
  \bibinfo{author}{Juvekar, P.}, \bibinfo{author}{Frisken, S.F.},
  \bibinfo{author}{Rigolo, L.}, \bibinfo{author}{Norton, I.},
  \bibinfo{author}{Kapur, T.}, \bibinfo{author}{Pujol, S.},
  \bibinfo{author}{Wells~III, W.}, \bibinfo{author}{Yarmarkovich, A.}, et~al.,
  \bibinfo{year}{2020}b.
\newblock \bibinfo{title}{{SlicerDMRI: Diffusion MRI and tractography research
  software for brain cancer surgery planning and visualization}}.
\newblock \bibinfo{journal}{JCO clinical cancer informatics}
  \bibinfo{volume}{4}, \bibinfo{pages}{299--309}.
\bibitem[{Zhang et~al.(2017)Zhang, Norton, Cai, Song, Wells and
  O'Donnell}]{zhang2017comparison}
\bibinfo{author}{Zhang, F.}, \bibinfo{author}{Norton, I.},
  \bibinfo{author}{Cai, W.}, \bibinfo{author}{Song, Y.},
  \bibinfo{author}{Wells, W.M.}, \bibinfo{author}{O'Donnell, L.J.},
  \bibinfo{year}{2017}.
\newblock \bibinfo{title}{Comparison between two white matter segmentation
  strategies: An investigation into white matter segmentation consistency}, in:
  \bibinfo{booktitle}{2017 IEEE 14th International Symposium on Biomedical
  Imaging (ISBI 2017)}, \bibinfo{organization}{IEEE}. pp.
  \bibinfo{pages}{796--799}.
\bibitem[{Zhang and O'Donnell(2020)}]{zhang2020support}
\bibinfo{author}{Zhang, F.}, \bibinfo{author}{O'Donnell, L.J.},
  \bibinfo{year}{2020}.
\newblock \bibinfo{title}{Support vector regression}, in:
  \bibinfo{booktitle}{Machine Learning}. \bibinfo{publisher}{Elsevier}, pp.
  \bibinfo{pages}{123--140}.
\bibitem[{Zhang et~al.(2018a)Zhang, Savadjiev, Cai, Song, Rathi, Tun{\c{c}},
  Parker, Kapur, Schultz, Makris et~al.}]{zhang2018whole}
\bibinfo{author}{Zhang, F.}, \bibinfo{author}{Savadjiev, P.},
  \bibinfo{author}{Cai, W.}, \bibinfo{author}{Song, Y.},
  \bibinfo{author}{Rathi, Y.}, \bibinfo{author}{Tun{\c{c}}, B.},
  \bibinfo{author}{Parker, D.}, \bibinfo{author}{Kapur, T.},
  \bibinfo{author}{Schultz, R.T.}, \bibinfo{author}{Makris, N.}, et~al.,
  \bibinfo{year}{2018}a.
\newblock \bibinfo{title}{Whole brain white matter connectivity analysis using
  machine learning: an application to autism}.
\newblock \bibinfo{journal}{Neuroimage} \bibinfo{volume}{172},
  \bibinfo{pages}{826--837}.
\bibitem[{Zhang et~al.(2018b)Zhang, Wu, Ning, McAnulty, Waber, Gagoski, Sarill,
  Hamoda, Song, Cai et~al.}]{zhang2018suprathreshold}
\bibinfo{author}{Zhang, F.}, \bibinfo{author}{Wu, W.}, \bibinfo{author}{Ning,
  L.}, \bibinfo{author}{McAnulty, G.}, \bibinfo{author}{Waber, D.},
  \bibinfo{author}{Gagoski, B.}, \bibinfo{author}{Sarill, K.},
  \bibinfo{author}{Hamoda, H.M.}, \bibinfo{author}{Song, Y.},
  \bibinfo{author}{Cai, W.}, et~al., \bibinfo{year}{2018}b.
\newblock \bibinfo{title}{Suprathreshold fiber cluster statistics: Leveraging
  white matter geometry to enhance tractography statistical analysis}.
\newblock \bibinfo{journal}{NeuroImage} \bibinfo{volume}{171},
  \bibinfo{pages}{341--354}.
\bibitem[{Zhang et~al.(2019)Zhang, Wu, Norton, Rathi, Golby and
  O'Donnell}]{zhang2019test}
\bibinfo{author}{Zhang, F.}, \bibinfo{author}{Wu, Y.}, \bibinfo{author}{Norton,
  I.}, \bibinfo{author}{Rathi, Y.}, \bibinfo{author}{Golby, A.J.},
  \bibinfo{author}{O'Donnell, L.J.}, \bibinfo{year}{2019}.
\newblock \bibinfo{title}{Test--retest reproducibility of white matter
  parcellation using diffusion {MRI} tractography fiber clustering}.
\newblock \bibinfo{journal}{Human brain mapping} \bibinfo{volume}{40},
  \bibinfo{pages}{3041--3057}.
\bibitem[{Zhang et~al.(2018c)Zhang, Wu, Norton, Rigolo, Rathi, Makris and
  O'Donnell}]{zhang2018anatomically}
\bibinfo{author}{Zhang, F.}, \bibinfo{author}{Wu, Y.}, \bibinfo{author}{Norton,
  I.}, \bibinfo{author}{Rigolo, L.}, \bibinfo{author}{Rathi, Y.},
  \bibinfo{author}{Makris, N.}, \bibinfo{author}{O'Donnell, L.J.},
  \bibinfo{year}{2018}c.
\newblock \bibinfo{title}{An anatomically curated fiber clustering white matter
  atlas for consistent white matter tract parcellation across the lifespan}.
\newblock \bibinfo{journal}{Neuroimage} \bibinfo{volume}{179},
  \bibinfo{pages}{429--447}.
\bibitem[{Zhang et~al.(2012)Zhang, Schneider, Wheeler-Kingshott and
  Alexander}]{Zhang2012}
\bibinfo{author}{Zhang, H.}, \bibinfo{author}{Schneider, T.},
  \bibinfo{author}{Wheeler-Kingshott, C.A.}, \bibinfo{author}{Alexander, D.C.},
  \bibinfo{year}{2012}.
\newblock \bibinfo{title}{Noddi: practical in vivo neurite orientation
  dispersion and density imaging of the human brain}.
\newblock \bibinfo{journal}{NeuroImage} \bibinfo{volume}{61},
  \bibinfo{pages}{1000--1016}.
\bibitem[{Zhang et~al.(2015)Zhang, Wei, Kang, Zalesky, Li, Xu, Li, Wang, Zheng,
  Wang et~al.}]{zhang2015disrupted}
\bibinfo{author}{Zhang, R.}, \bibinfo{author}{Wei, Q.}, \bibinfo{author}{Kang,
  Z.}, \bibinfo{author}{Zalesky, A.}, \bibinfo{author}{Li, M.},
  \bibinfo{author}{Xu, Y.}, \bibinfo{author}{Li, L.}, \bibinfo{author}{Wang,
  J.}, \bibinfo{author}{Zheng, L.}, \bibinfo{author}{Wang, B.}, et~al.,
  \bibinfo{year}{2015}.
\newblock \bibinfo{title}{Disrupted brain anatomical connectivity in
  medication-na{\"\i}ve patients with first-episode schizophrenia}.
\newblock \bibinfo{journal}{Brain Structure and Function}
  \bibinfo{volume}{220}, \bibinfo{pages}{1145--1159}.
\bibitem[{Zhang et~al.(2008)Zhang, Olivi, Hertig, Van~Zijl and
  Mori}]{zhang2008automated}
\bibinfo{author}{Zhang, W.}, \bibinfo{author}{Olivi, A.},
  \bibinfo{author}{Hertig, S.J.}, \bibinfo{author}{Van~Zijl, P.},
  \bibinfo{author}{Mori, S.}, \bibinfo{year}{2008}.
\newblock \bibinfo{title}{Automated fiber tracking of human brain white matter
  using diffusion tensor imaging}.
\newblock \bibinfo{journal}{Neuroimage} \bibinfo{volume}{42},
  \bibinfo{pages}{771--777}.
\bibitem[{Zhang et~al.(2010)Zhang, Zhang, Oishi, Faria, Jiang, Li, Akhter,
  Rosa-Neto, Pike, Evans et~al.}]{zhang2010atlas}
\bibinfo{author}{Zhang, Y.}, \bibinfo{author}{Zhang, J.},
  \bibinfo{author}{Oishi, K.}, \bibinfo{author}{Faria, A.V.},
  \bibinfo{author}{Jiang, H.}, \bibinfo{author}{Li, X.},
  \bibinfo{author}{Akhter, K.}, \bibinfo{author}{Rosa-Neto, P.},
  \bibinfo{author}{Pike, G.B.}, \bibinfo{author}{Evans, A.}, et~al.,
  \bibinfo{year}{2010}.
\newblock \bibinfo{title}{Atlas-guided tract reconstruction for automated and
  comprehensive examination of the white matter anatomy}.
\newblock \bibinfo{journal}{Neuroimage} \bibinfo{volume}{52},
  \bibinfo{pages}{1289--1301}.
\bibitem[{Zhang et~al.(2018d)Zhang, Descoteaux, Zhang, Girard, Chamberland,
  Dunson, Srivastava and Zhu}]{zhang2018mapping}
\bibinfo{author}{Zhang, Z.}, \bibinfo{author}{Descoteaux, M.},
  \bibinfo{author}{Zhang, J.}, \bibinfo{author}{Girard, G.},
  \bibinfo{author}{Chamberland, M.}, \bibinfo{author}{Dunson, D.},
  \bibinfo{author}{Srivastava, A.}, \bibinfo{author}{Zhu, H.},
  \bibinfo{year}{2018}d.
\newblock \bibinfo{title}{Mapping population-based structural connectomes}.
\newblock \bibinfo{journal}{NeuroImage} \bibinfo{volume}{172},
  \bibinfo{pages}{130--145}.
\bibitem[{Zhao et~al.(2015)Zhao, Cao, Niu, Zuo, Evans, He, Dong and
  Shu}]{zhao2015age}
\bibinfo{author}{Zhao, T.}, \bibinfo{author}{Cao, M.}, \bibinfo{author}{Niu,
  H.}, \bibinfo{author}{Zuo, X.N.}, \bibinfo{author}{Evans, A.},
  \bibinfo{author}{He, Y.}, \bibinfo{author}{Dong, Q.}, \bibinfo{author}{Shu,
  N.}, \bibinfo{year}{2015}.
\newblock \bibinfo{title}{Age-related changes in the topological organization
  of the white matter structural connectome across the human lifespan}.
\newblock \bibinfo{journal}{Human brain mapping} \bibinfo{volume}{36},
  \bibinfo{pages}{3777--3792}.
\bibitem[{Zhao et~al.(2019a)Zhao, Mishra, Jeon, Ouyang, Peng, Chalak,
  Wisnowski, Heyne, Rollins, Shu et~al.}]{zhao2019structural}
\bibinfo{author}{Zhao, T.}, \bibinfo{author}{Mishra, V.},
  \bibinfo{author}{Jeon, T.}, \bibinfo{author}{Ouyang, M.},
  \bibinfo{author}{Peng, Q.}, \bibinfo{author}{Chalak, L.},
  \bibinfo{author}{Wisnowski, J.L.}, \bibinfo{author}{Heyne, R.},
  \bibinfo{author}{Rollins, N.}, \bibinfo{author}{Shu, N.}, et~al.,
  \bibinfo{year}{2019}a.
\newblock \bibinfo{title}{Structural network maturation of the preterm human
  brain}.
\newblock \bibinfo{journal}{NeuroImage} \bibinfo{volume}{185},
  \bibinfo{pages}{699--710}.
\bibitem[{Zhao et~al.(2017a)Zhao, Sheng, Bi, Niu, Shu and Han}]{zhao2017age}
\bibinfo{author}{Zhao, T.}, \bibinfo{author}{Sheng, C.}, \bibinfo{author}{Bi,
  Q.}, \bibinfo{author}{Niu, W.}, \bibinfo{author}{Shu, N.},
  \bibinfo{author}{Han, Y.}, \bibinfo{year}{2017}a.
\newblock \bibinfo{title}{Age-related differences in the topological efficiency
  of the brain structural connectome in amnestic mild cognitive impairment}.
\newblock \bibinfo{journal}{Neurobiology of aging} \bibinfo{volume}{59},
  \bibinfo{pages}{144--155}.
\bibitem[{Zhao et~al.(2019b)Zhao, Xu and He}]{zhao2019graph}
\bibinfo{author}{Zhao, T.}, \bibinfo{author}{Xu, Y.}, \bibinfo{author}{He, Y.},
  \bibinfo{year}{2019}b.
\newblock \bibinfo{title}{Graph theoretical modeling of baby brain networks}.
\newblock \bibinfo{journal}{NeuroImage} \bibinfo{volume}{185},
  \bibinfo{pages}{711--727}.
\bibitem[{Zhao et~al.(2017b)Zhao, Tian, Yan, Yue, Yan and
  Zhang}]{zhao2017abnormal}
\bibinfo{author}{Zhao, X.}, \bibinfo{author}{Tian, L.}, \bibinfo{author}{Yan,
  J.}, \bibinfo{author}{Yue, W.}, \bibinfo{author}{Yan, H.},
  \bibinfo{author}{Zhang, D.}, \bibinfo{year}{2017}b.
\newblock \bibinfo{title}{Abnormal rich-club organization associated with
  compromised cognitive function in patients with schizophrenia and their
  unaffected parents}.
\newblock \bibinfo{journal}{Neuroscience bulletin} \bibinfo{volume}{33},
  \bibinfo{pages}{445--454}.
\bibitem[{Ziyan et~al.(2009)Ziyan, Sabuncu, Grimson and
  Westin}]{ziyan2009consistency}
\bibinfo{author}{Ziyan, U.}, \bibinfo{author}{Sabuncu, M.R.},
  \bibinfo{author}{Grimson, W.E.L.}, \bibinfo{author}{Westin, C.F.},
  \bibinfo{year}{2009}.
\newblock \bibinfo{title}{Consistency clustering: a robust algorithm for
  group-wise registration, segmentation and automatic atlas construction in
  diffusion mri}.
\newblock \bibinfo{journal}{International journal of computer vision}
  \bibinfo{volume}{85}, \bibinfo{pages}{279--290}.
\bibitem[{Z{\"o}llei et~al.(2019)Z{\"o}llei, Jaimes, Saliba, Grant and
  Yendiki}]{zollei2019tracts}
\bibinfo{author}{Z{\"o}llei, L.}, \bibinfo{author}{Jaimes, C.},
  \bibinfo{author}{Saliba, E.}, \bibinfo{author}{Grant, P.E.},
  \bibinfo{author}{Yendiki, A.}, \bibinfo{year}{2019}.
\newblock \bibinfo{title}{Tracts constrained by underlying infant anatomy
  (traculina): An automated probabilistic tractography tool with anatomical
  priors for use in the newborn brain}.
\newblock \bibinfo{journal}{NeuroImage} \bibinfo{volume}{199},
  \bibinfo{pages}{1--17}.

\end{thebibliography}
	
\end{document}